\documentclass[preprint2,9]{aastex}
\pdfoutput=1

\usepackage{caption}
\usepackage{subcaption}
\usepackage{fixltx2e}
\usepackage[usenames,dvipsnames]{color}
\usepackage{amsmath}
\usepackage{multirow}
\usepackage{graphics}
\usepackage{txfonts}
\usepackage{enumitem}

\newcommand{\vect}[1]{\textbf{\textit{#1}}}

\begin{document}

%%%%%%%%%%%%%%%%%%%%%%%%%%%%%%%%%%%%%%%
%TITLE PAGE
%%%%%%%%%%%%%%%%%%%%%%%%%%%%%%%%%%%%%%%

\title{
SIMULATING THREE-DIMENSIONAL NONTHERMAL HIGH-ENERGY PHOTON EMISSION IN COLLIDING-WIND BINARIES
}

\author{K.~Reitberger, R.~Kissmann, A.~Reimer, and O.~Reimer}
\affil{Institut f\"ur Astro- und Teilchenphysik and Institut f\"ur Theoretische Physik, Leopold-Franzens-Universit\"at Innsbruck, A-6020 Innsbruck, Austria} 
\email{klaus.reitberger@uibk.ac.at}

%%%%%%%%%%%%%%%%%%%%%%%%%%%%%%%%%%%%%%%
%ABSTRACT and SUBJECT KEYWORDS
%%%%%%%%%%%%%%%%%%%%%%%%%%%%%%%%%%%%%%%

\begin{abstract}

Massive stars in binary systems have long been regarded as potential sources of high-energy $\gamma$~rays. The emission is principally thought to arise in the region where the stellar winds collide and accelerate relativistic particles which subsequently emit $\gamma$~rays.
On the basis of a three-dimensional distribution function of high-energy particles in the wind collision region -- as obtained by a numerical hydrodynamics \& particle transport model -- we present the computation of the three-dimensional nonthermal photon emission for a given line of sight. 
Anisotropic inverse Compton emission is modelled using the target radiation field of both stars. Photons from relativistic bremsstrahlung and neutral pion decay are computed on the basis of local wind plasma densities. We also consider photon photon opacity effects due to the dense radiation fields of the stars.
Results are shown for different stellar separations of a given binary system comprising of a B star and a Wolf--Rayet star. The influence of orbital orientation with respect to the line of sight is also studied by using different orbital viewing angles. For the chosen electron-proton injection ratio of 10$^{-2}$, we present the ensuing photon emission in terms of two-dimensional projections maps, spectral energy distributions and integrated photon flux values in various energy bands. Here, we find a transition from hadron-dominated to lepton-dominated high-energy emission with increasing stellar separations. In addition, we confirm findings from previous analytic modeling that the spectral energy distribution varies significantly with orbital orientation.
\end{abstract}


%%%%%%%%%%%%%%%%%%%%%%%%%%%%%%%%%%%%%%%
%TEXT
%%%%%%%%%%%%%%%%%%%%%%%%%%%%%%%%%%%%%%%

%%%%%%%%%%%%%%%%%%%%%%%%%%%%%%%%%%%%%%%%%%%%%%%%%%%%%%%%%%%%%%
\section{INTRODUCTION} 
%%%%%%%%%%%%%%%%%%%%%%%%%%%%%%%%%%%%%%%%%%%%%%%%%%%%%%%%%%%%%%

The idea that binary systems of massive stars without compact objects provide suitable environments for efficient particle acceleration and subsequent nonthermal emission has been thoroughly discussed in literature \citep[e.g.,][]{Pollock1987, White1994}. Electrons and protons are thought to be accelerated at the shock fronts tracing the region where the stellar winds collide with supersonic velocities. Analytical models have predicted that the ensuing population of high-energy particles is liable to yield considerable emission of nonthermal radio waves, hard X~rays and $\gamma$~rays via various emission channels \citep[see e.g.;][]{Eichler1993, Benaglia2003, Reimer2006}.

The observational evidence concerning nonthermal photon emission from these particle-accelerating colliding-wind binaries (CBWs) was recently summarized by \cite{Becker2013} who provide a unified census of these systems. Their list of 43 confirmed or suspected binary systems together with their properties helps to pinpoint some important discrepancies between observations and model predictions, such as Fermi-LAT limits on the gamma-ray flux towards WR~140 or WR~147 \citep[see also ][]{Werner2013}, contrasted by the enigmatic properties of the $\eta$~Carinae binary system. The latter shows a comparably strong high-energy $\gamma$-ray signal that has been found to exhibit variability on orbital time scales and a peculiar two-component spectrum \citep[see \textit{Fermi} Large Area Telescope observation in][]{Reitberger2012}.

A better understanding of CWB systems can be obtained by dedicated hydrodynamical (HD) simulations. Studies such as presented in \cite{Pittard2009} have explored the highly dynamical nature of the WCR and its strong dependence on stellar and orbital parameters. The complex density, velocity and temperature structure of the colliding winds have further been used to model the thermal radio and X-ray emission in such systems \citep{Pittard2010,Pittard2010b}. Using magnetohydrodynamic (MHD) simulations \cite{Falceta2012} have recently explored the important role of the magnetic field in the wind collision region (WCR) of these systems and its impact on nonthermal radio emission. Focusing on the properties of a specific binary system, \cite{Madura2013} used three-dimensional (3D) smoothed particle hydrodynamics (SPH) to investigate the turbulent WCR structure in $\eta$~Carinae.

The present study provides simulations of high-energy nonthermal photon emission of CWBs, based on 3D distributions of high-energy particles. The latter has been obtained by combining a 3D HD description of the WCR with the solution of the transport equation of high-energy particles.

In \cite{Reitberger2014} (paper~I) we presented a method to numerically compute the spectral energy distribution (SED) of charged particles within a numerical HD model of the WCR in a binary system of two massive stars. By solving the transport equation including spatial convection, diffusion, diffusive shock acceleration (DSA) and various cooling processes for electrons and protons at every grid point and time step of the HD simulation, we derived the time-dependent 3D spatial distribution of particles at different energies.

In this work, we present the resulting nonthermal photon emission along a given line of sight on the basis of the previously derived electron and proton spectra. The interactions of high-energy particles with the stellar radiation fields and the dense wind material in the WCR give rise to several mechanisms of photon emission, including anisotropic inverse Compton (IC) scattering, relativistic bremsstrahlung and neutral pion decay. We compute the emission throughout the chosen computational domain and present results in terms of two-dimensional (2D) projection maps, spectral energy distributions and total integrated photon flux values. 

As not unexpected, the emission is very sensible to the conditions in the WCR, the position of the stars and orbital orientation. We present a case study of three binary systems differing in stellar separation. Non-thermal high-energy photon emission is computed for these systems with respect to different lines of sight.  

In Section 2 we provide a detailed description of our calculation of various emission processes, as well as the application of photon photon opacity effects. We investigate the ensuing photon emission in Section~3, first for two stationary stars, then including orbital motion. Sections 4 and 5 provide a summary of our findings as well as an outlook on future developments.

%%%%%%%%%%%%%%%%%%%%%%%%%%%%%%%%%%%%%%%%%%%%%%%%%%%%%%%%%%%%%%
\section{NON-THERMAL HIGH-ENERGY PHOTON EMISSION} 
%%%%%%%%%%%%%%%%%%%%%%%%%%%%%%%%%%%%%%%%%%%%%%%%%%%%%%%%%%%%%%
\label{gamma}

In order to estimate the nonthermal high-energy photon emission of a CWB system, we consider the three principal continuum emission processes for energies $E\gtrsim$10 MeV. These are IC scattering of the photons in the dense stellar radiation field to high energies, relativistic bremsstrahlung by the interaction of the energetic electrons with the ions in the wind, and the decay of neutral pions that are produced in hadronic nucleon nucleon collisions.

The computation of all three processes is based on a 3D distribution of high-energy electrons (IC scattering, bremsstrahlung) and protons (neutral pion decay) which we obtain by the method described in paper I. There, all relevant details concerning the hydrodynamics, the transport equation and the numerical implementation are provided.

In principle, the distribution of high-energy electrons can also be used to compute nonthermal low-energy emission such as synchrotron emission. This is highly relevant in modeling specific binary systems, where such observations of a nonthermal signal at radio wavelengths exists \citep[see e.g;][ for the case of WR140]{Williams1994}. A comparison of measured and modelled synchrotron emission yields constraints concerning the magnetic field in conjunction with the electron injection fraction in the WCR and hence the level of IC emission. As it is not the aim of the present study to investigate a specific binary system with fixed astronomical properties, we restrict ourselves to the high-energy part of the nonthermal emission spectrum considering the three emission channels as detailed below.
\begin{enumerate}
\item Anisotropic IC scattering of high-energy electrons on the photons in the radiation field of both stars: This process depends on the local energy density of radiation, the spatial and energetic distribution of electrons and the scattering angle, being the angle between the line of sight to the observer and the direction of the incoming stellar photons. The stars are approximated as monochromatic point sources. The high-energy electrons are considered to have an isotropic distribution function.
\item Relativistic bremsstrahlung of high-energy electrons passing through the wind plasma. This process depends on the number density of the wind plasma and the spatial and energetic distribution of high-energy electrons. We assume a typical interstellar medium (ISM) metallicity of 90\% H, 10\% He. 
\item The decay of neutral pions produced in collisions of high-energy protons with nucleons of the wind plasma. This process depends on the number density of the wind plasma and the spatial and energetic distribution of high-energy protons. 
\end{enumerate}

We compute the above nonthermal emission components for each numerical grid cell of the computational domain. For the calculation of IC scattering, additional information about the orientation of the system with respect to the observer (in terms of inclination $i$ and argument of periastron of the binary system $\Phi$) must be provided. Bremsstrahlung and neutral pion decay give isotropically distributed radiation.

In a second step we use the resulting 3D distribution of nonthermal photon emission per energy and volume and the information of direction and distance to the observer, in order to compute
\begin{enumerate}
\item 2D projection maps of the nonthermal emission for a given energy interval and line of sight,
\item the SED of the various nonthermal emission components integrated over the whole emission region, and
\item the total integrated flux for various energy bands.
\end{enumerate}
Additionally, we consider effects of photon photon opacity due to the dense radiation fields of the stars in the CWB system. Details are given in Section \ref{opac}. Other possible absorption effects (e.g., via interaction with the ISM) are, presently, not considered since they also depend on the specific system.

In the following, we give a detailed account of how we calculate the individual emission components and derive their 2D projections, SEDs and integrated flux values.

\subsection{Inverse Compton Scattering}
\label{ICs}
The computation of the emission by IC scattering requires information on the high-energy electrons, as well as on the radiation fields involved in the scattering process. For consistency with paper I, we use the approximation of monochromatic radiation from pointlike stars. The differential target photon field of star $j$ can then be expressed as
\begin{equation}
\label{ndelta}
\frac{dn_{\mathrm{ph},j}}{dE_\mathrm{ph}}(\vect{r},E_\mathrm{ph} ,\mu)=n_{0,j}(\vect{r})\delta(E_\mathrm{ph}-E_{T,j})\delta(\mu-\mu_{\mathrm{sc},j}) 
\end{equation}
where $E_{T,j}=2.7k_BT_{\ast,j}$ is the energy of a single photon from star $j$ with surface temperature $T_{\ast,j}$. The cosine of the scattering angle $\mu_\mathrm{sc}=\cos{\theta_\mathrm{sc}}$ is determined via the inner product of the unit vectors pointing from the position of the cell at $\vect{r}$ toward the observer and toward star $j$. $n_{0,j}(\vect{r})$ is the local photon number density at the position of the cell $\vect{r}$ given by 
\begin{equation}
n_{0,j}(\vect{r})=\frac{u_{\mathrm{ph},j}}{E_{T,j}}=\frac{L_{\ast,j}}{4\pi r_j^2cE_{T,j}}
\end{equation} with $r_j$ the distance to the star and $L_{\ast,j}$ its stellar luminosity.

The local differential number density of the scattering electrons is directly obtained from the output of the simulations described in paper~I. 
Differential number densities of target photons and high-energy electrons enter into the expression for the IC photon production rate $\frac{d\dot{n}_j(\vect{r},E_\gamma)}{dE_\gamma}$ at energy $E_\gamma$. The complete formula in the context of CWB systems for the case of an isotropic particle distribution is given by \cite{Reimer2006}.
It involves integration over all target photon energies, all target photon incoming directions $\mu$ and all electron energies. By applying the approximations of monochromacity and pointlike stars, only the integral over the energy of the electrons remains. Specific care has to be taken in determining the lower integration limit which marks the lowest energy at which electrons are liable to scatter photons to an energy $E_\gamma$ with regard to the scattering angle $\mu_{\mathrm{sc},j}$. The upper limit is determined by the maximum energy of the electrons. The integration is performed numerically. 

The IC flux from a single cell with volume $V_\mathrm{cell}$ reaching an observer along the line of sight at distance $d_l$ (neglecting absorption effects) then is
\begin{equation}
\frac{dF_\mathrm{IC}}{dE_\gamma}(\vect{r},E_\gamma)=\frac{V_\mathrm{cell}}{4\pi d_l^2}
\sum_{j=1}^2\frac{d\dot{n}_j(\vect{r},E_\gamma)}{dE_\gamma}
\end{equation}
where the sum accounts for the contribution of both stars. The integrated IC flux of the total emission region can be determined by a sum over all grid cells and an integration over the photon energy $E_\gamma$.

\subsection{Relativistic Bremsstrahlung}
According to \citet{Blumenthal}, the bremsstrahlung spectrum emitted by a single electron of energy $E_e$ passing through a medium containing various species of ions with number densities $n_i$ is
\begin{equation}
\label{single}
\frac{d\dot{N}_\gamma}{dE_\gamma}(E_e)=c\sum_in_iZ_i^2\frac{d\sigma_i}{dE_\gamma}(E_e)
\end{equation}
where the sum is over all elements relevant for the scattering process.

The differential cross-section $\frac{d\sigma_s}{dE_\gamma}$ is greatly simplified if we assume that all contributing species are fully ionized. This is approximately true in the high temperature plasma inside the WCR. With this assumption (the case of ``weak shielding'') the differential cross-section can be expressed as
\begin{equation}
\frac{d\sigma_i}{dE_\gamma}(E_e)=\alpha r_0^2\frac{\frac{4}{3}E_e^2-\frac{4}{3}E_eE_\gamma+E_\gamma^2}{E_\gamma E_e^2}\phi_\mathrm{weak}
\end{equation}
with
\begin{equation}
\phi_\mathrm{weak}=\left.
\begin{cases}
4\Big[\ln\Big(\frac{2E_e(E_e-E_\gamma)}{E_\gamma m_ec^2}\Big)-\frac{1}{2}\Big],\qquad &\mbox{for }E_\gamma < E_e\\
4\Big[\ln(4E_\gamma)-\frac{1}{2}\Big],\qquad &\mbox{for }E_\gamma \gtrsim E_e\\
\end{cases}
\right.
\end{equation}
We further make the assumption of only dealing with hydrogen and helium (90\% H, 10\% He), allowing to rewrite Equation (\ref{single}) as
\begin{equation}
\frac{d\dot{N}_\gamma}{dE_\gamma}=c(0.9\times1^2+0.1\times2^2)n_\mathrm{H}\frac{d\sigma}{dE_\gamma}
\end{equation}
where the number density $n_\mathrm{H}$ of particles in the wind plasma is obtained via $n_H = \frac{\rho}{m_H}$ from the output of the  hydrodynamic simulations described in paper~I. 
Computing the emission for isotropically distributed electrons $n_e(E_e)$ requires an integration over their energy spectrum.
\begin{equation}
\frac{d\dot{n}_\gamma}{dE_\gamma}=\int dE_e\frac{dn_e}{dE_e}(E_e)\frac{d\dot{N}_\gamma}{dE_\gamma}=1.3cn_\mathrm{H}\int dE_e\frac{dn_e}{dE_e}(E_e)\frac{d\sigma}{dE_\gamma}
\end{equation}
The integral is solved with the same numerical scheme as discussed in Section \ref{ICs}. 
To obtain the resulting flux emitted from a single computational cell at a distance $d_L$, we multiply by the cell volume $V_\mathrm{cell}$ and divide by the respective spherical distance surface.
\begin{eqnarray}
\frac{dF_\mathrm{brems}}{dE_\gamma}=\frac{V_\mathrm{cell}}{4\pi d_l^2}\frac{d\dot{n}_\gamma}{dE_\gamma}
\end{eqnarray}

\subsection{Neutral Pion Decay}
In computing the $\gamma$-ray emission due to the decay of neutral pions, we follow \cite{Kelner2006}  
who use a $\delta$-functional approximation for the cross-section $\sigma_{pp}^\pi(E_\pi,E_p)$ with $E_p$ being the total energy of the incident proton (the other one is assumed to be at rest). $E_\pi$ is the energy of the produced pion. For a distribution of protons $\frac{dn_{p}}{dE_{p}}(E_{p})$ the omnidirectional differential neutral pion spectrum can then be expressed as
\begin{align}
\label{pin}
\frac{dn_{\pi^0}}{dE_\pi}(E_\pi)&=
cn_\mathrm{H}\sigma_{pp}^\pi(m_pc^2+\frac{E_\pi}{K_\pi})\frac{dn_{p}}{dE_{p}}(m_pc^2+\frac{E_\pi}{K_\pi})
\end{align}
where $K_\pi$ is the mean fraction of the kinetic energy of the protons $E_\mathrm{kin}$ that is transferred to the secondary neutral pions.
According to \cite{Aharonian2000} the parameterization $K_\pi\approx0.17$ yields results of high accuracy in a broad region from GeV to TeV energies. It can be applied down to the threshold energy of neutral pion production. The total cross-section $ \sigma_{pp}^\pi(E_p)$ is approximated \citep[following][]{Kelner2006} as
\begin{equation}
\sigma_{pp}^\pi(E_p)=(34.3+1.88L+0.25L^2)\Bigg[1-\Big(\frac{E_\mathrm{th}}{E_p}\Big)^4\Bigg]^2\times10^{-31}\;\mathrm{m}^2
\end{equation}
with $L=\ln(E_p/1\mathrm{TeV})$ and $E_\mathrm{th}=m_p+2m_\pi+m_\pi^2/2m_p$.

Equation (\ref{pin}) allows to directly convert the proton spectra (obtained as described in paper I) to the resulting pion spectra. Via interpolation, $\frac{dn_{\pi^0}}{dE_\pi}(E_\pi)$ can be obtained for any given energy $E_\pi$.
The omnidirectional differential photon production rate can then be computed via the integral
\begin{equation}
\frac{d\dot{n}}{dE_\gamma}(E_\gamma)=2\int_{E_\gamma+\frac{m_\pi^2c^4}{4E_\gamma}}^{\infty}dE_\pi\frac{\frac{dn_{\pi^0}}{dE_\pi}(E_\pi)}{\sqrt{E_\pi^2-m_\pi^2c^4}}
\end{equation}
As in case of the bremsstrahlung component, the flux reaching an observer at a distance $d_l$ then is
\begin{equation}
\frac{dF_{\pi^0\mathrm{decay}}}{dE_\gamma}=\frac{V_\mathrm{cell}}{4\pi d_l^2}\frac{d\dot{n}}{dE_\gamma}.
\end{equation}

\subsection{Photon-Photon Opacity in the Stellar Radiation Fields}
\label{opac}
Due to the intense stellar radiation field close to the stars, the emitted nonthermal photon flux above $\sim$~100 GeV may be attenuated significantly via the electron pair production process 
\begin{equation}
\gamma+\gamma^\prime\rightarrow e^++e^-.
\end{equation}
This attenuation can be expressed as 
\begin{equation}
\frac{dF}{dE_\gamma}(\vect{r},E_\gamma)\longrightarrow\frac{dF}{dE_\gamma}(\vect{r},E_\gamma) \exp\Big(-\tau_{\gamma\gamma}(\vect{r},E_\gamma)\Big)
\end{equation}
with $\tau_{\gamma\gamma}$ being the optical depth due to photon-photon pair production.
For each position within the emission region, the computation of $\tau_{\gamma\gamma}$ requires an integration along the path $\ell$ toward the observer. The optical depth is proportional to the cross-section $\sigma_{\gamma\gamma}$ as well as to the differential photon number density $\frac{dn_{\mathrm{ph}}}{dE_\mathrm{ph}}(E_\mathrm{ph} ,\mu_j)$ of the stellar radiation field. The general expression can be written as
\begin{align}
&\tau_{\gamma\gamma}(\vect{r},E_\gamma)=\nonumber \\ 
\int d\ell\int dE_\mathrm{ph}\int d\mu \frac{dn_{\mathrm{ph}}}{dE_\mathrm{ph}}&(\vect{x}(\ell),E_\mathrm{ph} ,\mu)\sigma_{\gamma\gamma}(E_\mathrm{ph},E_\gamma,\mu)(1-\mu)\nonumber \\
\;
\end{align}
where the integrals run along the path $\ell$, over the energy of the photons in the stellar radiation fields $E_\mathrm{ph}$, and over the stellar surface described by the cosine of the angle between the two interacting photons $\mu$. 

Analogous to the computation of the IC emission in Section \ref{ICs}, we assume the stars to be pointlike and their radiation field to be monochromatic. Thus, the differential photon number density can be expressed as in Equation (\ref{ndelta}) and
\begin{align}
&\tau_{\gamma\gamma}(\vect{r},E_\gamma)=\nonumber\\
\int{d\ell}\sum_j^\mathrm{stars}
n_{0,j}(\vect{x}(\ell))& \sigma_{\gamma\gamma}(E_\mathrm{ph},E_{T,j},\mu_j(\vect{x}(\ell)))(1-\mu_j(\vect{x}(\ell)))
\end{align} 
 
where $\mu_j$ is the angle between the line of sight (LOS) and the direction toward the star j.
The integral along path $\ell$ is solved numerically.

According to \cite{Gould1967} the cross-section of the process $\gamma+\gamma^\prime\rightarrow e^++e^-$ is 
\begin{equation}
\sigma_{\gamma\gamma}(E_T,E_\gamma,\mu) = \frac{1}{2}\pi r_0^2(1-\beta^2)\Big[(3-\beta^4)\ln\frac{1+\beta}{1-\beta}-2\beta(2-\beta^2)\Big]
\end{equation}
with 
\begin{equation}
\beta=\sqrt{1-s^{-1}}\quad\mathrm{and}\quad s=\frac{E_T E_\gamma}{2m_e^2c^4}(1-\mu)
\end{equation}
The threshold for photon photon absorption is at $E_TE_\gamma\geq(m_ec^2)^2$ (in the most favourable case of $\mu=-1$). For a star of temperature $T = 3\cdot10^4 K$, opacity effects do occur for $E_\gamma\gtrsim$ 40 GeV.

\begin{figure*}
	\setlength{\unitlength}{0.001\textwidth}
	\begin{subfigure}[c]{500\unitlength}
		\begin{picture}(500,370)
			\put(15,15){
				\includegraphics[width=\textwidth]{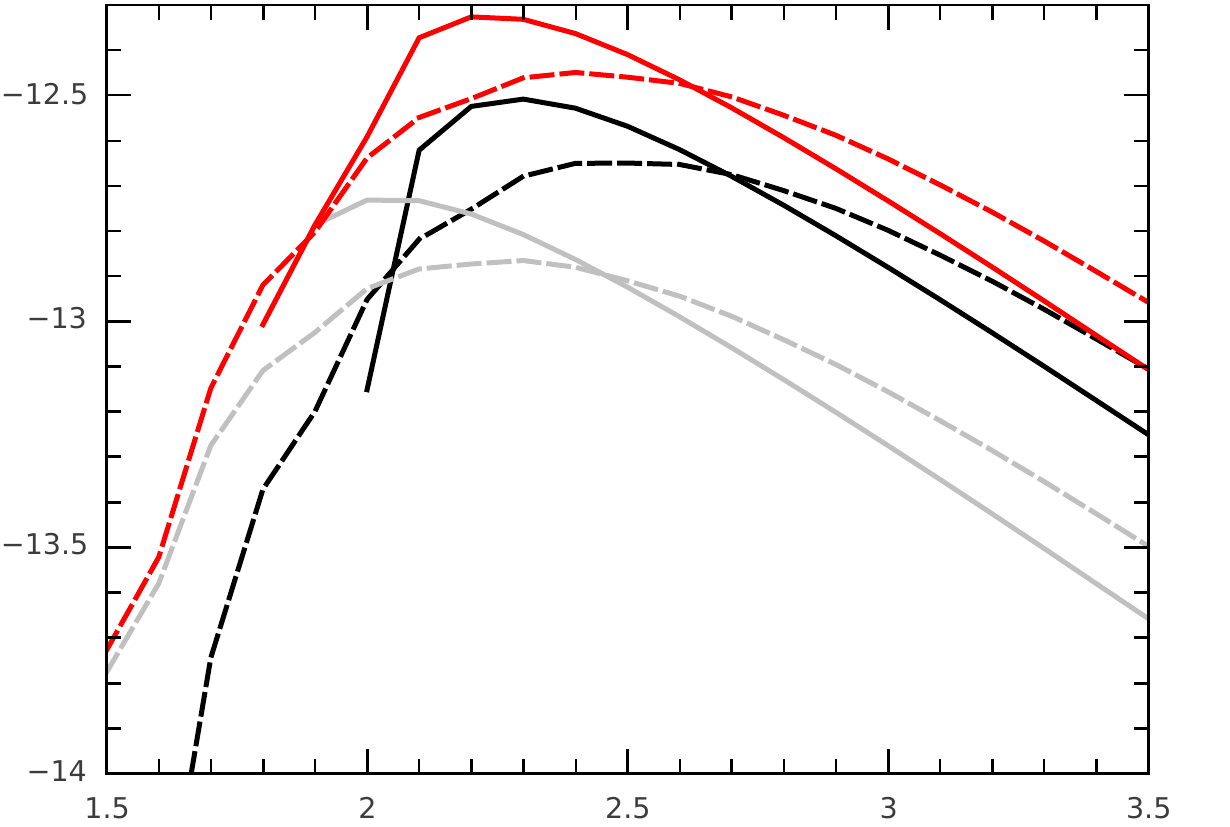}
			}
			\put(200,0){\footnotesize{log( E in GeV)}}
			\put(0,160){\rotatebox{90}{\footnotesize{log($\tau$)}}}
			\put(450,60){a)}
		\end{picture}
	\end{subfigure}
	\begin{subfigure}[c]{500\unitlength}
		\begin{picture}(500,370)
			\put(15,15){
				\includegraphics[width=0.96\textwidth]{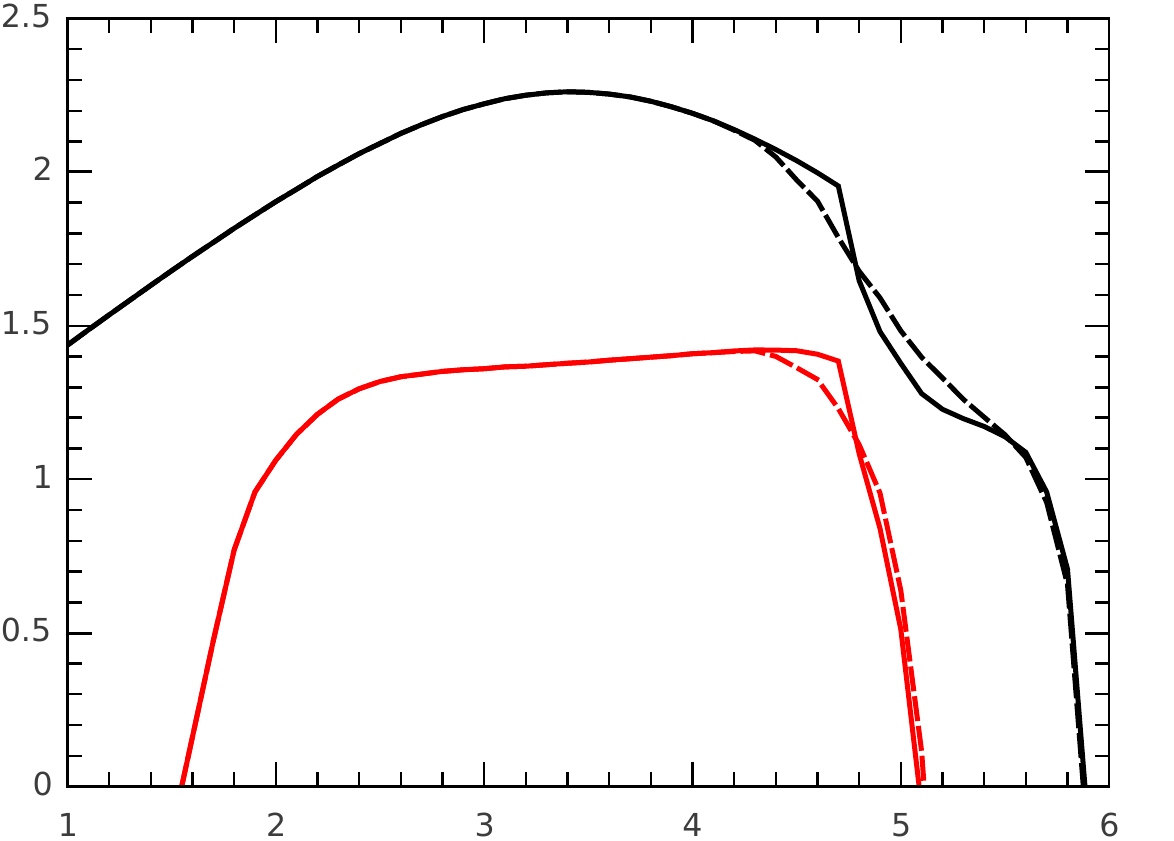}
			}
			\put(200,0){\footnotesize{log( E in MeV)}}
			\put(0,120){\rotatebox{90}{\footnotesize{log( $E^2N$ in MeV cm$^{-3}$ )}}}
			\put(440,60){b)}
		\end{picture}
	\end{subfigure}
		\caption{a) Optical depth $\tau$ from photon photon absorption for a single grid cell as a function of the energy of the high-energy photon scattering with the mono-chromatic (solid) or black-body type (dashed) radiation field of the B star (black), the WR star (gray) and both stars (red). \\
		b) Photon spectra emitted by a significant fraction of the computational domain for IC scattering (black) and neutral pion decay (red) for a viewing angle where opacity effects are maximal ($i=90^\circ$, $\Phi=0^\circ$). The spectra have been computing using mono-chromatic (solid) and black-body type (dashed) radiation fields. Stellar and orbital parameters are the same as for case C in Section \ref{results}. \label{opacity}}
\end{figure*}
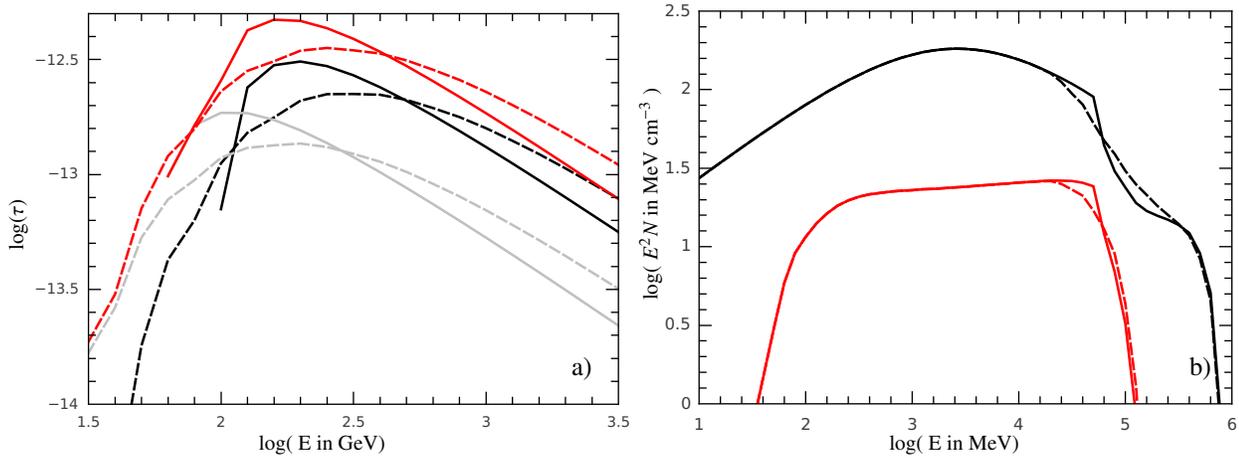

In order to test the validity of the monochromatic approximation, we also computed the absorption assuming a black-body spectrum for both stars. This requires an additional integration over the energies of the stellar photons. We approximate the spectrum by 20 energy bins (equally spaced on a logarithmic scale) in the interval from 0.1 to 100 eV and  integrate numerically. 
A comparison between the two cases is given in Figure \ref{opacity} where we show the optical depth $\tau$ for a single computational cell as a function of $E_\gamma$ for either the WR star, the B star or both stars (a), as well as the impact of the opacity effects on the SED emitted by a representative fraction of the computational domain (b). The monochromatic approximation is found to overestimate the optical depth near its maximum and underestimates it for lower and higher energies. However, the differences found in the SED are sufficiently small (see Figure \ref{opacity} b)) to warrant the use of the monochromatic approximation, keeping in mind that it has significantly lower computational cost, as the black-body description of the radiation field (with 20 energy bins) increases the overall computation time by a factor of 20.

%%%%%%%%%%%%%%%%%%%%%%%%%%%%%%%%%%%%%%%%%%%%%%%%%%%%%%%%%%%%%%
\section{RESULTS} 
%%%%%%%%%%%%%%%%%%%%%%%%%%%%%%%%%%%%%%%%%%%%%%%%%%%%%%%%%%%%%%
\label{results}

In the following, we will investigate three different binary systems with different stellar separations regarding their wind plasma properties, their distribution of charged particles and their ensuing nonthermal high-energy emission. The latter is explored using 2D projection maps, SEDs and integrated flux values for various energy bands. We also present the results for an additional case including effects of orbital motion.
% tau curve

\subsection{HD Model Parameters}

\begin{figure*}
\begin{center}
\includegraphics[trim=0cm 1cm 0cm 0cm, clip=true,width=0.7\textwidth]{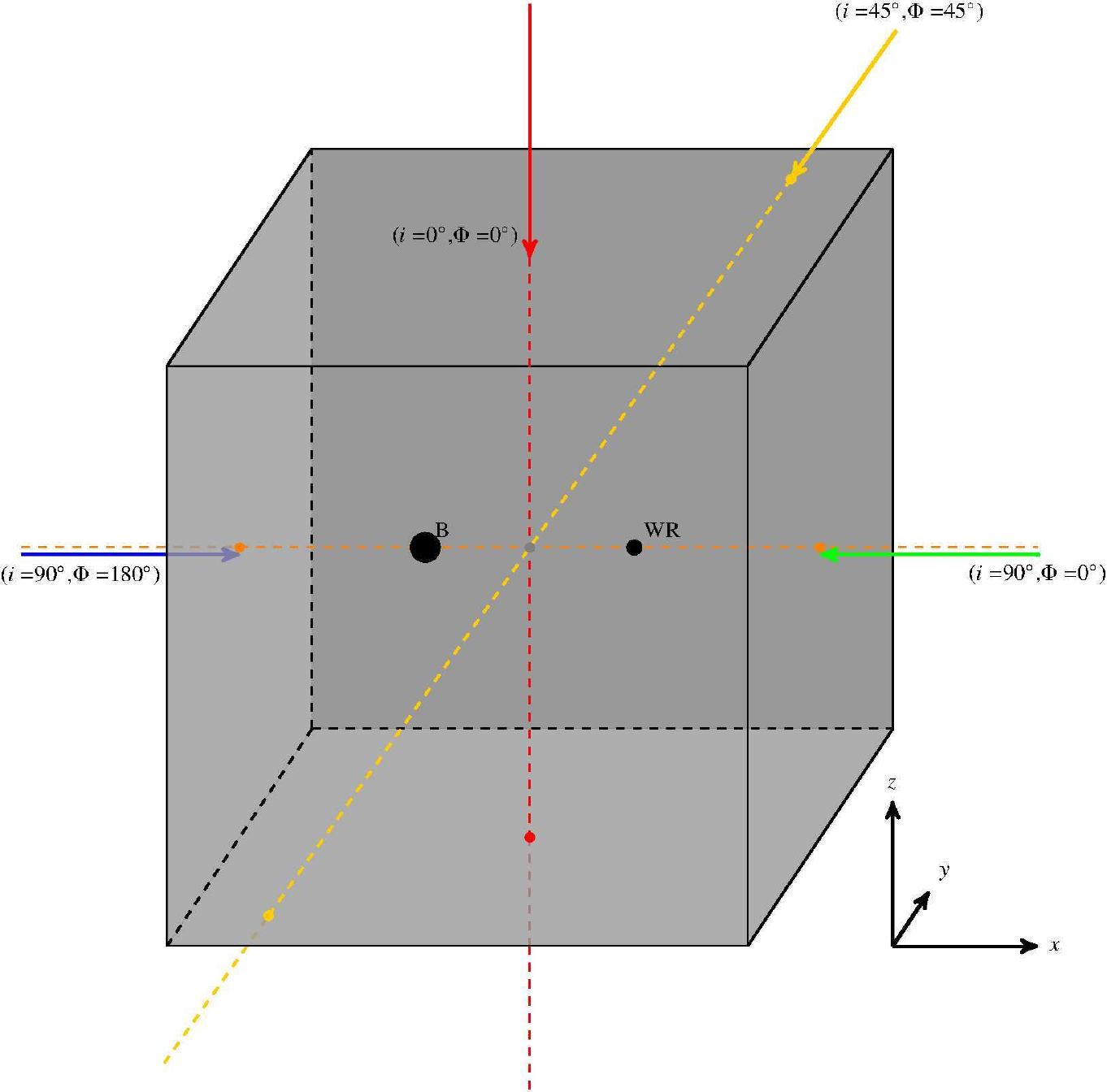}
\end{center}
\caption{Schematic view of the two stars within the computational domain. The line of centers is represented by the horizontal dashed orange line. The various viewing angles (lines-of-sight) in respect to which the emission is computed are indicated as well.
\label{orient}}
\end{figure*}

To allow for quick comparison, we consider three binary systems (Wolf--Rayet (WR) + B star) that differ only in their stellar separation $d$ for which we chose 720 (case A), 1440 (case B) and 2880 (case C) R$\odot$. All stellar and stellar wind parameters (listed in Table \ref{params}) are the same as in the parameter studies in \cite{Reimer2006} and in paper I. The important parameter $\eta = \frac{\dot{M}_\mathrm{B}v_{\infty,\mathrm{B}}}{\dot{M}_\mathrm{WR}v_{\infty,\mathrm{WR}}}$ which indicates the point of ram pressure balance between the two stars has a value of 0.1 for the chosen parameters.
In a first approach, we neglect orbital motion and study the three systems in a converged state after all traces of the initial conditions have vanished. 
For all three cases we chose the identical computational domain in the shape of a cube with side length 4000 $R_\odot$. The two stars are located on the $x$ axis.
Figure \ref{orient} schematically depicts the computational domain containing the two stars. It also shows the line of centers and various arrows indicating the different viewing angles in respect to which we investigate the nonthermal high-energy emission of the system.

In Section \ref{orbital} we repeat the analysis of case A, now including effects of orbital motion. Again, the two stars are located initially on the $x$ axis and move on a Keplerian orbit in the $x$--$y$ plane.

\subsection{Properties of the Wind Plasma}
\label{hydrod}
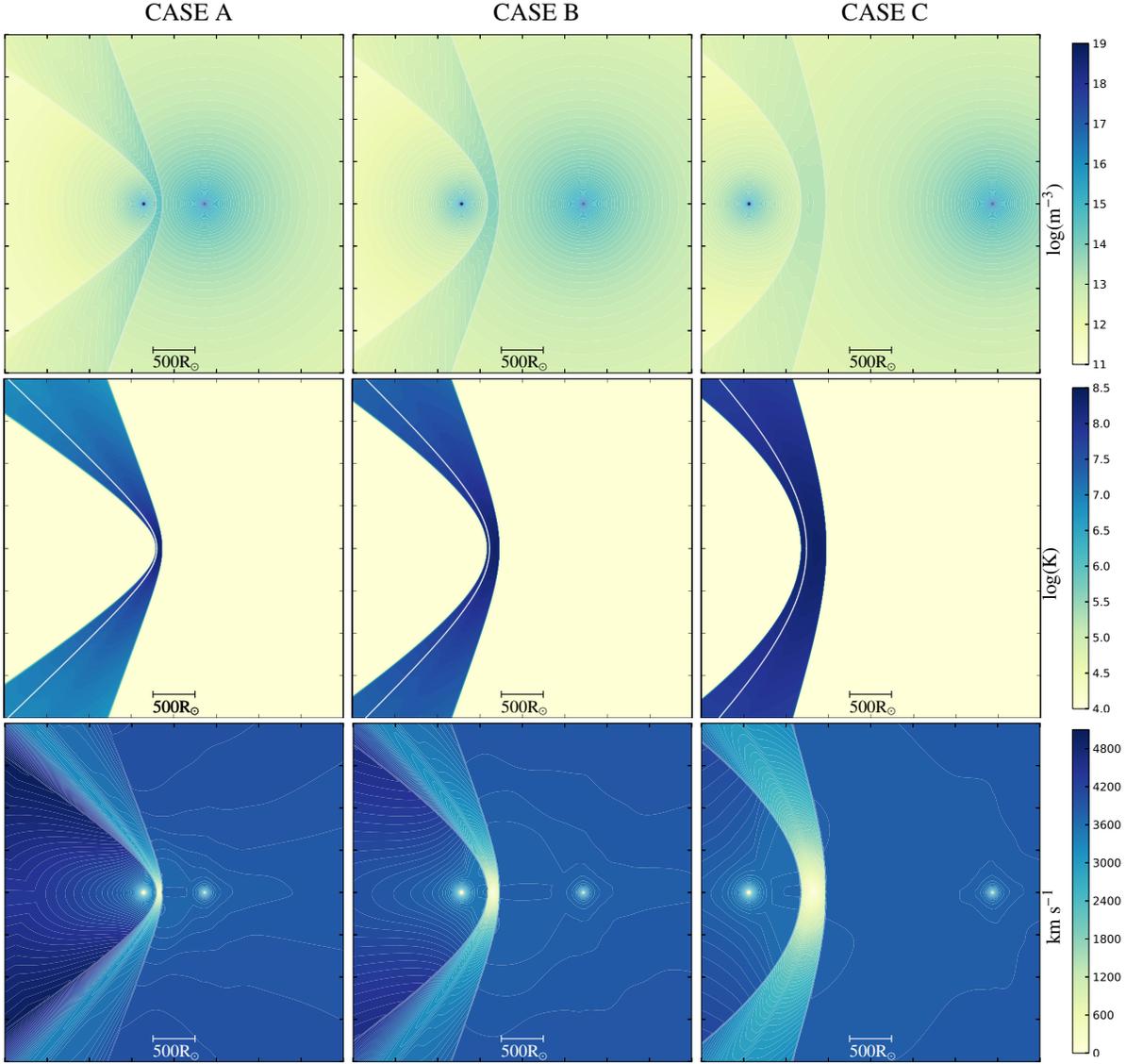
\begin{figure*}
	\setlength{\unitlength}{0.001\textwidth}
		\begin{subfigure}{290\unitlength}
			\begin{picture}(290,290)
					\put(0,0){\includegraphics[trim=2.9cm 1.5cm 5.1cm 1.5cm, clip=true,width=290\unitlength]{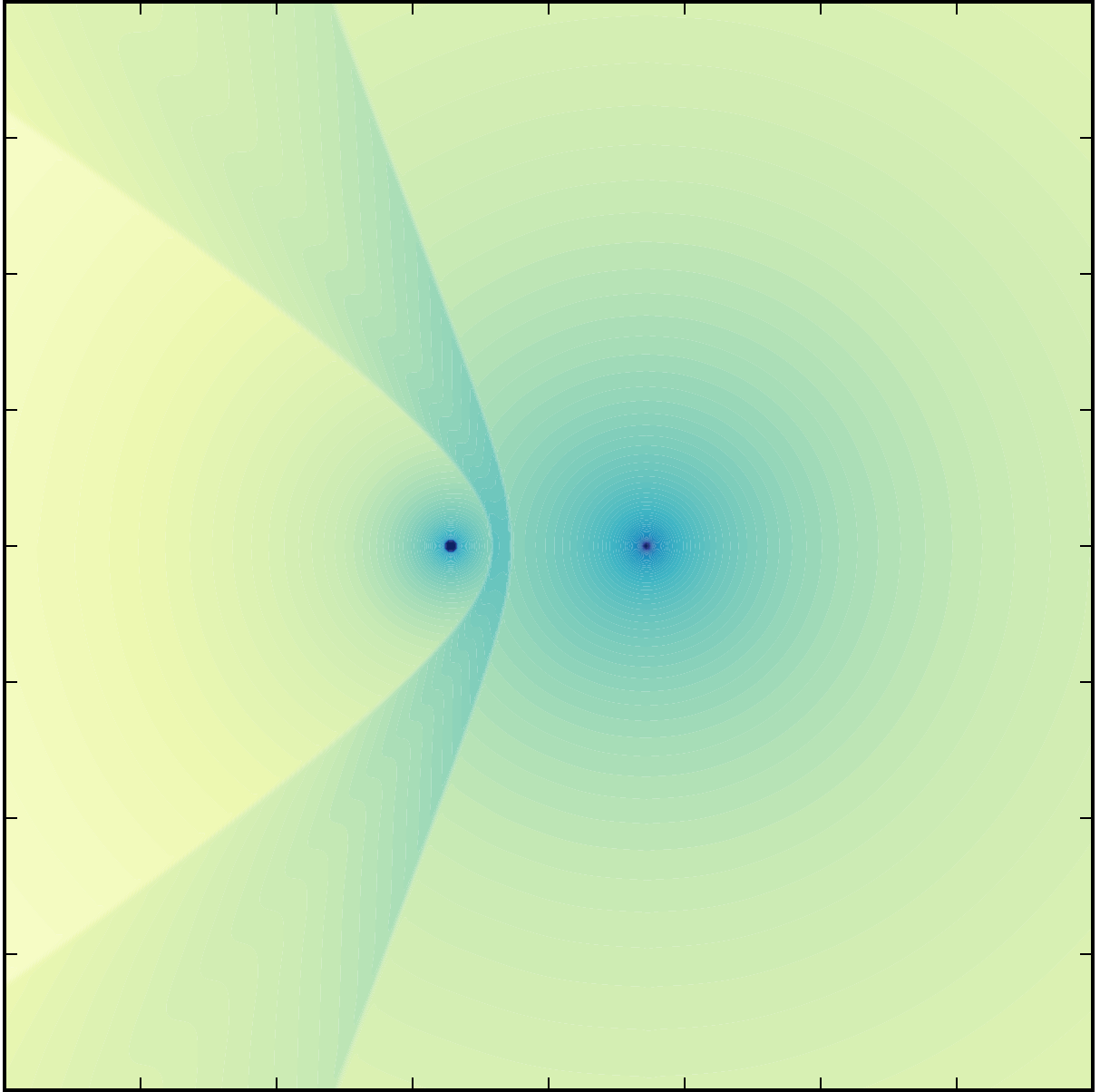}}
				\put(120,300){CASE A}
				\put(126.7,20){\line(1,0){35.625}}
				\put(126.7,17){\line(0,1){6}}
				\put(162.325,17){\line(0,1){6}}
				\put(126.7,5){\scriptsize{500R$_\odot$}}
			\end{picture}
		\end{subfigure}
		\begin{subfigure}{290\unitlength}
			\begin{picture}(290,290)
					\put(0,0){\includegraphics[trim=2.9cm 1.5cm 5.1cm 1.5cm, clip=true,width=290\unitlength]{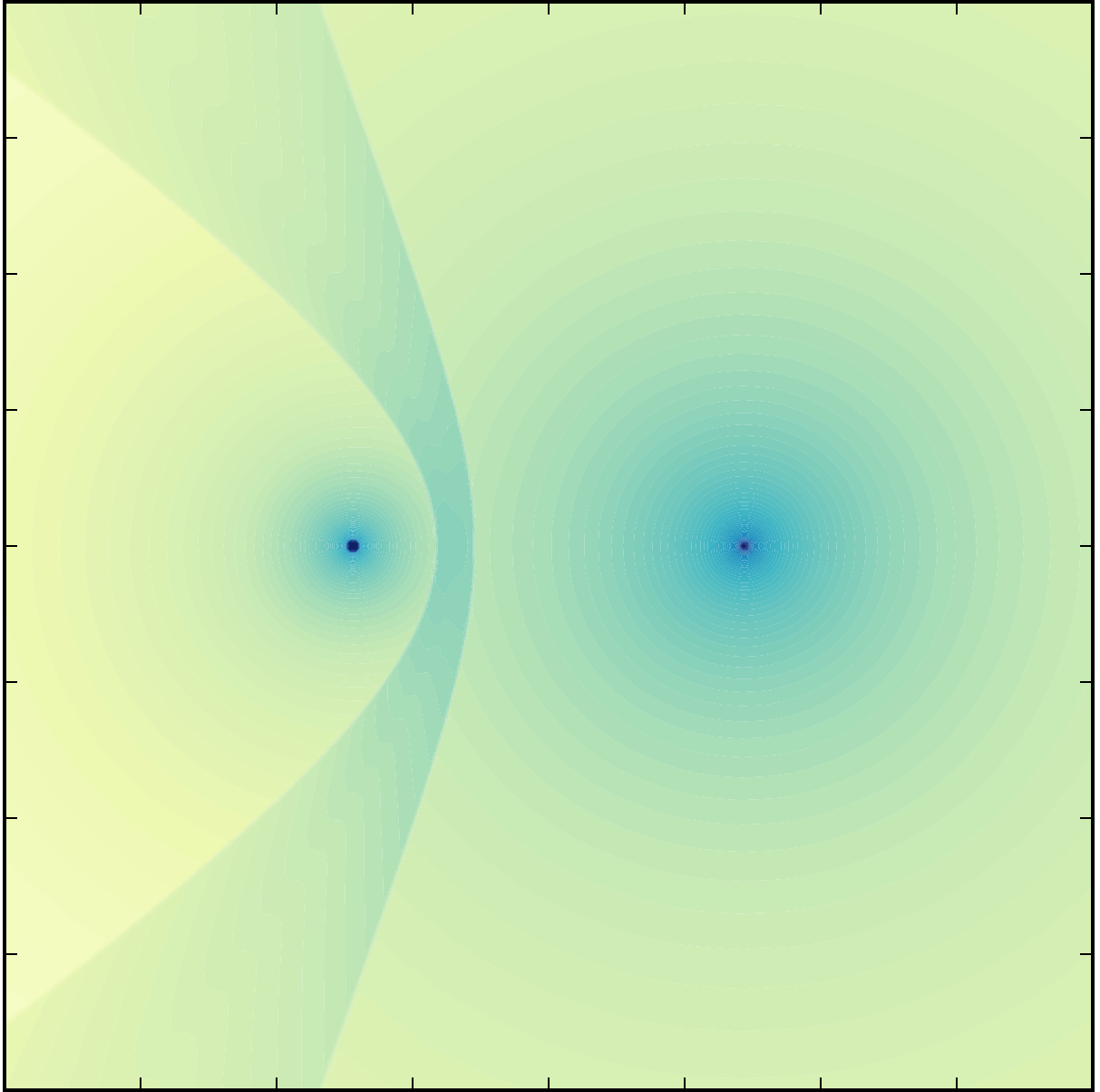}}
				\put(120,300){CASE B}
				\put(126.7,20){\line(1,0){35.625}}
				\put(126.7,17){\line(0,1){6}}
				\put(162.325,17){\line(0,1){6}}
				\put(126.7,5){\scriptsize{500R$_\odot$}}
			\end{picture}
		\end{subfigure}
		\begin{subfigure}{290\unitlength}
			\begin{picture}(290,290)
			\put(0,0){\includegraphics[trim=2.9cm 1.5cm 5.1cm 1.5cm, clip=true,width=290\unitlength]{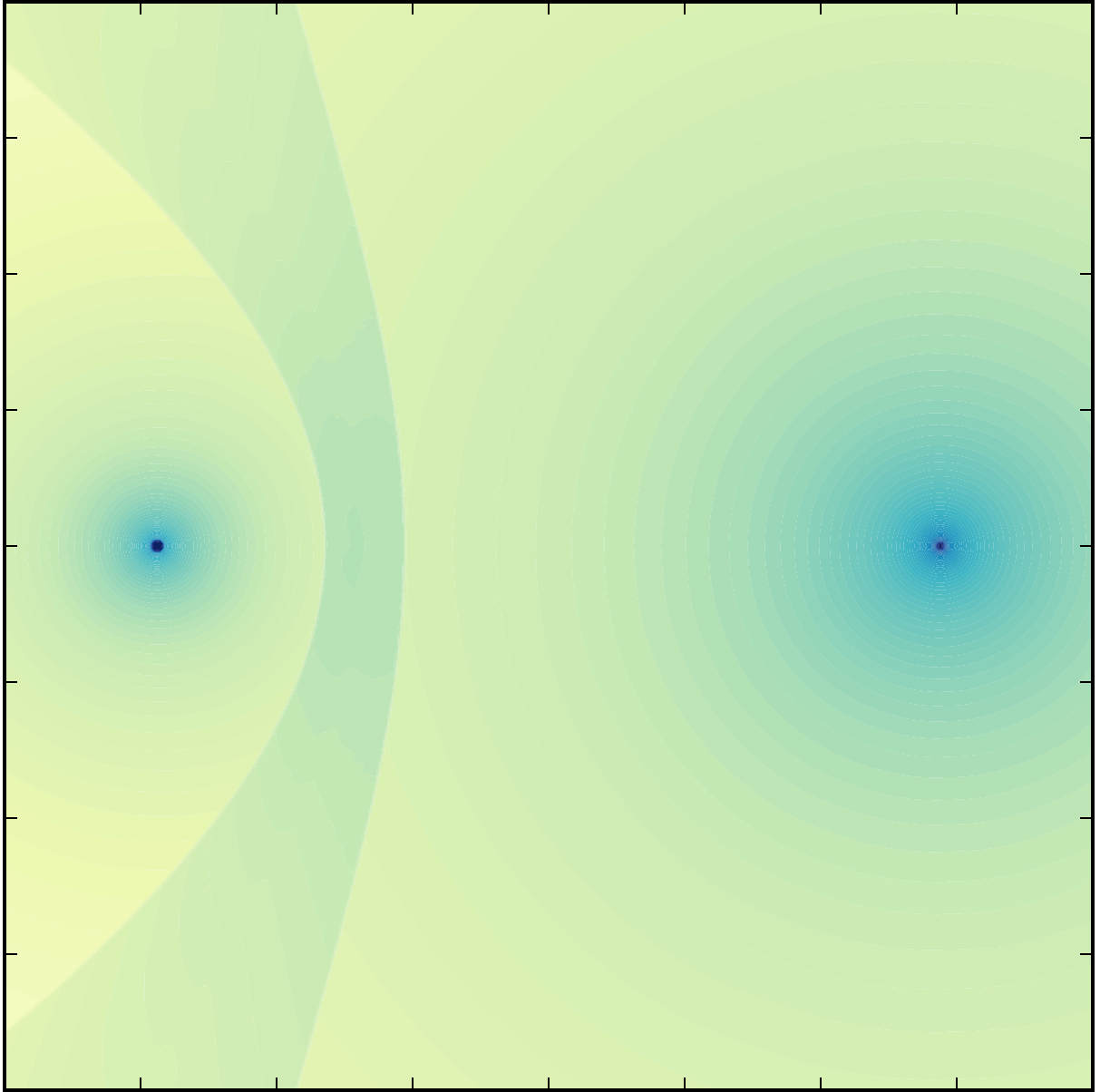}}
				\put(120,300){CASE C}
				\put(126.7,20){\line(1,0){35.625}}
				\put(126.7,17){\line(0,1){6}}
				\put(162.325,17){\line(0,1){6}}
				\put(126.7,5){\scriptsize{500R$_\odot$}}
			\end{picture}
		\end{subfigure}
		\begin{subfigure}{95\unitlength}
		\begin{picture}(95,290)
				\put(20,5){\includegraphics[trim=15.9cm 1.4cm 2.4cm 1.3cm, clip=true,height=280\unitlength]{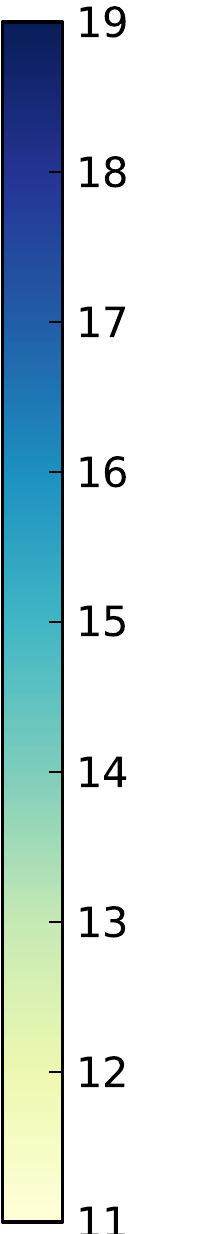}}
			\put(-5,100){\rotatebox{90}{\footnotesize{log(m$^{-3}$)}}}
		\end{picture}
		\end{subfigure}\\
		\begin{subfigure}{290\unitlength}
			\begin{picture}(290,290)
				\put(0,0){\includegraphics[trim=2.9cm 1.5cm 5.1cm 1.5cm, clip=true,width=290\unitlength]{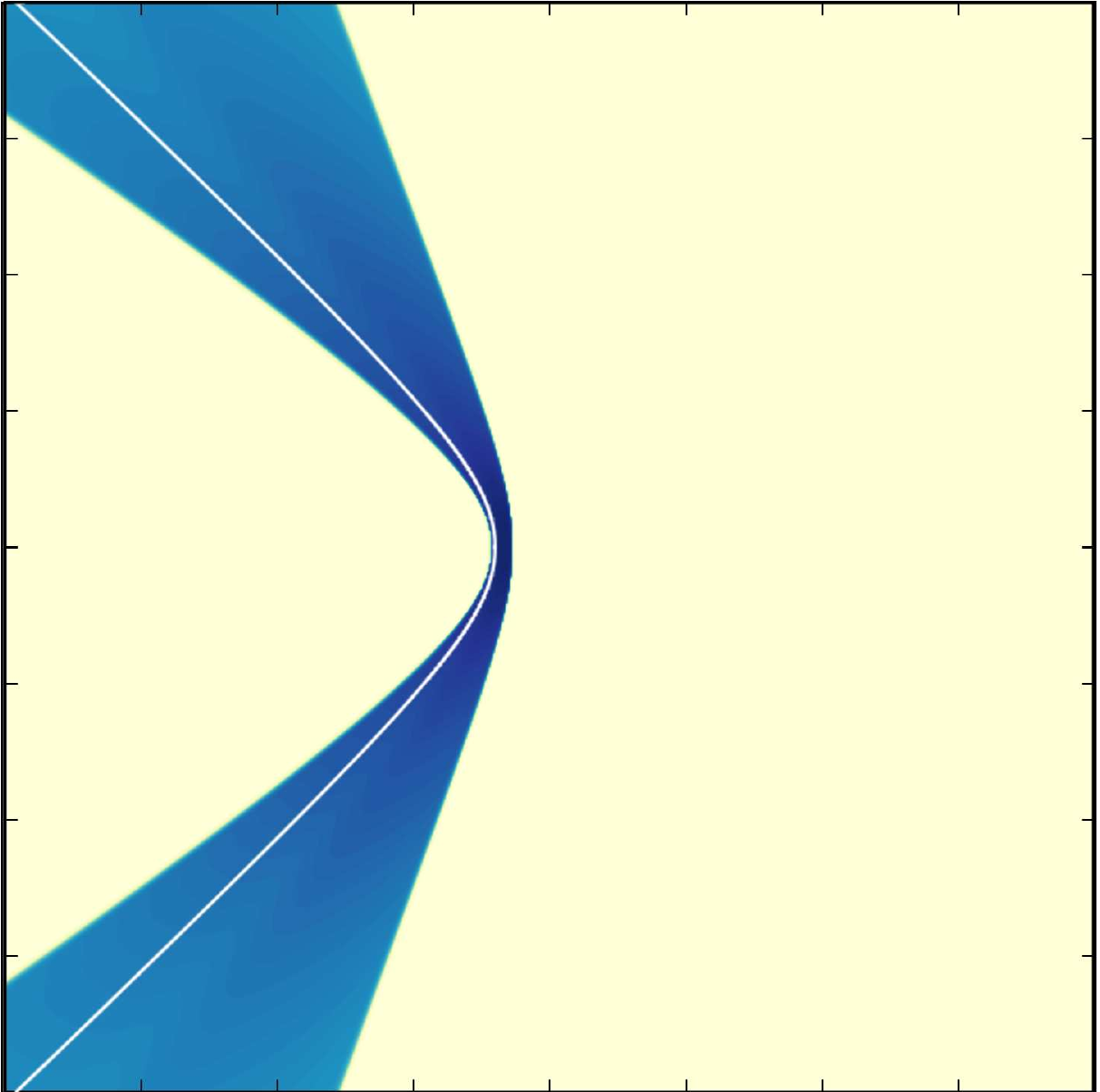}}
				\put(126.7,20){\line(1,0){35.625}}
				\put(126.7,17){\line(0,1){6}}
				\put(162.325,17){\line(0,1){6}}
				\put(126.7,5){\scriptsize{500R$_\odot$}}
				\put(126.7,5){\scriptsize{500R$_\odot$}}
			\end{picture}
		\end{subfigure}
		\begin{subfigure}{290\unitlength}
			\begin{picture}(290,290)
				\put(0,0){\includegraphics[trim=2.9cm 1.5cm 5.1cm 1.5cm, clip=true,width=290\unitlength]{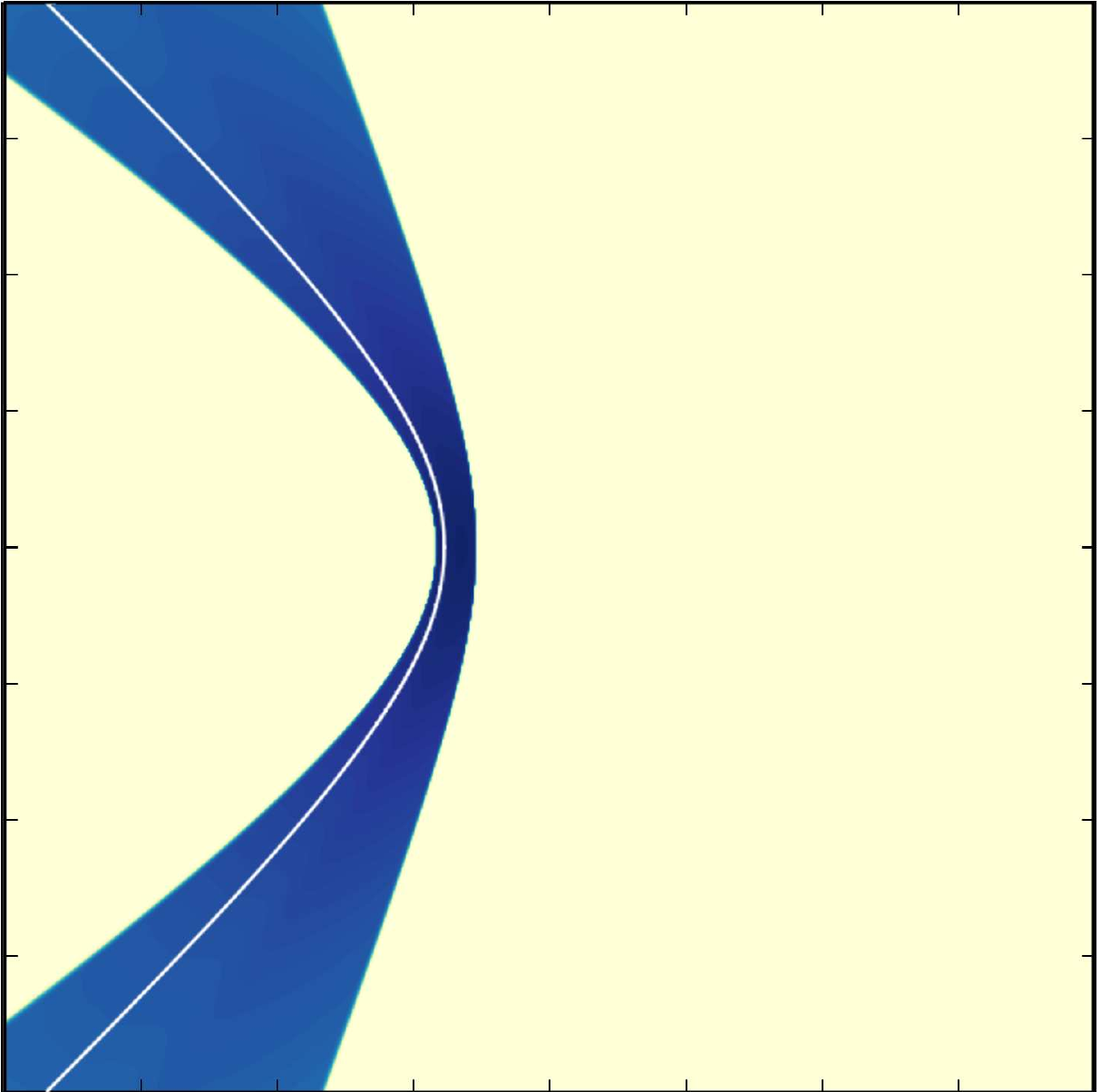}}
				\put(126.7,20){\line(1,0){35.625}}
				\put(126.7,17){\line(0,1){6}}
				\put(162.325,17){\line(0,1){6}}
				\put(126.7,5){\scriptsize{500R$_\odot$}}
			\end{picture}
		\end{subfigure}
		\begin{subfigure}{290\unitlength}	
			\begin{picture}(290,290)
				\put(0,0){\includegraphics[trim=2.9cm 1.5cm 5.1cm 1.5cm, clip=true,width=290\unitlength]{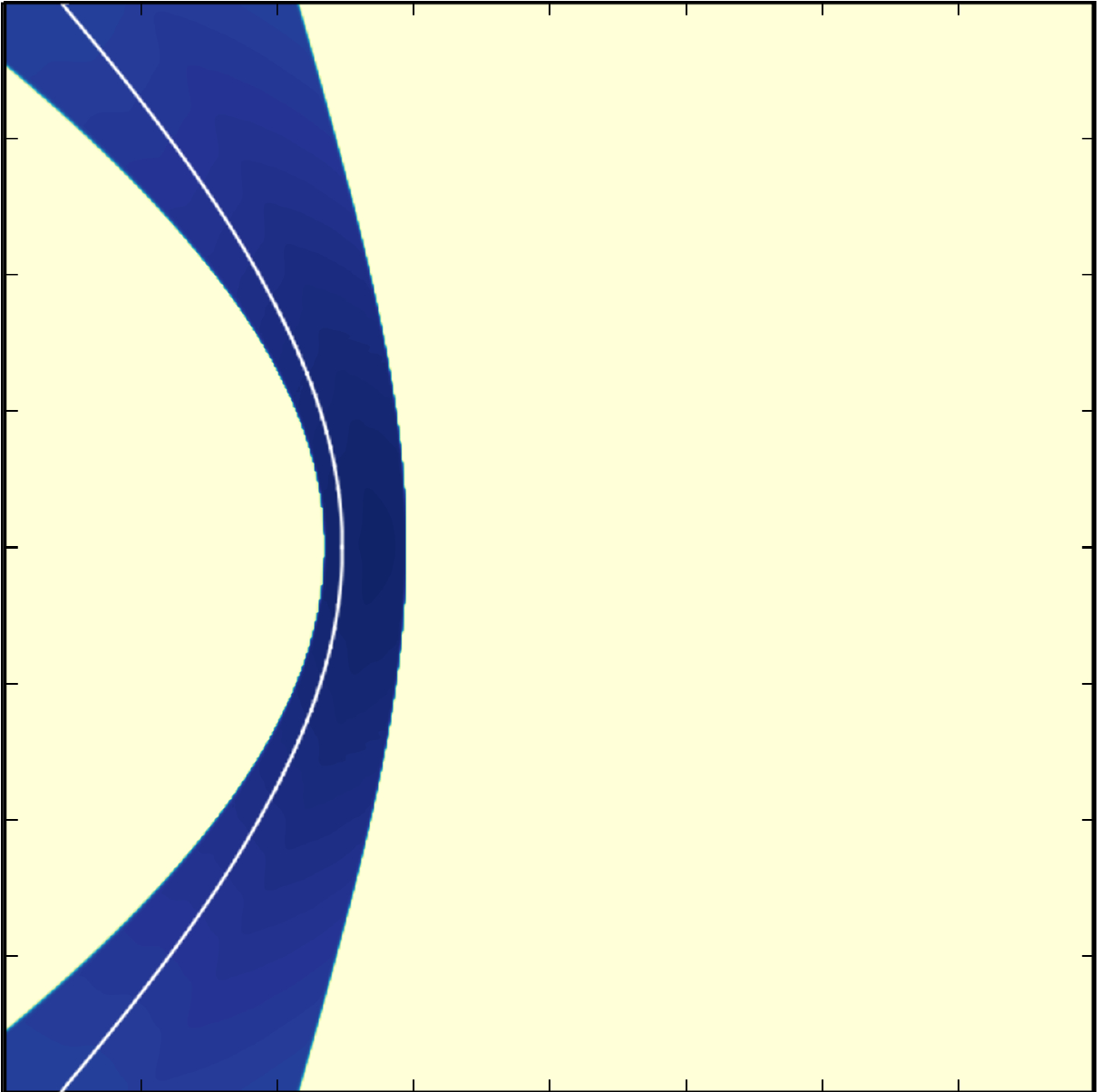}}
				\put(126.7,20){\line(1,0){35.625}}
				\put(126.7,17){\line(0,1){6}}
				\put(162.325,17){\line(0,1){6}}
				\put(126.7,5){\scriptsize{500R$_\odot$}}
			\end{picture}
		\end{subfigure}
		\begin{subfigure}{85\unitlength}
		\begin{picture}(95,290)
				\put(20,5){\includegraphics[trim=15.9cm 1.4cm 2.4cm 1.3cm, clip=true,height=280\unitlength]{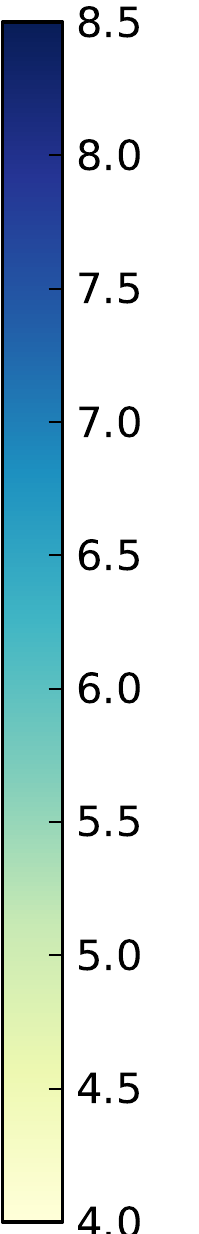}}
			\put(-5,100){\rotatebox{90}{\footnotesize{log(K)}}}
		\end{picture}
	\end{subfigure}\\
		\begin{subfigure}{290\unitlength}
			\begin{picture}(290,290)
				\put(0,0){\includegraphics[trim=2.9cm 1.5cm 5.1cm 1.5cm, clip=true,width=290\unitlength]{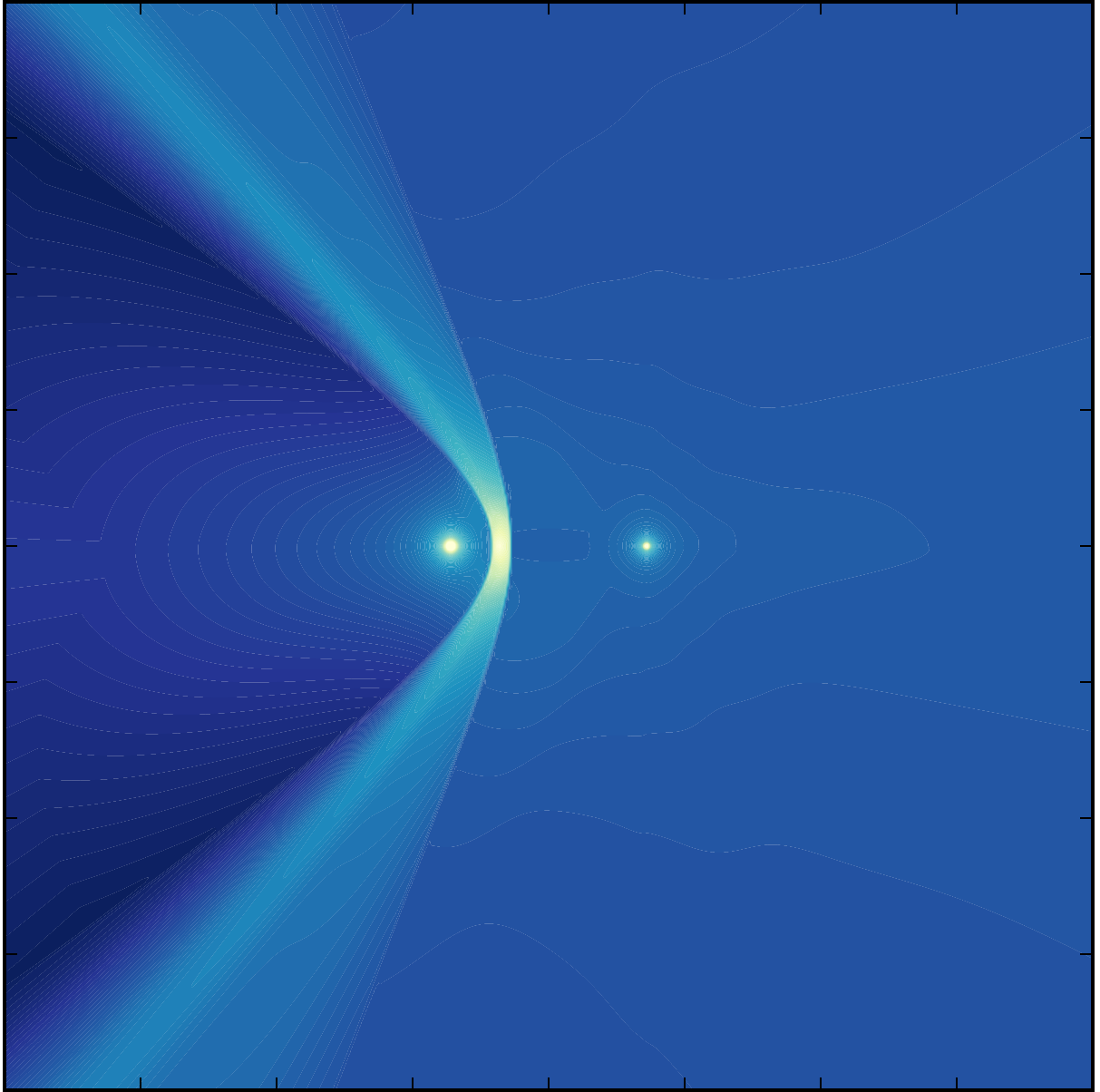}}
				\put(126.7,20){{\color{white}\line(1,0){35.625}}}
				\put(126.7,17){{\color{white}\line(0,1){6}}}
				\put(162.325,17){{\color{white}\line(0,1){6}}}
				\put(126.7,5){\scriptsize{{\color{white}500R$_\odot$}}}
			\end{picture}
		\end{subfigure}
		\begin{subfigure}{290\unitlength}
			\begin{picture}(290,290)
				\put(0,0){\includegraphics[trim=2.9cm 1.5cm 5.1cm 1.5cm, clip=true,width=290\unitlength]{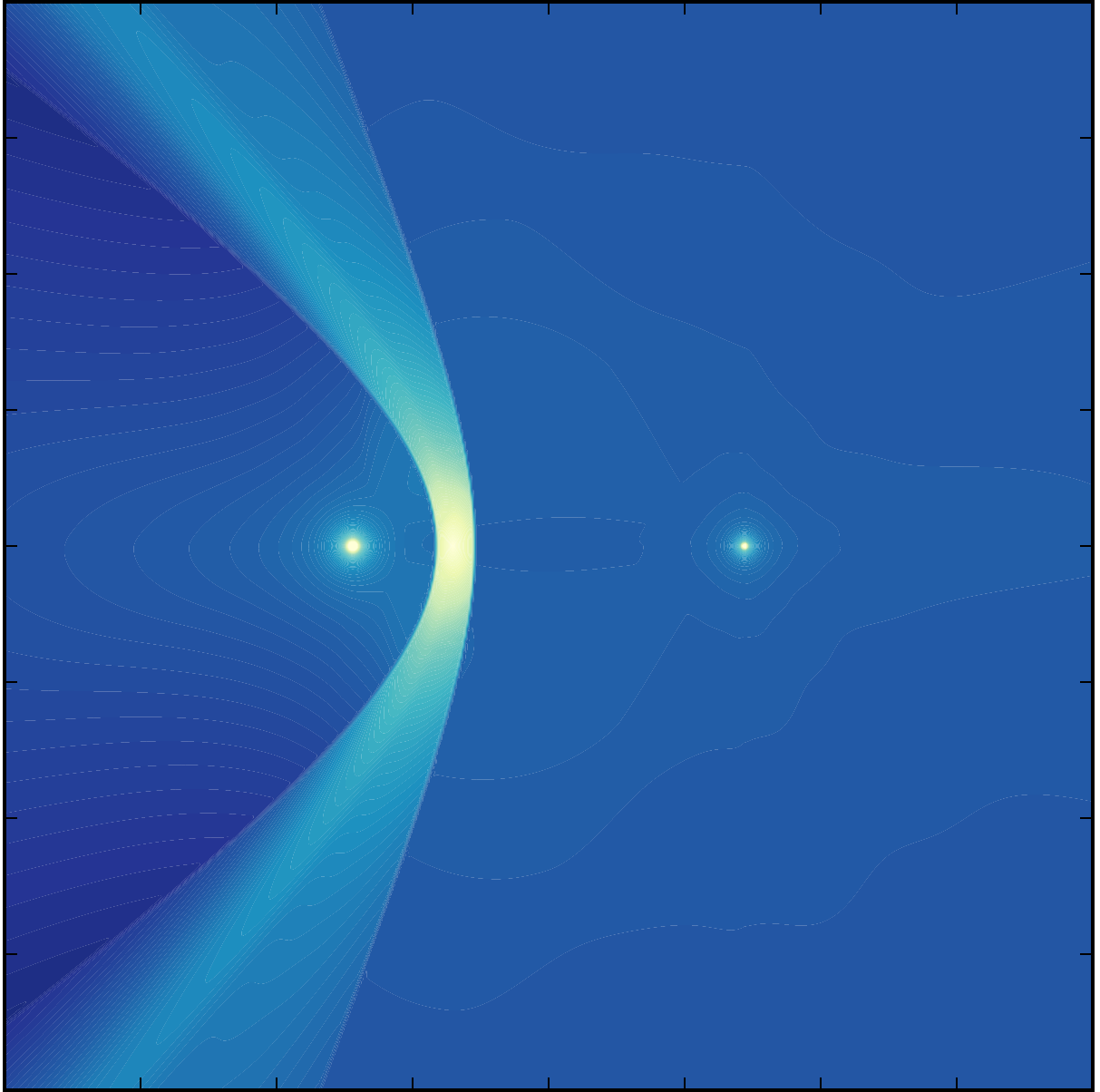}}		
				\put(126.7,20){{\color{white}\line(1,0){35.625}}}
				\put(126.7,17){{\color{white}\line(0,1){6}}}
				\put(162.325,17){{\color{white}\line(0,1){6}}}
				\put(126.7,5){\scriptsize{{\color{white}500R$_\odot$}}}
			\end{picture}
		\end{subfigure}
		\begin{subfigure}{290\unitlength}	
			\begin{picture}(290,290)
				\put(0,0){\includegraphics[trim=2.9cm 1.5cm 5.1cm 1.5cm, clip=true,width=290\unitlength]{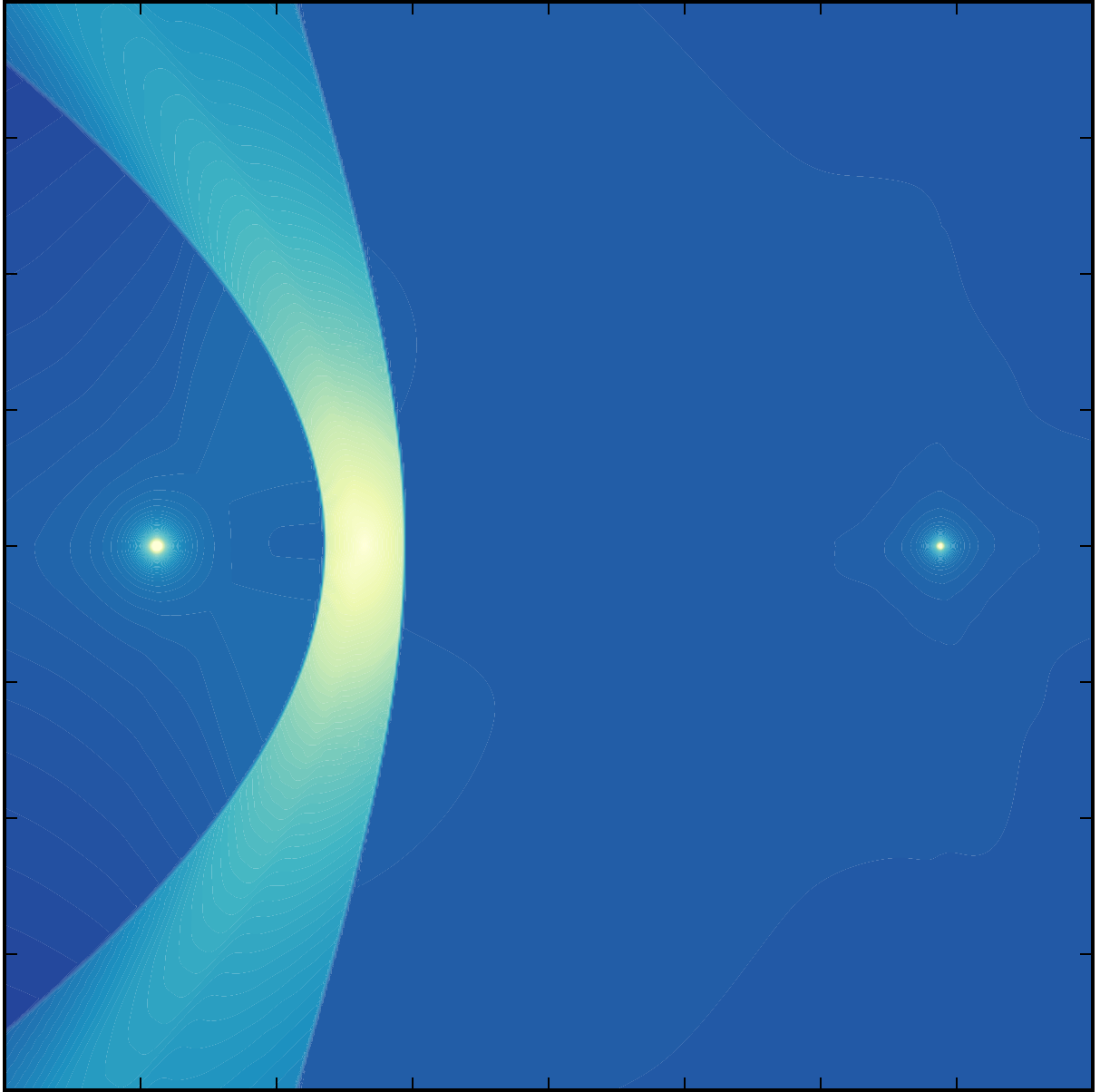}}
				\put(126.7,20){{\color{white}\line(1,0){35.625}}}
				\put(126.7,17){{\color{white}\line(0,1){6}}}
				\put(162.325,17){{\color{white}\line(0,1){6}}}
				\put(126.7,5){\scriptsize{{\color{white}500R$_\odot$}}}
			\end{picture}
		\end{subfigure}
	\begin{subfigure}{85\unitlength}
		\begin{picture}(95,290)
				\put(20,5){\includegraphics[trim=15.9cm 1.4cm 2.4cm 1.4cm, clip=true,height=280\unitlength]{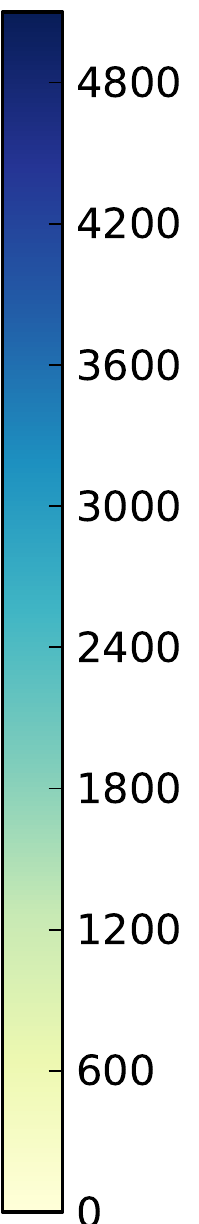}}
			\put(-5,100){\rotatebox{90}{\footnotesize{km s$^{-1}$}}}
		\end{picture}
	\end{subfigure}
\caption{HD quantities in the converged state for three different stellar separations. The plots show the $x$--$y$ plane of a $512\times512\times512$ simulation  at z=0. 
The WR star is located on the right-hand side of the WCR, and the B star is on its left-hand side. The depicted quantities are particle density (first row) in log(m$^{-3}$), temperature in log(K) (second row), and absolute velocity in km s$^{-1}$ (third row). In the second row, we also plot the analytical approximation  for the WRC position following by \cite{Stevens1992} (white solid line).
\label{hydros}}
\end{figure*}

As expected, the three different stellar separations show considerable contrasts in terms of properties and structure of the WCR. In Figure \ref{hydros} we show density, absolute velocity and temperature in the $x$-$y$ plane at $z=0$ for cases A, B, and C. 

\begin{table}[t]
\begin{center}
\textbf{Stellar and stellar-wind parameters}\\
\end{center}
\centering
\begin{tabular}{r ||c |c |c| l}
  & \textbf{B} & \textbf{WR}  & \textit{unit}\\ \hline \hline
 $M_\ast$ &30 & 30  & M$_{\odot}$\\
 $R_\ast$ &20 & 10 & R$_{\odot}$\\
 $T_\ast$ &23000 & 40000 & K\\
 $L_\ast$  & $10^{5}$ & $2.3\times 10^{5}$   & L$_{\odot}$\\
 $\dot{M}$ &$10^{-6}$ & $10^{-5}$  & M$_{\odot}$ yr\textsuperscript{-1}\\
 $v_\infty$ & 4000 & 4000  & km s$^{-1}$\\
 \end{tabular}
\caption{Stellar and stellar-wind parameters of a typical binary system as considered in this work.\label{params}}
\end{table}

\paragraph{Shape of WCR}
The thickness of the WCR at the apex widens considerably with increasing stellar separation. Along the line of centers, its width is $\sim$100, $\sim$170 and $\sim$310 $R_\odot$ for cases A, B, and C, respectively. Analogously, the distance between the WCR and the B star (which is generally closer because of $\eta=0.1$) increases. The star is located $\sim$140, $\sim$290 and $\sim$600 $R_\odot$ from the edge, and $\sim$190, $\sim$380 and $\sim$760 $R_\odot$ from the center of the WCR for cases A, B, and C. Positions and curvatures of the WCR in all models agree well with the analytical approximation by \cite{Stevens1992} in which the distance of the apex of the WCR from the B star is determined by $d_B=\frac{\sqrt{\eta}}{1+\sqrt{\eta}}d$ and the curvature of the downstream WCR is approximated by solving the differential equation
\begin{equation}
\frac{dy}{dx}=\frac{\left(\sqrt{\eta}d_\mathrm{WR}^2+d_\mathrm{B}^2\right)y}{\sqrt{\eta}d_\mathrm{WR}^2x+d_\mathrm{B}^2(x-d)}.
\end{equation}
The resulting curves are indicated in the second row of Figure \ref{hydros}.
Contrasting the effect of increasing WCR thickness with increasing stellar separation, the opening angle (or curvature) of the collision region significantly decreases. Among the three systems under investigation, case A has the narrowest WCR at the apex, but the widest WCR close to the edge of the computational domain.

\paragraph{Density}
The maximum number density values within the WCR are 33.1, 7.6 and 1.8 $\times10^{13}$ m$^{-3}$ for cases A, B, and C. The approximate inverse proportionality to $d^2$ is understandable from the fact that, for a strong shock, $\rho_\mathrm{postshock}\sim4\rho_\mathrm{preshock}$ and $\rho_\mathrm{preshock}=\frac{\dot{M}}{4\pi r^2 v_\mathrm{wind}}$. The density contrast of the unshocked winds manifests itself in  comparable contrasts within the WCR where the side toward the WR star shows considerable higher density values of the shocked wind plasma.

\paragraph{Temperature}
The maximum temperature reached at the apex of the WCR shows only little variation. It is 2.14, 2.17 and 2.21 $\times$10$^8$ K for cases A, B, and C. In contrast to that the temperature gradient is far more affected by the change of stellar separation. The closer the stars, the larger the drop in temperature with increasing distance from the apex of the WCR. The minimum temperature of the shocked wind plasma reached at the edge of the computational domain varies accordingly, with values of 6.9, 20.9, 61.7 $\times10^6$ K for cases A, B, and C.
\paragraph{Absolute velocity}
In all three systems, the wind is slowed down to near zero velocity at the apex. We observe a considerable difference concerning the size of the region in which the velocity remains comparatively low. The distance downstream from the apex at which the shocked wind plasma is efficiently re-accelerated is significantly shorter for the case of smallest stellar separation. 
Another notable difference is the maximum velocity reached within the computational domain in the unshocked wind outside the WCR. For case A, the B wind close to the WCR at the edge of the domain reaches values up to 5100 km s$^{-1}$ in contrast to merely 4700 and 4300 km s$^{-1}$ for case B and C. The reason for these velocity values which lie all above the B wind's terminal velocity of $v_\infty=$4000 km s$^{-1}$ is the radiative wind acceleration mechanism which allows additional acceleration of the unshocked B star's wind due to the radiation of the WR star. The effect increases with increasing proximity of the WR star, thus it is greatest for case A.

\paragraph{Other properties}
The local magnetic field strength and the energy density of radiation are two important properties that critically determine the energy distribution of the particles accelerated at the shocks by influencing energy loss by synchrotron emission and IC cooling. In estimating the magnetic field strength in dependence of the distance of the stars we rely on the approximations following \cite{Usov1992} (details also in paper I). The respective values at the center of the WCR for cases A, B and C are $\sim$0.64 G, $\sim$0.16 G, and $\sim$0.04 G. \\
The energy density of radiation is proportional to the luminosity of the stars and the inverse square of their distance. Its values at the center of the WCR are $\sim$0.77, $\sim$0.19, and $\sim$0.05 J~m$^{-3}$ for cases A, B and C, respectively.

%************** HYDRO PLOT *********************

\subsection{High-Energy Particles}
\label{partd}
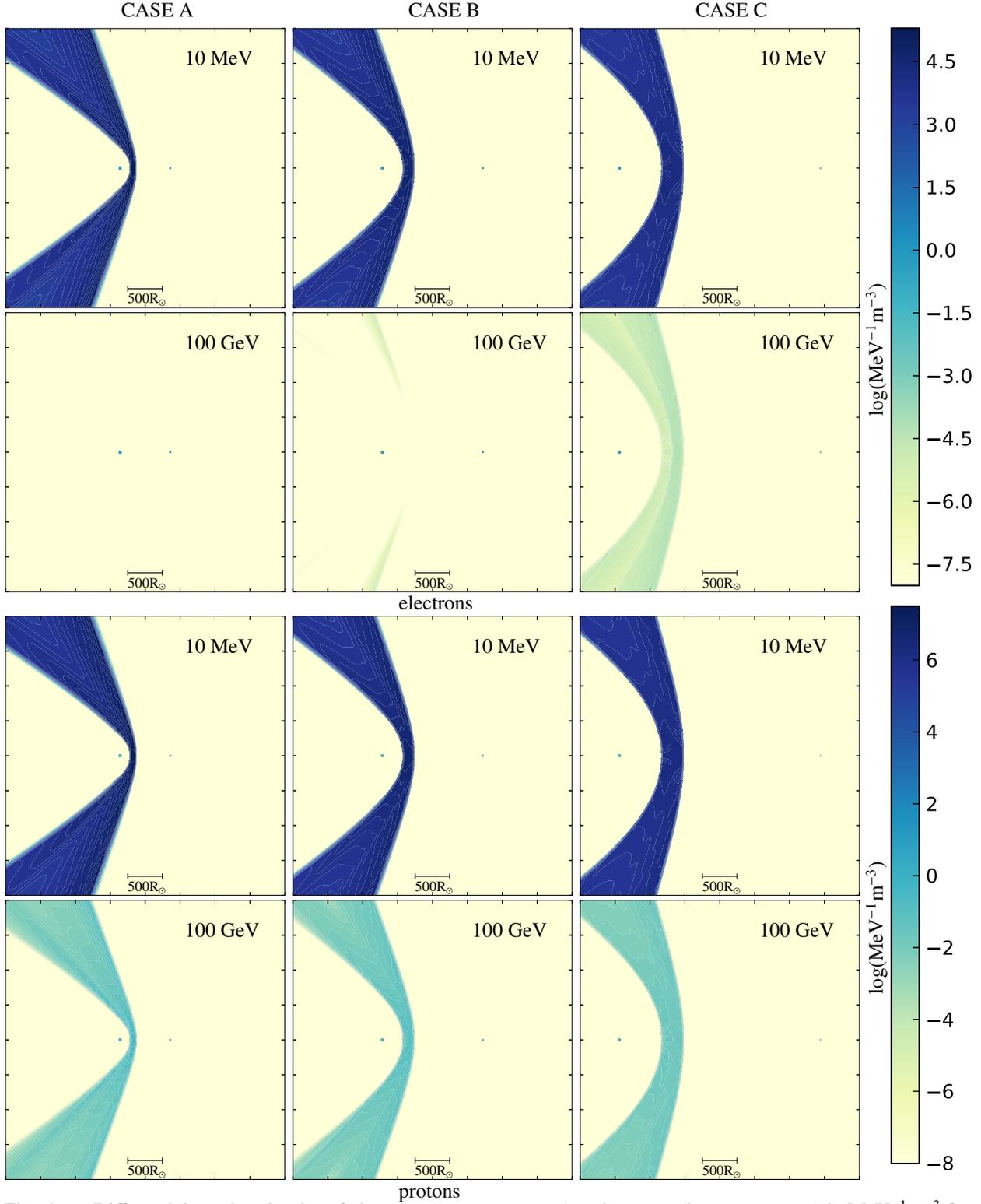
\begin{figure*}
	\setlength{\unitlength}{0.001\textwidth}
	\begin{subfigure}[l]{900\unitlength}
		\begin{subfigure}{290\unitlength}
			\begin{picture}(290,290)
					\put(0,0){\includegraphics[trim=2.9cm 1.5cm 5.1cm 1.5cm, clip=true,width=290\unitlength]{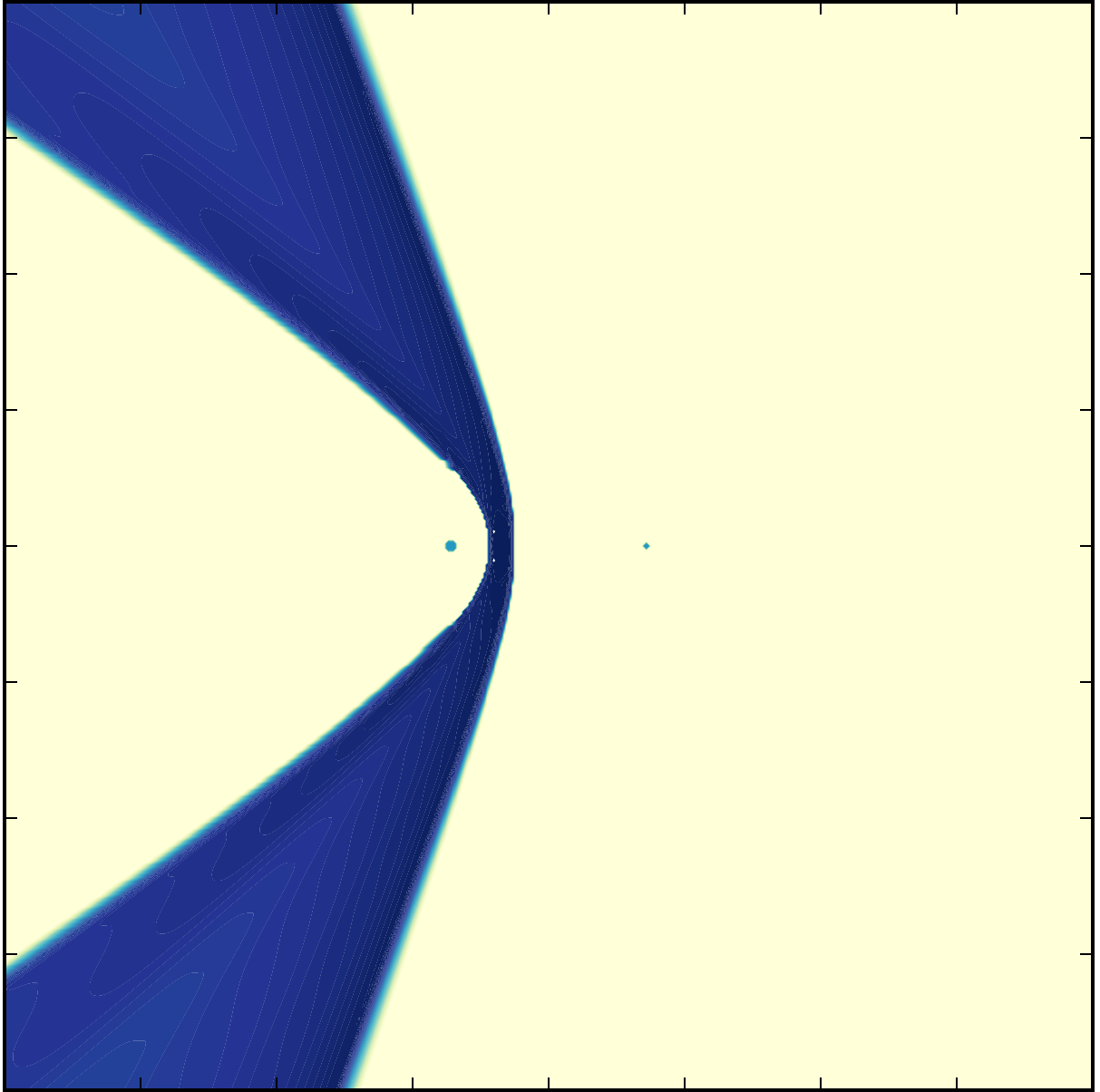}}
				\put(120,300){CASE A}
				\put(180,250){ 10 MeV}
				\put(126.7,20){\line(1,0){35.625}}
				\put(126.7,17){\line(0,1){6}}
				\put(162.325,17){\line(0,1){6}}
				\put(126.7,5){\scriptsize{500R$_\odot$}}
			\end{picture}
		\end{subfigure}
		\begin{subfigure}{290\unitlength}
			\begin{picture}(290,290)
					\put(0,0){\includegraphics[trim=2.9cm 1.5cm 5.1cm 1.5cm, clip=true,width=290\unitlength]{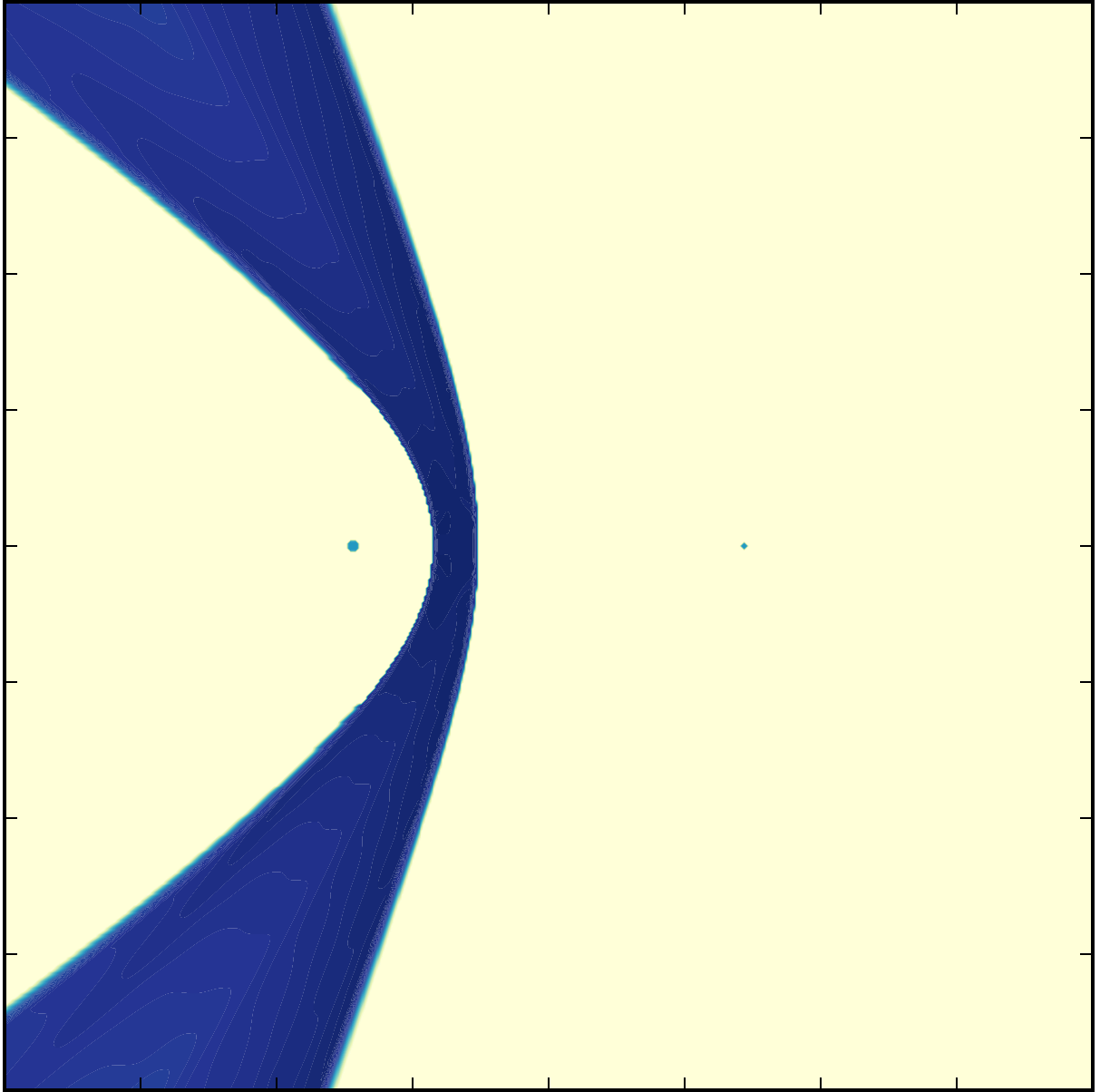}}
				\put(120,300){CASE B}
				\put(180,250){ 10 MeV}
				\put(126.7,20){\line(1,0){35.625}}
				\put(126.7,17){\line(0,1){6}}
				\put(162.325,17){\line(0,1){6}}
				\put(126.7,5){\scriptsize{500R$_\odot$}}
			\end{picture}
				\end{subfigure}
		\begin{subfigure}{290\unitlength}
			\begin{picture}(290,290)
			\put(0,0){\includegraphics[trim=2.9cm 1.5cm 5.1cm 1.5cm, clip=true,width=290\unitlength]{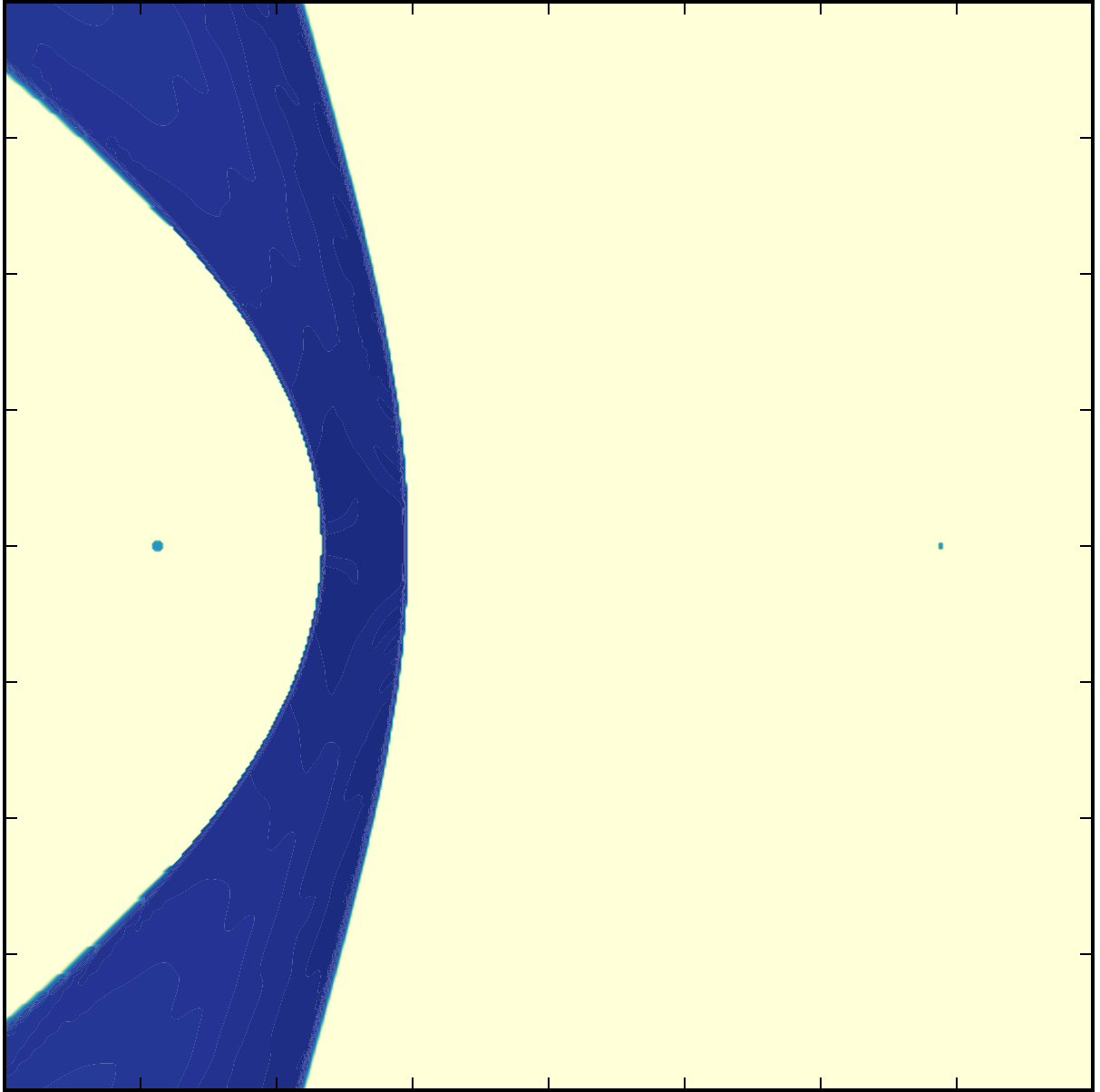}}
				\put(120,300){CASE C}
				\put(180,250){ 10 MeV}
				\put(126.7,20){\line(1,0){35.625}}
				\put(126.7,17){\line(0,1){6}}
				\put(162.325,17){\line(0,1){6}}
				\put(126.7,5){\scriptsize{500R$_\odot$}}
			\end{picture}
		\end{subfigure}\\
		\begin{subfigure}{290\unitlength}
			\begin{picture}(290,290)
				\put(0,0){\includegraphics[trim=2.9cm 1.5cm 5.1cm 1.5cm, clip=true,width=290\unitlength]{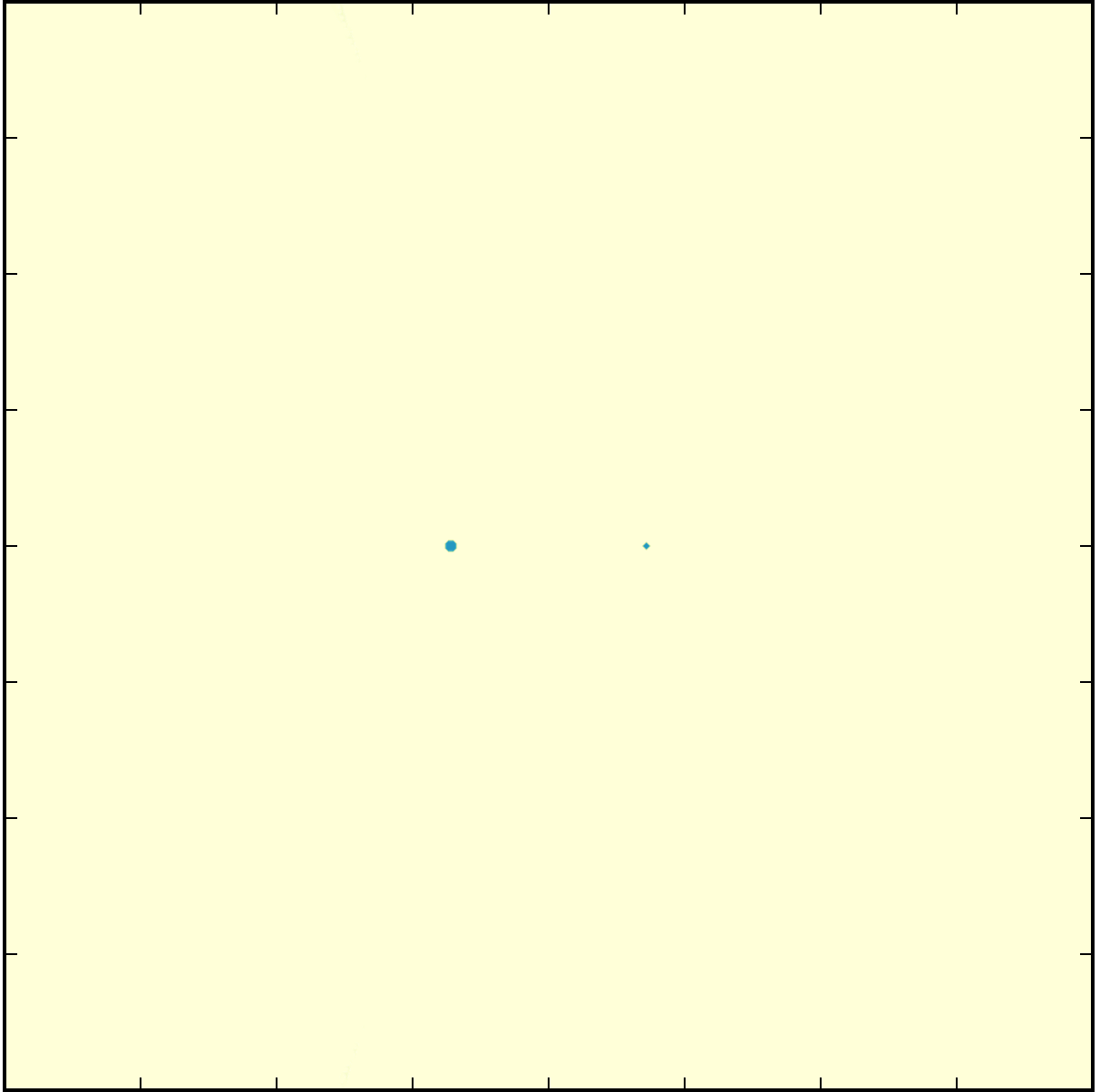}}
				\put(180,250){ 100 GeV}
				\put(126.7,20){\line(1,0){35.625}}
				\put(126.7,17){\line(0,1){6}}
				\put(162.325,17){\line(0,1){6}}
				\put(126.7,5){\scriptsize{500R$_\odot$}}
			\end{picture}
		\end{subfigure}
		\begin{subfigure}{290\unitlength}
			\begin{picture}(290,290)
				\put(0,0){\includegraphics[trim=2.9cm 1.5cm 5.1cm 1.5cm, clip=true,width=290\unitlength]{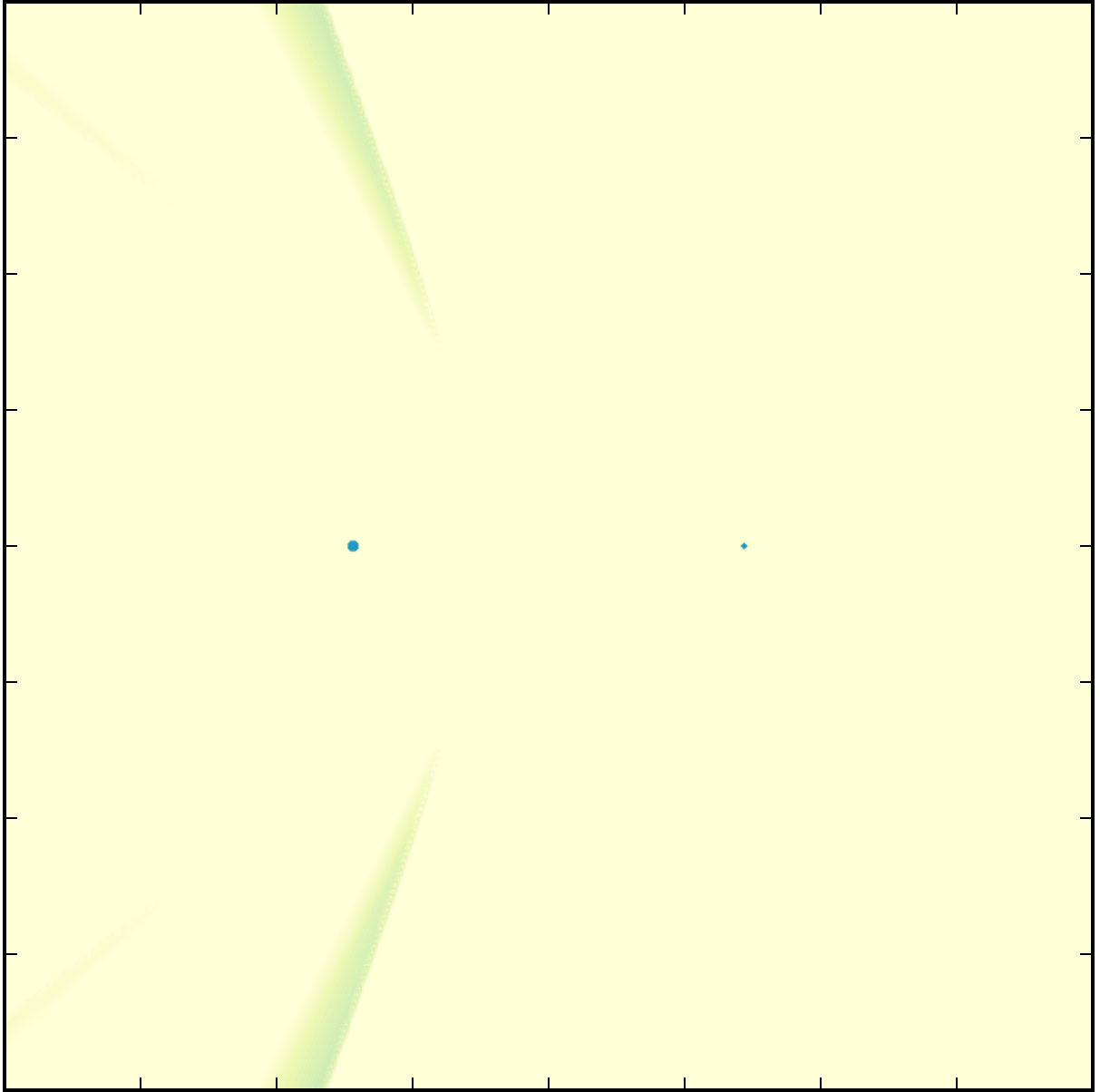}}
				\put(180,250){ 100 GeV}
				\put(126.7,20){\line(1,0){35.625}}
				\put(126.7,17){\line(0,1){6}}
				\put(162.325,17){\line(0,1){6}}
				\put(126.7,5){\scriptsize{500R$_\odot$}}
				\put(110,-18){electrons}
			\end{picture}
		\end{subfigure}
		\begin{subfigure}{290\unitlength}	
			\begin{picture}(290,290)
				\put(0,0){\includegraphics[trim=2.9cm 1.5cm 5.1cm 1.5cm, clip=true,width=290\unitlength]{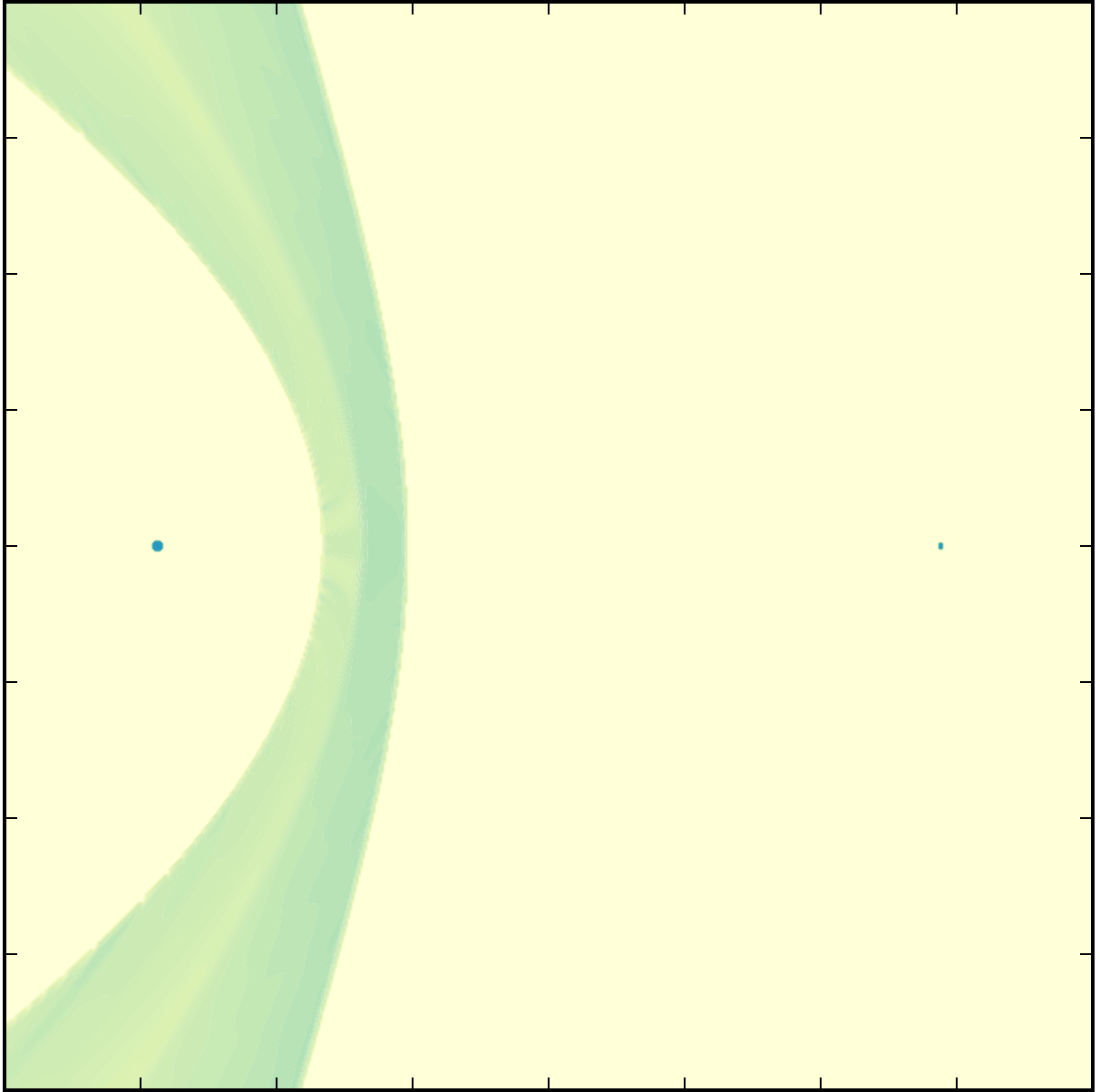}}
				\put(180,250){ 100 GeV}
				\put(126.7,20){\line(1,0){35.625}}
				\put(126.7,17){\line(0,1){6}}
				\put(162.325,17){\line(0,1){6}}
				\put(126.7,5){\scriptsize{500R$_\odot$}}
			\end{picture}
		\end{subfigure}
	\end{subfigure}
		\begin{subfigure}{85\unitlength}
		\begin{picture}(85,580)
				\put(5,0){\includegraphics[trim=15.9cm 1.4cm 2.3cm 1.5cm, clip=true,height=580\unitlength]{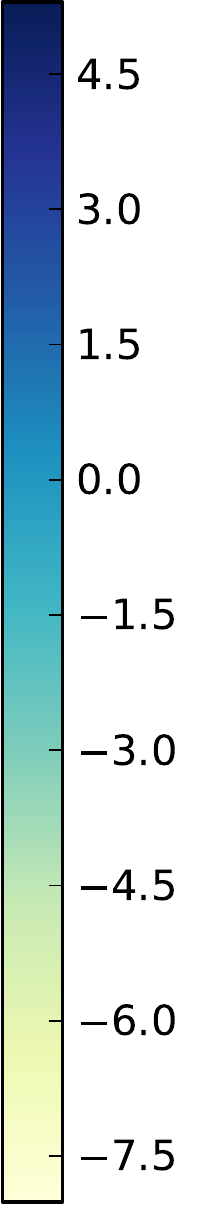}}
			\put(-20,180){\rotatebox{90}{log(MeV$^{-1}$m$^{-3}$)}}
		\end{picture}
	\end{subfigure}\\
	\begin{subfigure}[l]{900\unitlength}
		\begin{subfigure}{290\unitlength}
			\begin{picture}(290,310)
				\put(0,0){\includegraphics[trim=2.9cm 1.5cm 5.1cm 1.5cm, clip=true,width=290\unitlength]{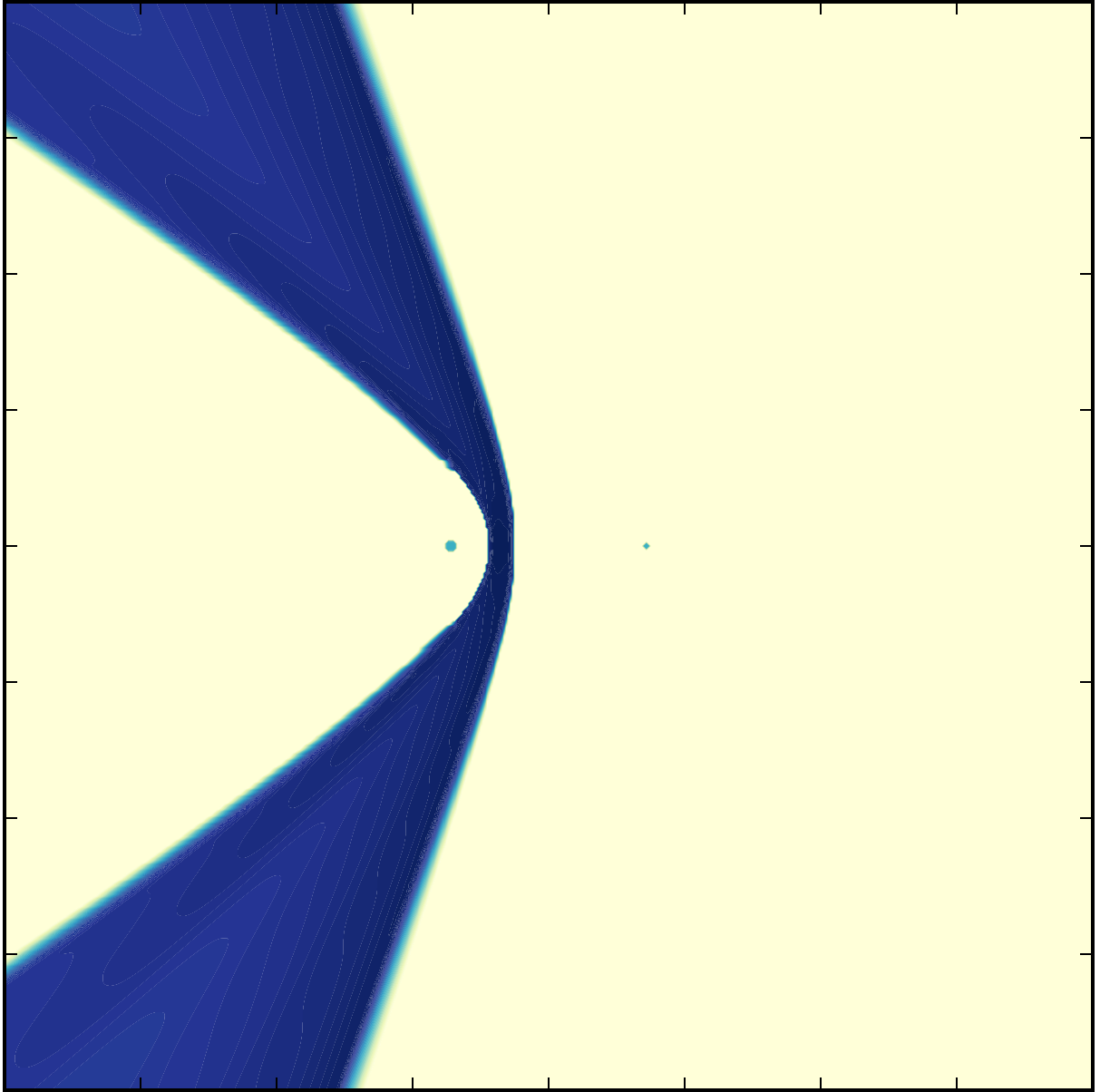}}
				\put(180,250){ 10 MeV}
				\put(126.7,20){\line(1,0){35.625}}
				\put(126.7,17){\line(0,1){6}}
				\put(162.325,17){\line(0,1){6}}
				\put(126.7,5){\scriptsize{500R$_\odot$}}
			\end{picture}
		\end{subfigure}
		\begin{subfigure}{290\unitlength}
			\begin{picture}(290,310)
				\put(0,0){\includegraphics[trim=2.9cm 1.5cm 5.1cm 1.5cm, clip=true,width=290\unitlength]{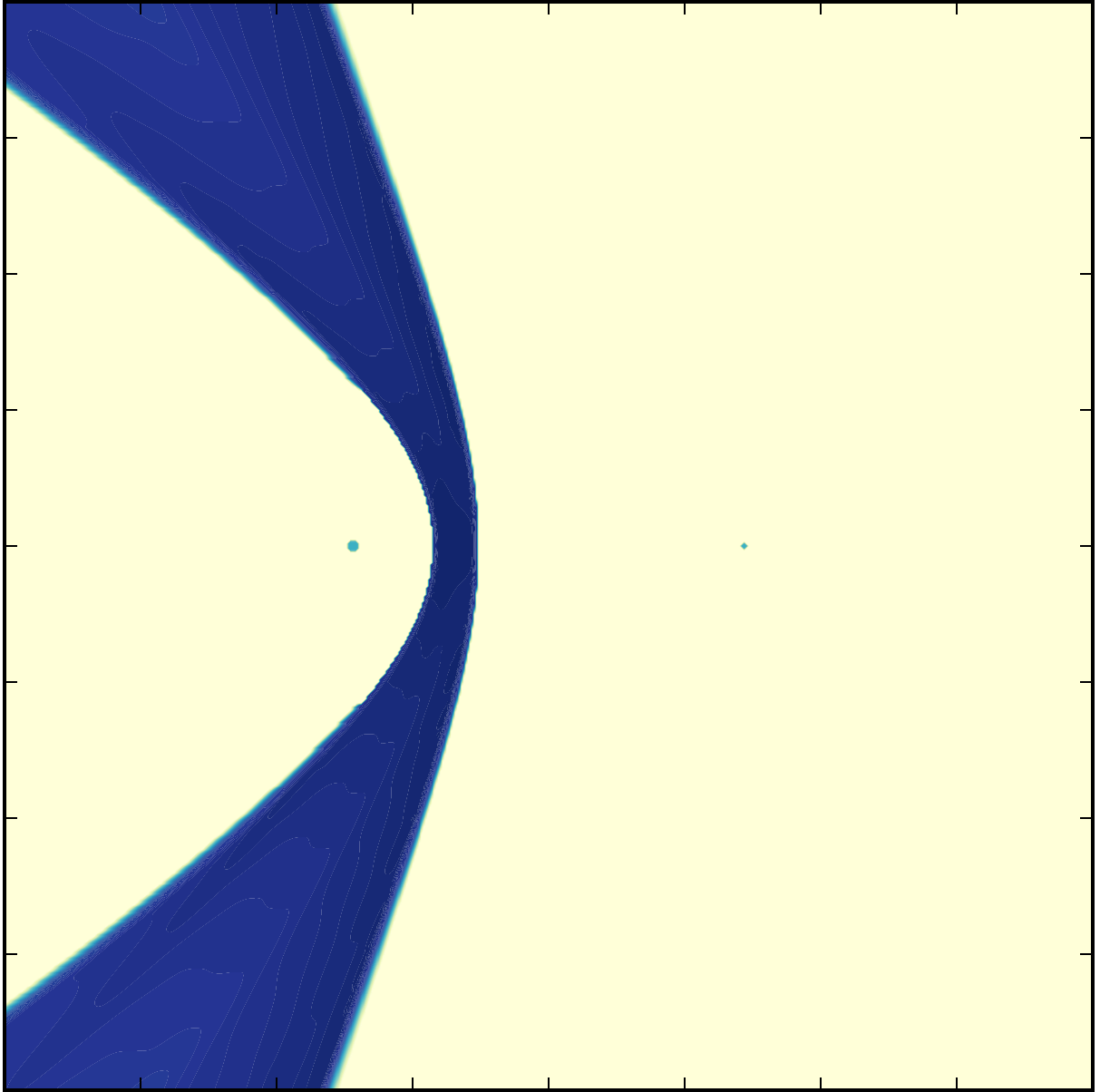}}
				\put(180,250){ 10 MeV}
				\put(126.7,20){\line(1,0){35.625}}
				\put(126.7,17){\line(0,1){6}}
				\put(162.325,17){\line(0,1){6}}
				\put(126.7,5){\scriptsize{500R$_\odot$}}
			\end{picture}
		\end{subfigure}
		\begin{subfigure}{290\unitlength}	
			\begin{picture}(290,310)
				\put(0,0){\includegraphics[trim=2.9cm 1.5cm 5.1cm 1.5cm, clip=true,width=290\unitlength]{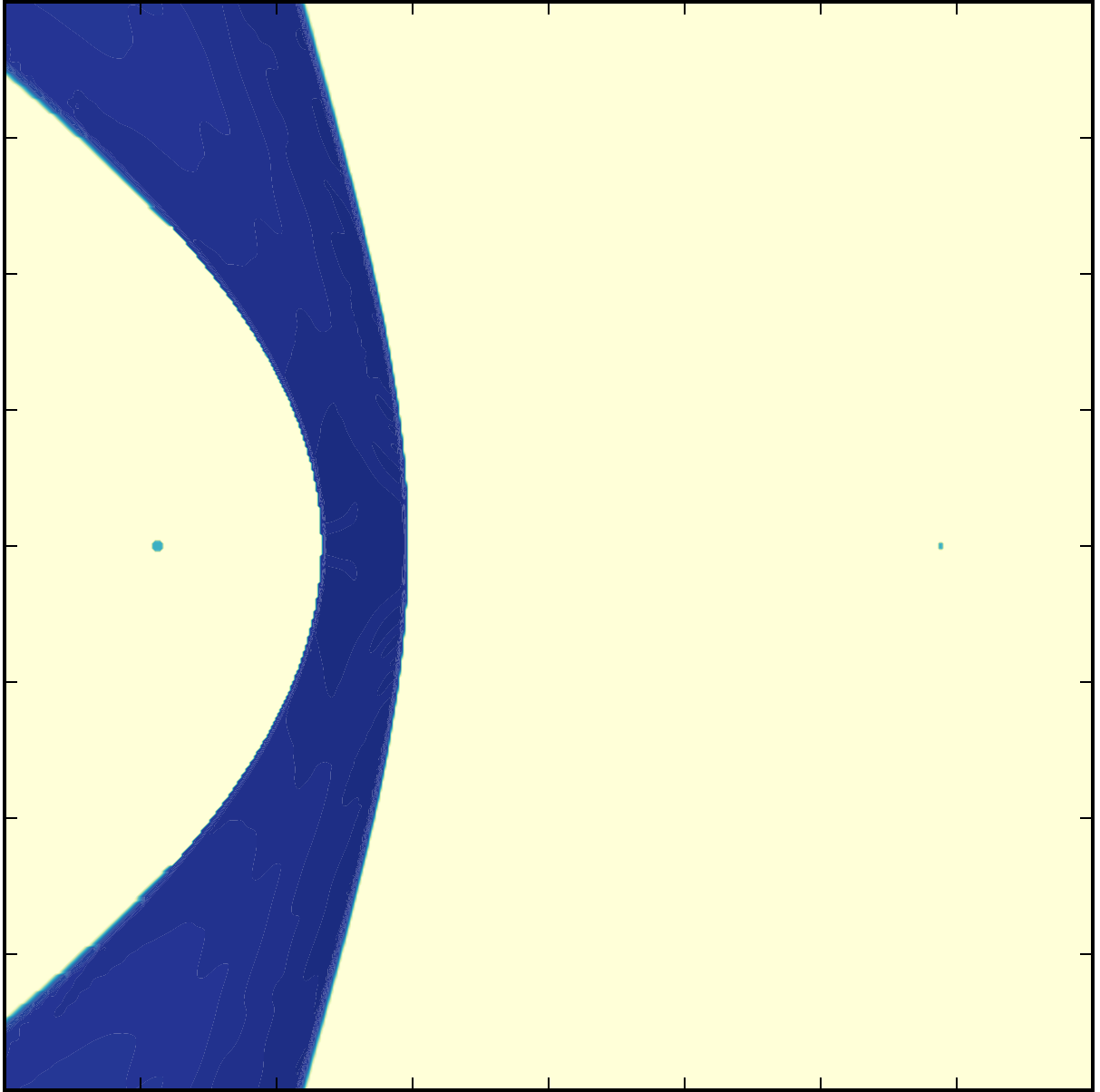}}
				\put(180,250){ 10 MeV}
				\put(126.7,20){\line(1,0){35.625}}
				\put(126.7,17){\line(0,1){6}}
				\put(162.325,17){\line(0,1){6}}
				\put(126.7,5){\scriptsize{500R$_\odot$}}
			\end{picture}
		\end{subfigure}\\
		\begin{subfigure}{290\unitlength}
			\begin{picture}(290,290)
				\put(0,0){\includegraphics[trim=2.9cm 1.5cm 5.1cm 1.5cm, clip=true,width=290\unitlength]{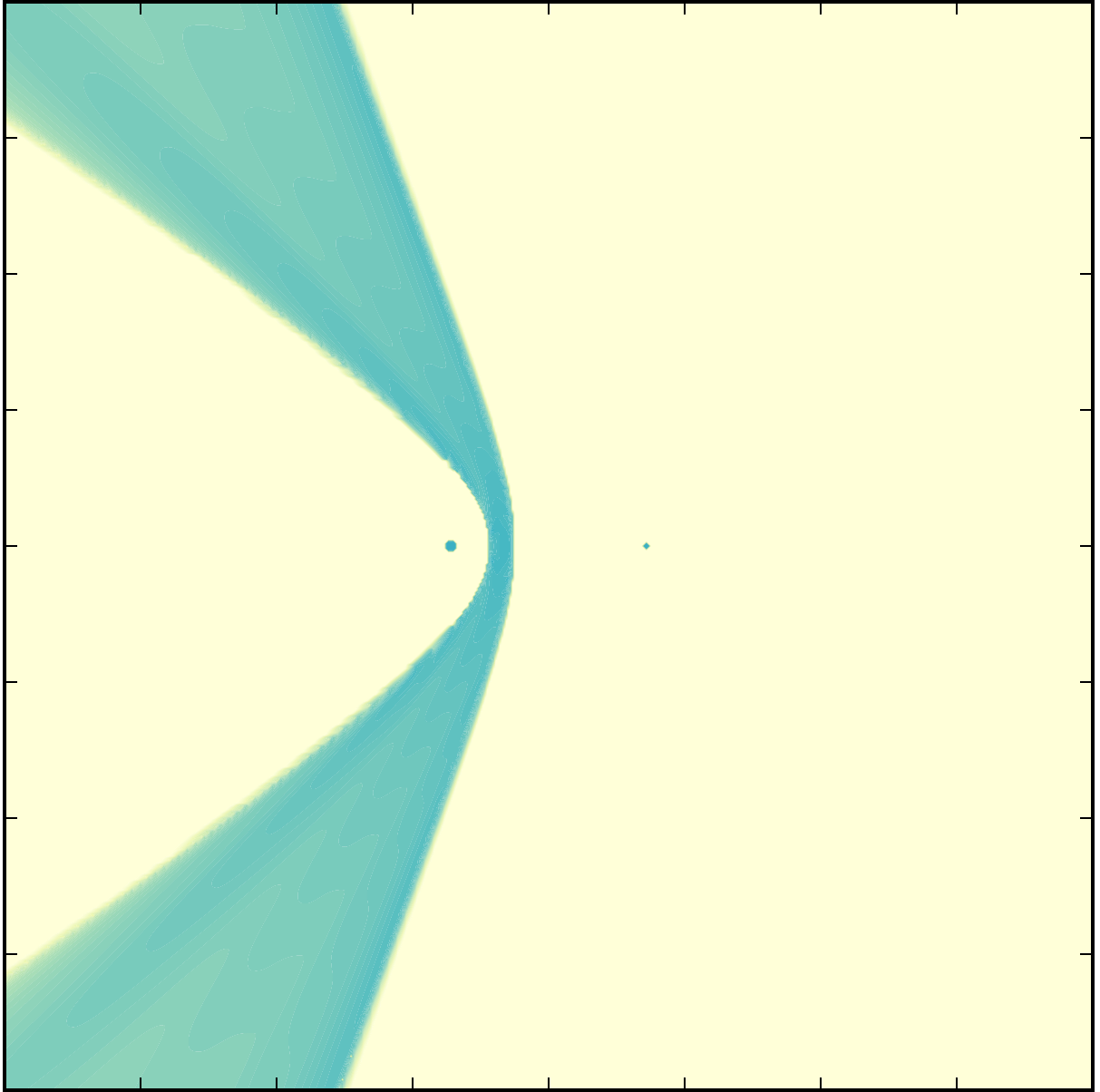}}
				\put(180,250){ 100 GeV}
				\put(126.7,20){\line(1,0){35.625}}
				\put(126.7,17){\line(0,1){6}}
				\put(162.325,17){\line(0,1){6}}
				\put(126.7,5){\scriptsize{500R$_\odot$}}
			\end{picture}
		\end{subfigure}
		\begin{subfigure}{290\unitlength}
			\begin{picture}(290,290)
				\put(0,0){\includegraphics[trim=2.9cm 1.5cm 5.1cm 1.5cm, clip=true,width=290\unitlength]{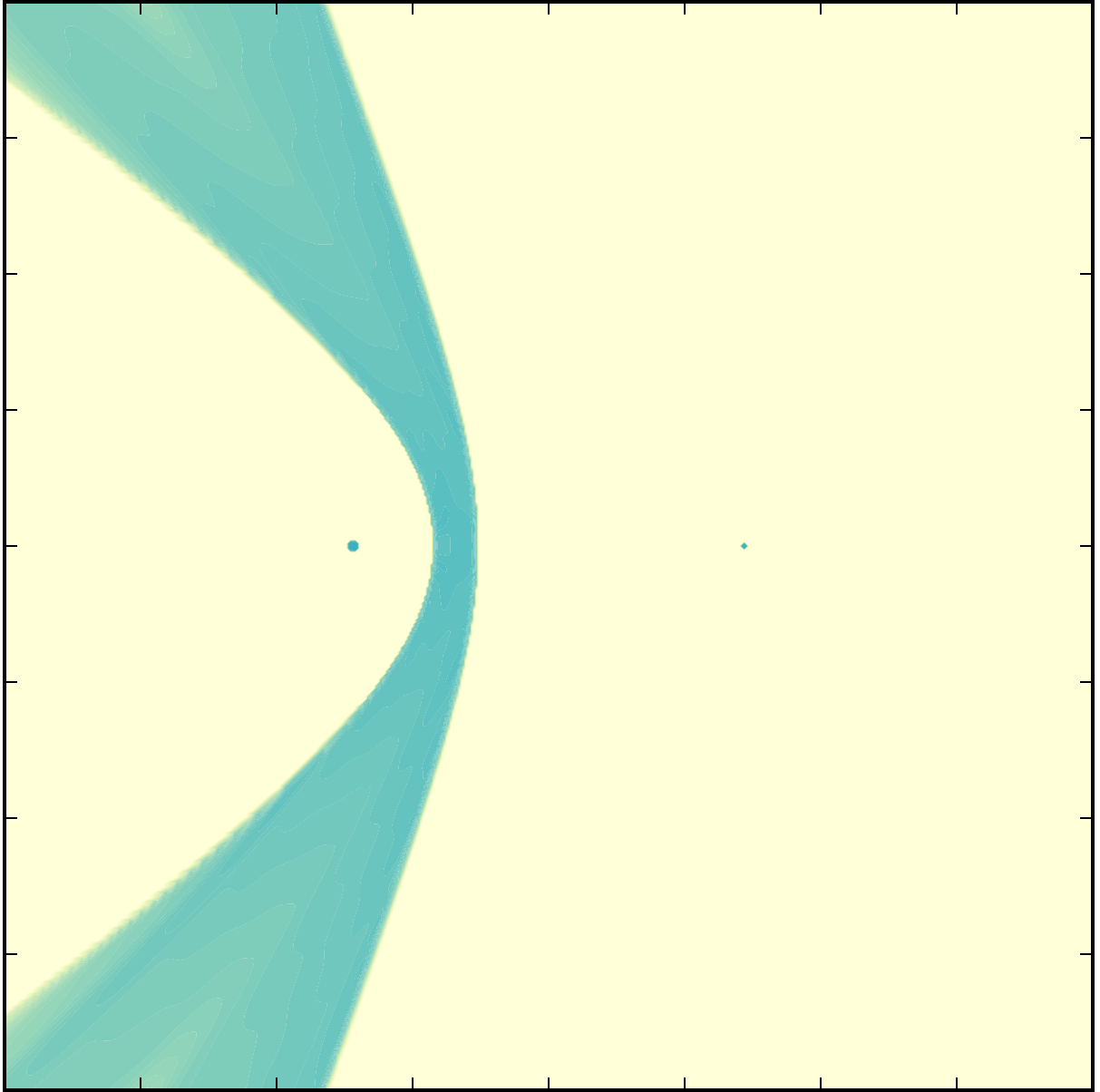}}
				\put(180,250){ 100 GeV}
				\put(126.7,20){\line(1,0){35.625}}
				\put(126.7,17){\line(0,1){6}}
				\put(162.325,17){\line(0,1){6}}
				\put(126.7,5){\scriptsize{500R$_\odot$}}
				\put(110,-18){protons}
			\end{picture}
		\end{subfigure}
		\begin{subfigure}{290\unitlength}	
			\begin{picture}(290,290)
				\put(0,0){\includegraphics[trim=2.9cm 1.5cm 5.1cm 1.5cm, clip=true,width=290\unitlength]{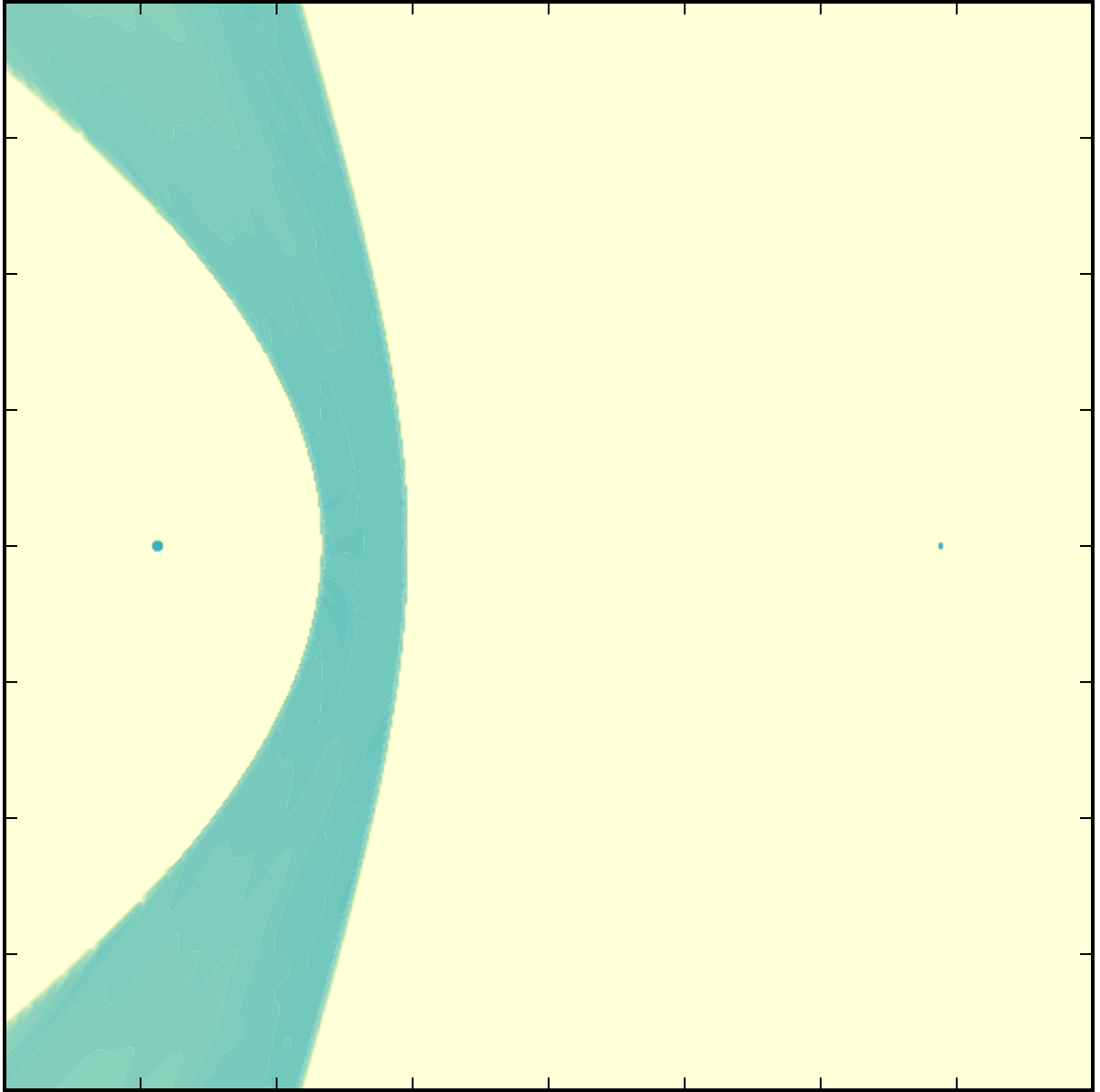}}
				\put(180,250){ 100 GeV}
				\put(126.7,20){\line(1,0){35.625}}
				\put(126.7,17){\line(0,1){6}}
				\put(162.325,17){\line(0,1){6}}
				\put(126.7,5){\scriptsize{500R$_\odot$}}
			\end{picture}
		\end{subfigure}
	\end{subfigure}
	\begin{subfigure}{85\unitlength}
		\begin{picture}(85,580)
				\put(5,0){\includegraphics[trim=15.9cm 1.4cm 2.3cm 1.5cm, clip=true,height=580\unitlength]{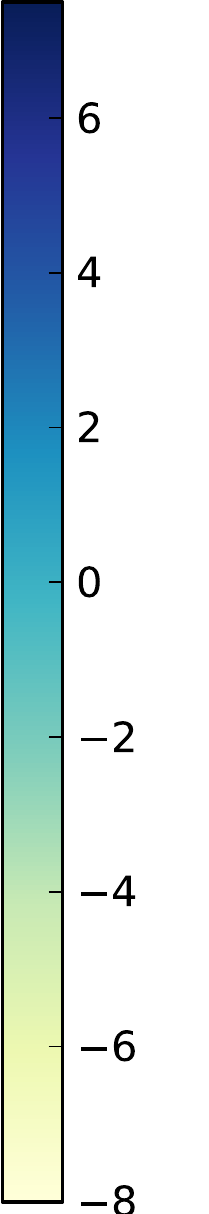}}
			\put(-20,180){\rotatebox{90}{log(MeV$^{-1}$m$^{-3}$)}}
		\end{picture}
	\end{subfigure}
\caption{Differential number density of electrons (upper two rows) and protons (lower two rows) in MeV$^{-1}$m$^{-3}$ for different values of kinetic particle energy and different stellar separations. The colour maps show the $x$--$y$ plane of a 512$\times$512$\times$\textbf{512} simulation at z=0.   
\label{els}}
\end{figure*}

Along with the hydrodynamic variables of density, velocity and temperature, we consider 200 advected scalar fields containing the number densities of high-energy electrons and protons at different energies accelerated at the shock fronts of the WCR. A transport equation for both electrons and protons in energy space is solved after every hydrodynamical time step. The injection rate of electrons and protons at the shock front is treated proportional to the number density of particles in the wind. An important free parameter is the electron-proton injection ratio for which we chose a typical value of 10$^{-2}$ (for details, see paper I).
Figure \ref{els} depicts the spatial distribution of electrons and protons at two different energies for all three cases A, B, and C. 

%************** PARTICLE *********************

At energies of $\sim$10 MeV, electrons as well as protons reach higher densities for smaller stellar separation. This is due to the proportionality of the electron and proton injection rate to the wind density. The most important difference between the three cases lies in the maximum energy attained by electrons. Whereas electrons of 100 GeV are distributed throughout most of the WCR for case C, they are confined to large stellar distances for case B or even vanish completely for case A. This has two reasons. For smaller stellar separations, the winds collide long before reaching their terminal velocity. This produces lower shock velocities $V_\mathrm{shock}$. As the diffusive shock acceleration term depends on $V_\mathrm{shock}^2$, the maximum attainable energies are significantly reduced for small stellar separations. In addition, close proximity to the stars leads to much higher energy densities of radiation and magnetic field. Together with higher wind plasma densities in the WCR, this greatly increases IC losses, synchrotron losses and bremsstrahlung losses. 

As protons are not affected by these loss terms, energies of $\sim$100 GeV are reached for all three cases. However -- mainly due to higher downstream velocities --  they cannot efficiently propagate into the inner downstream regions of the WCR for low stellar separations. This leads to a notable drop in number density of high-energy protons from the shock fronts toward the inner WCR which is most notable for case A. For case C, low downstream velocities allow the protons to fill the WCR homogeneously.

\subsection{The 3D Opacity Structure}
\begin{figure*}
	\setlength{\unitlength}{0.001\textwidth}
	\begin{subfigure}{490\unitlength}
\includegraphics[trim=0cm 0cm 0cm 0cm, clip=true,width=490\unitlength]{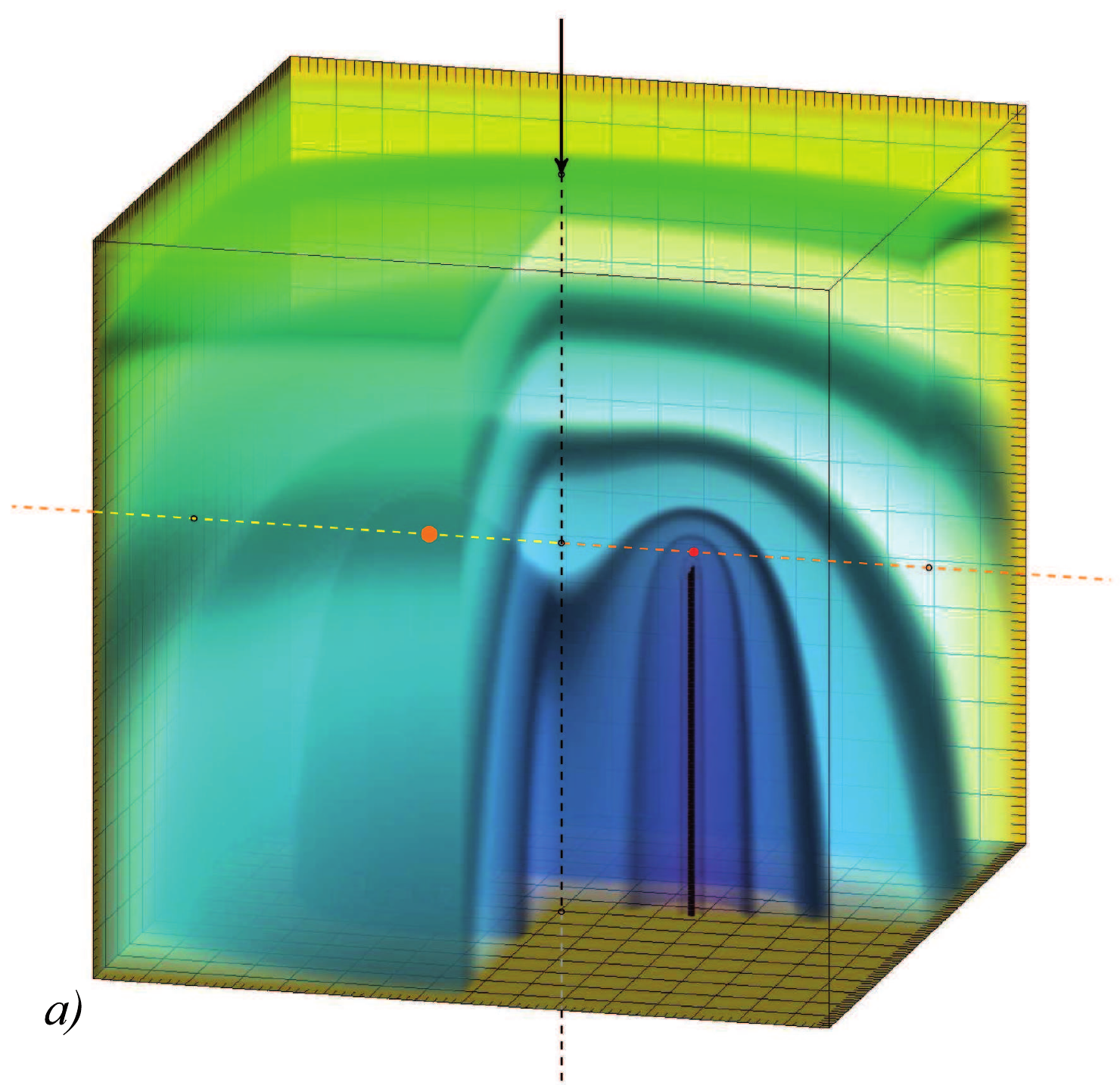}
\end{subfigure}
\begin{subfigure}{490\unitlength}
\includegraphics[trim=0cm 0cm 0cm 0cm, clip=true,width=490\unitlength]{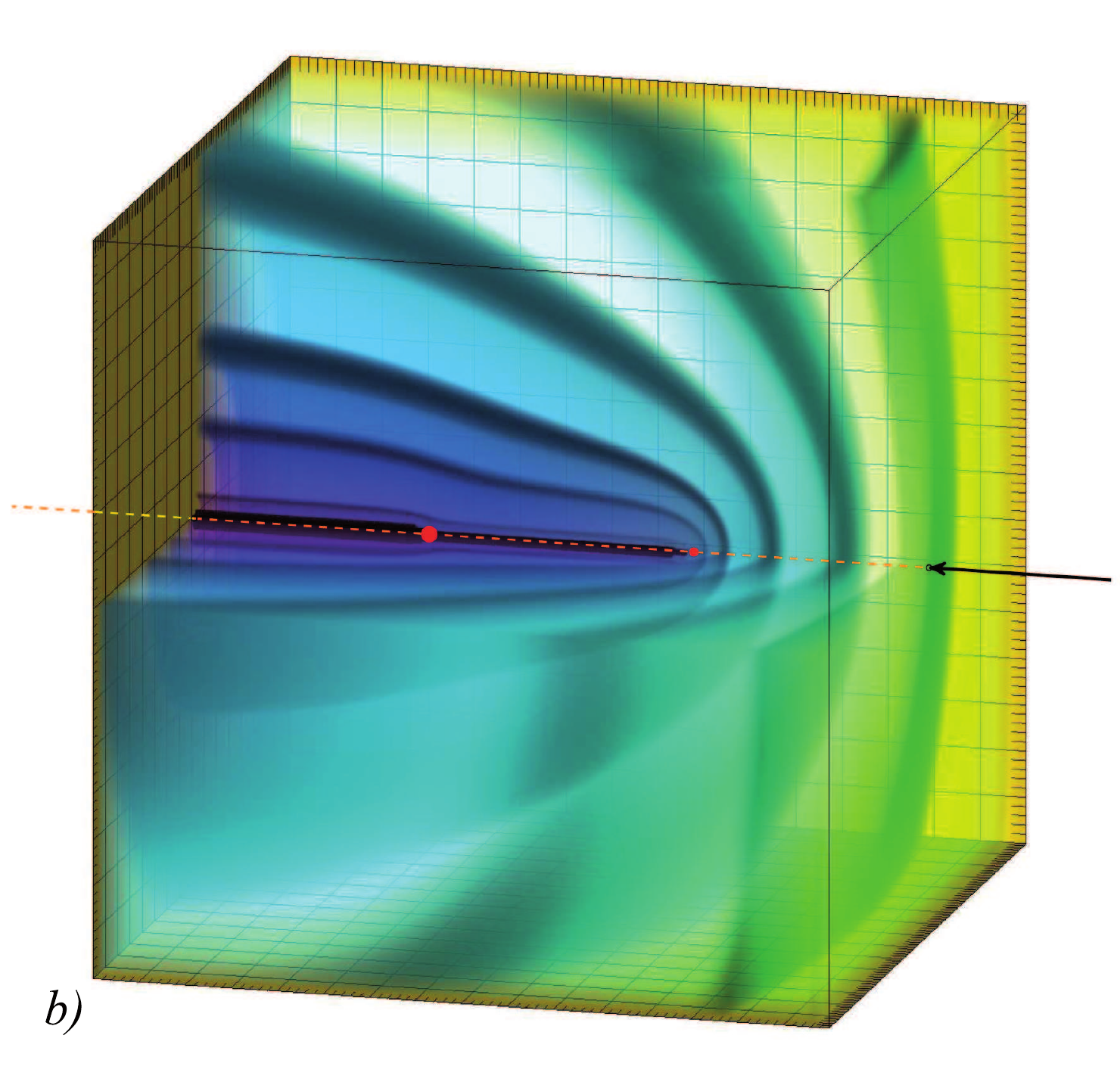}
\end{subfigure}
\begin{subfigure}{490\unitlength}
\includegraphics[trim=0cm 0cm 0cm 0cm, clip=true,width=490\unitlength]{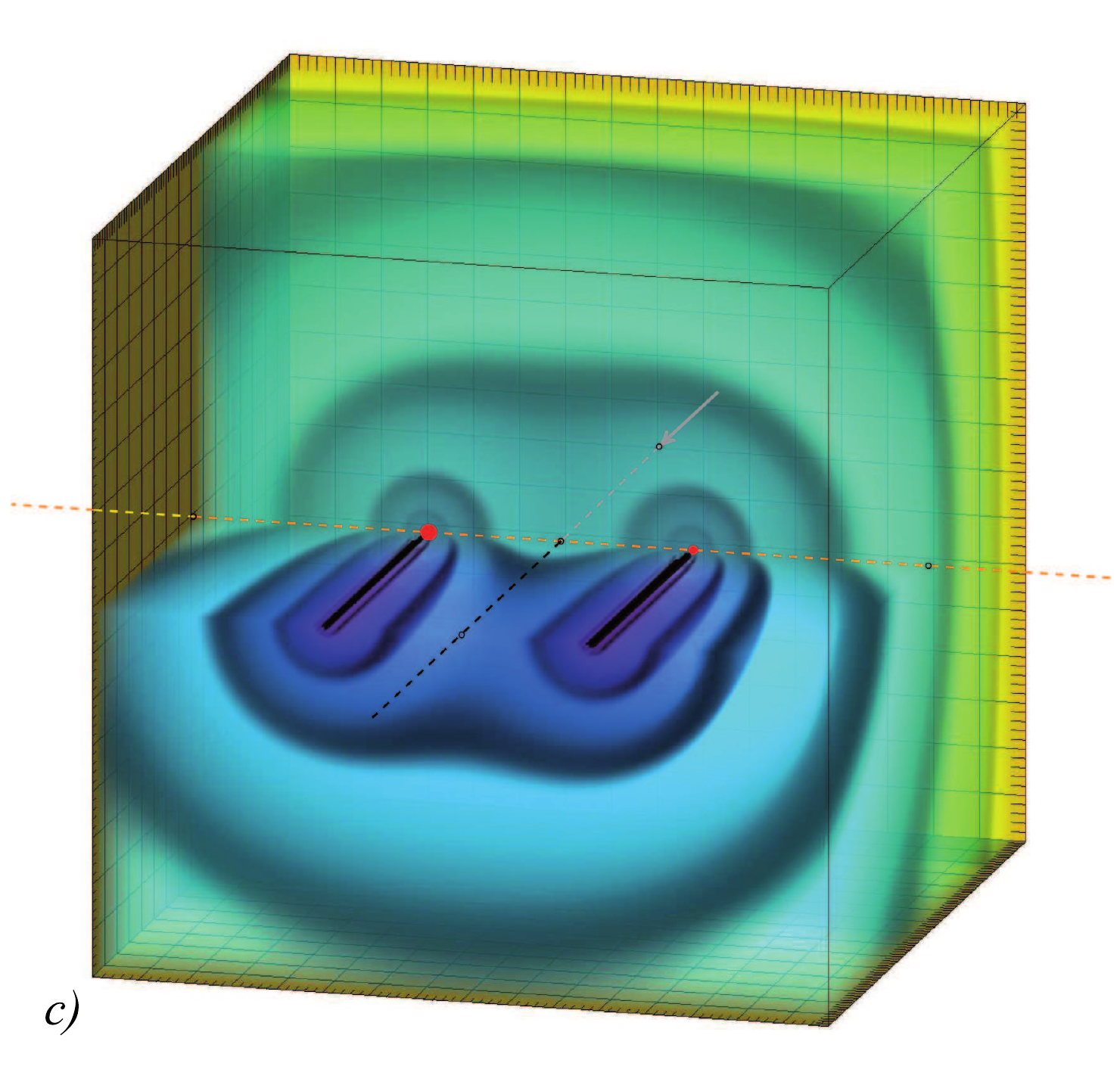}
\end{subfigure}
\begin{subfigure}{490\unitlength}
\includegraphics[trim=0cm 0cm 0cm 0cm, clip=true,width=490\unitlength]{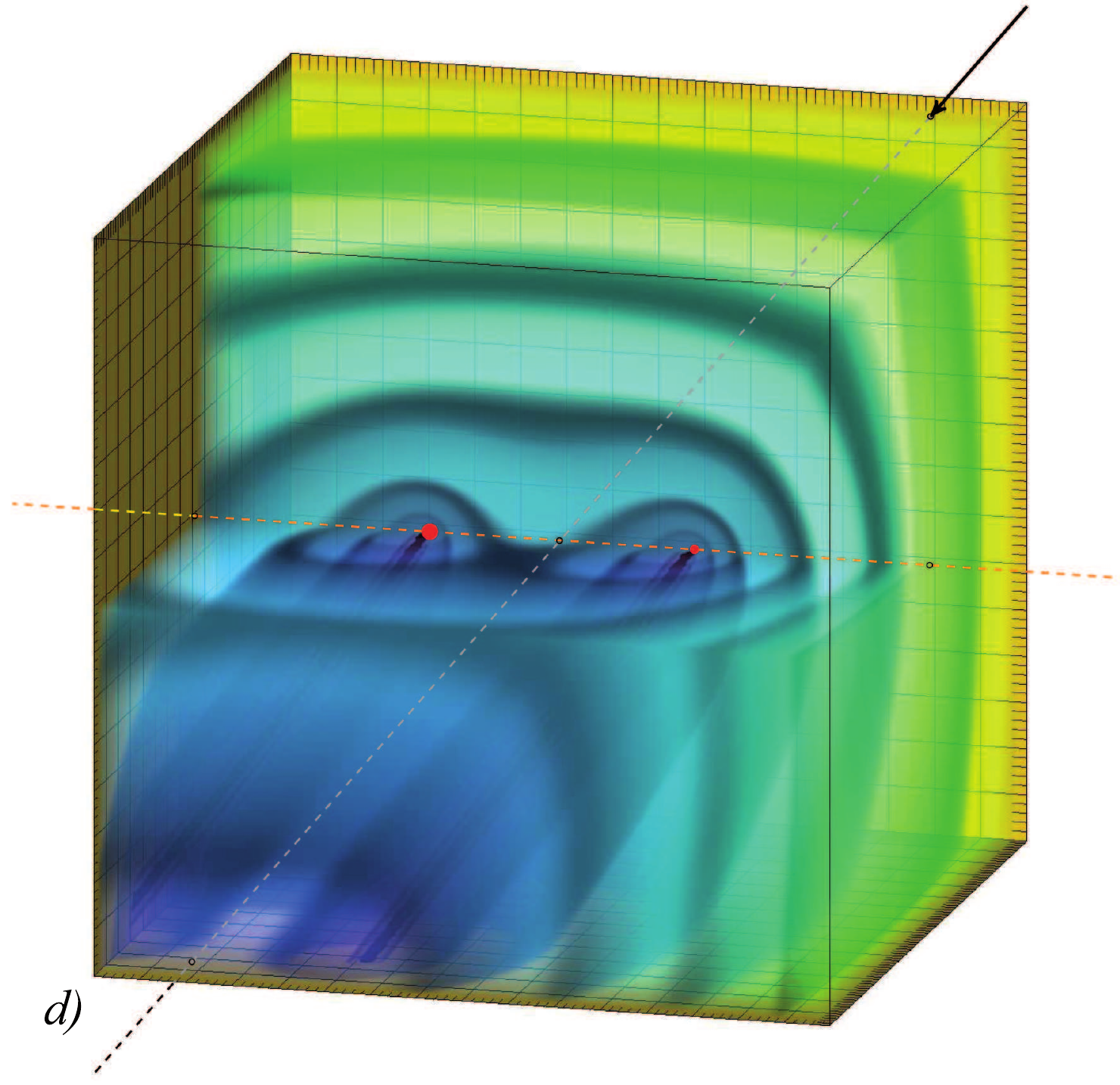}
\end{subfigure}
\begin{subfigure}{990\unitlength}
\centering
\includegraphics[trim=0cm 7cm 0cm 11cm, clip=true,width=990\unitlength]{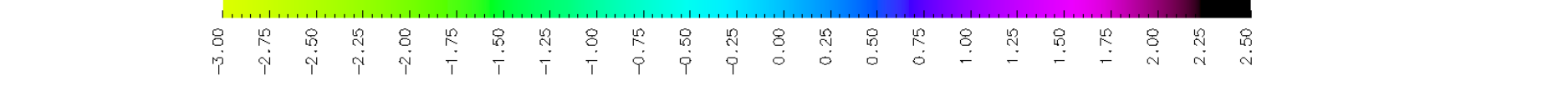}
\end{subfigure}

	\caption{Photon-photon opacity log($\tau$) at $\sim$200 GeV integrated for each position along the line of sight through the computational domain for different orientations. a) $i=$0$^\circ$, $\Phi=$0$^\circ$, b) $i=$90$^\circ$, $\Phi=$0$^\circ$, c) $i=$90$^\circ$, $\Phi=$90$^\circ$, d) $i=$45$^\circ$, $\Phi=$45$^\circ$. The box size is $4000\times4000\times4000$ R$_\odot$. A segment is cut out to allow a better view of the structure. The two stars are indicated in red. The line of centers is indicated in orange and yellow, the line of sight in black and gray.	
\label{tau}		}
\end{figure*}

We illustrate the spatial variation of the optical depth due to photon photon absorption caused by the radiation fields of the two stars in Figure \ref{tau} which depicts 3D illustrations of $\log_{10}(\tau_{\gamma\gamma})$ for an incident photon energy of $\sim$200 GeV. All four images represent case B for different orientations of the binary system relative to the observer as indicated in the figure. They cover the entire computational domain. 

Each point shows the total value of $\tau$ integrated from the point itself along the line of sight toward the observer. Thus, the region that lies precisely behind a star with respect to the line of sight has a very high opacity. Its emitted flux is reduced to very low levels according to $\frac{dF}{dE}\rightarrow\frac{dF}{dE}e^{-\tau}$.

Photon photon absorption is most effective if the line of centers and the line of sight are aligned such that the opacity contributions of both stars add up. In Figure \ref{tau} b) we see that the high-$\tau$ region notably widens at the $x$-position of the B star.

%The higher temperature of the WR star leads to higher luminosity and therefore a higher density of target photons. This results in a significantly larger region of $\tau>1$ surrounding the WR star. The difference is clearly visible for the face-on orientation (a), but also in the edge-on case with perpendicular orientation of line of centers and line of sight (c).

We find that the photon photon absorption in the radiation field of the stars remains inefficient 
below incident photon energies of $\sim$100 GeV. The value of 200 GeV -- chosen for the illustration -- is close to the energy where absorption is at maximum for the given binary system. 

 \subsection{Photon Emission}
Applying the formalism described in Section \ref{gamma}, we can now compute 2D projection maps, SEDs and integrated flux values of the nonthermal photon emission of the three discussed binary systems. Here, we start with the former of those.
\subsubsection{2D Projection Maps}
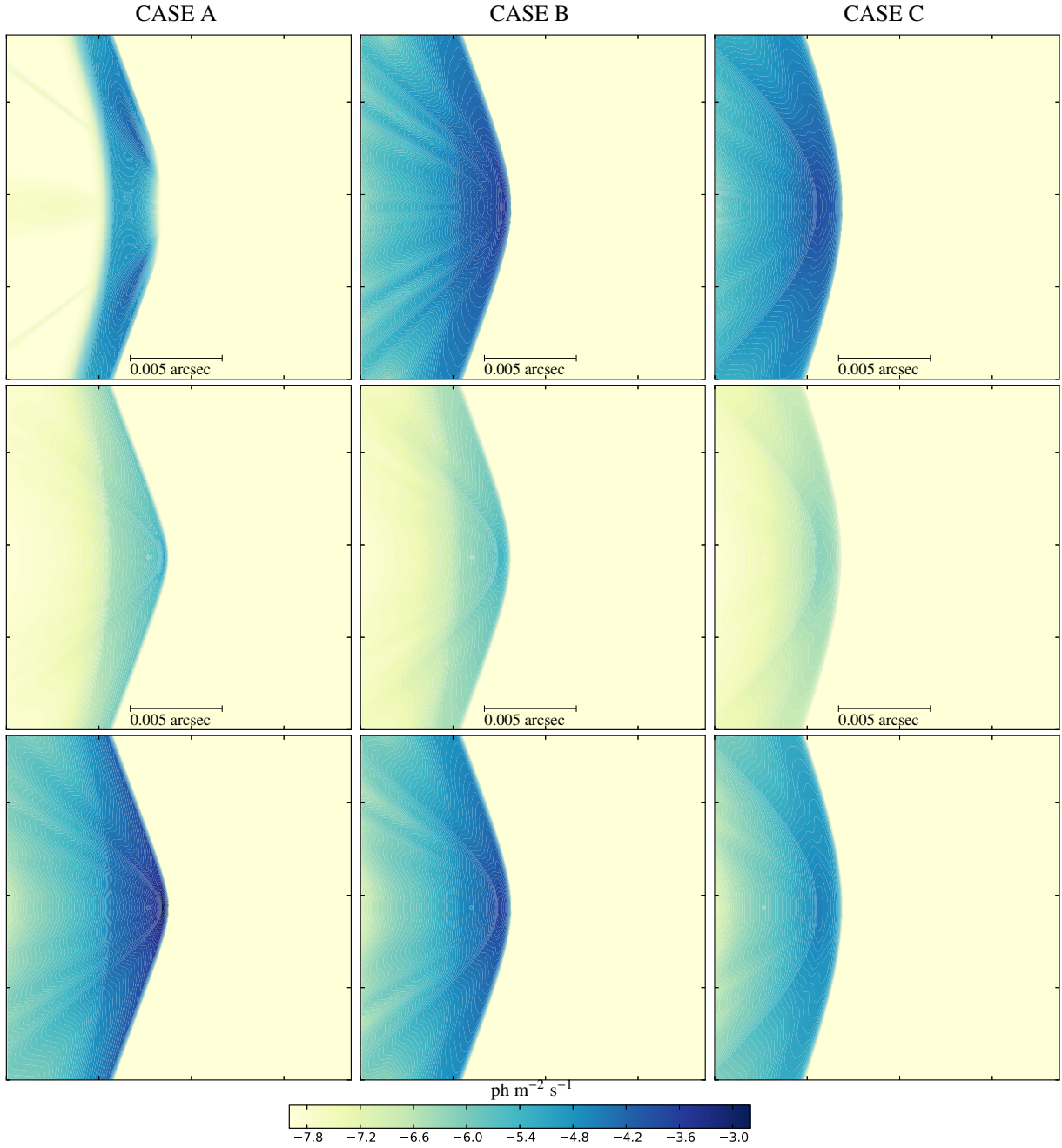
\begin{figure*}
	\setlength{\unitlength}{0.001\textwidth}
		\begin{subfigure}{320\unitlength}
			\begin{picture}(320,320)
					\put(0,0){\includegraphics[trim=2.9cm 1.5cm 5.1cm 1.5cm, clip=true,width=320\unitlength]{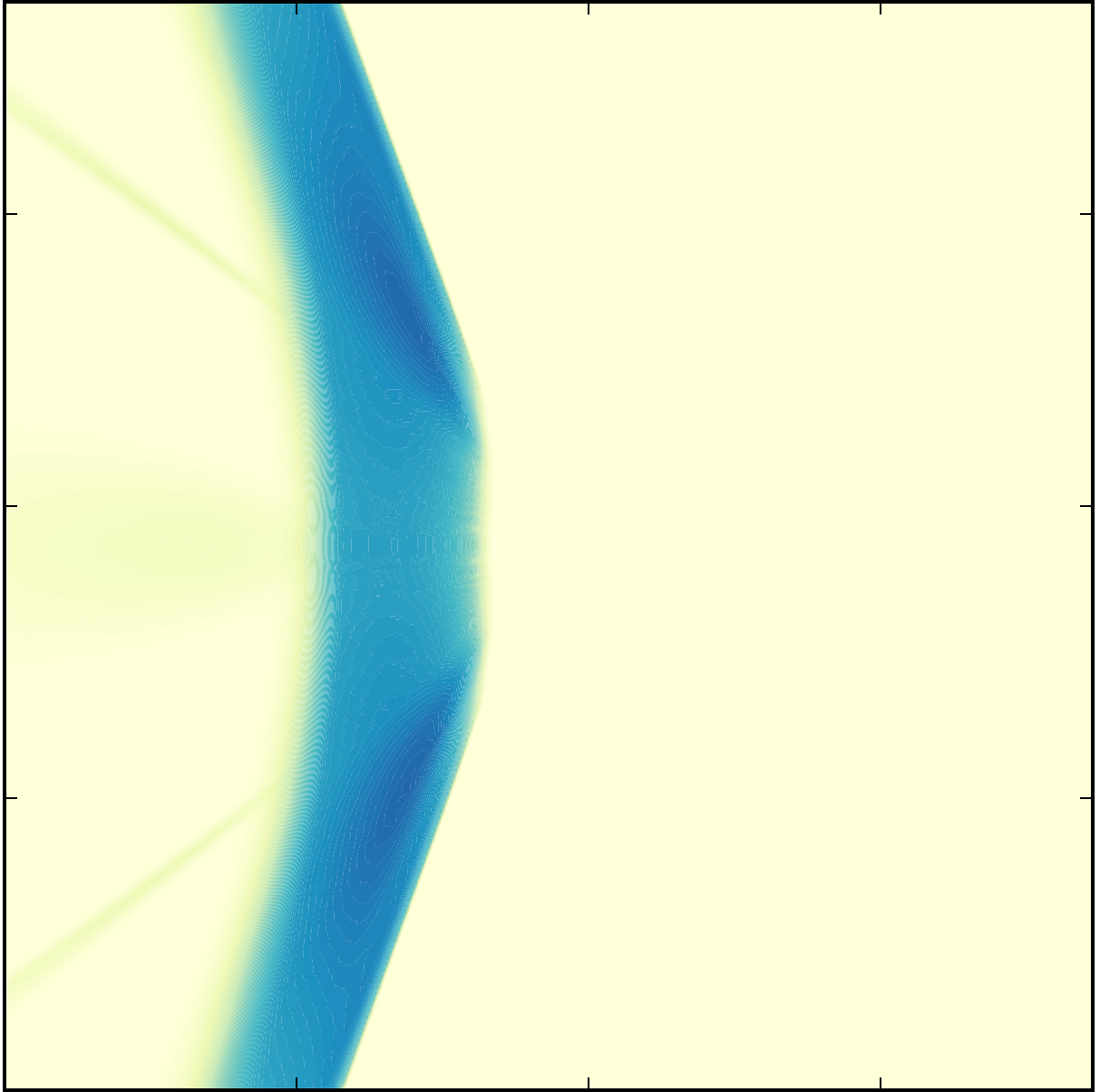}}
				\put(120,330){CASE A}
				\put(115,20){\line(1,0){84.4}}
				\put(115,17){\line(0,1){6}}
				\put(199.4,17){\line(0,1){6}}
				\put(115,5){\scriptsize{0.005 arcsec}}
			\end{picture}
		\end{subfigure}
		\begin{subfigure}{320\unitlength}
			\begin{picture}(320,320)
					\put(0,0){\includegraphics[trim=2.9cm 1.5cm 5.1cm 1.5cm, clip=true,width=320\unitlength ,]{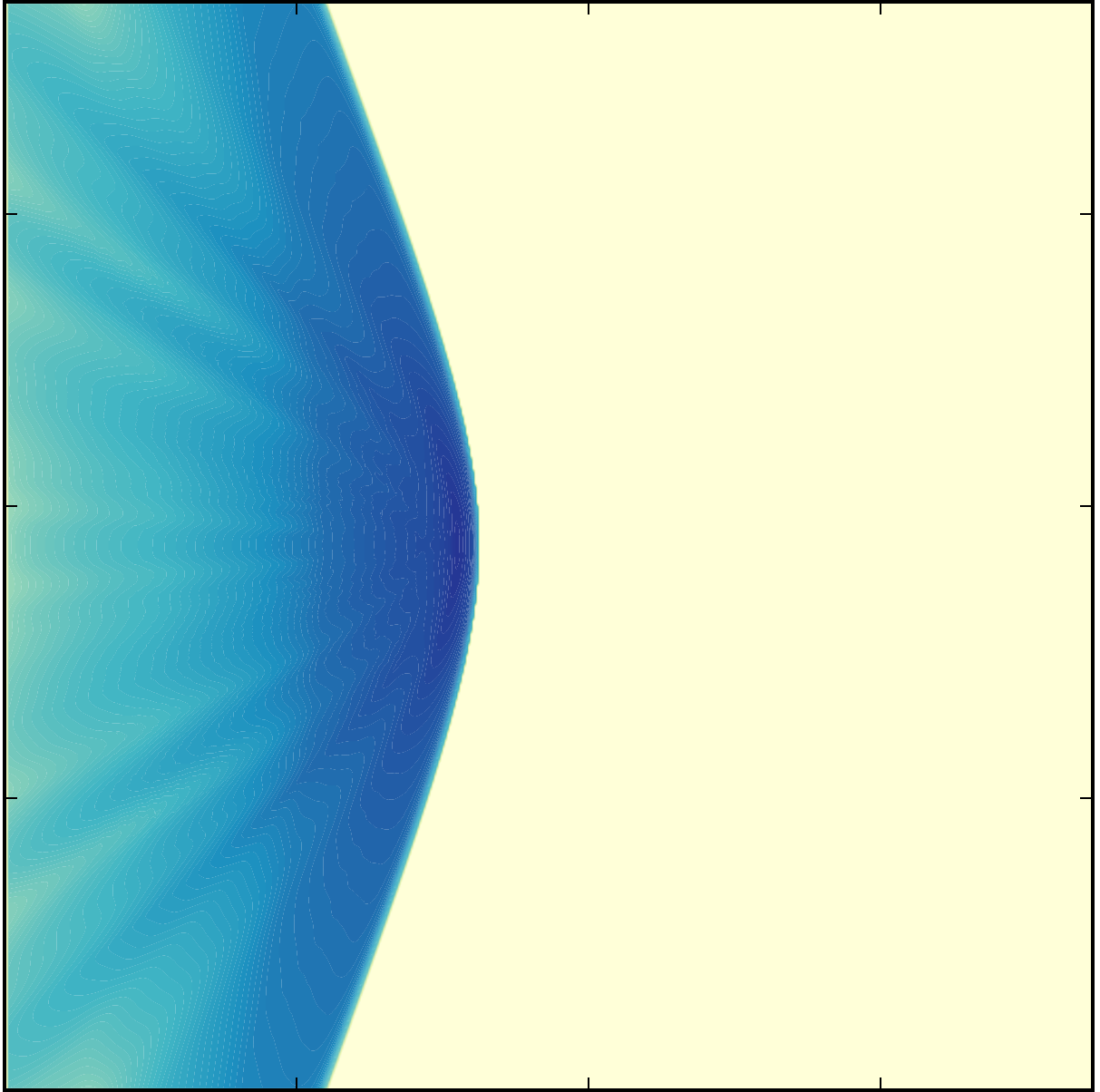}}
				\put(120,330){CASE B}
				\put(115,20){\line(1,0){84.4}}
				\put(115,17){\line(0,1){6}}
				\put(199.4,17){\line(0,1){6}}
				\put(115,5){\scriptsize{0.005 arcsec}}
			\end{picture}
				\end{subfigure}
		\begin{subfigure}{320\unitlength}
			\begin{picture}(320,320)
			\put(0,0){\includegraphics[trim=2.9cm 1.5cm 5.1cm 1.5cm, clip=true,width=320\unitlength]{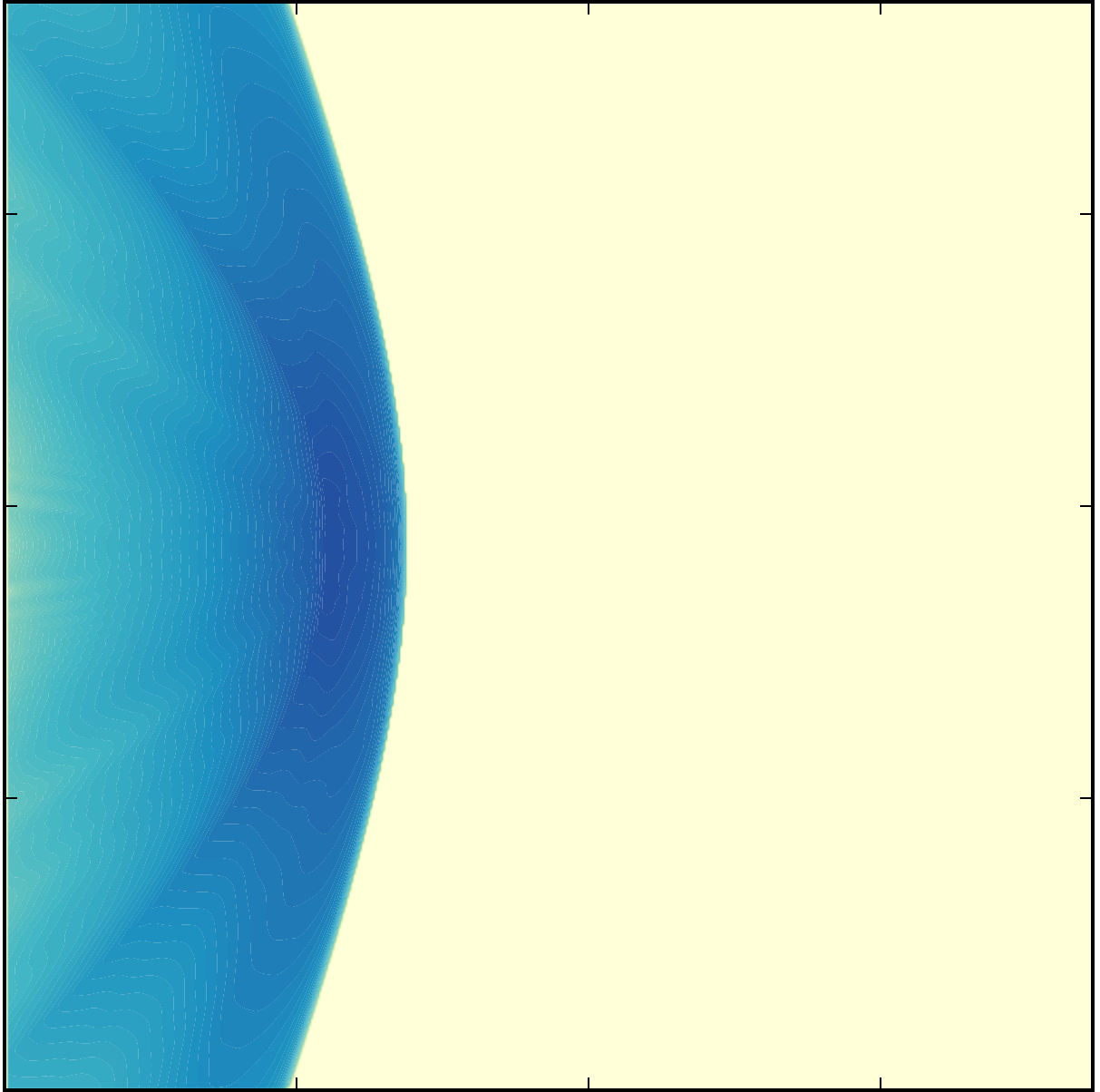}}
				\put(120,330){CASE C}
				\put(115,20){\line(1,0){84.4}}
				\put(115,17){\line(0,1){6}}
				\put(199.4,17){\line(0,1){6}}
				\put(115,5){\scriptsize{0.005 arcsec}}
			\end{picture}
		\end{subfigure}\\
		\begin{subfigure}{320\unitlength}
			\begin{picture}(320,320)
				\put(0,0){\includegraphics[trim=2.9cm 1.5cm 5.1cm 1.5cm, clip=true,width=320\unitlength]{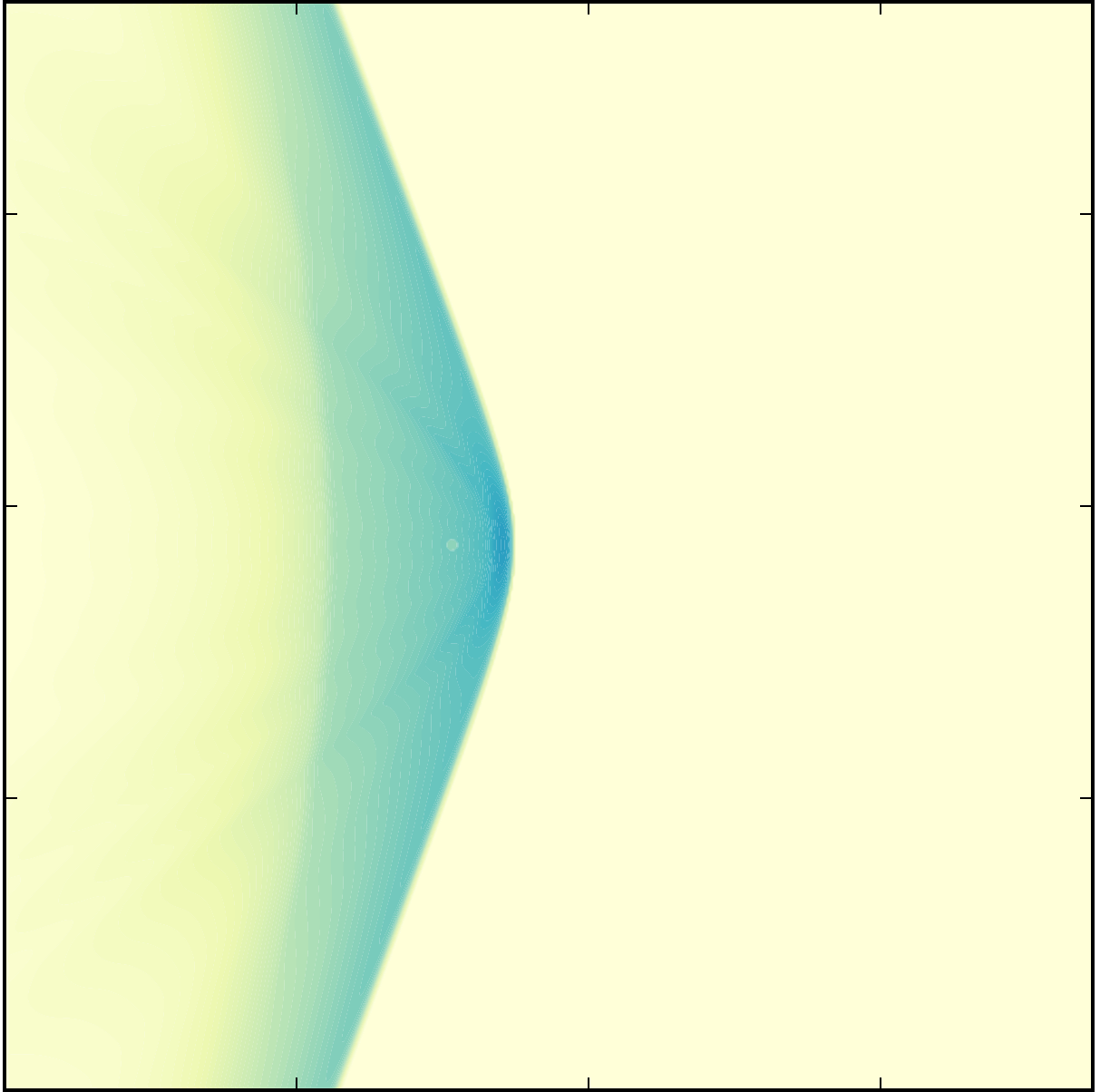}}
				\put(115,20){\line(1,0){84.4}}
				\put(115,17){\line(0,1){6}}
				\put(199.4,17){\line(0,1){6}}
				\put(115,5){\scriptsize{0.005 arcsec}}
			\end{picture}
		\end{subfigure}
		\begin{subfigure}{320\unitlength}
			\begin{picture}(320,320)
				\put(0,0){\includegraphics[trim=2.9cm 1.5cm 5.1cm 1.5cm, clip=true,width=320\unitlength]{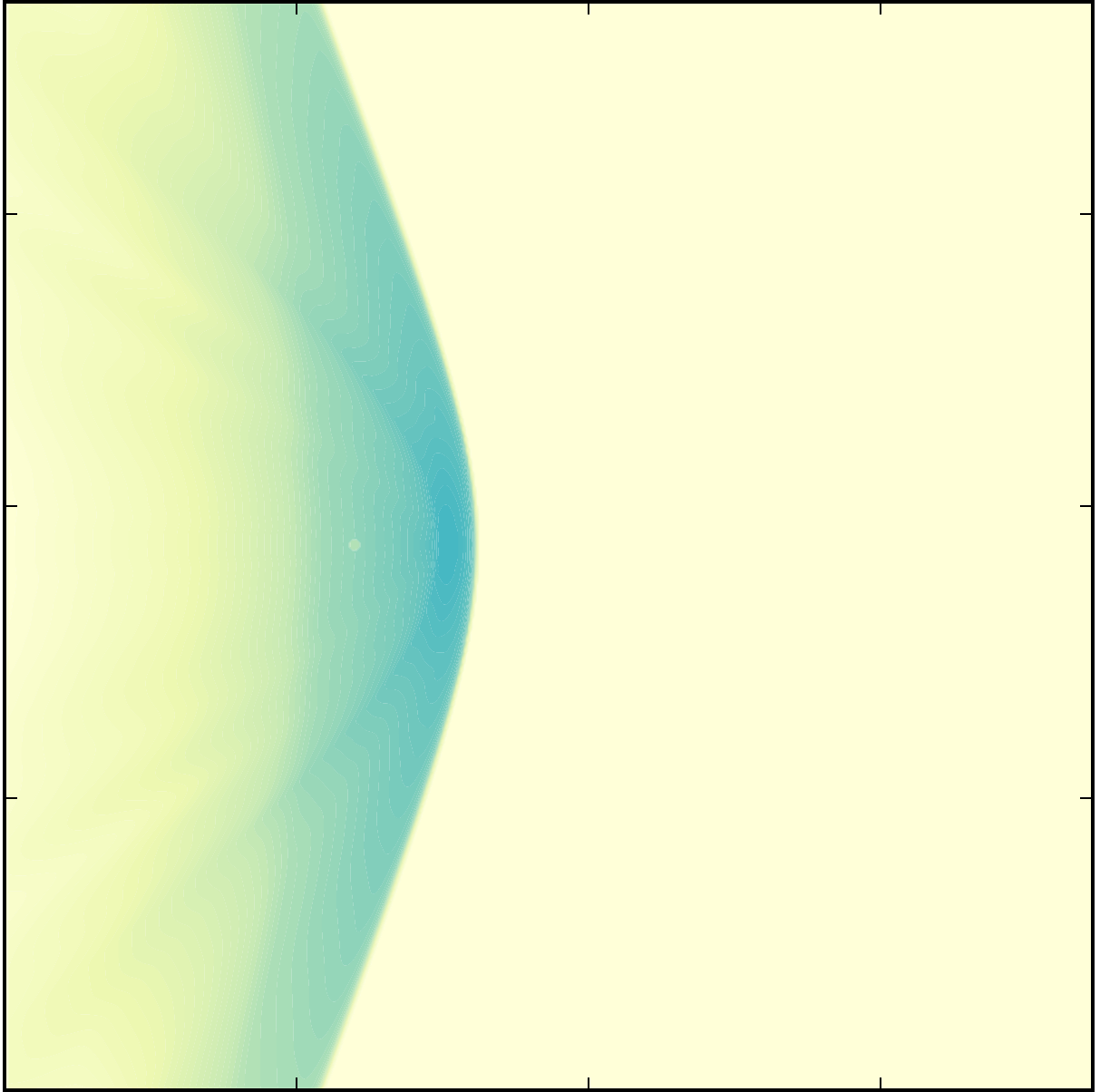}}
				\put(115,20){\line(1,0){84.4}}
				\put(115,17){\line(0,1){6}}
				\put(199.4,17){\line(0,1){6}}
				\put(115,5){\scriptsize{0.005 arcsec}}
			\end{picture}
		\end{subfigure}
		\begin{subfigure}{320\unitlength}	
			\begin{picture}(320,320)
				\put(0,0){\includegraphics[trim=2.9cm 1.5cm 5.1cm 1.5cm, clip=true,width=320\unitlength]{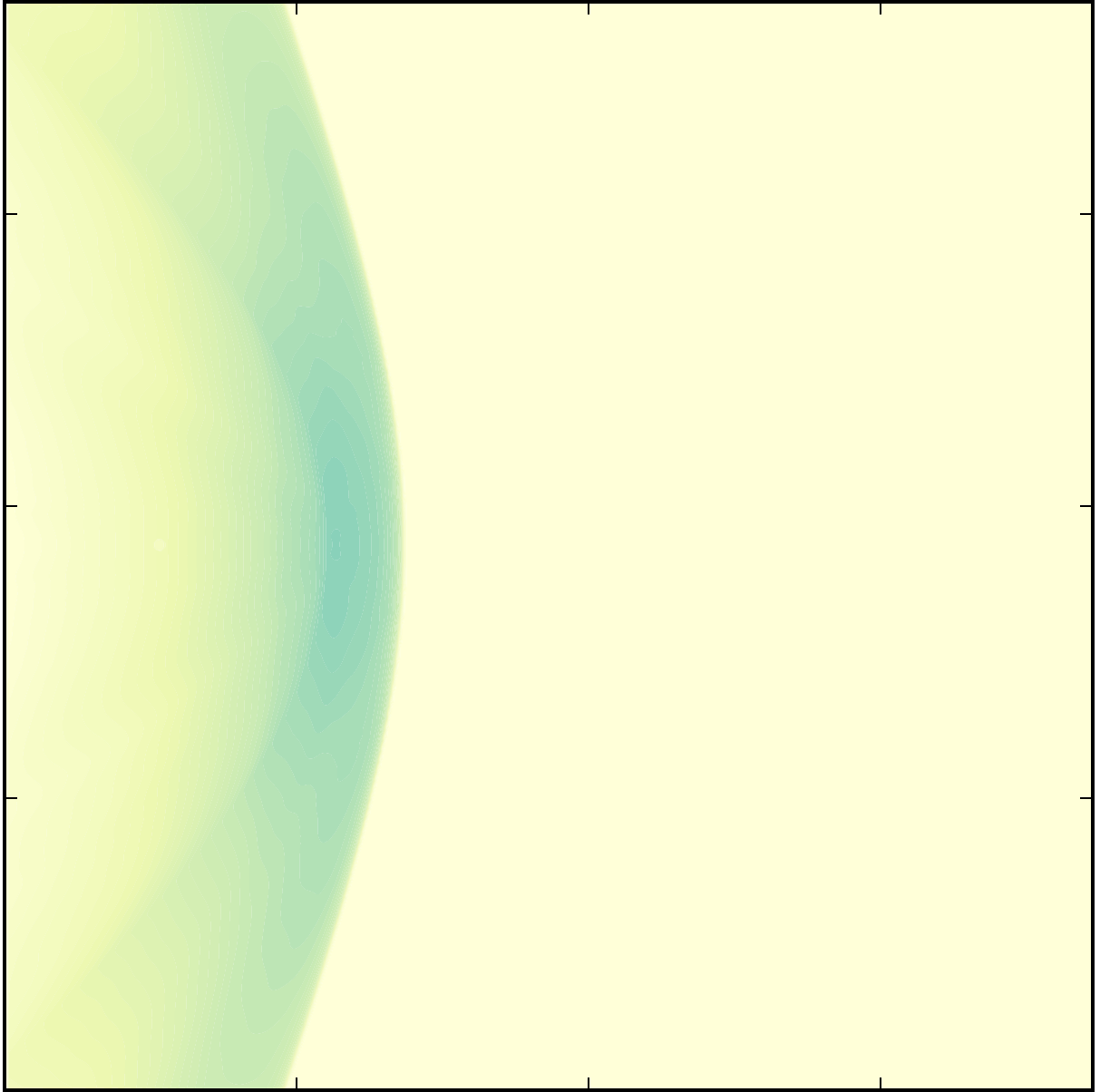}}
				\put(115,20){\line(1,0){84.4}}
				\put(115,17){\line(0,1){6}}
				\put(199.4,17){\line(0,1){6}}
				\put(115,5){\scriptsize{0.005 arcsec}}
			\end{picture}
		\end{subfigure}\\
		\begin{subfigure}{320\unitlength}
			\begin{picture}(320,320)
				\put(0,0){\includegraphics[trim=2.9cm 1.5cm 5.1cm 1.5cm, clip=true,width=320\unitlength]{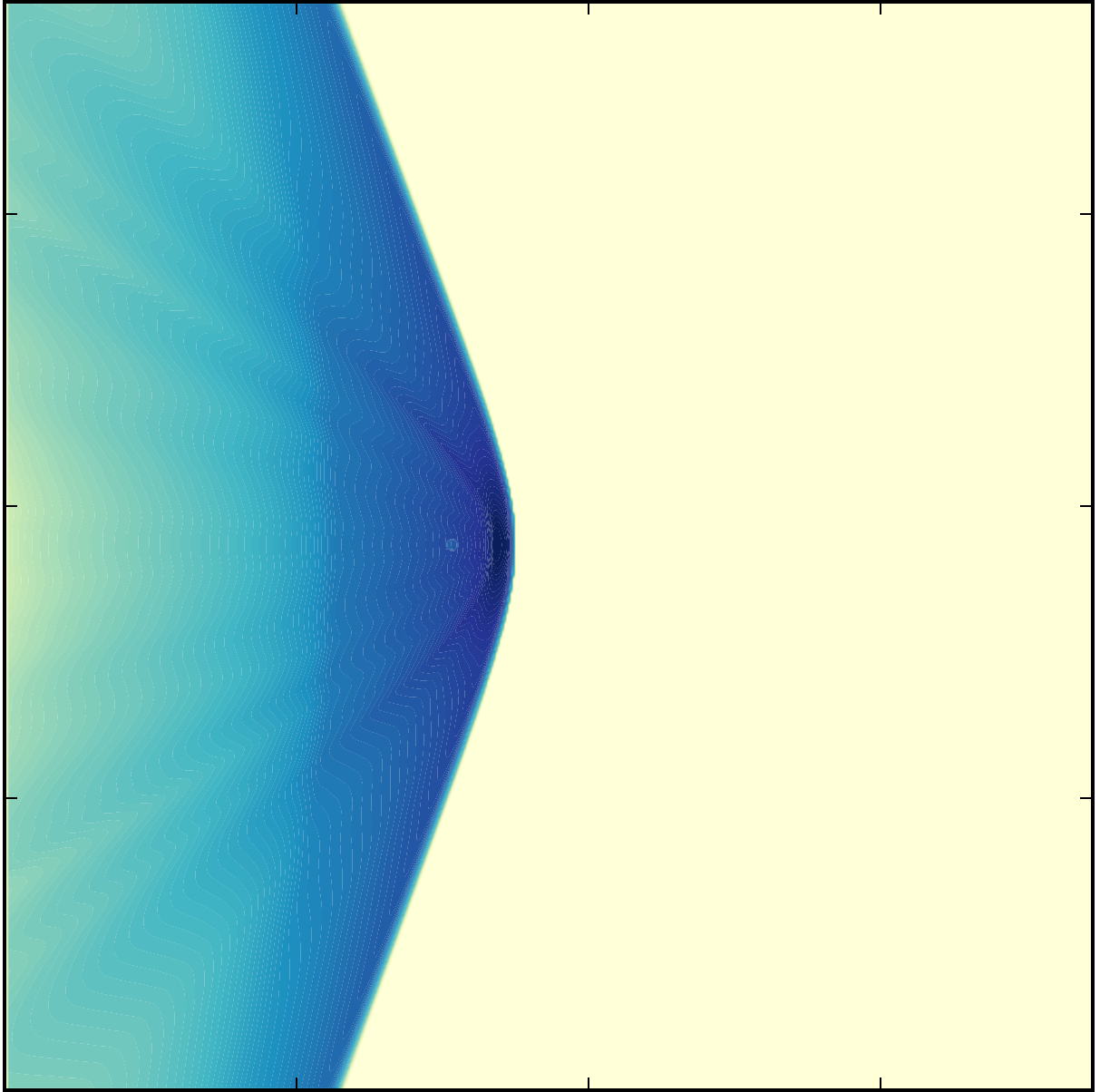}}
			%	\put(115,20){\line(1,0){84.4}}
				%\put(115,17){\line(0,1){6}}
				%\put(199.4,17){\line(0,1){6}}
				%\put(115,5){\scriptsize{0.005 arcsec}}
			\end{picture}
		\end{subfigure}
		\begin{subfigure}{320\unitlength}
			\begin{picture}(320,320)
				\put(0,0){\includegraphics[trim=2.9cm 1.5cm 5.1cm 1.5cm, clip=true,width=320\unitlength]{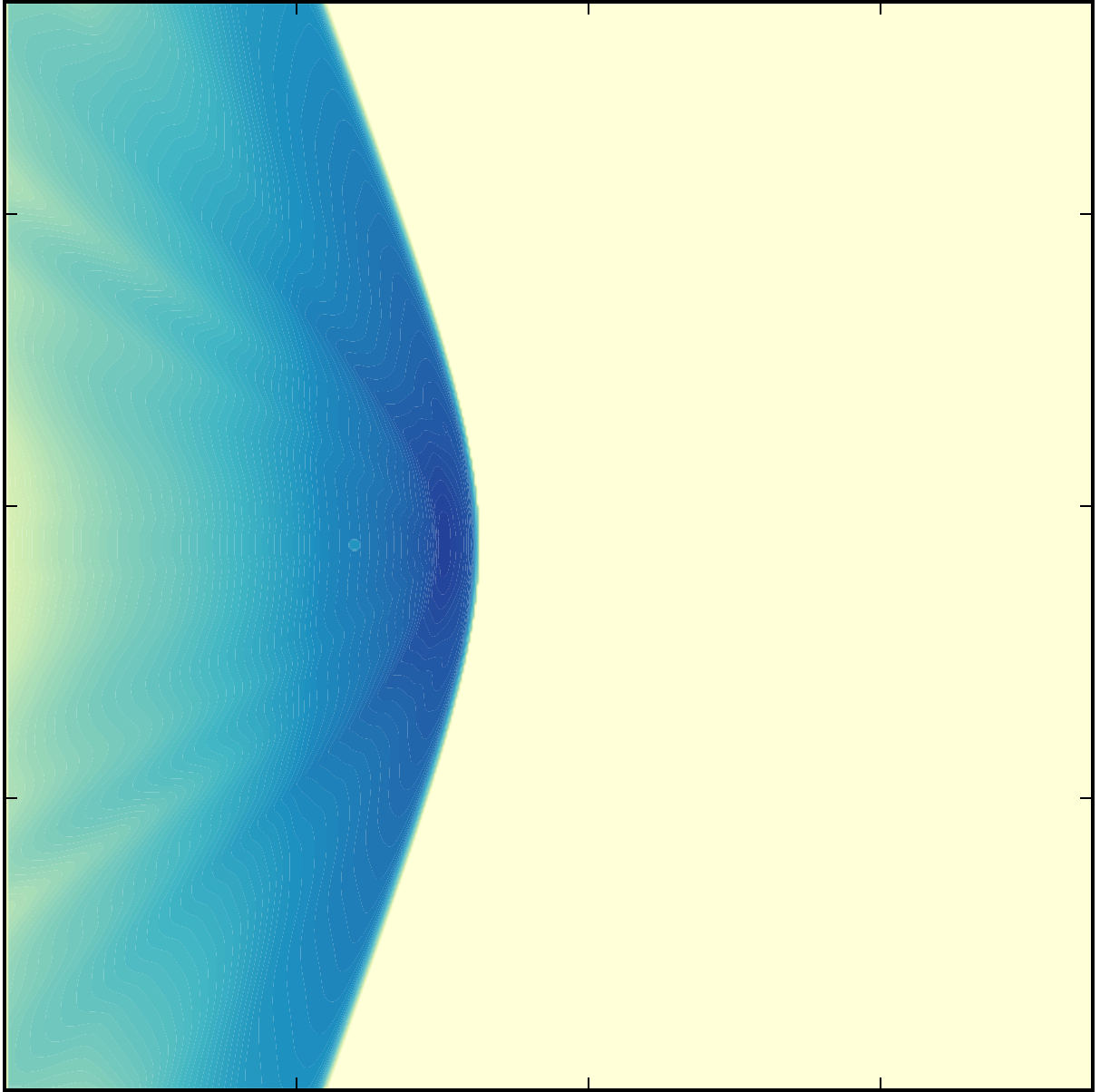}}
				%\put(115,20){\line(1,0){84.4}}
				%\put(115,17){\line(0,1){6}}
				%\put(199.4,17){\line(0,1){6}}
				%\put(115,5){\scriptsize{0.005 arcsec}}
			\end{picture}
		\end{subfigure}
		\begin{subfigure}{320\unitlength}	
			\begin{picture}(320,320)
				\put(0,0){\includegraphics[trim=2.9cm 1.5cm 5.1cm 1.5cm, clip=true,width=320\unitlength]{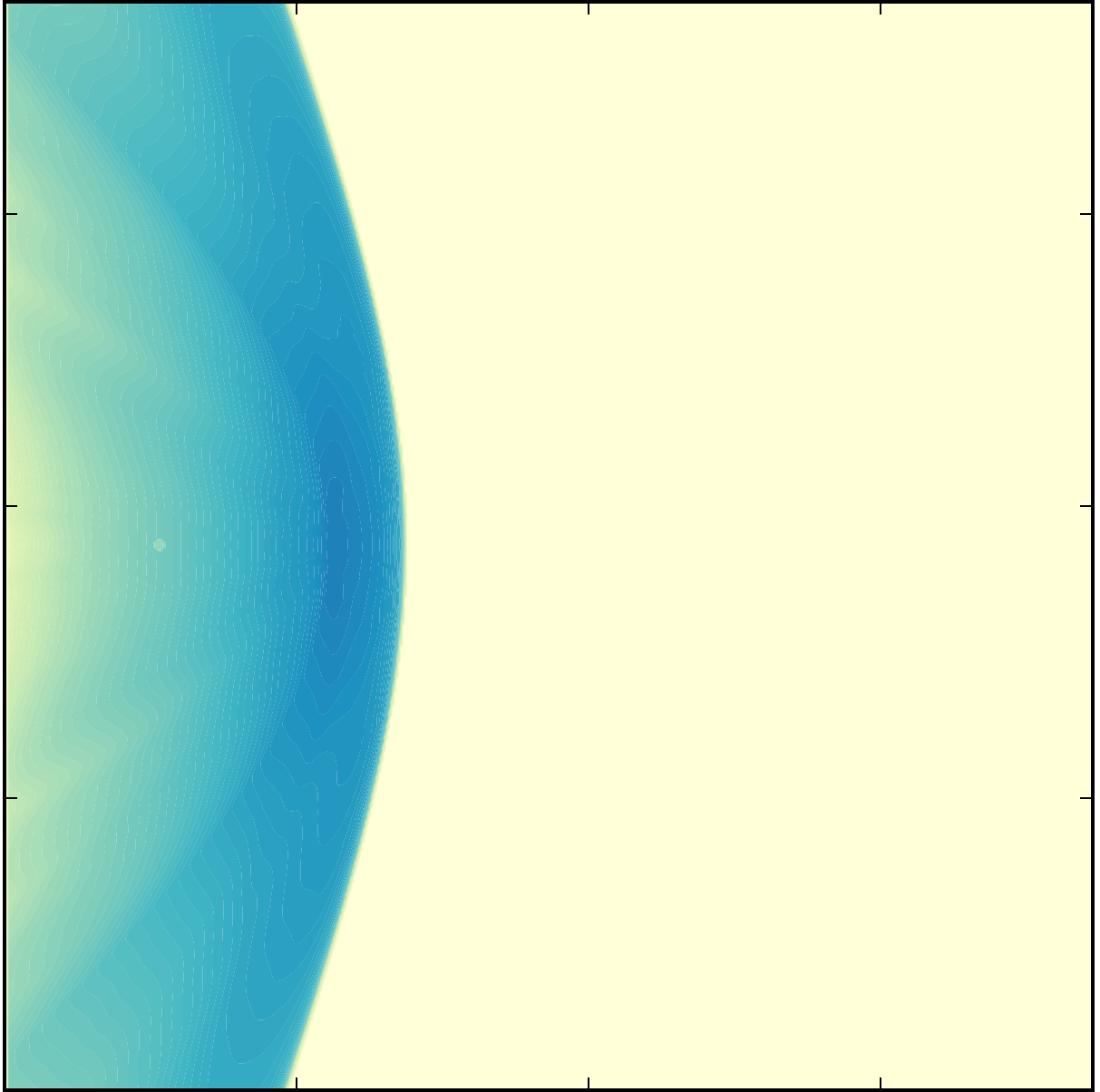}}
				%\put(115,20){\line(1,0){84.4}}
				%\put(115,17){\line(0,1){6}}
				%\put(199.4,17){\line(0,1){6}}
				%\put(115,5){\scriptsize{0.005 arcsec}}
			\end{picture}
		\end{subfigure}\\
		\begin{center}
		\begin{subfigure}{500\unitlength}	
			\begin{picture}(500,0)
				\put(0,-20){\includegraphics[trim=2cm 2cm 1cm 24cm, clip=true,width=500\unitlength]{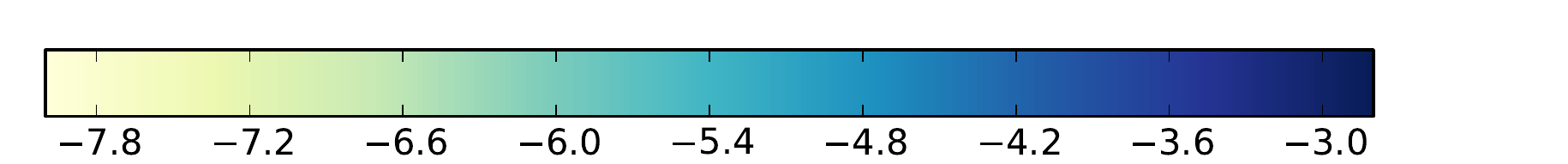}}
				\put(200,25){\footnotesize{ph m$^{-2}$ s$^{-1}$}}
			\end{picture}
		\end{subfigure}
		\end{center}	
\caption{Photon flux above 100 MeV at 1 kpc distance with face-on orientation $i=0^\circ$, $\Phi=0^\circ$ for IC-emission (first row), bremsstrahlung (second row) and neutral pion decay (third row). 
\label{em_0_0}}
\end{figure*}

Figure \ref{em_0_0} illustrates how the integrated flux above 100 MeV from the individual nonthermal emission channels appears in the face-on configuration ($i=$0$^\circ$,$\Phi=$0$^\circ$) for an ideal observatory at 1 kpc distance with infinite angular resolution and without absorption in the ISM. Several features noted previously in Sections \ref{hydrod} and \ref{partd} concerning the wind structure and the high-energy particle distribution have an impact on the resulting photon emission maps.
The dominant emission channel shifts from IC-scattering to neutral pion decay as the stellar separation decreases. This is because of the lack of high-energy electrons when the WCR is close to the stars (case A). 

\begin{figure*}
	\setlength{\unitlength}{0.001\textwidth}
		\begin{subfigure}{320\unitlength}
			\begin{picture}(320,320)
					\put(0,0){\includegraphics[trim=2.9cm 1.5cm 5.1cm 1.5cm, clip=true,width=320\unitlength]{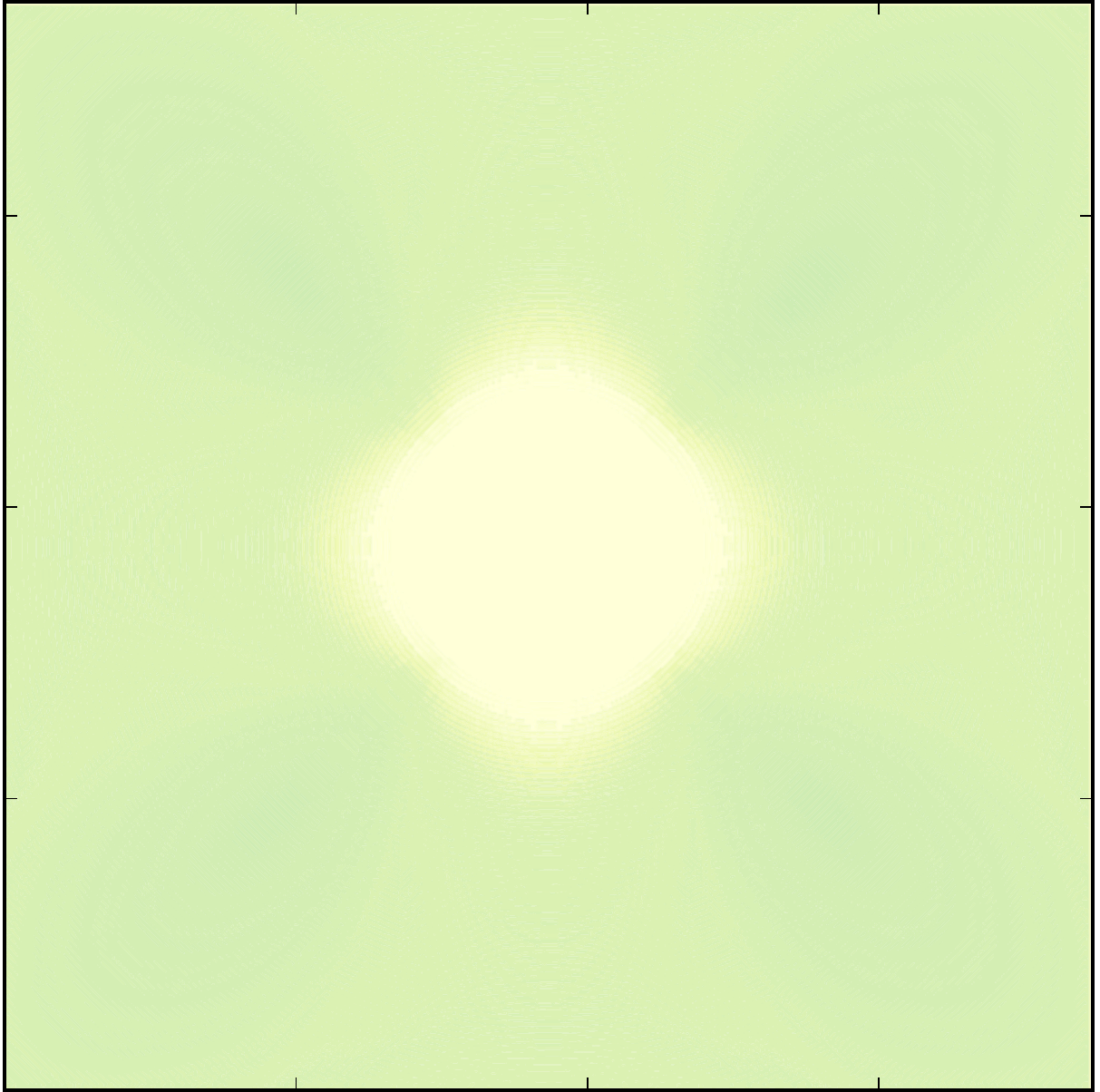}}
				\put(120,330){CASE A}
				\put(115,20){\line(1,0){84.4}}
				\put(115,17){\line(0,1){6}}
				\put(199.4,17){\line(0,1){6}}
				\put(115,5){\scriptsize{0.005 arcsec}}
			\end{picture}
		\end{subfigure}
		\begin{subfigure}{320\unitlength}
			\begin{picture}(320,320)
					\put(0,0){\includegraphics[trim=2.9cm 1.5cm 5.1cm 1.5cm, clip=true,width=320\unitlength ,]{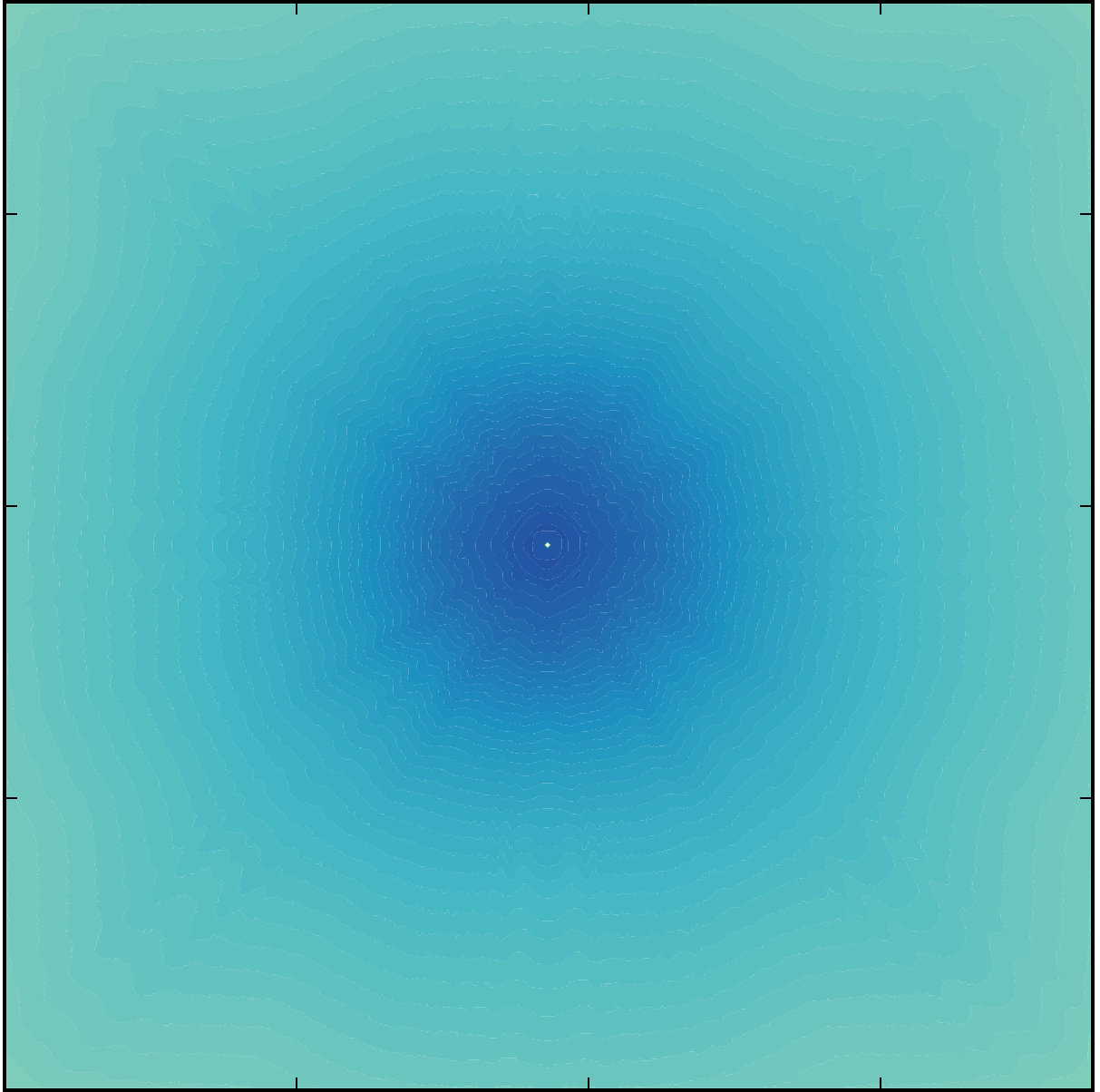}}
				\put(120,330){CASE B}
				\put(115,20){\line(1,0){84.4}}
				\put(115,17){\line(0,1){6}}
				\put(199.4,17){\line(0,1){6}}
				\put(115,5){\scriptsize{0.005 arcsec}}
			\end{picture}
				\end{subfigure}
		\begin{subfigure}{320\unitlength}
			\begin{picture}(320,320)
			\put(0,0){\includegraphics[trim=2.9cm 1.5cm 5.1cm 1.5cm, clip=true,width=320\unitlength]{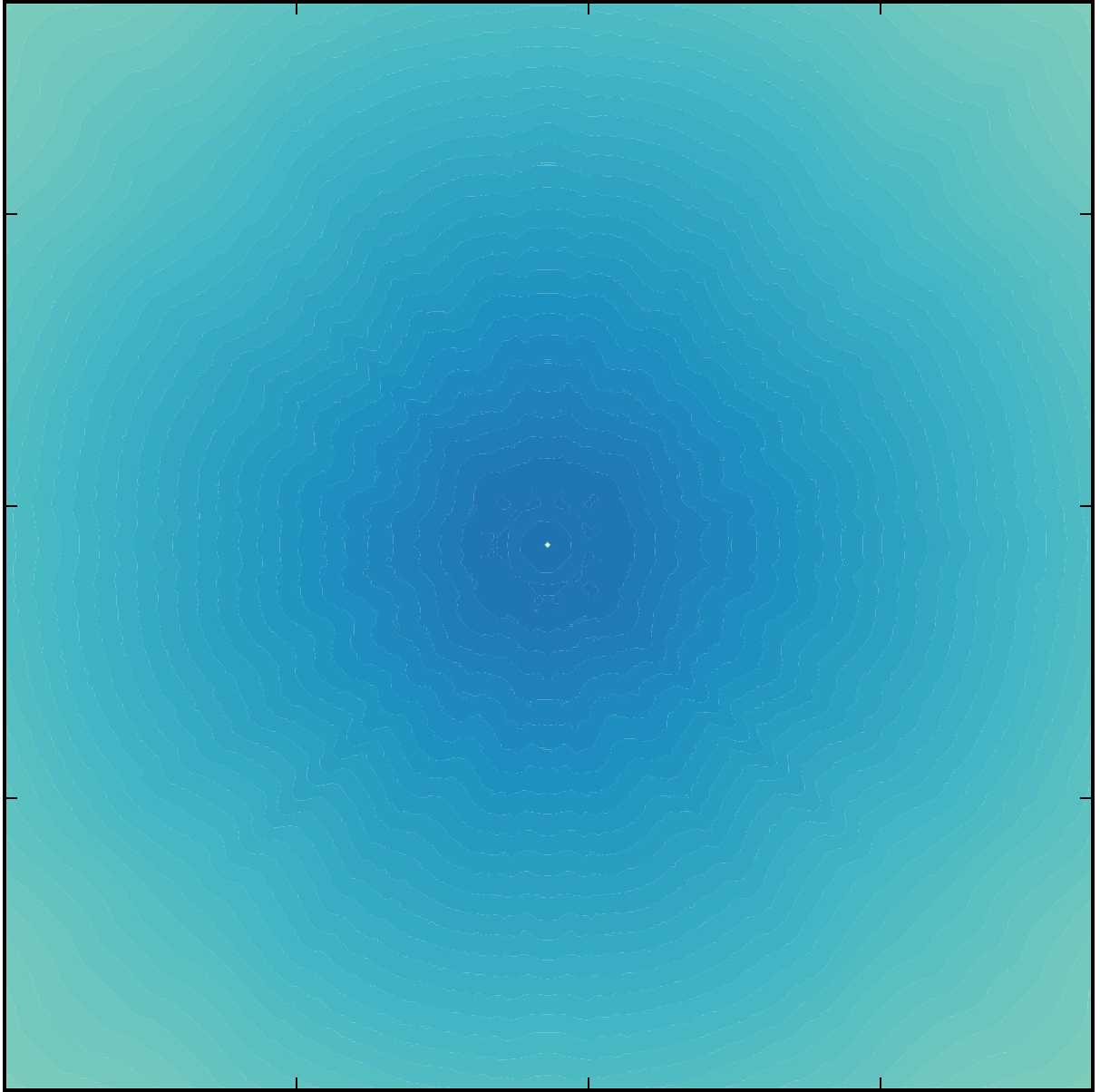}}
				\put(120,330){CASE C}
				\put(115,20){\line(1,0){84.4}}
				\put(115,17){\line(0,1){6}}
				\put(199.4,17){\line(0,1){6}}
				\put(115,5){\scriptsize{0.005 arcsec}}
			\end{picture}
		\end{subfigure}\\
		\begin{subfigure}{320\unitlength}
			\begin{picture}(320,320)
				\put(0,0){\includegraphics[trim=2.9cm 1.5cm 5.1cm 1.5cm, clip=true,width=320\unitlength]{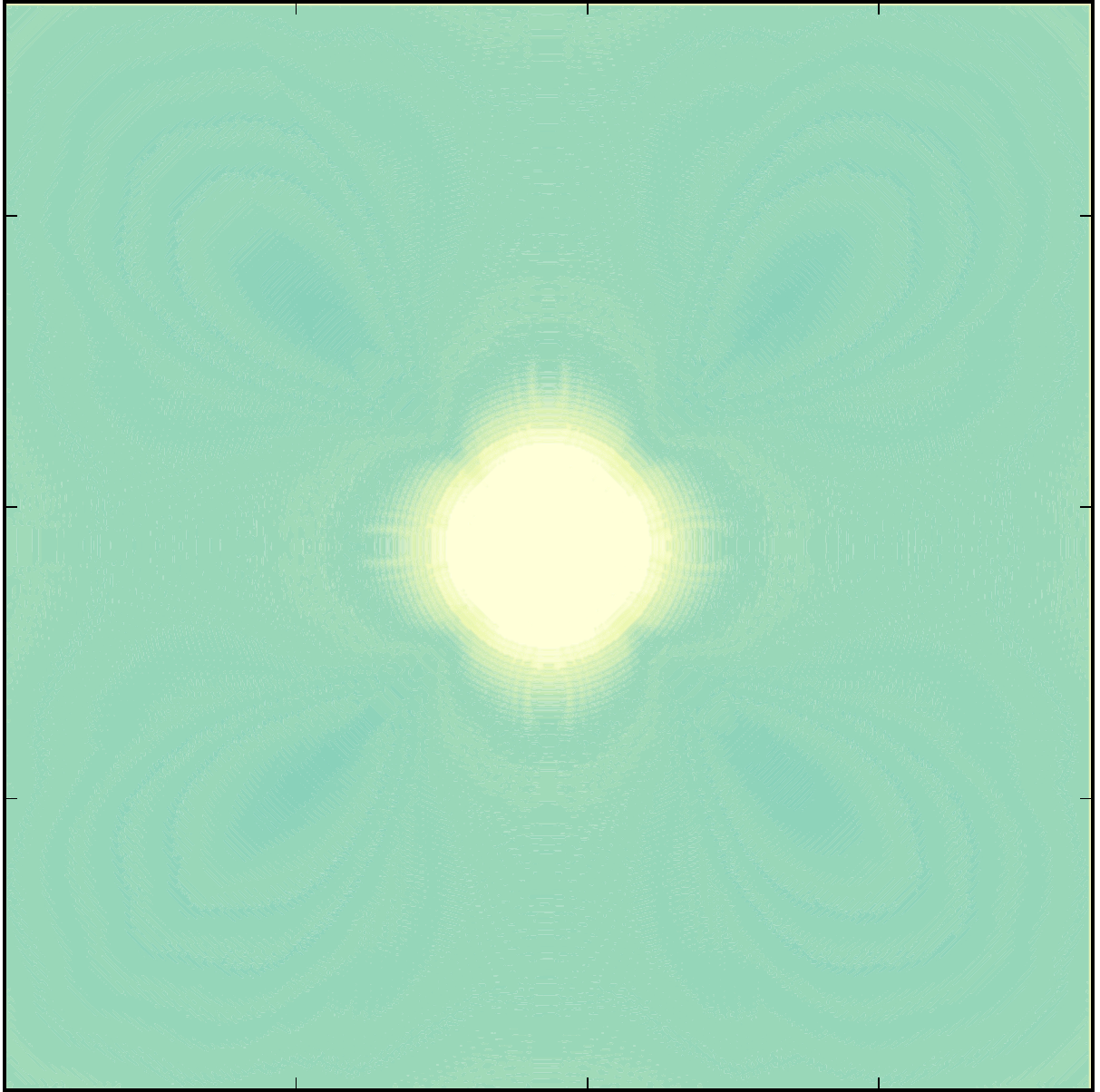}}
				\put(115,20){\line(1,0){84.4}}
				\put(115,17){\line(0,1){6}}
				\put(199.4,17){\line(0,1){6}}
				\put(115,5){\scriptsize{0.005 arcsec}}
			\end{picture}
		\end{subfigure}
		\begin{subfigure}{320\unitlength}
			\begin{picture}(320,320)
				\put(0,0){\includegraphics[trim=2.9cm 1.5cm 5.1cm 1.5cm, clip=true,width=320\unitlength]{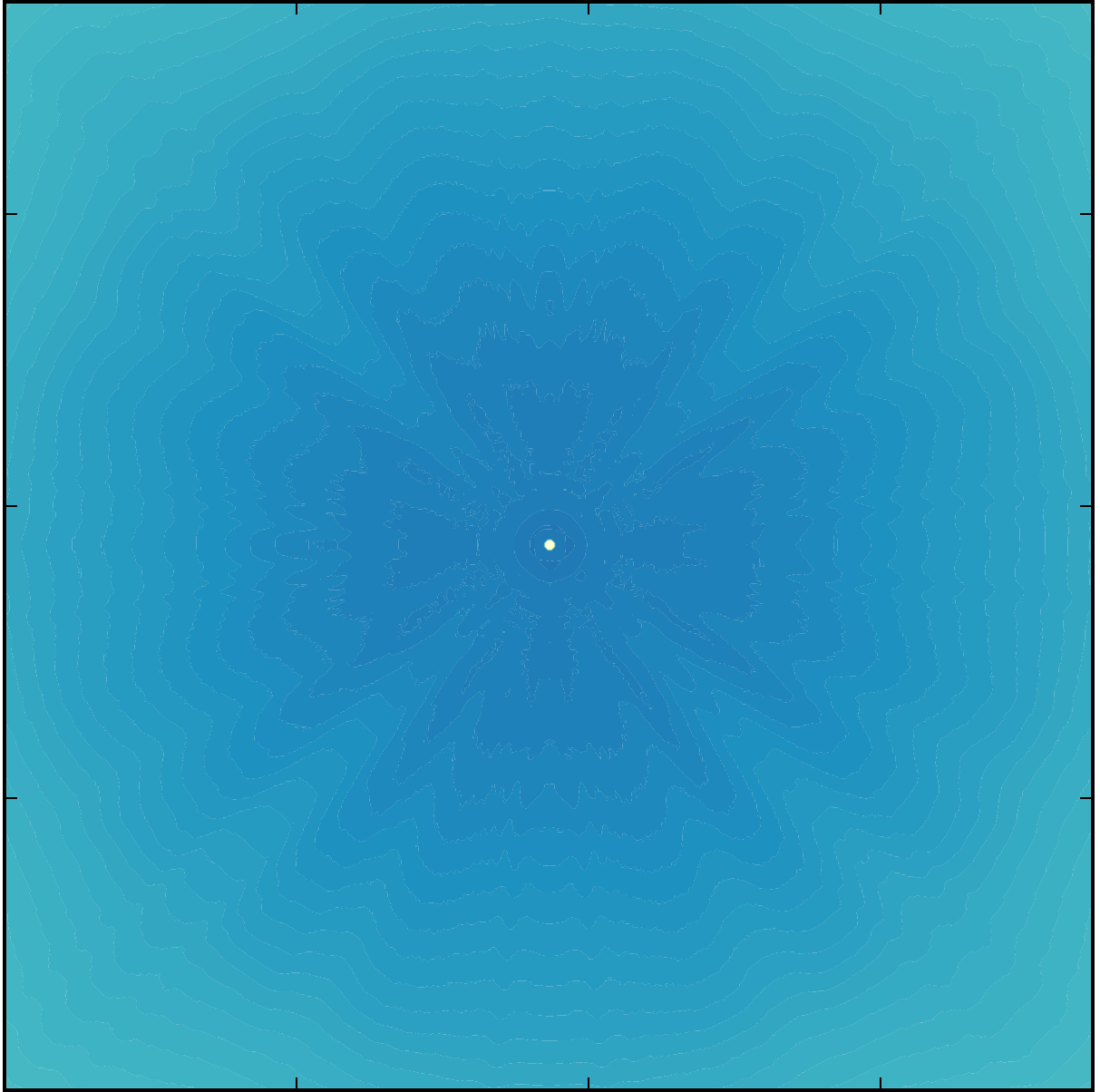}}
				\put(115,20){\line(1,0){84.4}}
				\put(115,17){\line(0,1){6}}
				\put(199.4,17){\line(0,1){6}}
				\put(115,5){\scriptsize{0.005 arcsec}}
			\end{picture}
		\end{subfigure}
		\begin{subfigure}{320\unitlength}	
			\begin{picture}(320,320)
				\put(0,0){\includegraphics[trim=2.9cm 1.5cm 5.1cm 1.5cm, clip=true,width=320\unitlength]{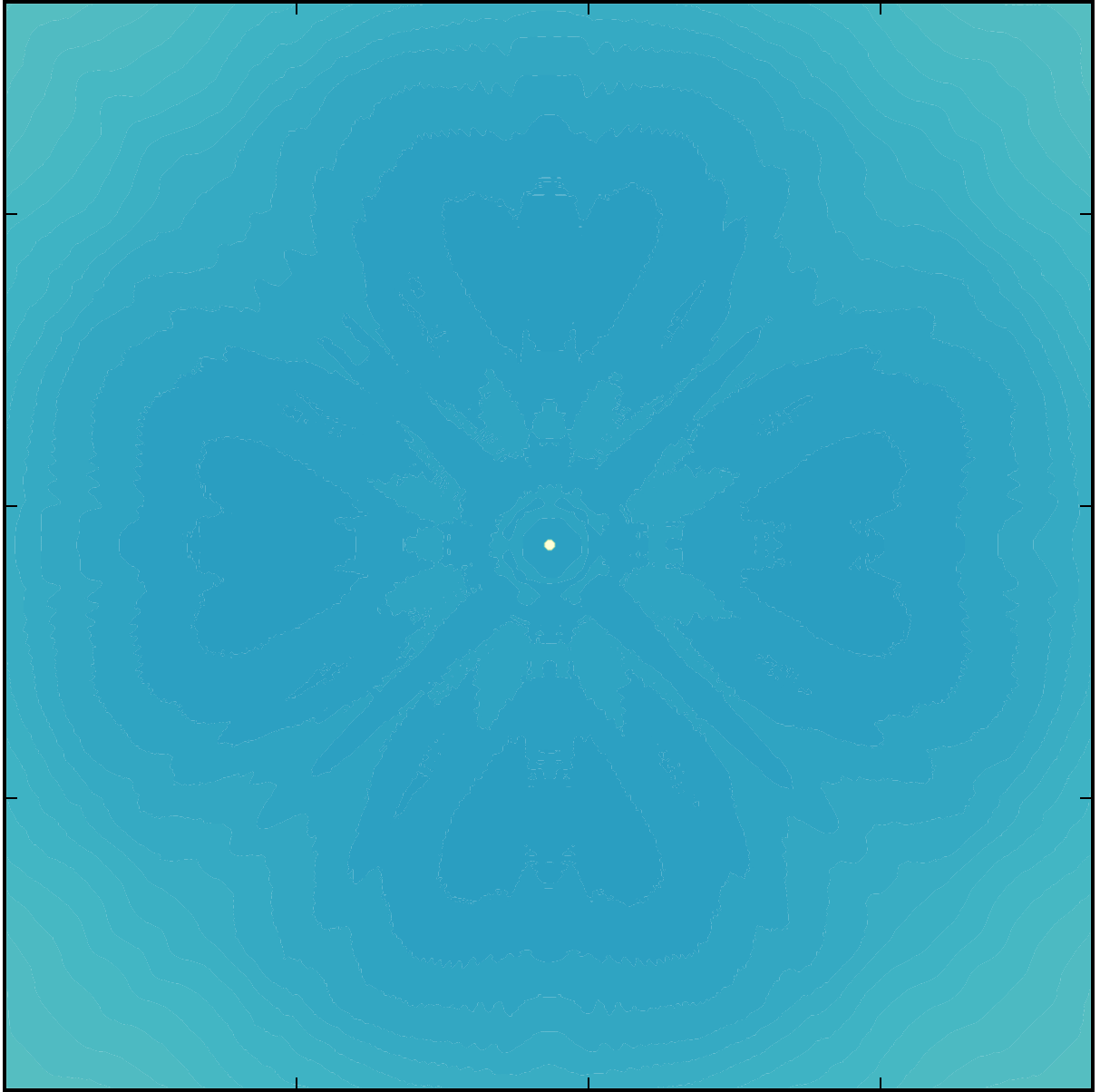}}
				\put(115,20){\line(1,0){84.4}}
				\put(115,17){\line(0,1){6}}
				\put(199.4,17){\line(0,1){6}}
				\put(115,5){\scriptsize{0.005 arcsec}}
			\end{picture}
		\end{subfigure}\\
		\begin{subfigure}{320\unitlength}
			\begin{picture}(320,358)
				\put(0,0){\includegraphics[trim=4.2cm 1.5cm 5.1cm 1.5cm, clip=true,width=320\unitlength]{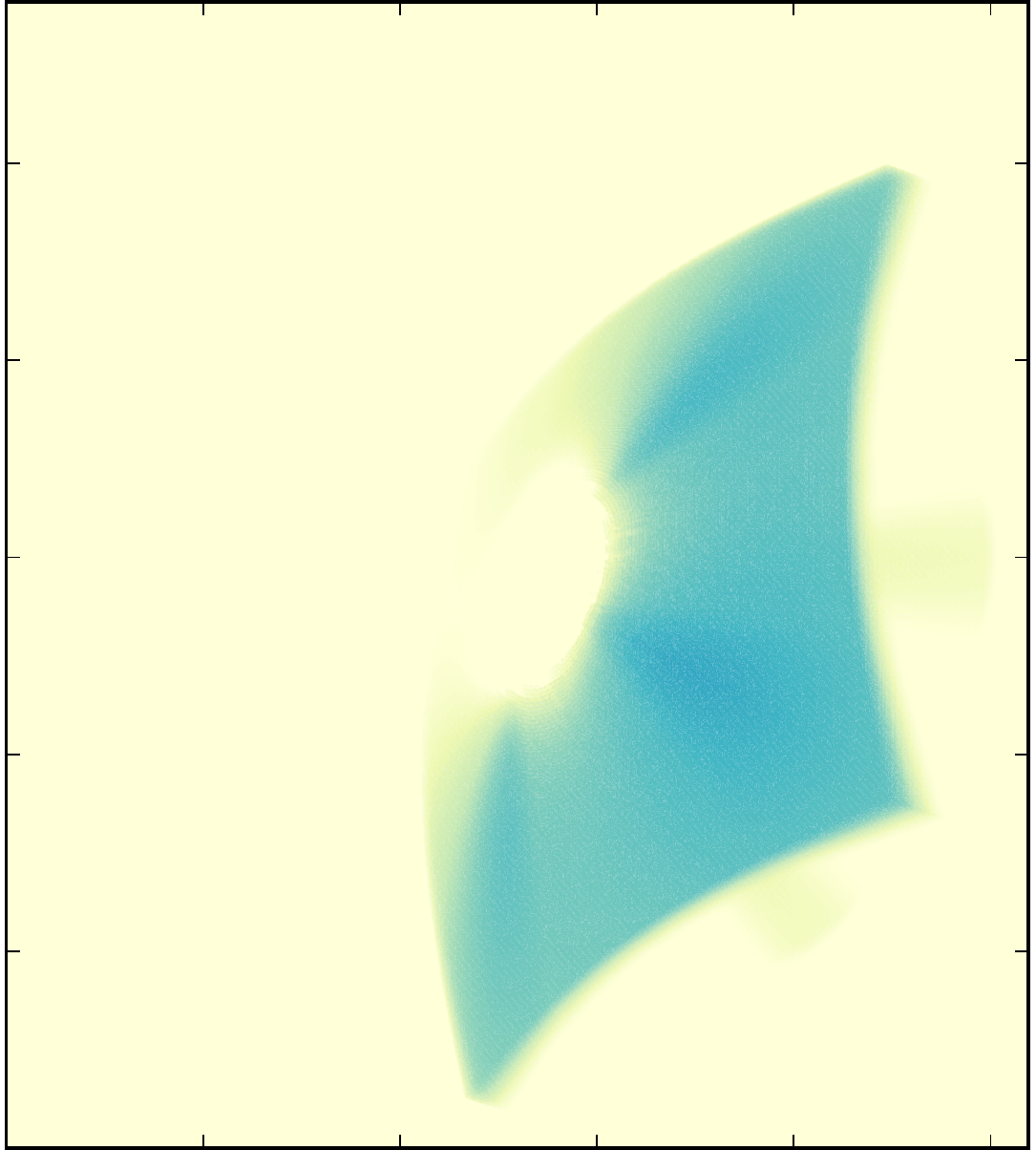}}
				\put(115,20){\line(1,0){84.4}}
				\put(115,17){\line(0,1){6}}
				\put(199.4,17){\line(0,1){6}}
				\put(115,5){\scriptsize{0.005 arcsec}}
			\end{picture}
		\end{subfigure}
		\begin{subfigure}{320\unitlength}
			\begin{picture}(320,358)
				\put(0,0){\includegraphics[trim=4.2cm 1.5cm 5.1cm 1.5cm, clip=true,width=320\unitlength]{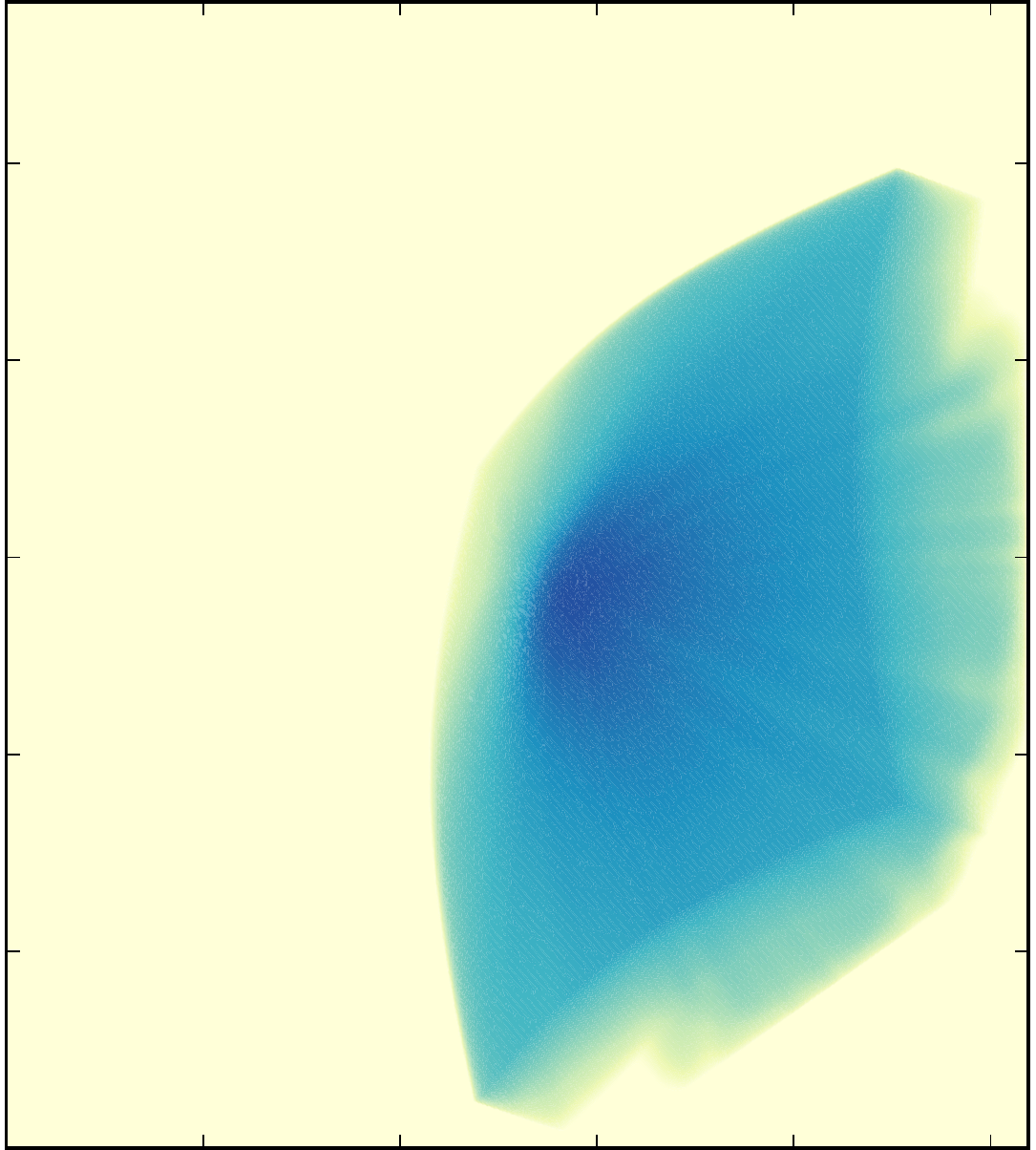}}
				\put(115,20){\line(1,0){84.4}}
				\put(115,17){\line(0,1){6}}
				\put(199.4,17){\line(0,1){6}}
				\put(115,5){\scriptsize{0.005 arcsec}}
			\end{picture}
		\end{subfigure}
		\begin{subfigure}{320\unitlength}	
			\begin{picture}(320,358)
				\put(0,0){\includegraphics[trim=4.2cm 1.5cm 5.1cm 1.5cm, clip=true,width=320\unitlength]{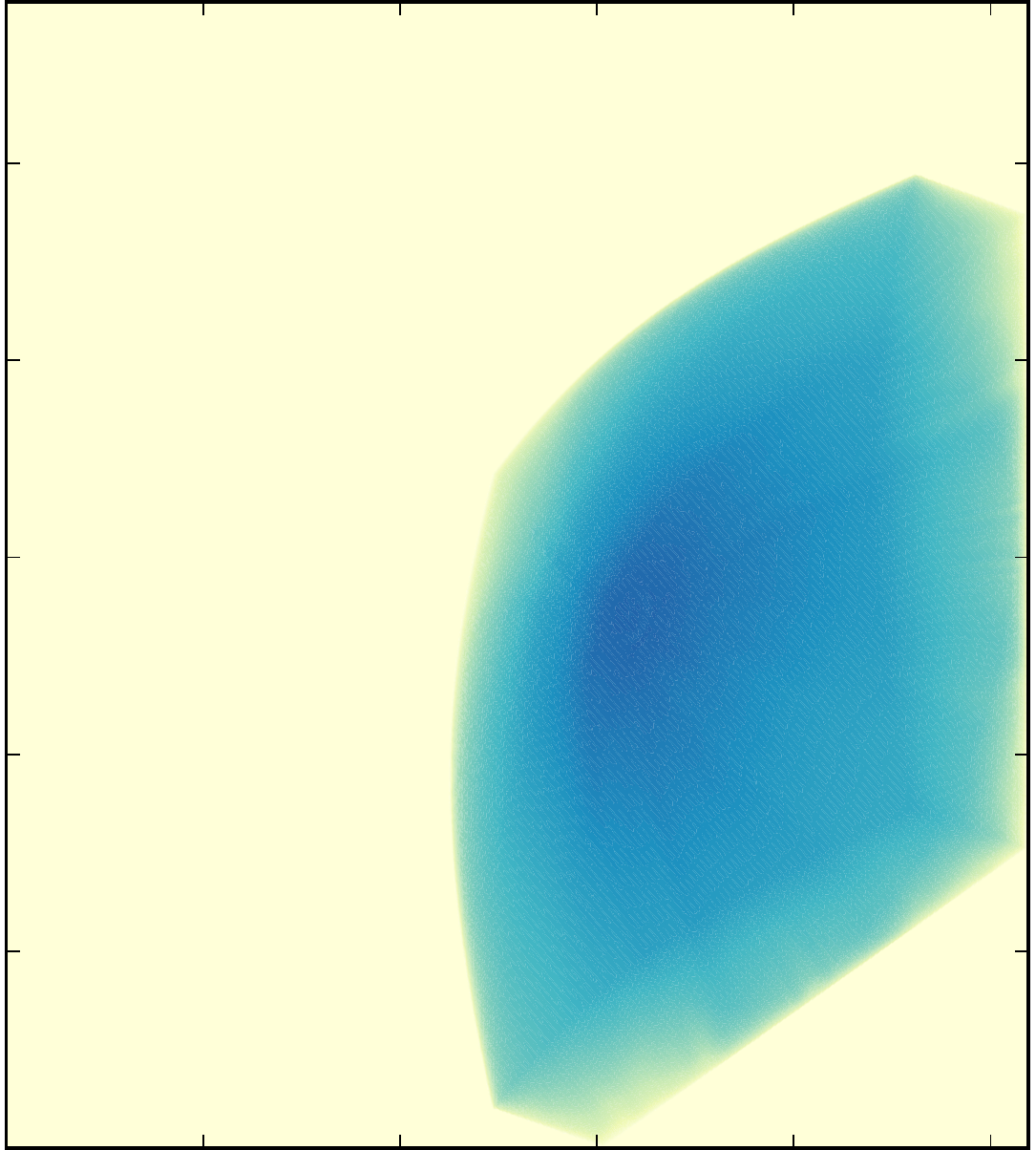}}
				\put(115,20){\line(1,0){84.4}}
				\put(115,17){\line(0,1){6}}
				\put(199.4,17){\line(0,1){6}}
				\put(115,5){\scriptsize{0.005 arcsec}}
			\end{picture}
		\end{subfigure}\\
		\begin{center}
		\begin{subfigure}{500\unitlength}	
			\begin{picture}(500,0)
				\put(0,-20){\includegraphics[trim=2cm 2cm 1cm 24cm, clip=true,width=500\unitlength]{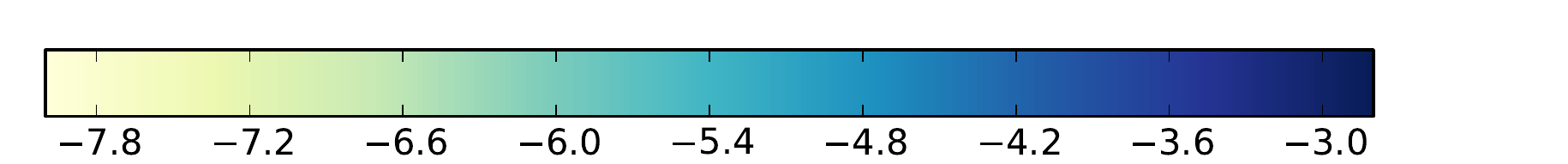}}
				\put(200,25){\footnotesize{ph m$^{-2}$ s$^{-1}$}}
			\end{picture}
		\end{subfigure}
		\end{center}	
\caption{Photon flux above 100 MeV at 1 kpc distance for IC-emission with different orientations. first row: $i=$90$^\circ$, $\Phi=$0$^\circ$ (edge-on, along line of centers), second row: $i=$90$^\circ$, $\Phi=$180$^\circ$ (edge-on, along line of centers, opposite of second row), 
third row: $i=$45$^\circ$, $\Phi=$45$^\circ$. The spatial dimensions are the same as in Figure \ref{em_0_0}.
\label{em_varic}}
\end{figure*}
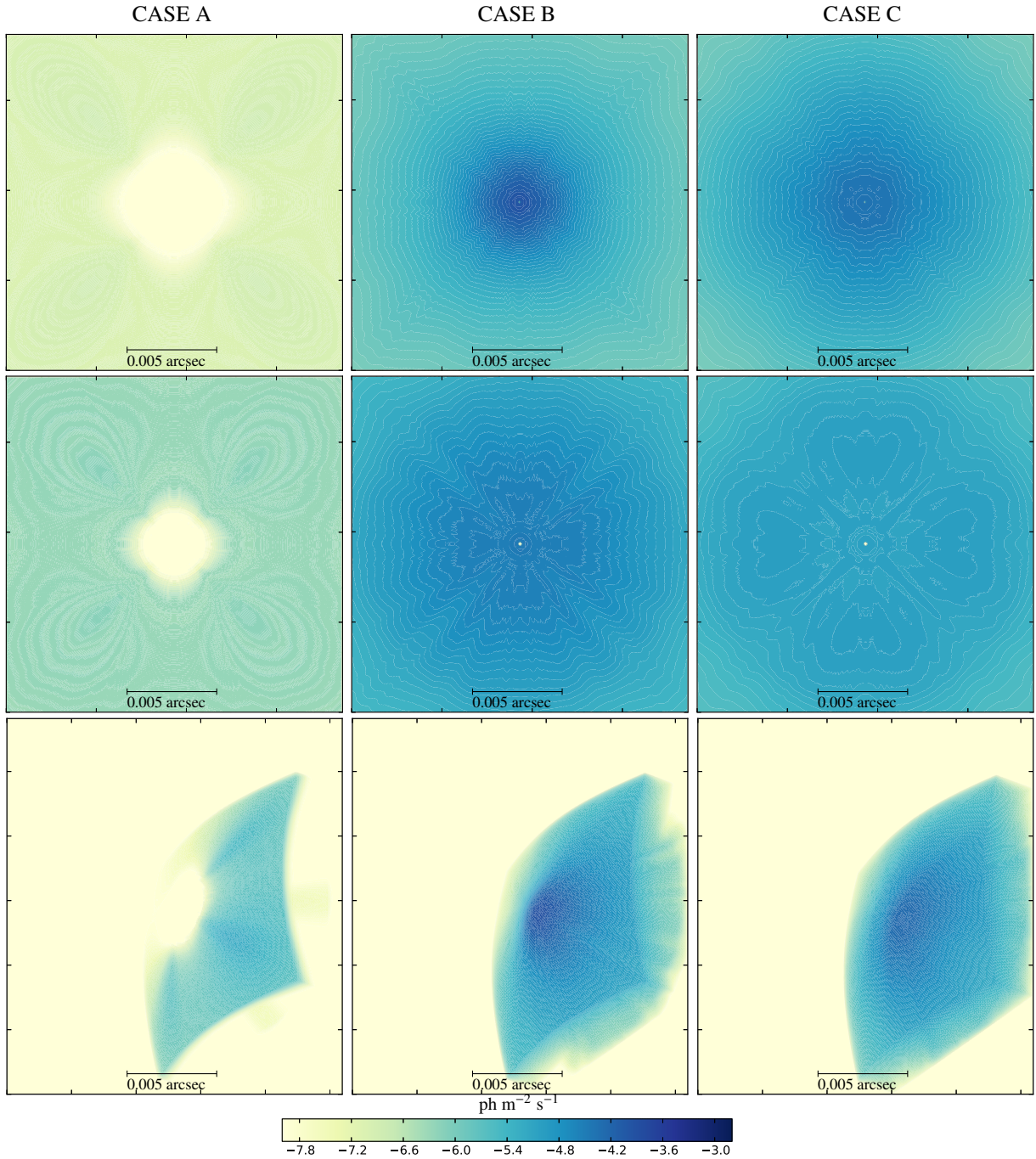

\begin{figure*}
	\setlength{\unitlength}{0.001\textwidth}
		\begin{subfigure}{320\unitlength}
			\begin{picture}(320,320)
					\put(0,0){\includegraphics[trim=2.9cm 1.5cm 5.1cm 1.5cm, clip=true,width=320\unitlength]{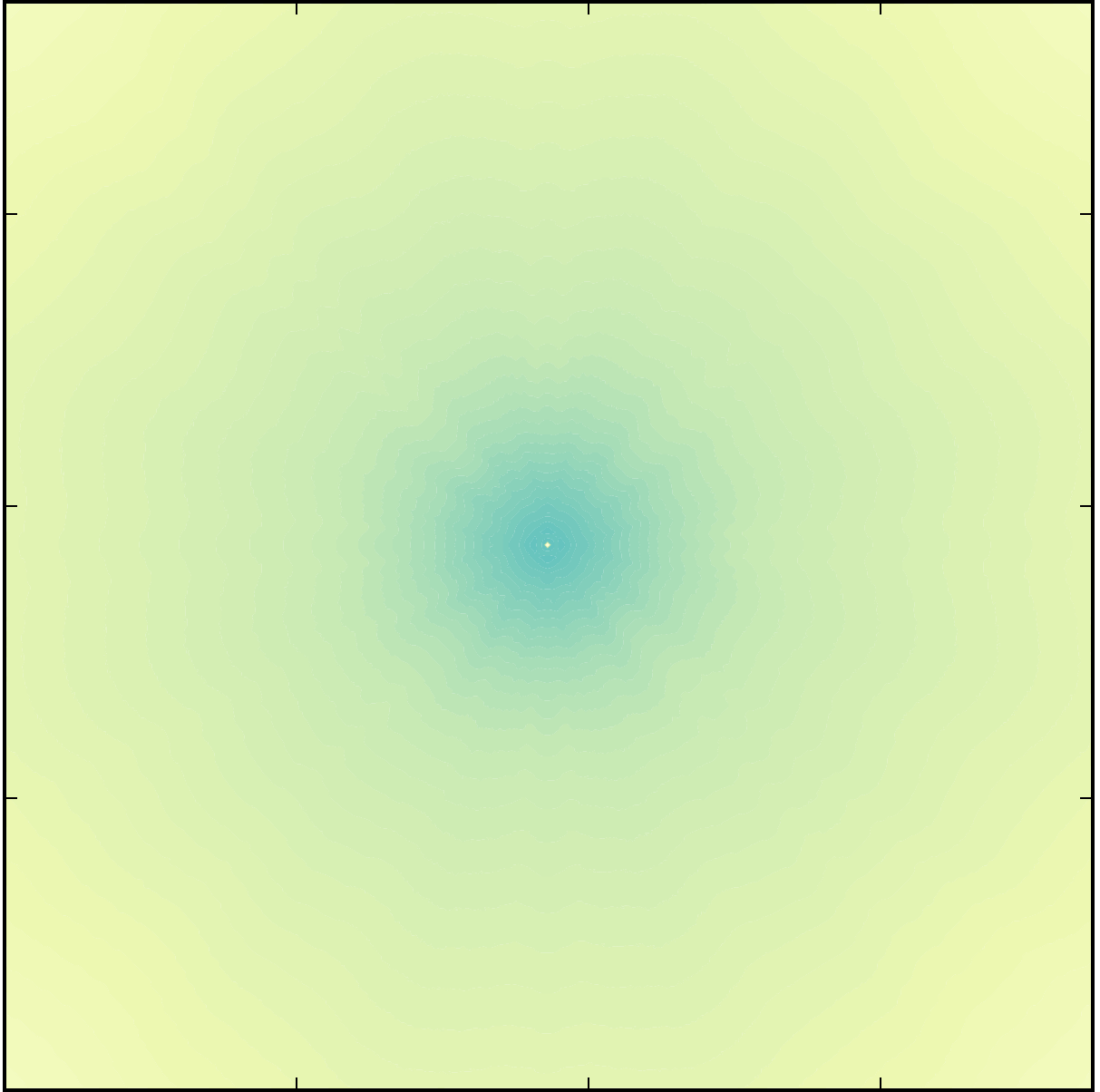}}
				\put(120,330){CASE A}
				\put(115,20){\line(1,0){84.4}}
				\put(115,17){\line(0,1){6}}
				\put(199.4,17){\line(0,1){6}}
				\put(115,5){\scriptsize{0.005 arcsec}}
			\end{picture}
		\end{subfigure}
		\begin{subfigure}{320\unitlength}
			\begin{picture}(320,320)
					\put(0,0){\includegraphics[trim=2.9cm 1.5cm 5.1cm 1.5cm, clip=true,width=320\unitlength ,]{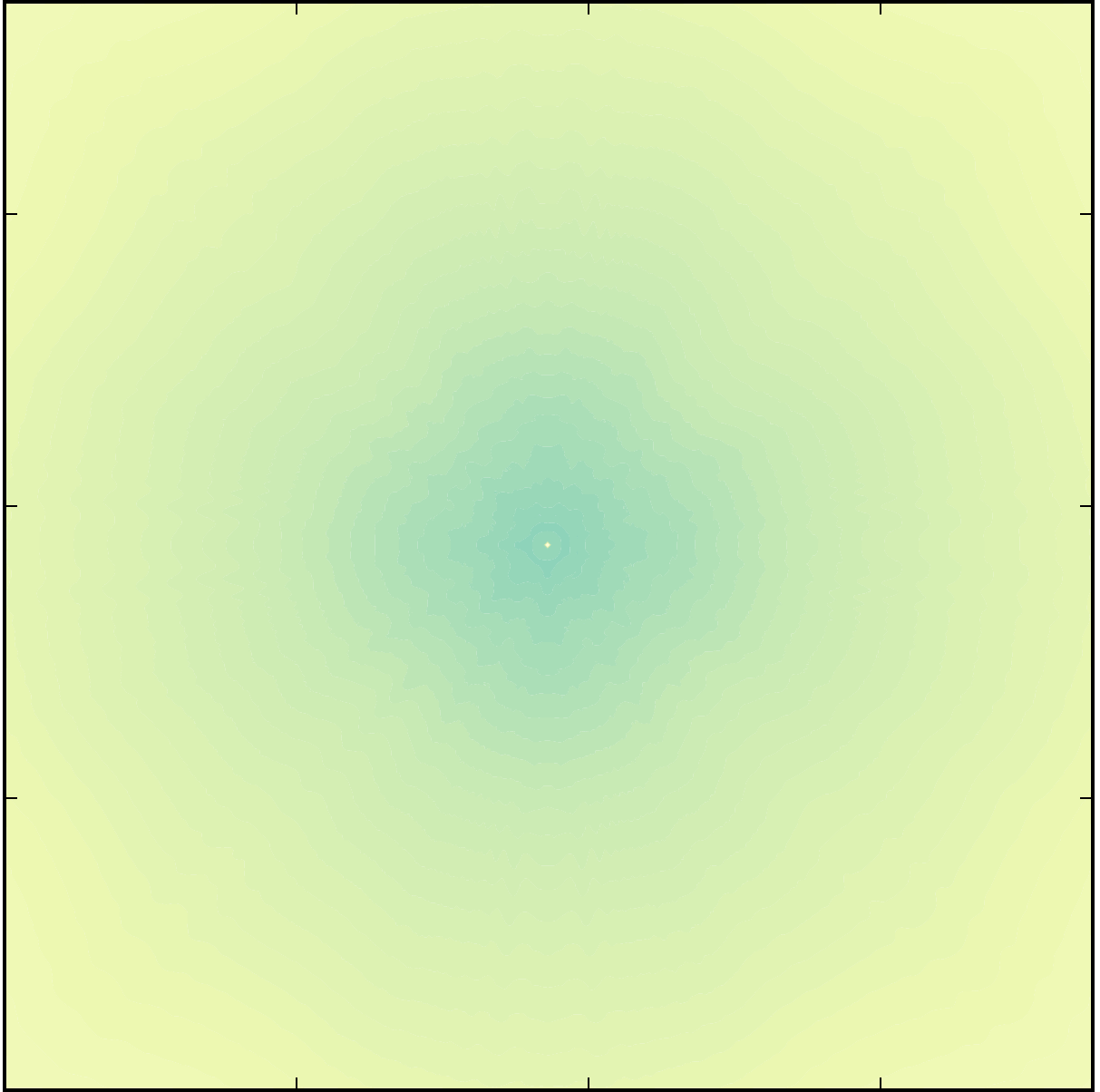}}
				\put(120,330){CASE B}
				\put(115,20){\line(1,0){84.4}}
				\put(115,17){\line(0,1){6}}
				\put(199.4,17){\line(0,1){6}}
				\put(115,5){\scriptsize{0.005 arcsec}}
			\end{picture}
				\end{subfigure}
		\begin{subfigure}{320\unitlength}
			\begin{picture}(320,320)
			\put(0,0){\includegraphics[trim=2.9cm 1.5cm 5.1cm 1.5cm, clip=true,width=320\unitlength]{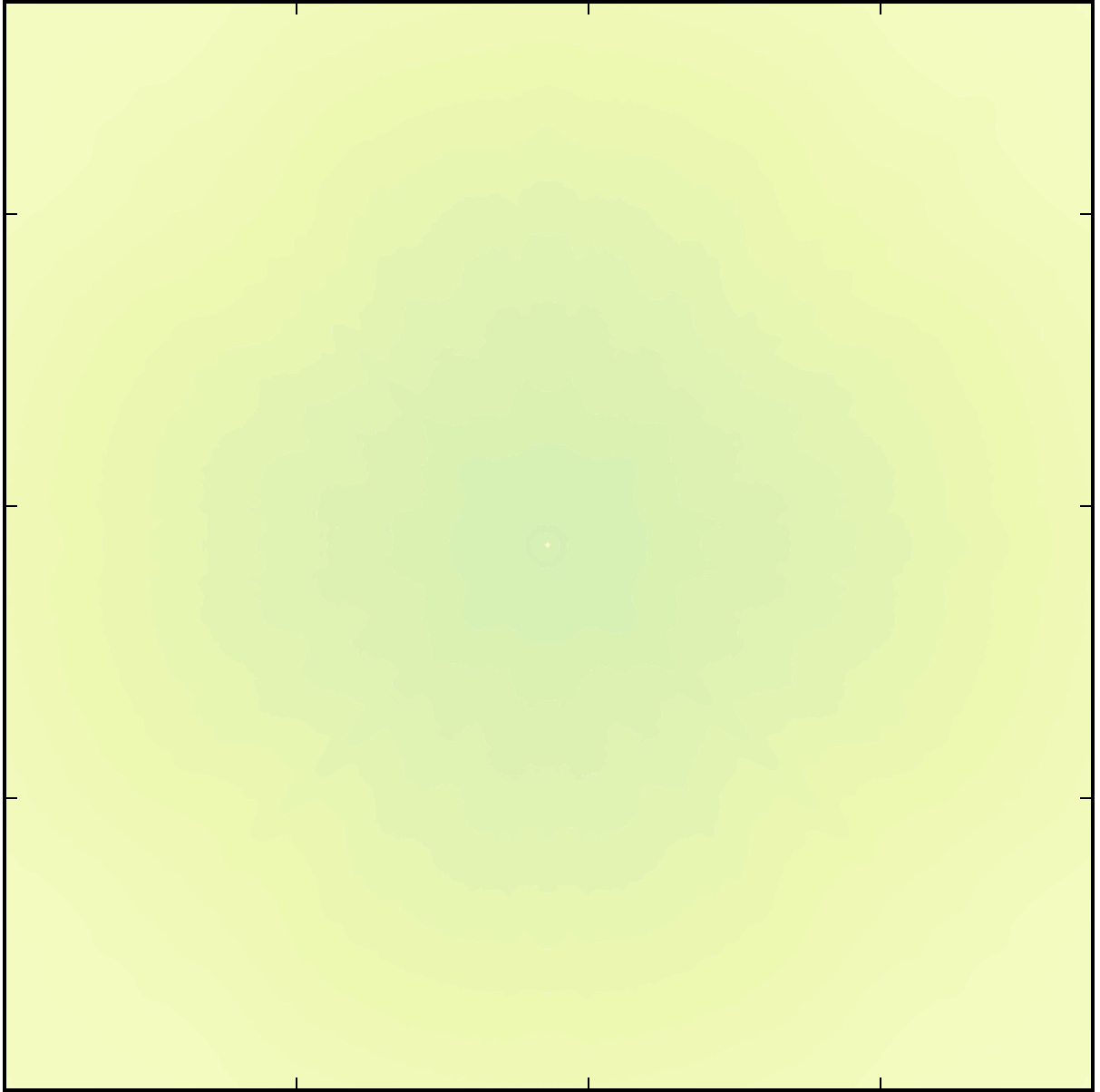}}
				\put(120,330){CASE C}
				\put(115,20){\line(1,0){84.4}}
				\put(115,17){\line(0,1){6}}
				\put(199.4,17){\line(0,1){6}}
				\put(115,5){\scriptsize{0.005 arcsec}}
			\end{picture}
		\end{subfigure}\\
		\begin{subfigure}{320\unitlength}
			\begin{picture}(320,320)
				\put(0,0){\includegraphics[trim=2.9cm 1.5cm 5.1cm 1.5cm, clip=true,width=320\unitlength]{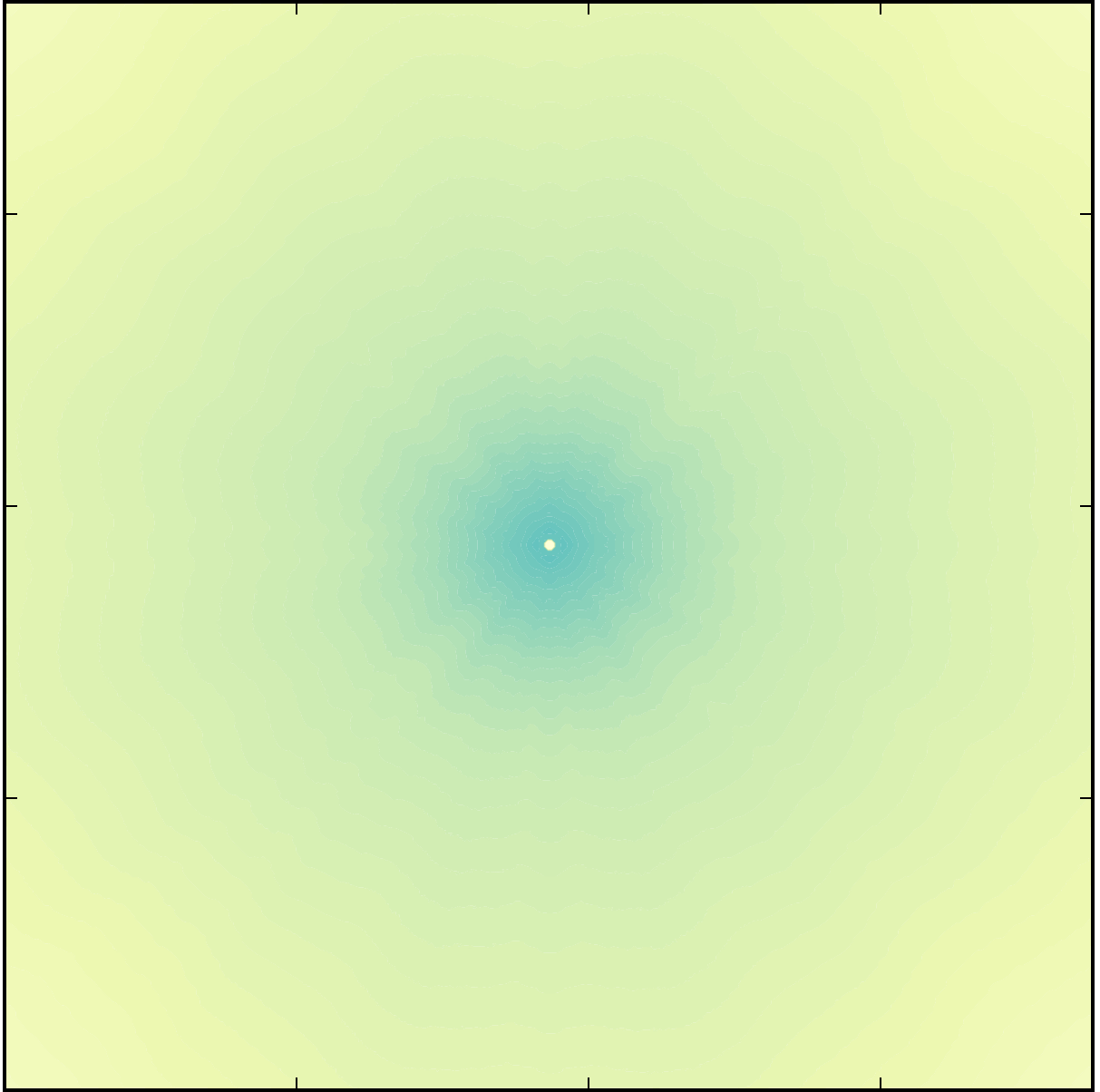}}
				\put(115,20){\line(1,0){84.4}}
				\put(115,17){\line(0,1){6}}
				\put(199.4,17){\line(0,1){6}}
				\put(115,5){\scriptsize{0.005 arcsec}}
			\end{picture}
		\end{subfigure}
		\begin{subfigure}{320\unitlength}
			\begin{picture}(320,320)
				\put(0,0){\includegraphics[trim=2.9cm 1.5cm 5.1cm 1.5cm, clip=true,width=320\unitlength]{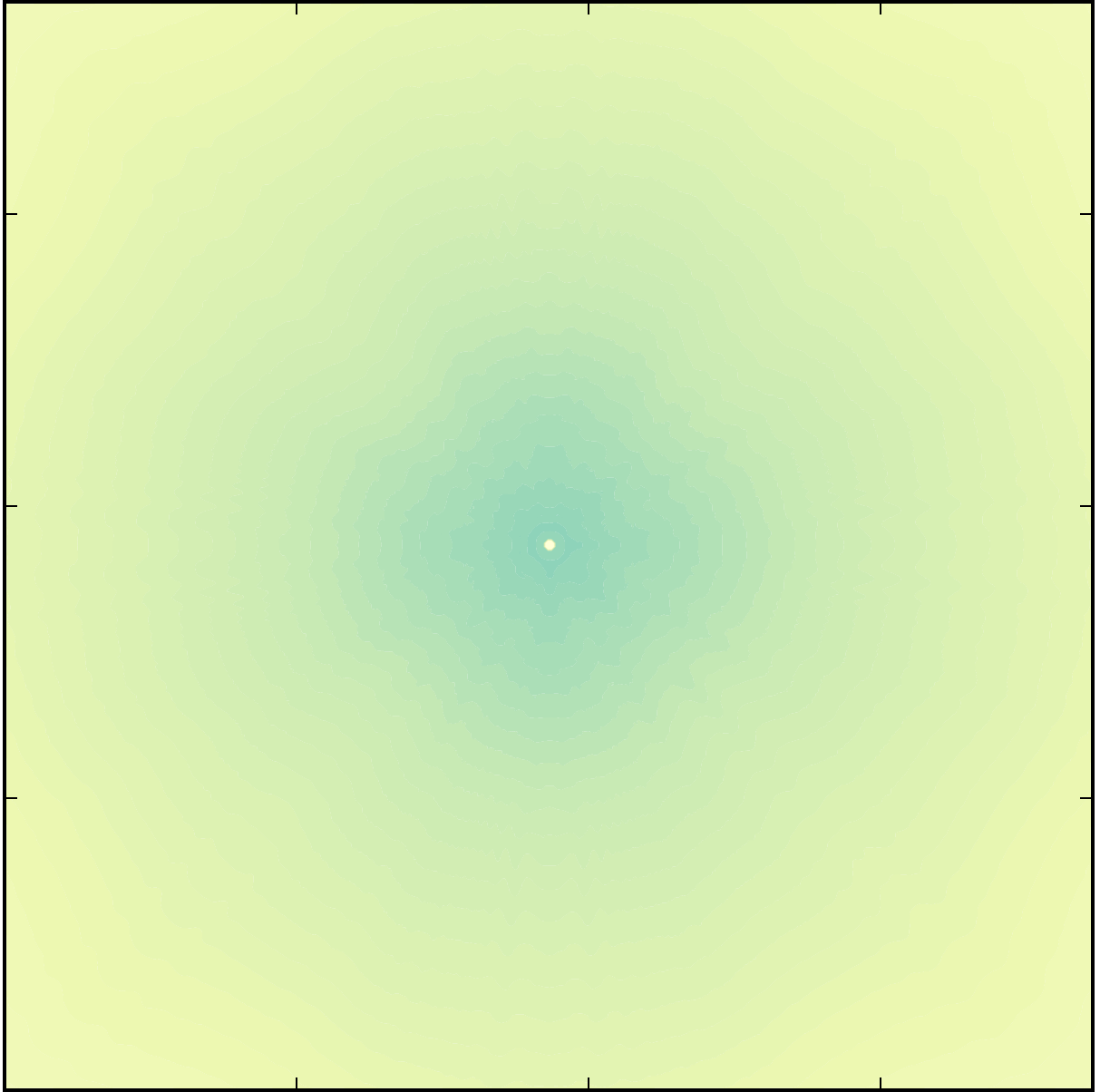}}
				\put(115,20){\line(1,0){84.4}}
				\put(115,17){\line(0,1){6}}
				\put(199.4,17){\line(0,1){6}}
				\put(115,5){\scriptsize{0.005 arcsec}}
			\end{picture}
		\end{subfigure}
		\begin{subfigure}{320\unitlength}	
			\begin{picture}(320,320)
				\put(0,0){\includegraphics[trim=2.9cm 1.5cm 5.1cm 1.5cm, clip=true,width=320\unitlength]{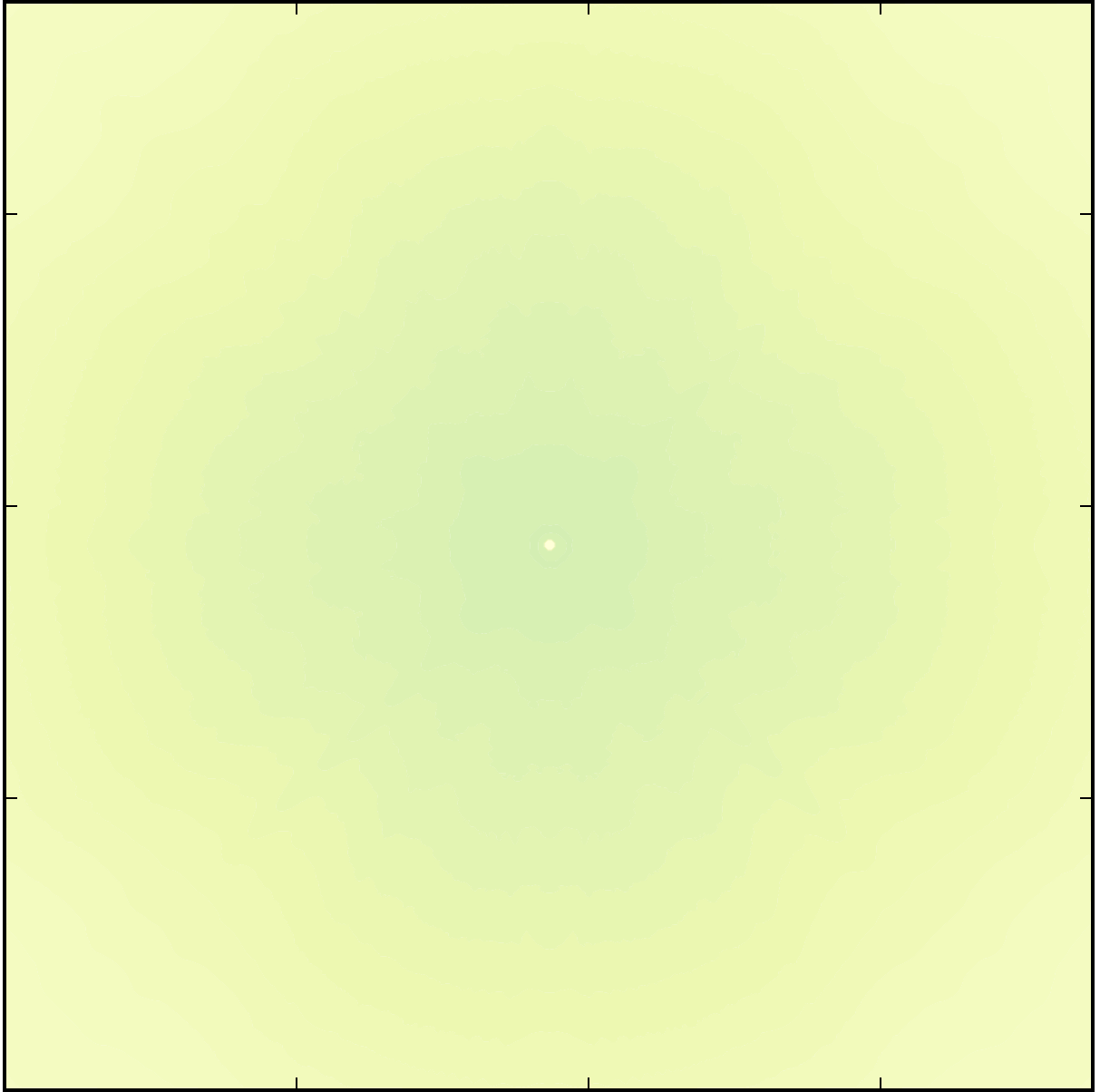}}
				\put(115,20){\line(1,0){84.4}}
				\put(115,17){\line(0,1){6}}
				\put(199.4,17){\line(0,1){6}}
				\put(115,5){\scriptsize{0.005 arcsec}}
			\end{picture}
		\end{subfigure}\\
		\begin{subfigure}{320\unitlength}
			\begin{picture}(320,358)
				\put(0,0){\includegraphics[trim=4.2cm 1.5cm 5.1cm 1.5cm, clip=true,width=320\unitlength]{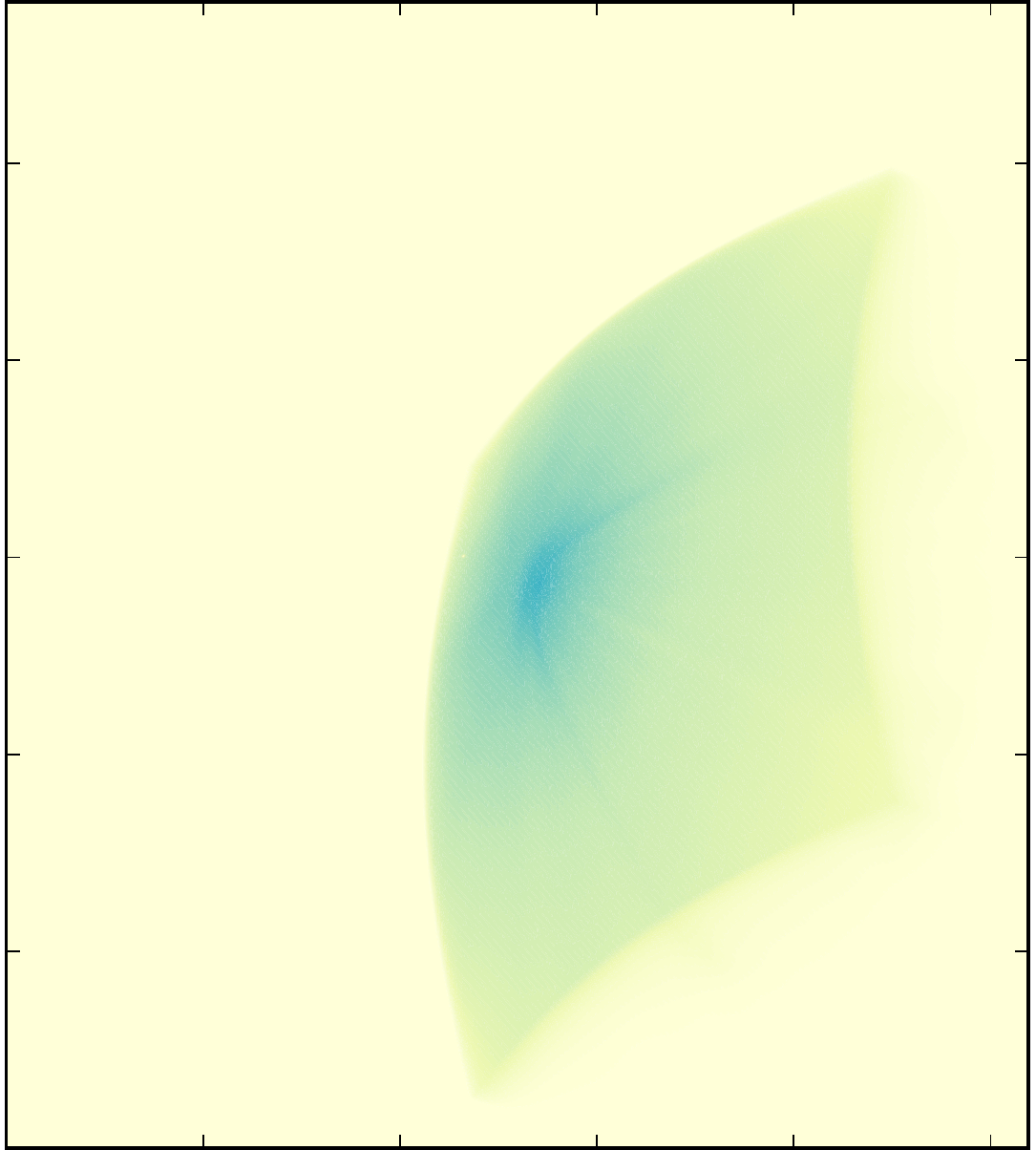}}
				%\put(115,20){\line(1,0){84.4}}
			%	\put(115,17){\line(0,1){6}}
			%	\put(199.4,17){\line(0,1){6}}
				%\put(115,5){\scriptsize{0.005 arcsec}}
			\end{picture}
		\end{subfigure}
		\begin{subfigure}{320\unitlength}
			\begin{picture}(320,358)
				\put(0,0){\includegraphics[trim=4.2cm 1.5cm 5.1cm 1.5cm, clip=true,width=320\unitlength]{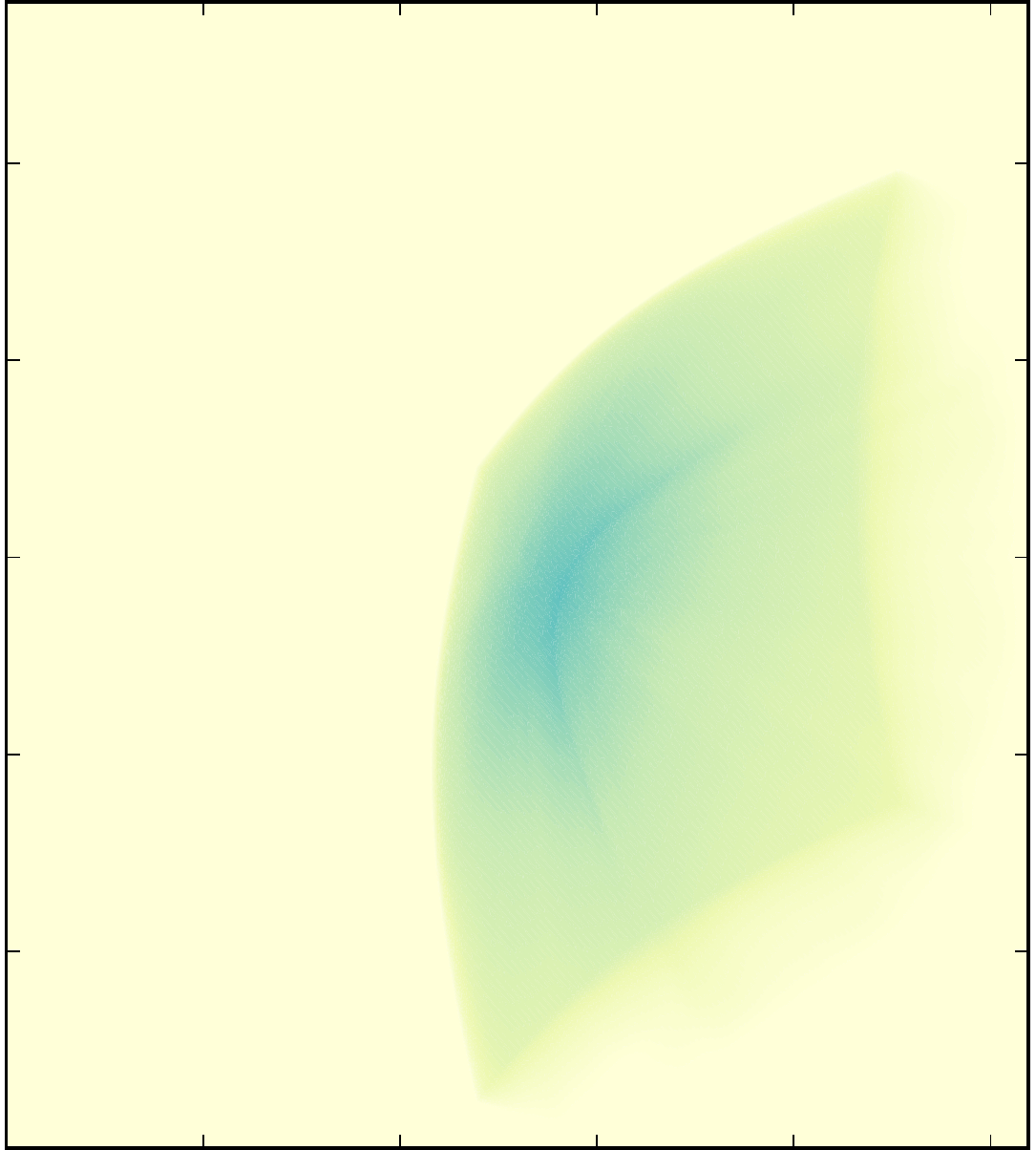}}
				%\put(115,20){\line(1,0){84.4}}
				%\put(115,17){\line(0,1){6}}
				%\put(199.4,17){\line(0,1){6}}
				%\put(115,5){\scriptsize{0.005 arcsec}}
			\end{picture}
		\end{subfigure}
		\begin{subfigure}{320\unitlength}	
			\begin{picture}(320,358)
				\put(0,0){\includegraphics[trim=4.2cm 1.5cm 5.1cm 1.5cm, clip=true,width=320\unitlength]{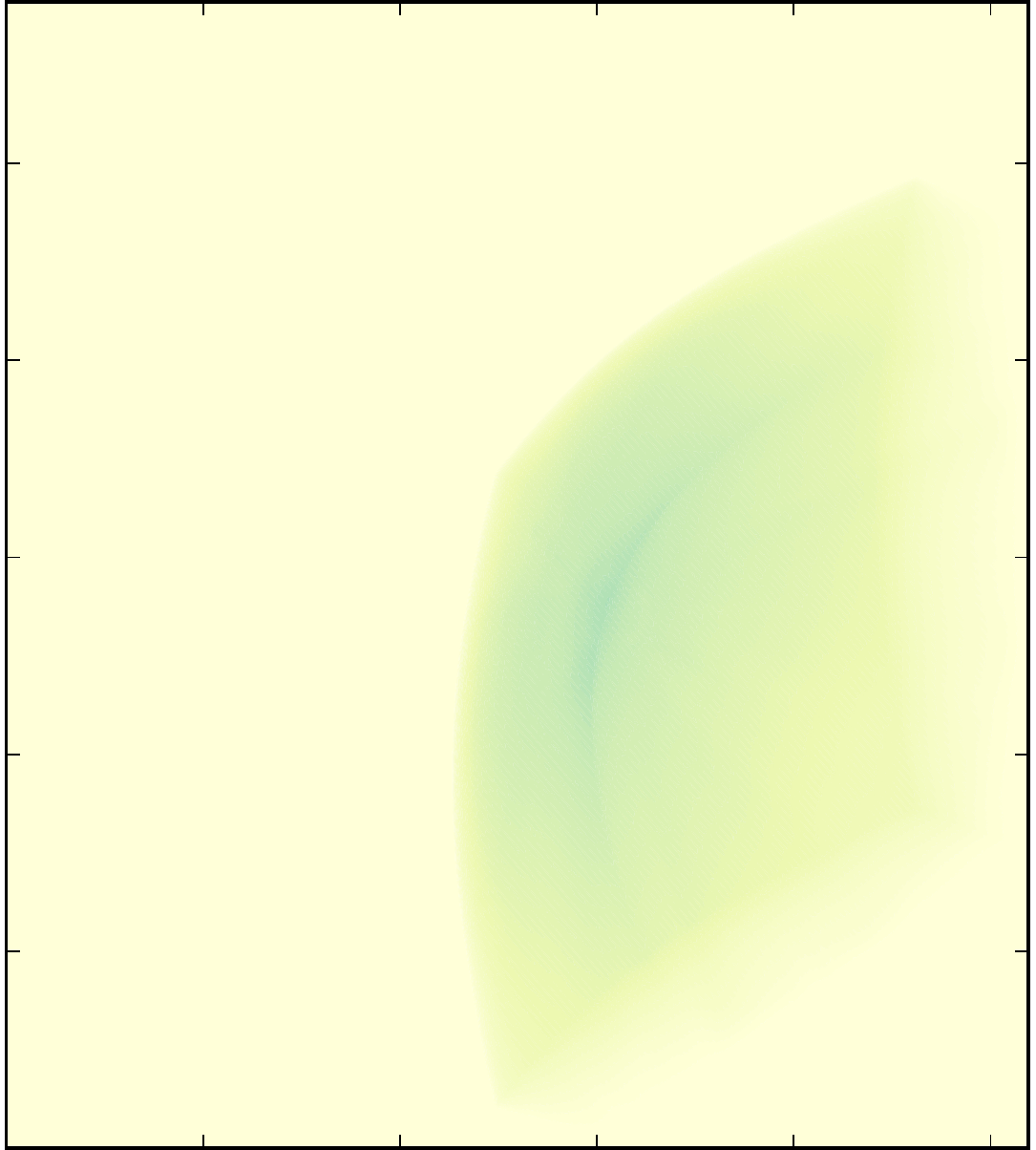}}
				%\put(115,20){\line(1,0){84.4}}
				%\put(115,17){\line(0,1){6}}
				%\put(199.4,17){\line(0,1){6}}
				%\put(115,5){\scriptsize{0.005 arcsec}}
			\end{picture}
		\end{subfigure}\\
		\begin{center}
		\begin{subfigure}{500\unitlength}	
			\begin{picture}(500,0)
				\put(0,-20){\includegraphics[trim=2cm 2cm 1cm 24cm, clip=true,width=500\unitlength]{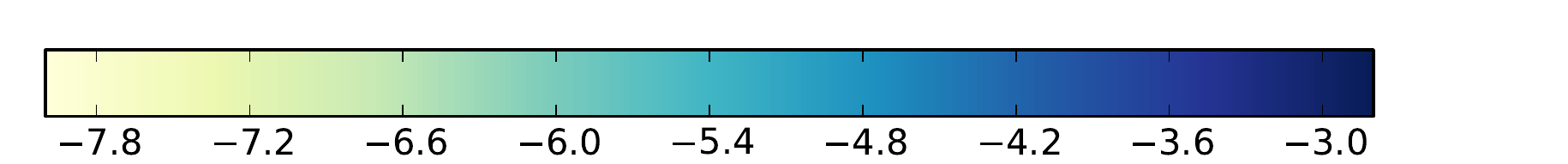}}
				\put(200,25){\footnotesize{ph m$^{-2}$ s$^{-1}$}}
			\end{picture}
		\end{subfigure}
		\end{center}	
\caption{Same as Figure \ref{em_varic} for bremsstrahlung.
\label{em_varbr}}
\end{figure*}
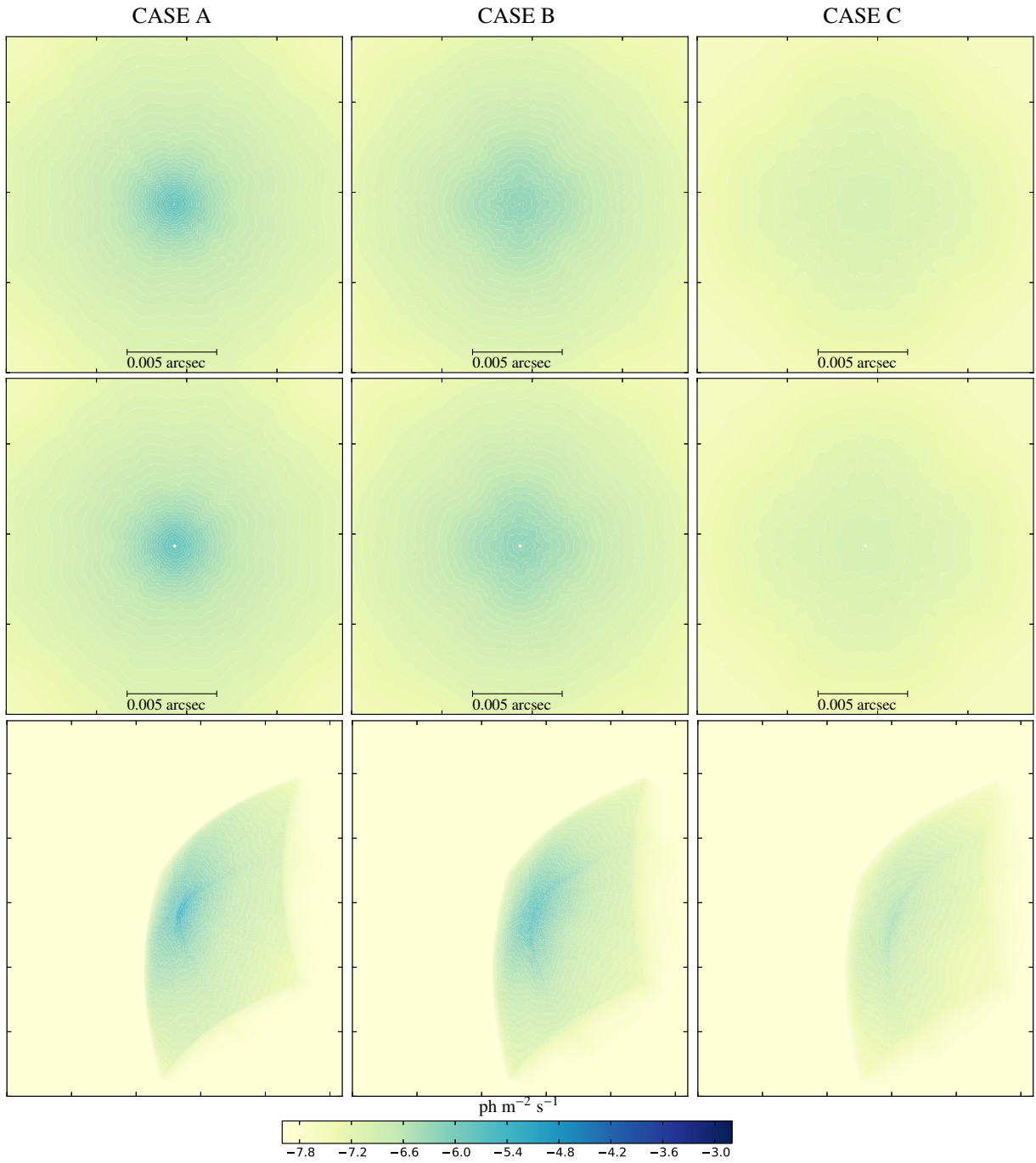

\begin{figure*}
	\setlength{\unitlength}{0.001\textwidth}
		\begin{subfigure}{320\unitlength}
			\begin{picture}(320,320)
					\put(0,0){\includegraphics[trim=2.9cm 1.5cm 5.1cm 1.5cm, clip=true,width=320\unitlength]{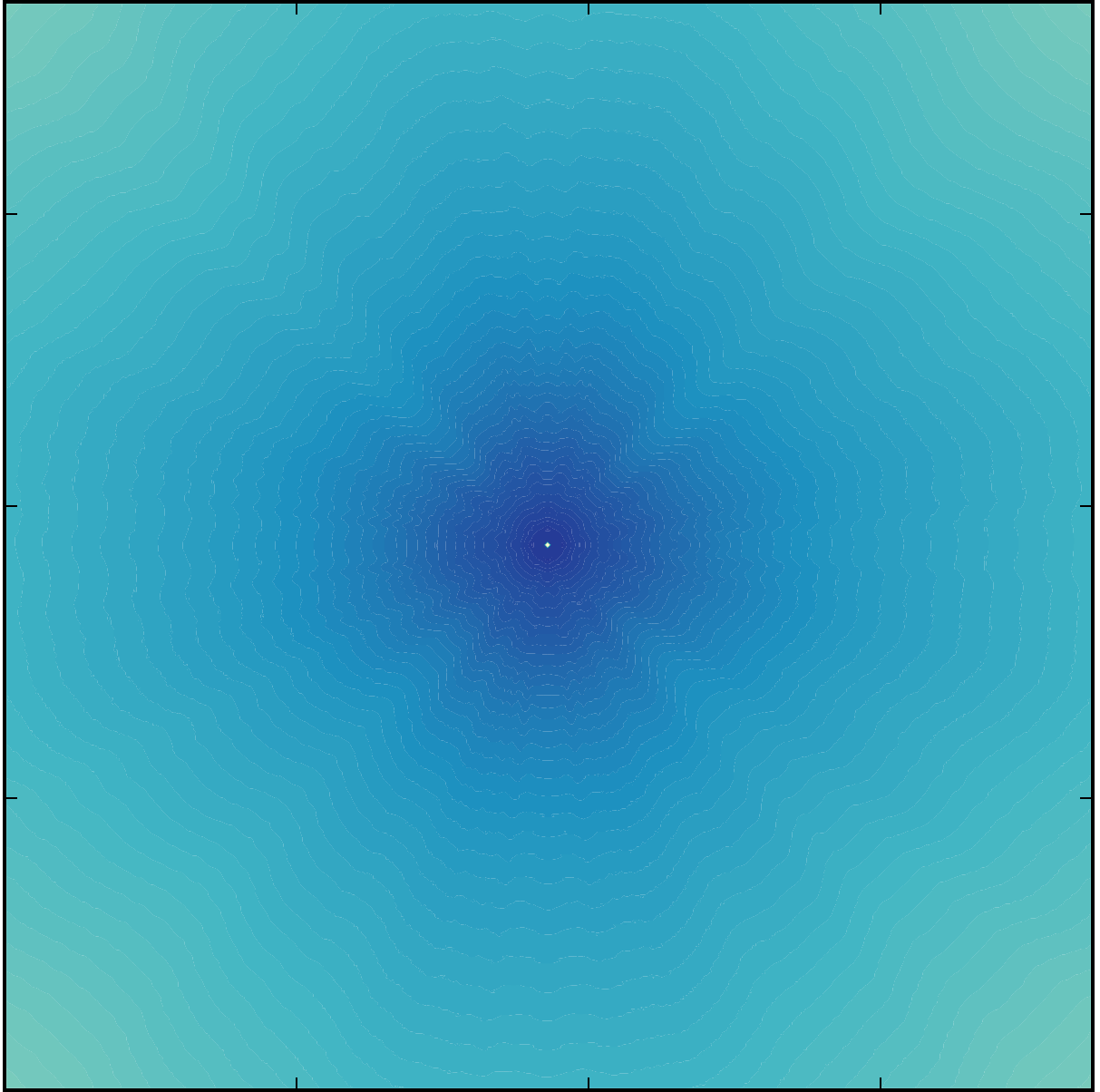}}
				\put(120,330){CASE A}
				\put(115,20){\line(1,0){84.4}}
				\put(115,17){\line(0,1){6}}
				\put(199.4,17){\line(0,1){6}}
				\put(115,5){\scriptsize{0.005 arcsec}}
			\end{picture}
		\end{subfigure}
		\begin{subfigure}{320\unitlength}
			\begin{picture}(320,320)
					\put(0,0){\includegraphics[trim=2.9cm 1.5cm 5.1cm 1.5cm, clip=true,width=320\unitlength ,]{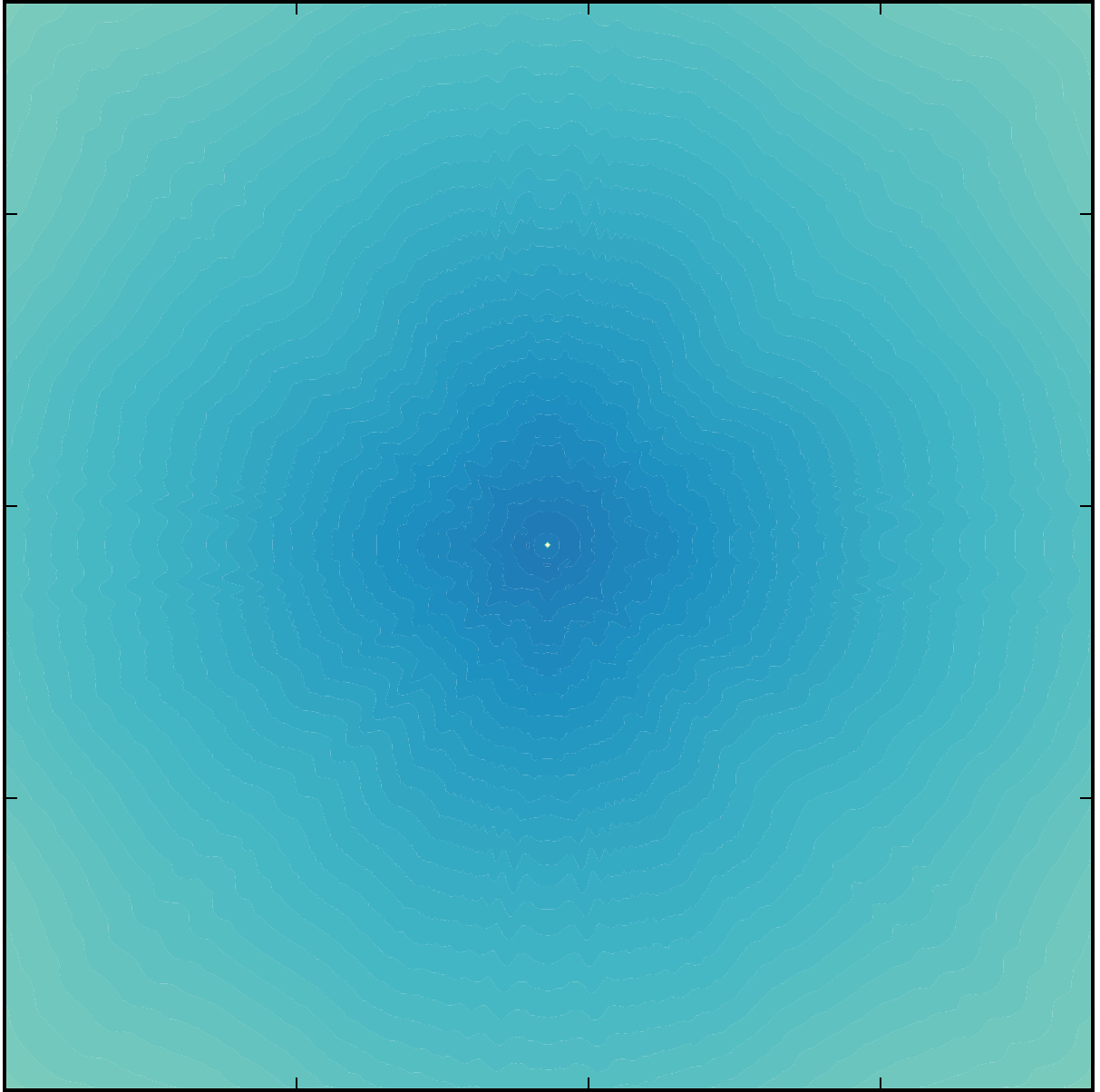}}
				\put(120,330){CASE B}
				\put(115,20){\line(1,0){84.4}}
				\put(115,17){\line(0,1){6}}
				\put(199.4,17){\line(0,1){6}}
				\put(115,5){\scriptsize{0.005 arcsec}}
			\end{picture}
				\end{subfigure}
		\begin{subfigure}{320\unitlength}
			\begin{picture}(320,320)
			\put(0,0){\includegraphics[trim=2.9cm 1.5cm 5.1cm 1.5cm, clip=true,width=320\unitlength]{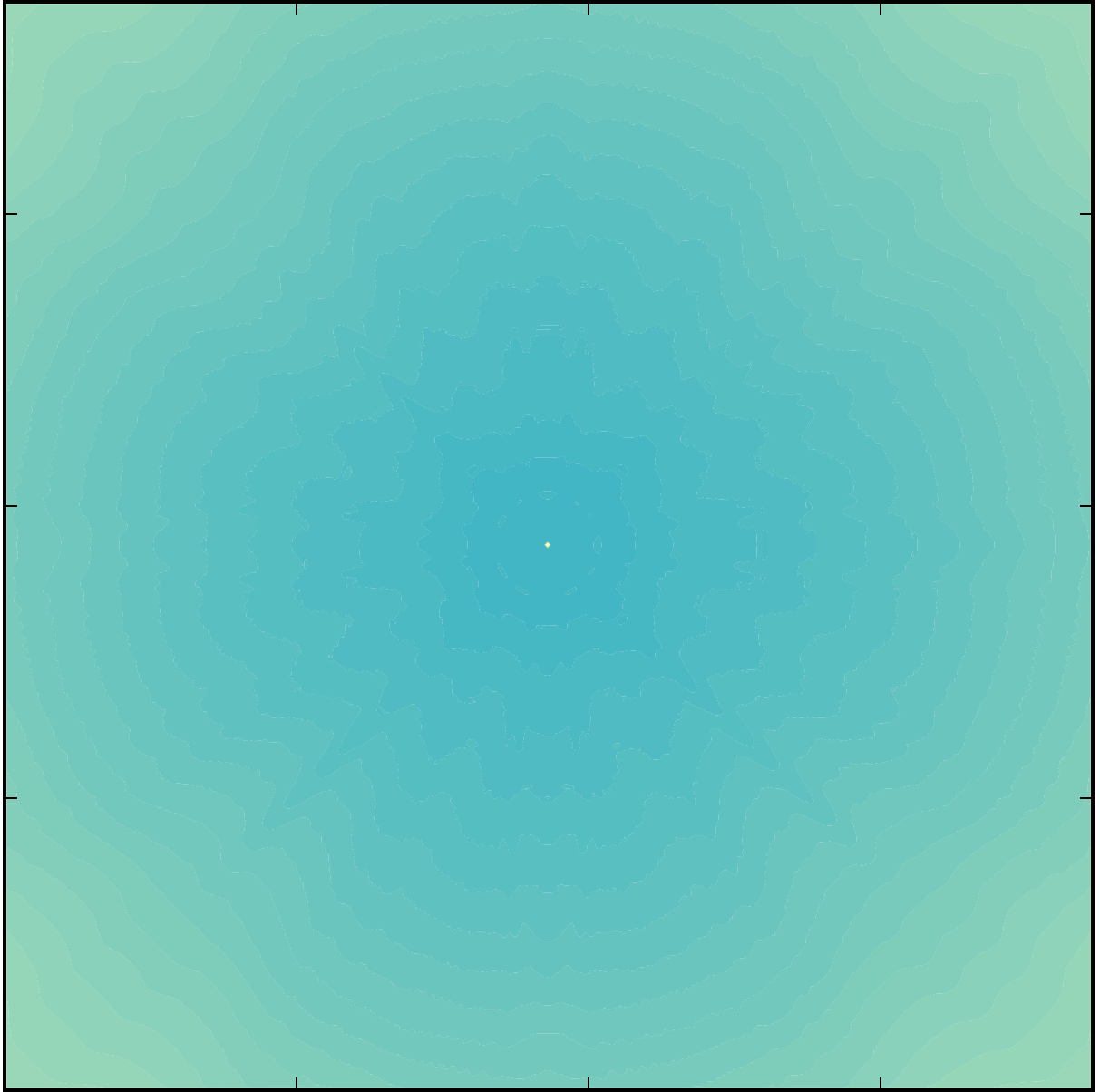}}
				\put(120,330){CASE C}
				\put(115,20){\line(1,0){84.4}}
				\put(115,17){\line(0,1){6}}
				\put(199.4,17){\line(0,1){6}}
				\put(115,5){\scriptsize{0.005 arcsec}}
			\end{picture}
		\end{subfigure}\\
		\begin{subfigure}{320\unitlength}
			\begin{picture}(320,320)
				\put(0,0){\includegraphics[trim=2.9cm 1.5cm 5.1cm 1.5cm, clip=true,width=320\unitlength]{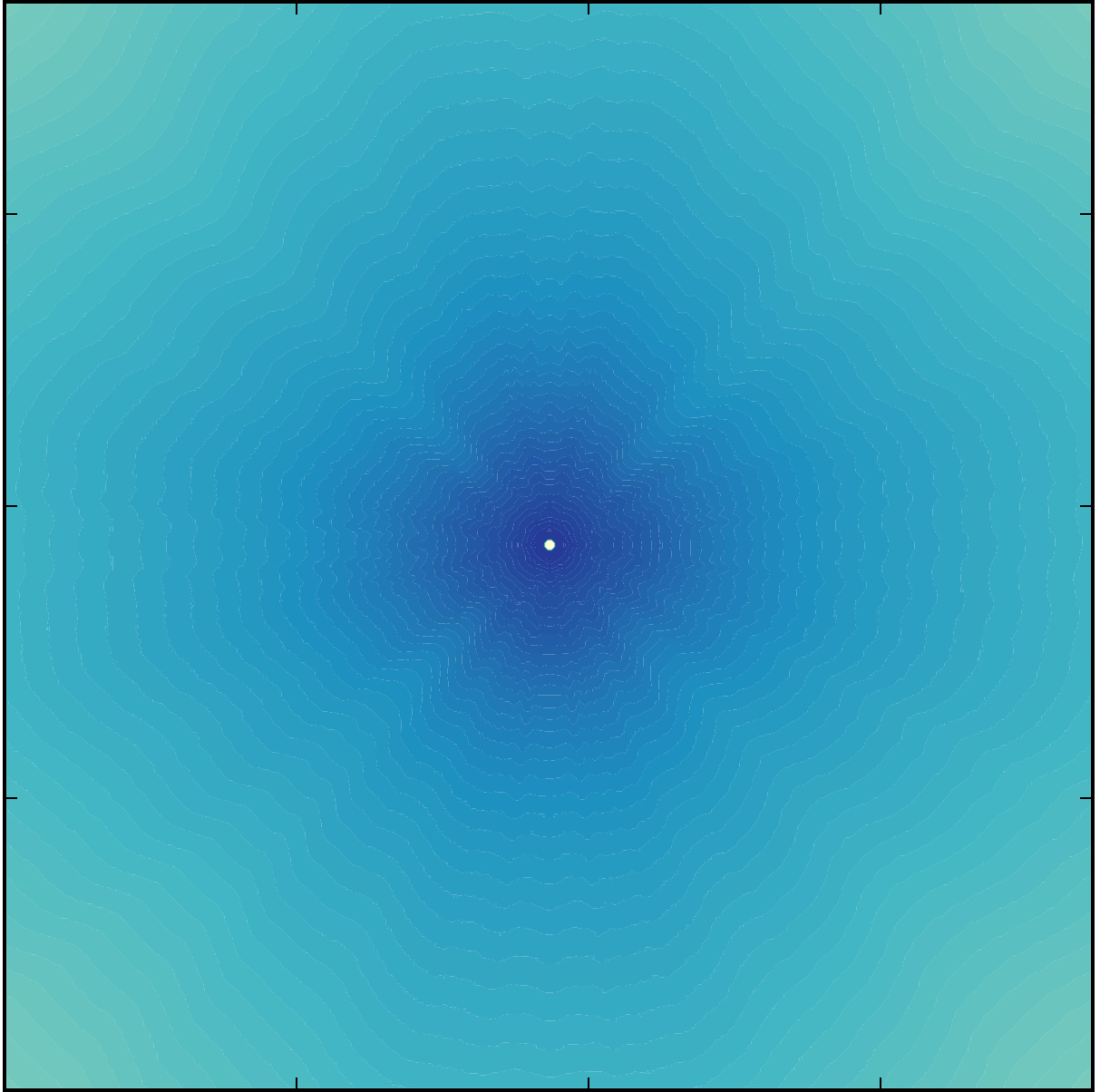}}
				\put(115,20){\line(1,0){84.4}}
				\put(115,17){\line(0,1){6}}
				\put(199.4,17){\line(0,1){6}}
				\put(115,5){\scriptsize{0.005 arcsec}}
			\end{picture}
		\end{subfigure}
		\begin{subfigure}{320\unitlength}
			\begin{picture}(320,320)
				\put(0,0){\includegraphics[trim=2.9cm 1.5cm 5.1cm 1.5cm, clip=true,width=320\unitlength]{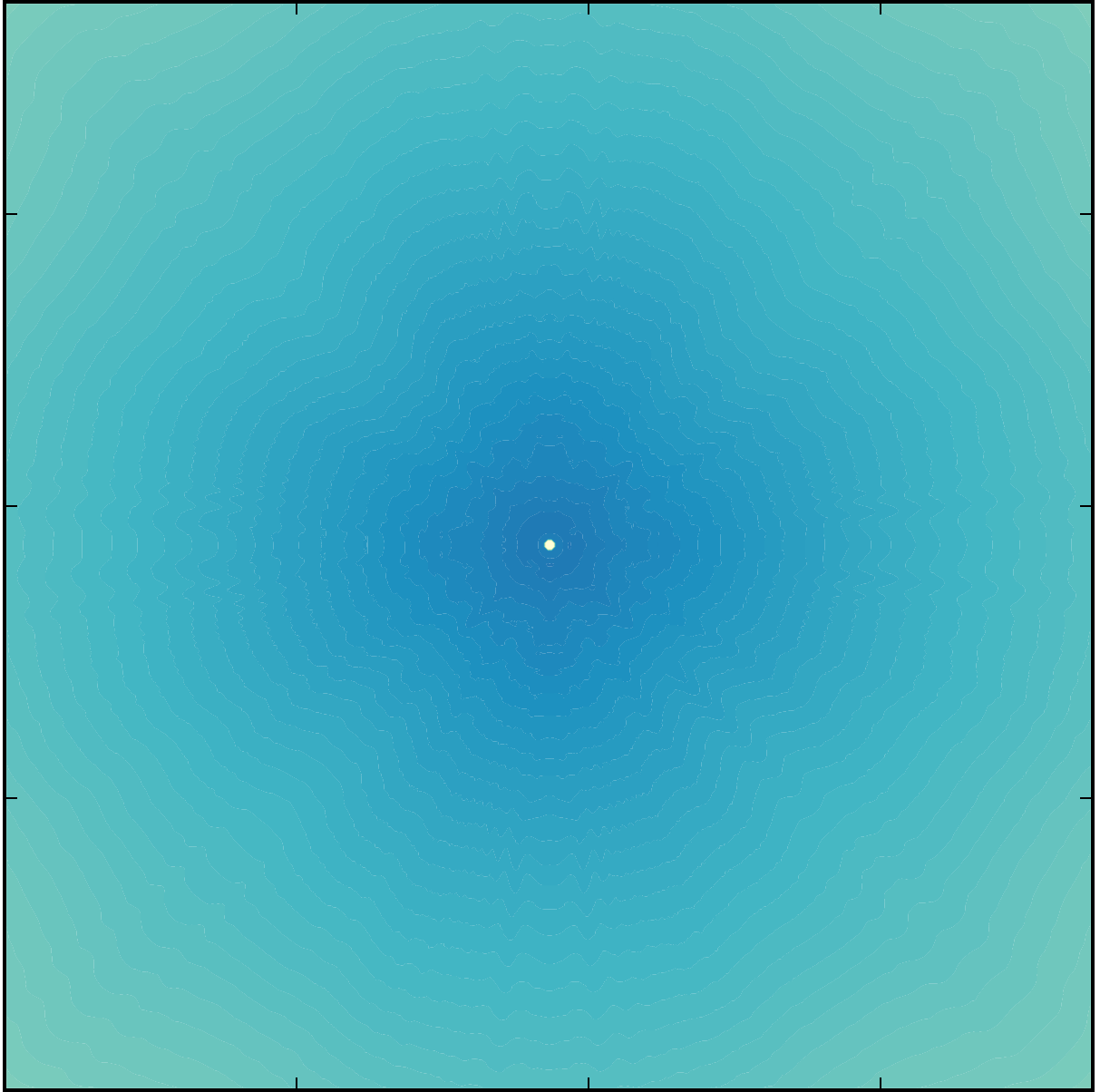}}
				\put(115,20){\line(1,0){84.4}}
				\put(115,17){\line(0,1){6}}
				\put(199.4,17){\line(0,1){6}}
				\put(115,5){\scriptsize{0.005 arcsec}}
			\end{picture}
		\end{subfigure}
		\begin{subfigure}{320\unitlength}	
			\begin{picture}(320,320)
				\put(0,0){\includegraphics[trim=2.9cm 1.5cm 5.1cm 1.5cm, clip=true,width=320\unitlength]{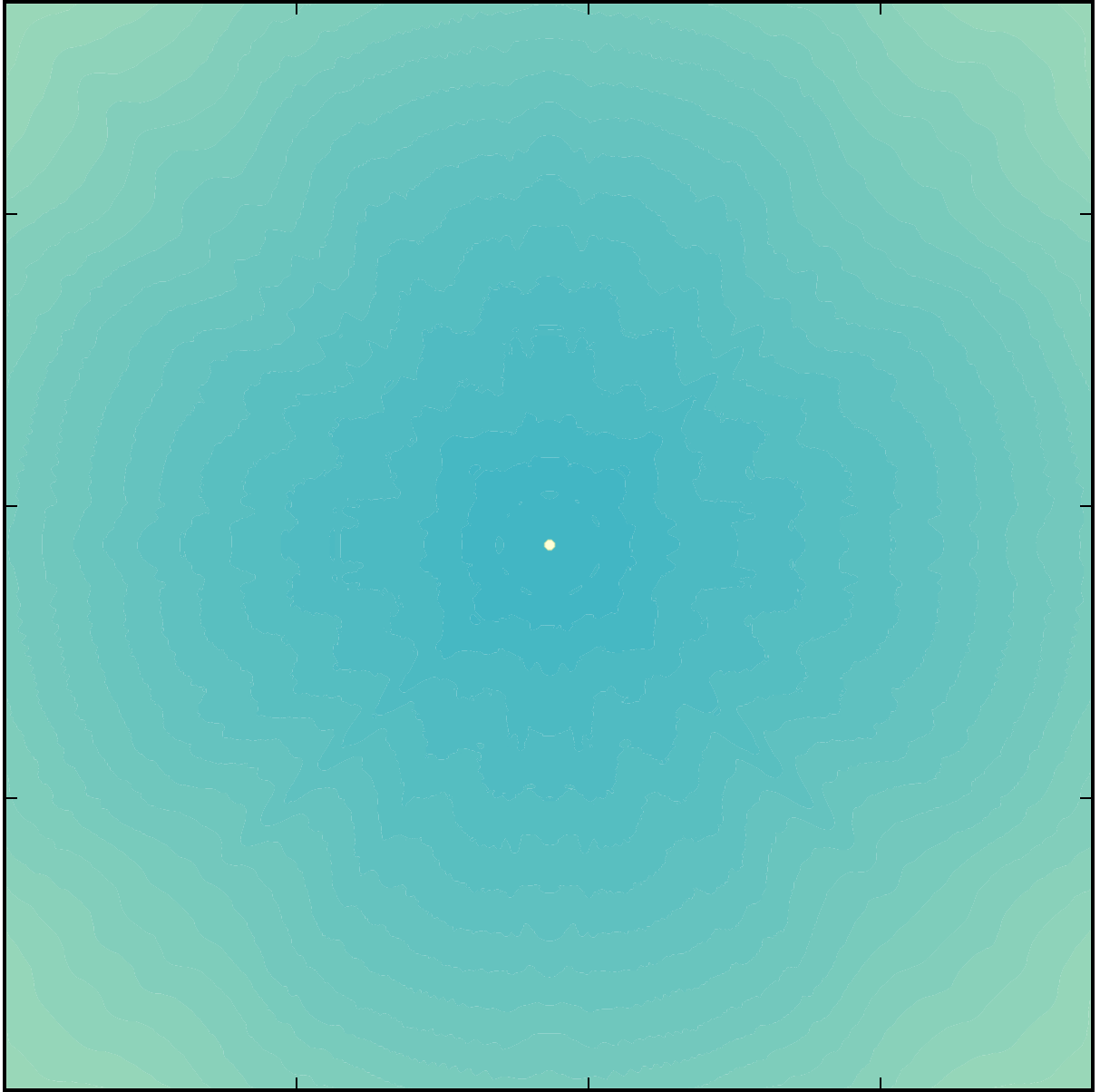}}
				\put(115,20){\line(1,0){84.4}}
				\put(115,17){\line(0,1){6}}
				\put(199.4,17){\line(0,1){6}}
				\put(115,5){\scriptsize{0.005 arcsec}}
			\end{picture}
		\end{subfigure}\\
		\begin{subfigure}{320\unitlength}
			\begin{picture}(320,358)
				\put(0,0){\includegraphics[trim=4.2cm 1.5cm 5.1cm 1.5cm, clip=true,width=320\unitlength]{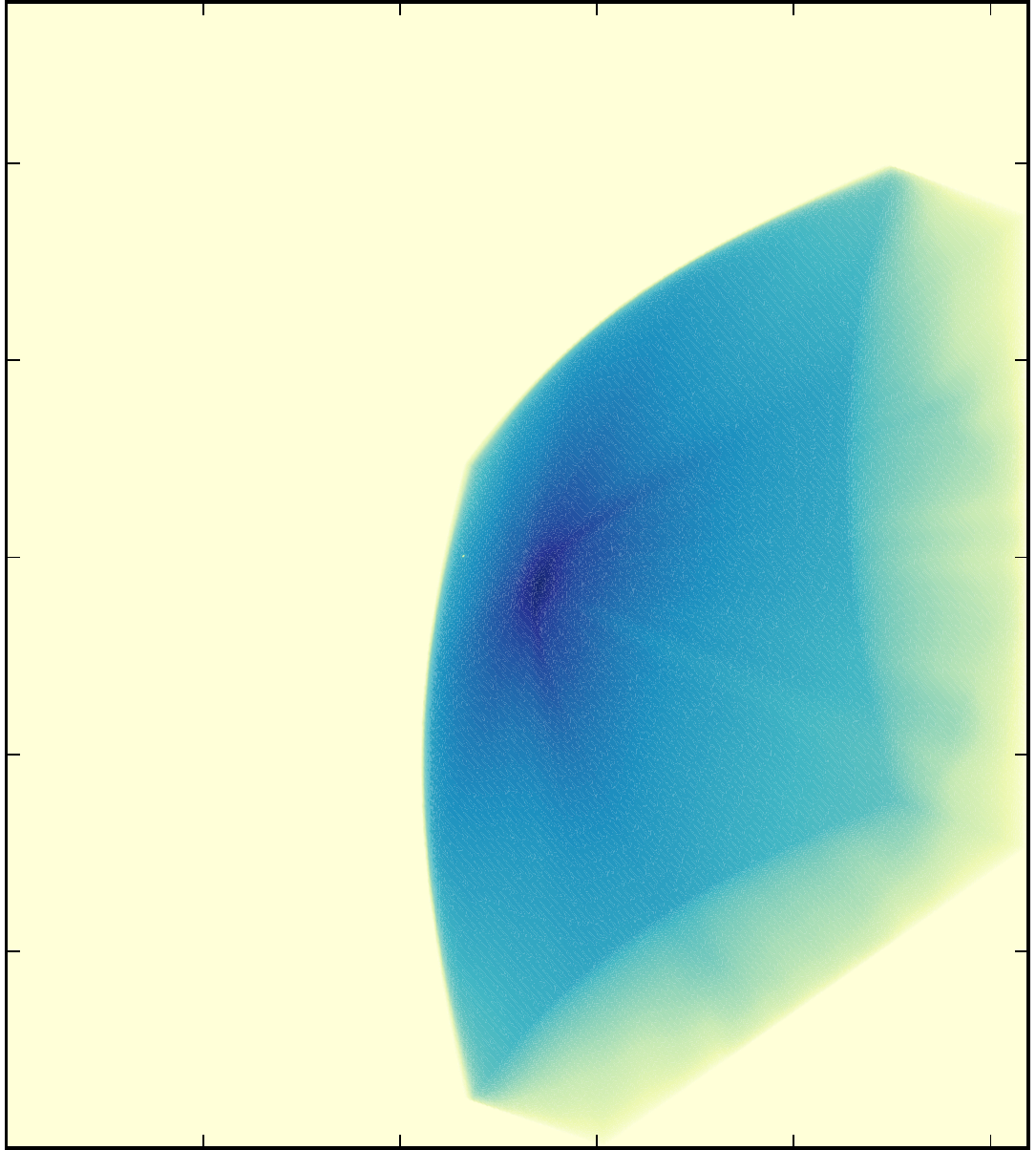}}
				%\put(115,20){\line(1,0){84.4}}
				%\put(115,17){\line(0,1){6}}
				%\put(199.4,17){\line(0,1){6}}
				%\put(115,5){\scriptsize{0.005 arcsec}}
			\end{picture}
		\end{subfigure}
		\begin{subfigure}{320\unitlength}
			\begin{picture}(320,358)
				\put(0,0){\includegraphics[trim=4.2cm 1.5cm 5.1cm 1.5cm, clip=true,width=320\unitlength]{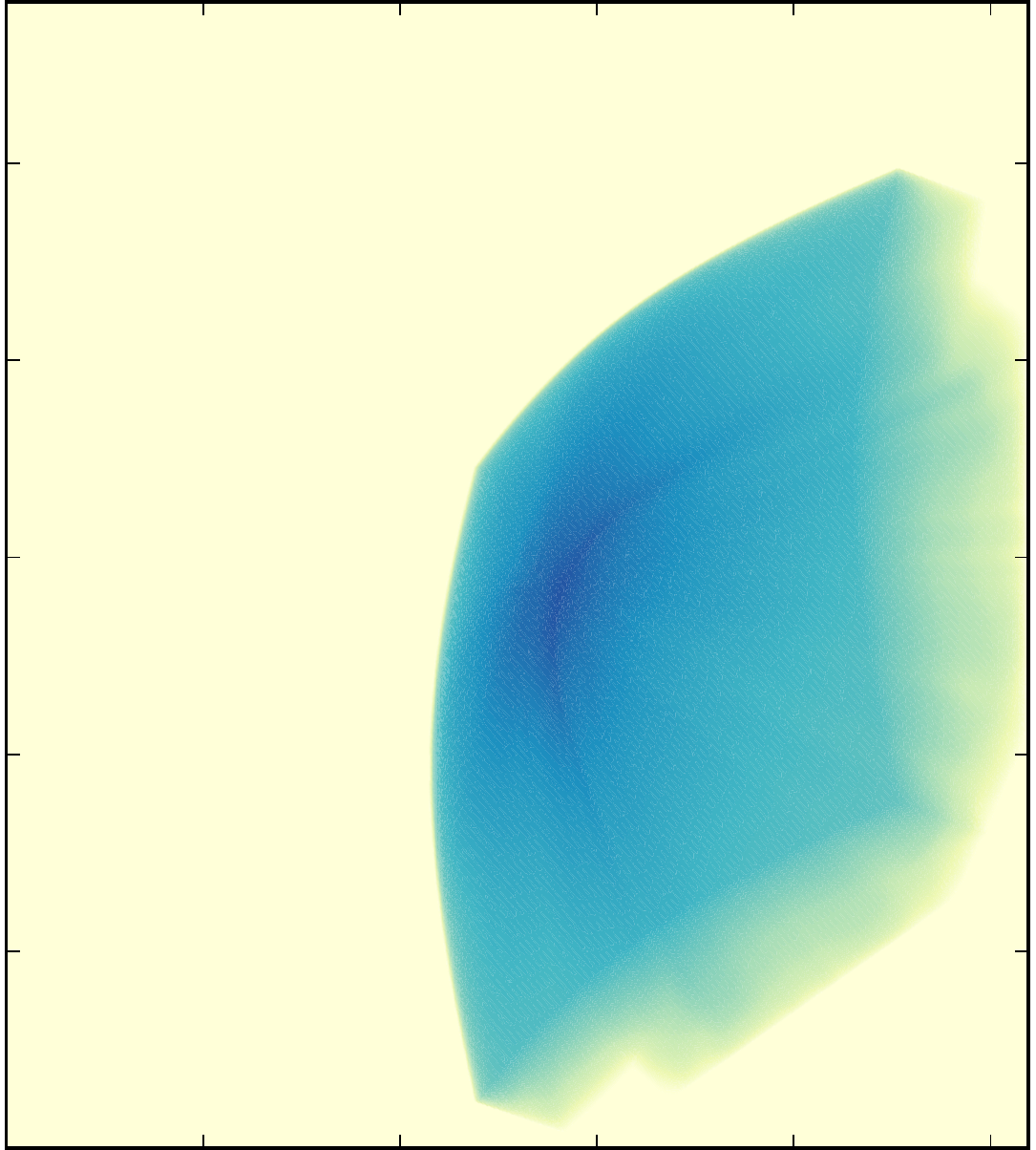}}
				%\put(115,20){\line(1,0){84.4}}
				%\put(115,17){\line(0,1){6}}
				%\put(199.4,17){\line(0,1){6}}
				%\put(115,5){\scriptsize{0.005 arcsec}}
			\end{picture}
		\end{subfigure}
		\begin{subfigure}{320\unitlength}	
			\begin{picture}(320,358)
				\put(0,0){\includegraphics[trim=4.2cm 1.5cm 5.1cm 1.5cm, clip=true,width=320\unitlength]{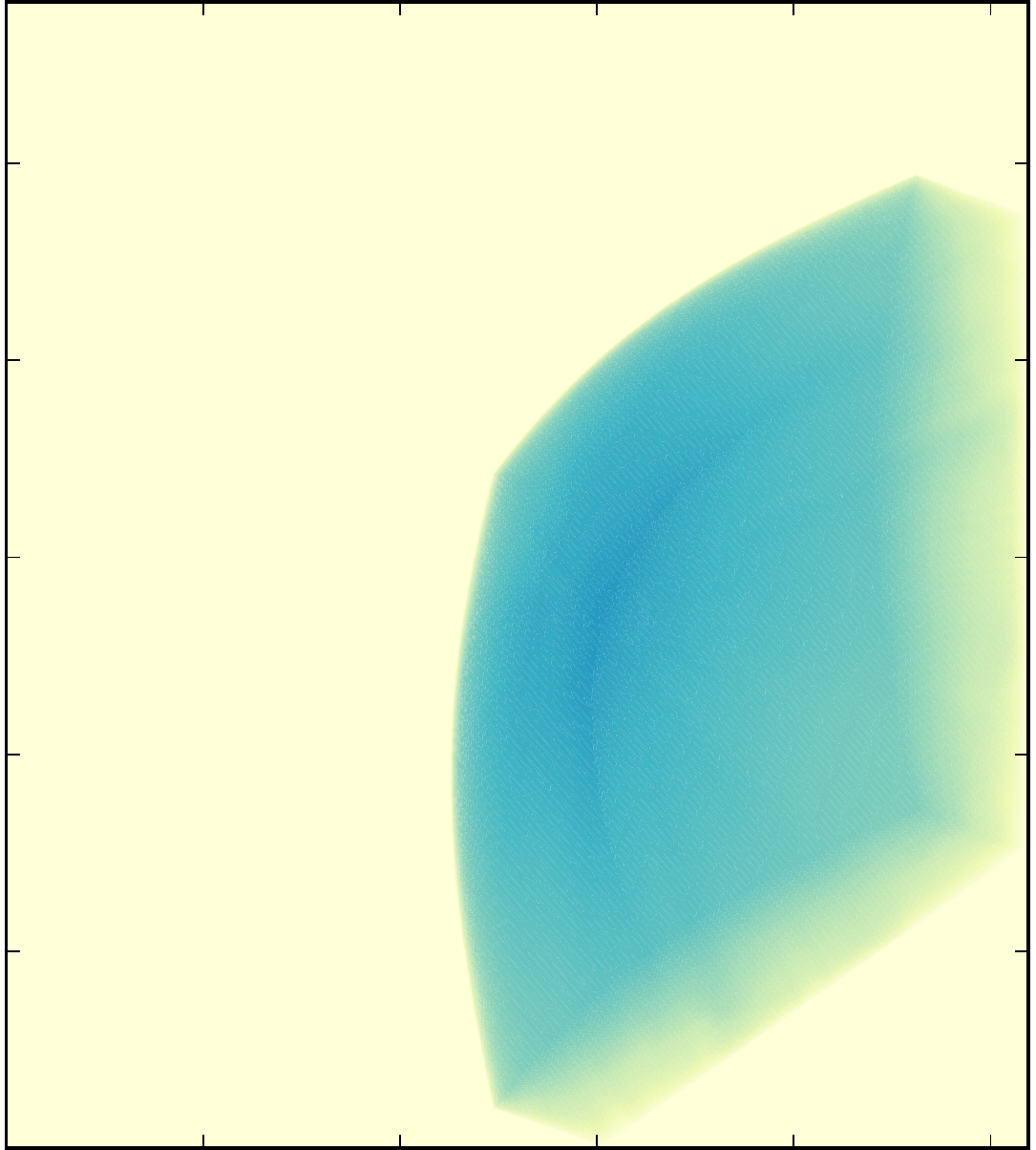}}
				%\put(115,20){\line(1,0){84.4}}
				%\put(115,17){\line(0,1){6}}
				%\put(199.4,17){\line(0,1){6}}
				%\put(115,5){\scriptsize{0.005 arcsec}}
			\end{picture}
		\end{subfigure}\\
		\begin{center}
		\begin{subfigure}{500\unitlength}	
			\begin{picture}(500,0)
				\put(0,-20){\includegraphics[trim=2cm 2cm 1cm 24cm, clip=true,width=500\unitlength]{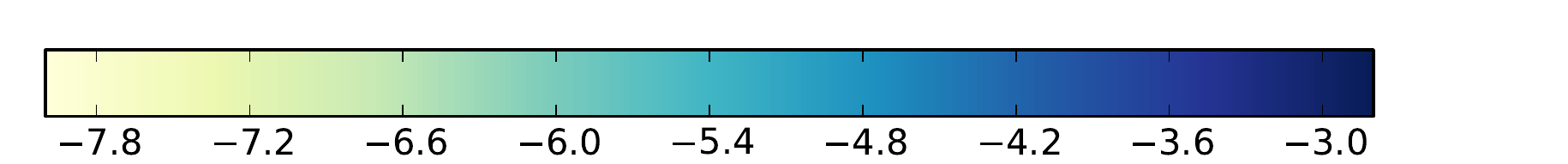}}
				\put(200,25){\footnotesize{ph m$^{-2}$ s$^{-1}$}}
			\end{picture}
		\end{subfigure}
		\end{center}	
\caption{Same as Figure \ref{em_varic} for for neutral pion decay.
\label{em_varp0}}
\end{figure*}

Figures \ref{em_varic} to \ref{em_varp0} provide further insight by illustrating the 2D projection maps of the individual nonthermal high-energy emission channels for different orientations of the system relative to the observer. The scaling of the colour bars has been kept identical in order to allow quick comparison.
The first and second rows show a case where the line of centers and the line of sight are aligned. Note the different sizes of the occultation region due to the stellar disk for $i=$90$^\circ$,$\Phi=$0$^\circ$ (smaller WR star in front of B star) and $i=$90$^\circ$,$\Phi=$180$^\circ$ (larger B star in front of WR star). 

Studying the first and second row clearly reveals the anisotropic character of the IC-component. Both orientations look identical (except for the occultation by the stellar disks) for photons from bremsstrahlung (Figure \ref{em_varbr}) and neutral pion decay (Figure \ref{em_varp0}). This is rather different for photons from IC-scattering where the relative angle of stellar positions and line of sight cause notable contrasts between the two orientations due to a different scattering angle (see first and second row in Figure \ref{em_varic}). 

The general shape and extent of the emission region becomes especially clear in the projection for an observer at $i=$45$^\circ$,$\Phi=$45$^\circ$ as it is shown in the third row of Figures \ref{em_varic} to \ref{em_varp0}. 

One of the most striking features is the missing photon emission from IC scattering close to the apex of case A, which is due to the lack of high-energy electrons in that region. Again, the dominance of the IC-component for case C in contrast to the dominance of the pion decay component for case A becomes apparent.

% ______________ABSORPTION

%_____________ EMISSION 0 0 ___________________
%******************* diff incl IC ***************************

%******************* diff incl brems ***************************
%******************* diff incl p0 ***************************

\subsubsection{Spectral Energy Distribution}
\begin{figure*}
	\setlength{\unitlength}{0.001\textwidth}
	\begin{subfigure}[c]{500\unitlength}
		\begin{picture}(500,370)
			\put(15,15){
				\includegraphics[width=\textwidth]{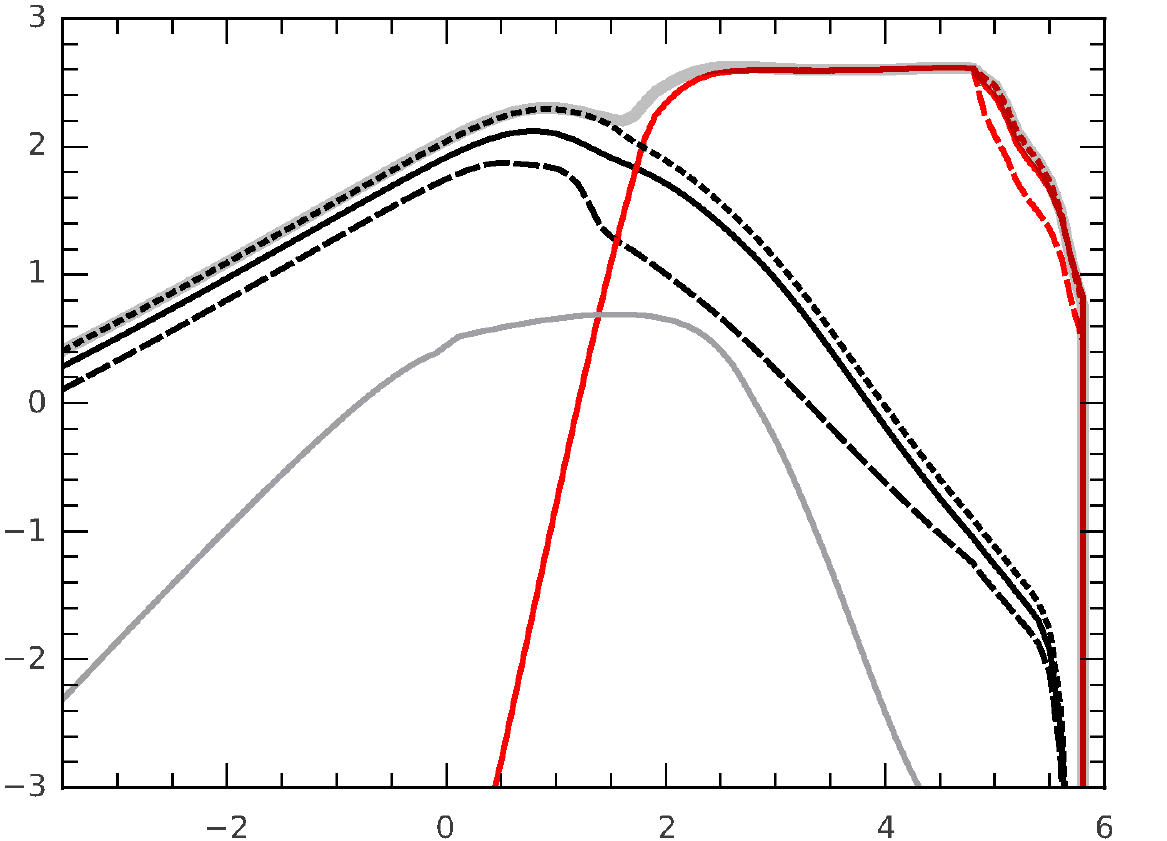}
			}
			\put(200,0){\footnotesize{log( E in MeV)}}
			\put(0,120){\rotatebox{90}{\footnotesize{log( $E^2N$ in MeV m$^{-2}$s$^{-1}$ )}}}
			\put(450,80){a)}
		\end{picture}
	\end{subfigure}
	\begin{subfigure}[c]{500\unitlength}
		\begin{picture}(500,370)
			\put(15,15){
				\includegraphics[width=\textwidth]{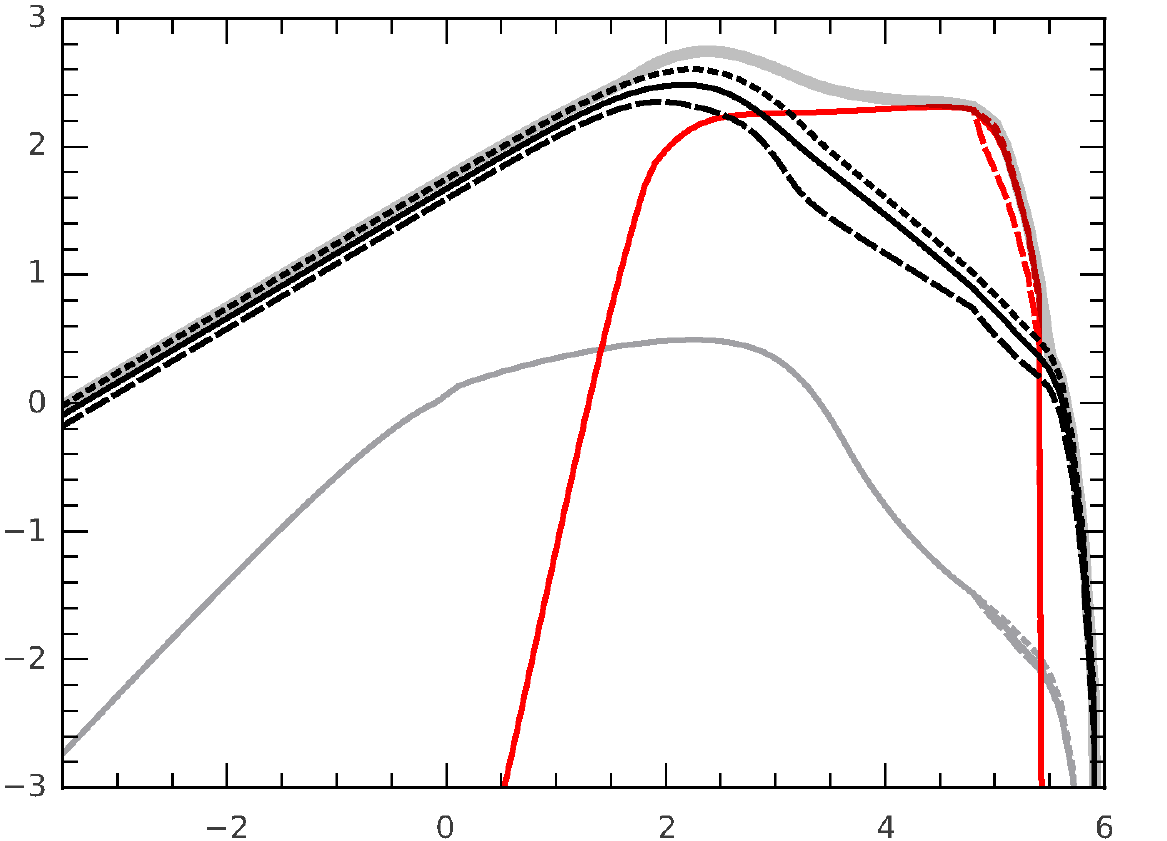}
			}
			\put(200,0){\footnotesize{log( E in MeV)}}
			\put(0,120){\rotatebox{90}{\footnotesize{log( $E^2N$ in MeV cm$^{-3}$ )}}}
			\put(450,80){b)}
		\end{picture}
	\end{subfigure}
	\begin{subfigure}[c]{500\unitlength}
		\begin{picture}(500,400)
			\put(15,15){
			\includegraphics[trim=0.cm 0.cm 0.cm 0cm,
				 clip=true,width=\textwidth]{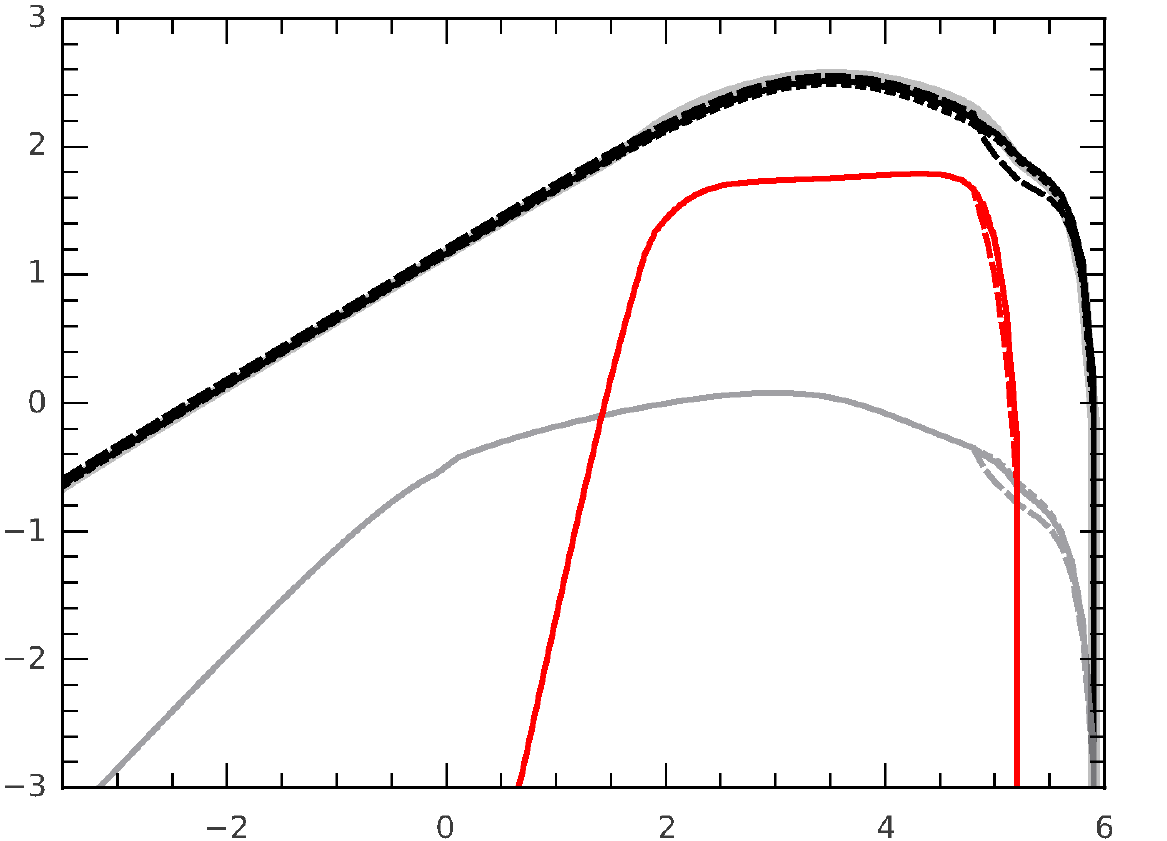}
			}
			\put(200,0){\footnotesize{log( E in MeV)}}
			\put(0,120){\rotatebox{90}{\footnotesize{log( $E^2N$ in MeV cm$^{-3}$ )}}}
			\put(450,80){c)}
		\end{picture}
	\end{subfigure}
	\begin{subfigure}[c]{500\unitlength}
		\begin{picture}(500,400)
			\put(15,15){
			\includegraphics[trim=0.cm 0.cm 0.cm 0cm,
				 clip=true,width=\textwidth]{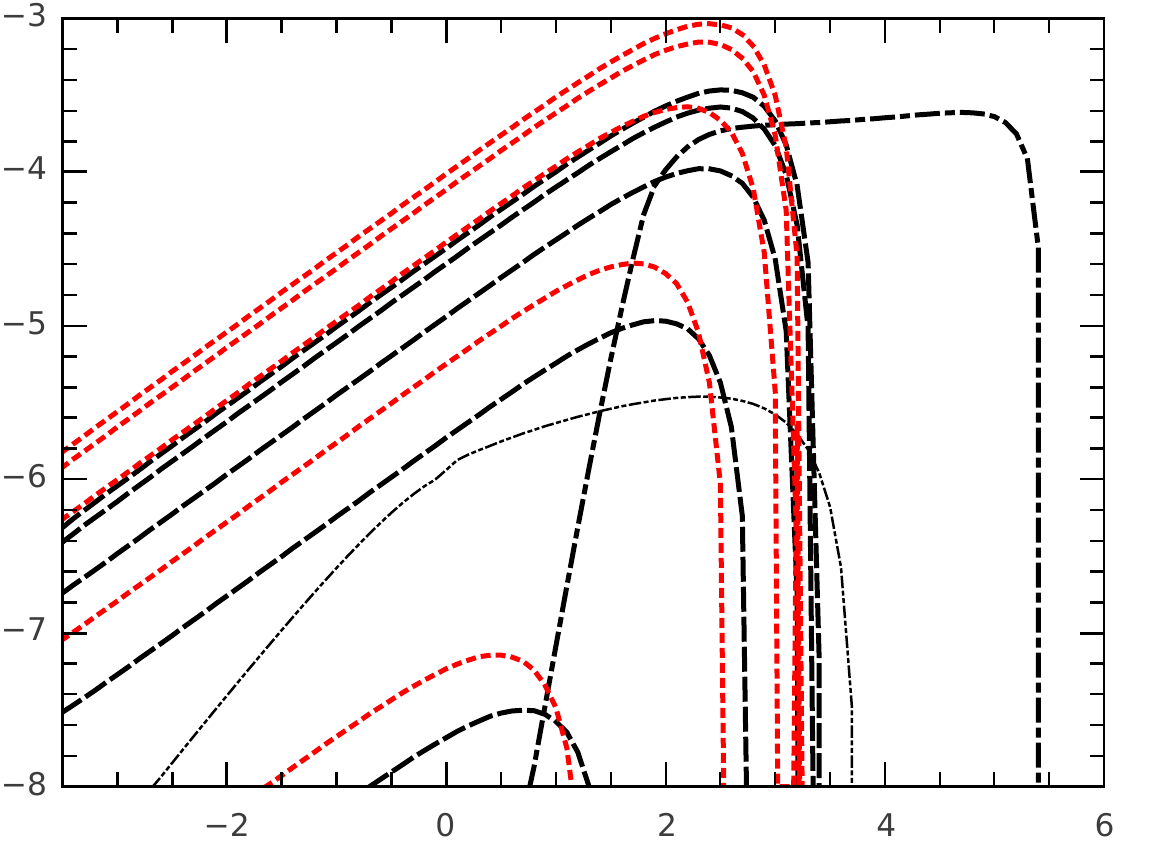}
			}
			\put(200,0){\footnotesize{log( E in MeV)}}
			\put(0,120){\rotatebox{90}{\footnotesize{log( $E^2N$ in MeV cm$^{-3}$ )}}}
			\put(450,80){d)}
		\end{picture}	
			\end{subfigure}
		\caption{a), b), c) Photon spectra emitted by all particles in the whole computational domain of cases A to C for IC scattering (black), bremsstrahlung (gray) and neutral pion decay (red) with $i=$0$^{\circ}$ (solid), $i=90^\circ$, $\Phi=0^\circ$ (dashed) and $i=90^\circ$, $\Phi=180^\circ$ (dotted). The sum of all three components (for $i=90^\circ$, $\Phi=180^\circ$) is indicated in shaded gray.\\
d) Photon spectra emitted by one single cell at the apex of case B. The black dashed lines show the IC flux from scattering in the radiation field of the WR star for an orientation of ($i=90^\circ$, $\Phi=180^\circ$), ($i=45^\circ$, $\Phi=180^\circ$), ($i=0^\circ$, $\Phi=0^\circ$), ($i=45^\circ$, $\Phi=0^\circ$) and ($i=80^\circ$, $\Phi=0^\circ$) in order of decreasing flux. The red dotted lines show the IC flux from scattering in the radiation field of the B star for an orientation of ($i=90^\circ$, $\Phi=0^\circ$), ($i=45^\circ$, $\Phi=0^\circ$), ($i=0^\circ$, $\Phi=0^\circ$), ($i=45^\circ$, $\Phi=180^\circ$) and ($i=80^\circ$, $\Phi=180^\circ$) in order of decreasing flux. The photon flux from neutral pion decay and bremsstrahlung is represented by the dash-dotted and double-dot-dashed line.
\label{gamspec1} }
\end{figure*}

We calculate the SED by summation over all emitting grid cells for a fixed energy throughout the computational domain. This is shown for case A, B, and C and various orientations in Figure \ref{gamspec1} a), b) and c). 
Figure \ref{gamspec1} d) explores the variation of the IC-component for a single cell at the apex of the WCR for case B as a function of orientation. It also discriminates between the two components of IC-emission, one being due to the scattering of electrons on the radiation field of the B star (red dotted curves), the other due to scattering on the radiation field of the WR star (dashed black). Whereas the first one is maximal at ($i=$90$^\circ$,$\Phi=$0$^\circ$) and disappears at ($i=$90$^\circ$,$\Phi=$180$^\circ$), the other behaves in the exact opposite manner. This is due to the dependence on the scattering angle $\theta_\mathrm{sc}$. For the case ($i=$90$^\circ$,$\Phi=$0$^\circ$) the scattering angle of photons from the B star is 180$^\circ$ . The corresponding IC emission from the apex of the  WCR is therefore at its maximum. For the photon field of the WR star the scattering angle is zero and there is no emission from the apex via IC scattering. 
The spectra of photons by neutral pion decay and bremsstrahlung remain the same, as they do not depend on the scattering angle. This last statement excludes the effects of photon photon opacity which is not taken into account in Figure \ref{gamspec1} d). 
Studying the spectra in Figure \ref{gamspec1} (a) to (c) (for case A to C) significant differences between small and large stellar separation become apparent. 

Case A shows a neutral pion component which dominates the spectrum at E $>$100 MeV for all orientations. The lack of high-energy electrons near the apex of the WCR leads to an early cutoff of the IC-component at $\sim$10 MeV, followed by a steady decline where high-energy electrons in the wings of the WCR contribute a low number of photons at higher energies. 
%At $\sim$100 MeV to $\sim$1 GeV, the high plasma densities in the WCR even allow the bremsstrahlung component to exceed the IC-component for some orientations. 
It is interesting to note that amongst the three depicted orientations, the IC flux is highest for ($i=90^\circ$, $\Phi=180^\circ$). The same can be seen in Figure \ref{em_varic} by comparing the first and second row for case A. Because of the lack of high-energy electrons at the apex, it is the wings of the WCR that mainly contribute to the IC emission at energies $>$100 MeV. In these regions, the scattering angle for the radiation field of the WR star is much more favourable for ($i=90^\circ$, $\Phi=180^\circ$) than for ($i=90^\circ$, $\Phi=0^\circ$), whereas the scattering angle for the B-star is unfavourable for both. This is an effect of the curvature of the WCR and the different distance from the WCR to the stars.  
Figure \ref{gamspec1} a) also shows effects of photon photon opacity in the spectrum of the neutral pion component which completely dominates the emission at high energies. Owing to the denser radiation field of the WR star, opacity effects are highest for the orientation ($i=90^\circ$, $\Phi=0^\circ$) in which emission from the apex has to pass by close to the star. The higher temperature of the star also causes the onset of photon photon absorption at lower energies than for the orientation ($i=90^\circ$, $\Phi=180^\circ$) for which high-energy photons from the apex come close to the cooler B star.
Due to the gap between the pion-bump and the cutoff of the IC spectrum, the total emitted photon spectrum shows a pronounced dip at $\sim$50 MeV.

Case B is already quite different from case A. Due to higher energies reached by the electrons, the cutoff in the IC component occurs at higher photon energies, causing a broad overlap with the neutral pion component. At the same time, lower plasma densities in the WCR lead to a lower flux for the bremsstrahlung and neutral pion component. The maximum energy of photons by bremsstrahlung increases with the maximum electron energy. As some regions in the wings of the WCR now produce sufficiently high electron energies, both, the IC and the bremsstrahlung component reach up to $\sim$1 TeV. Effects of photon-photon opacity become more pronounced for these components also. Again, they are largest for the case ($i=90^\circ$, $\Phi=0^\circ$) where the emission from the WCR has to pass by the luminous WR star. 

% ________________________ spectra  cell

For case C, a large population of high-energy electrons leads to complete dominance of the IC-component. Whereas lower plasma densities lead to still lower flux values for the neutral pion and the bremsstrahlung component, the latter along with the IC component again reaches energies of up to 1 TeV. As the IC flux also decreases with increasing stellar distance (and thus decreasing energy density of radiation), it shows lower flux values with respect to case B below $\sim$1 GeV. For higher energies, flux values unprecedented for lower stellar separations are reached. Effects of photon-photon opacity are visible for all three emission components. For the same reasons as above, they are strongest for ($i=90^\circ$, $\Phi=0^\circ$). Note that the IC component is now maximal for the case ($i=90^\circ$, $\Phi=0^\circ$) which can be understood by the increased density of high-energy electrons at the apex where the radiation field of the B star dominates. The latter has most favourable scattering angles for ($i=90^\circ$, $\Phi=180^\circ$) . Because of the effects of photon photon opacity, the face-on orientation shows the highest resulting IC flux at E$>$100 GeV.

Comparing all three cases, we find that (for an electron-proton injection ratio of $10^{-2}$) growing stellar separation leads to a transition from hadron-dominated emission (neutral pion decay) to lepton-dominated emission (IC) at energies E$>$100 MeV. The latter is inhibited for close stellar separations because of a lack of high-energy electrons at the apex of the WCR. As an additional effect, different viewing angles produce significant variation in the IC emission spectra. These dependencies on stellar separation and viewing angle can be quantified using integrated flux values for various energy bands.

\subsubsection{Integrated Fluxes}
By integrating over energy and space, we obtain the total emitted flux, which we computed for three energy intervals. Table \ref{fluxtable} provides flux values for three different orientations for case A, B, and C. The main characteristics in this table will be further discussed below.
\paragraph{The flux above 10 GeV}
As suggested by the spectral shapes discussed in the previous section, the dominant radiation process at E $>$ 10 GeV changes with stellar separation. From Table \ref{fluxtable} one can assess that, roughly, the total IC and bremsstrahlung fluxes increase by $\sim$ 3 orders of magnitude from case A to C, whereas the neutral pion component decreases on a much smaller scale. 
Only for case C, the IC-component dominates over the neutral pion component. Studying effects of changing orientation, we find that the variations in the bremsstrahlung and neutral pion component due to changing magnitude of photon photon opacity are very low. 
For the IC component, its anisotropic nature as well as the larger effect of absorption cause flux variations due to the orientation of the system of a factor of 3 for case A and lower factors for cases B and C. 

\paragraph{The flux above 100 MeV}
Including also lower energies down to 100 MeV, we find that growing stellar separation has the opposite effect than for the previous case. For E$>$10 GeV, higher maximum energies for electrons for large stellar separation led to an \textit{increase} of the IC and bremsstrahlung photon flux with increasing stellar separation. At lower energies, higher plasma densities in the WCR, along with denser radiation fields for smaller stellar separations, reverse this order and cause IC and bremsstrahlung photon fluxes to \textit{decrease} with increasing stellar separation. For E$>$100 MeV this transition is only partly realised, as we still find an increase in flux from case A to B for the IC component. However, all components decrease in their average flux from case B to C. There is no such reversal of proportionality for the neutral pion component.
We also find that the IC component is now dominant for cases B and C, albeit for some orientations of case B that are least favourable for IC emission, it is just a factor of 1.2 above the neutral pion component. For case A, the latter still dominates all possible orientations.
Variations due to changes in the viewing angle cause differences of a factor of $\sim$8 in the IC flux component for case A with a smaller contrast for cases B and C. 

\paragraph{The flux above 1 keV}
Including still lower energies, all components now show the identical trend that increasing stellar separation causes decreasing flux values. Furthermore, the IC component dominates for all three cases. The total flux of the neutral pion component is even below the total flux of the bremsstrahlung component. This can be understood in terms of the lack of photons from pion decay below a few MeV due to the shape of the pion bump.
Effects of varying orientation are still lower (factor $\sim$2) . However the previously observed order of larger fluxes for orientation ($i=$90$^\circ$,$\Phi=$180$^\circ$) than ($i=$90$^\circ$,$\Phi=$0$^\circ$) in case A and the opposite trend in case C remains in place.

\subsection{Effects of Orbital Motion}
\label{orbital}
We finally address the impact of circular orbital motion and the ensuing deformation of the WCR. As the effects increase with orbital velocity, we  choose the case of smallest stellar separation (case A) and let it evolve until the WCR  adjusts to the new situation and there is no further change in its curvature. We analyse the system after 1.5 orbits when the stars are once more on the $x$ axis. For the given stellar masses and a distance of 720 R$_\odot$, a Keplerian orbit has a period of $\sim$290 days. The orbital velocity of the stars is $\sim$263 km s$^{-1}$. With orbital motion, the two arms of the WCR (henceforth forward arm and trailing arm) develop considerable differences.

\begin{table*}
\begin{center}
\textbf{Integrated flux for various energy bands}\\
\end{center}
\centering
\begin{tabular}{r ccc|ccc|ccc}

\textit{($i$,$\Phi$)}&
\textit{$F_\mathrm{IC}^\mathrm{>10GeV}$}          & \textit{$F_\mathrm{br}^\mathrm{>10GeV}$}  &
\textit{$F_\mathrm{\pi^0}^\mathrm{>10GeV}$}          & \textit{$F_\mathrm{IC}^\mathrm{>100MeV}$}    &
\textit{$F_\mathrm{br}^\mathrm{>100MeV}$}  & \textit{$F_\mathrm{\pi^0}^\mathrm{>100MeV}$}  &
\textit{$F_\mathrm{IC}^\mathrm{>1keV}$} & \textit{$F_\mathrm{br}^\mathrm{>1keV}$} & \textit{$F_\mathrm{\pi^0}^\mathrm{>1keV}$}\\

\textbf{Case A}\\
($0^\circ$, $0^\circ$)   &3.2	&1.4				 &3.8 
    &33 &3.1  &3.4 
    &6.2 &6.5 &5.2 \\
($90^\circ$, $0^\circ$)  &1.4 &1.4				 &3.7 
    &6.3 &3.1  &3.4
    &4.1 &6.5  &5.2 \\
($90^\circ$, $180^\circ$) &4.6 &1.4				 &3.9 
    &49 &3.1  &3.4
    &8.2  &6.5	 &5.2 \\
    &$\times10^{-5}$ &$\times10^{-7}$ &$\times10^{-2}$ 
    &$\times10^{-2}$ &$\times10^{-2}$  & 
    &$\times10^{3}$  &$\times10^{1}$	 & \\
    
\textbf{Case B     }\\
($0^\circ$, $0^\circ$)   &18 &8.4 &1.9 
    &2.7 &2.9 &1.5 
    &3.0 &2.5 				&2.3 \\
($90^\circ$, $0^\circ$)  &9.8 &8.4	&1.8
    &1.8 &2.9 &1.5 
    &2.5 &2.5 				&2.3\\
($90^\circ$, $180^\circ$) &24 &8.5	&1.9 
    &3.6 &2.9 &1.5 
    &3.6  &2.5 				&2.3 \\
    &$\times10^{-4}$ &$\times10^{-6}$	&$\times10^{-2}$ 
    & &$\times10^{-2}$ & 
    &$\times10^{3}$  &$\times10^{1}$ &  \\
\textbf{Case C\qquad}\\
($0^\circ$, $0^\circ$)   &2.4	&6.3 &5.5 
    &2.0 &1.1 &4.4 
    &8.9 &6.8 				&6.6 \\
($90^\circ$, $0^\circ$)  &2.5 &6.3	&5.4 
    &2.2 &1.1 &4.4 
    &9.8 &6.8 				&6.6 \\
($90^\circ$, $180^\circ$) &2.2 &6.4	&5.5 
    &1.9 &1.1 &4.4 
    &8.7  &6.8 				&6.6 \\
    &$\times10^{-2}$ &$\times10^{-5}$	&$\times10^{-3}$ 
    & &$\times10^{-2}$ &$\times10^{-1}$ 
    &$\times10^{2}$  & 				&$\times10^{-1}$ \\
\end{tabular}
\caption{Integrated flux values in units of m$^{-2}$s$^{-1}$ as obtained for the chosen electron-proton injection ratio of $10^{-2}$ for three different stellar separations (case A, B, C), as well as three different viewing angles. \label{fluxtable}}
\end{table*}

%********** ORBIT case A HD + particles ***********

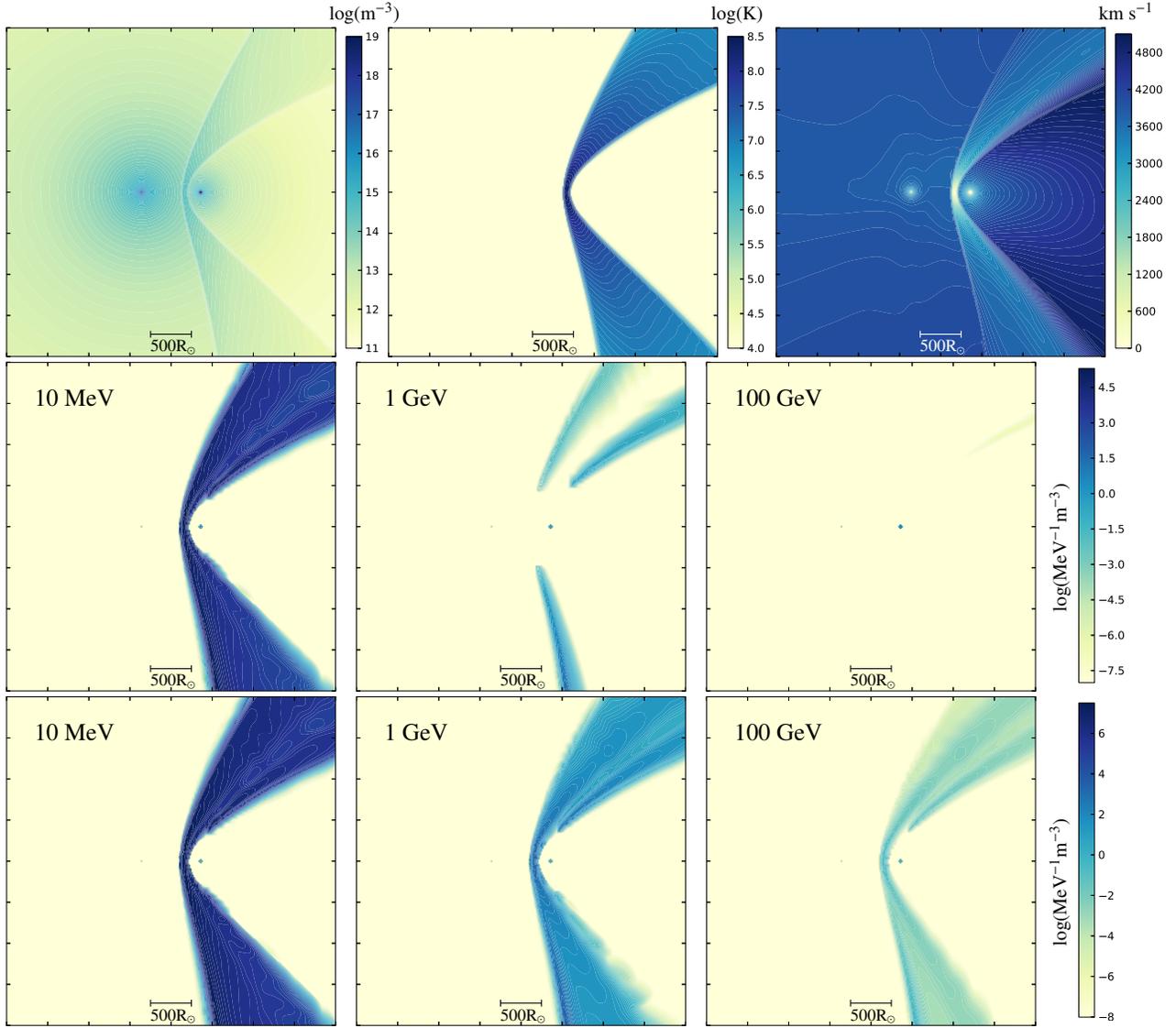
\begin{figure*}
	\setlength{\unitlength}{0.001\textwidth}
		\begin{subfigure}{290\unitlength}
			\begin{picture}(290,290)
					\put(0,0){\includegraphics[trim=2.9cm 1.5cm 5.1cm 1.5cm, clip=true,width=290\unitlength]{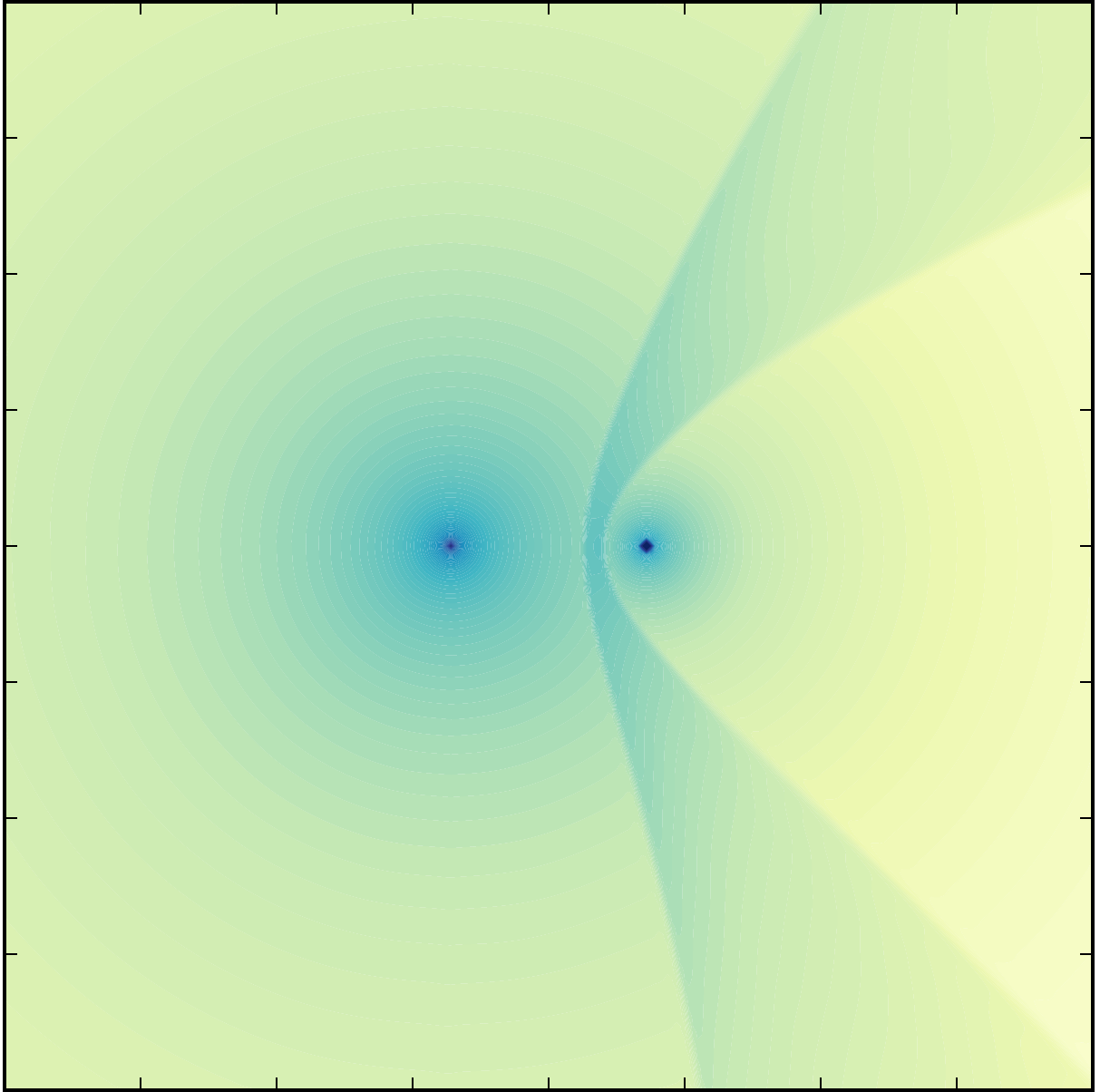}}
				\put(126.7,20){\line(1,0){35.625}}
				\put(126.7,17){\line(0,1){6}}
				\put(162.325,17){\line(0,1){6}}
				\put(126.7,5){\scriptsize{500R$_\odot$}}
			\end{picture}
		\end{subfigure}
		\begin{subfigure}{20\unitlength}
		\begin{picture}(20,290)
				\put(-5,5){\includegraphics[trim=15.9cm 1.4cm 2.4cm 1.3cm, clip=true,height=280\unitlength]{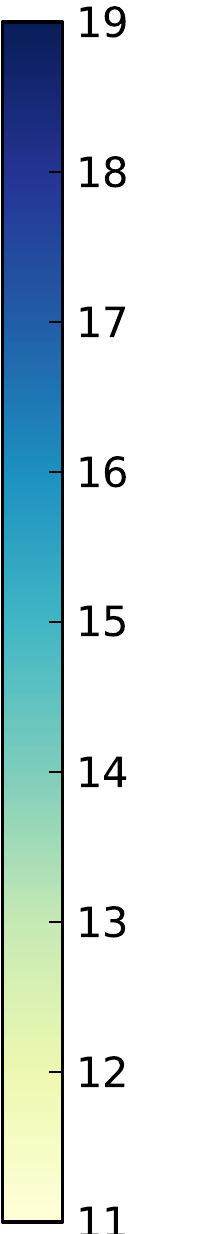}}
			\put(-19,295){{\footnotesize{log(m$^{-3}$)}}}
		\end{picture}
		\end{subfigure}
		\begin{subfigure}{290\unitlength}
			\begin{picture}(290,290)
				\put(0,0){\includegraphics[trim=2.9cm 1.5cm 5.1cm 1.5cm, clip=true,width=290\unitlength]{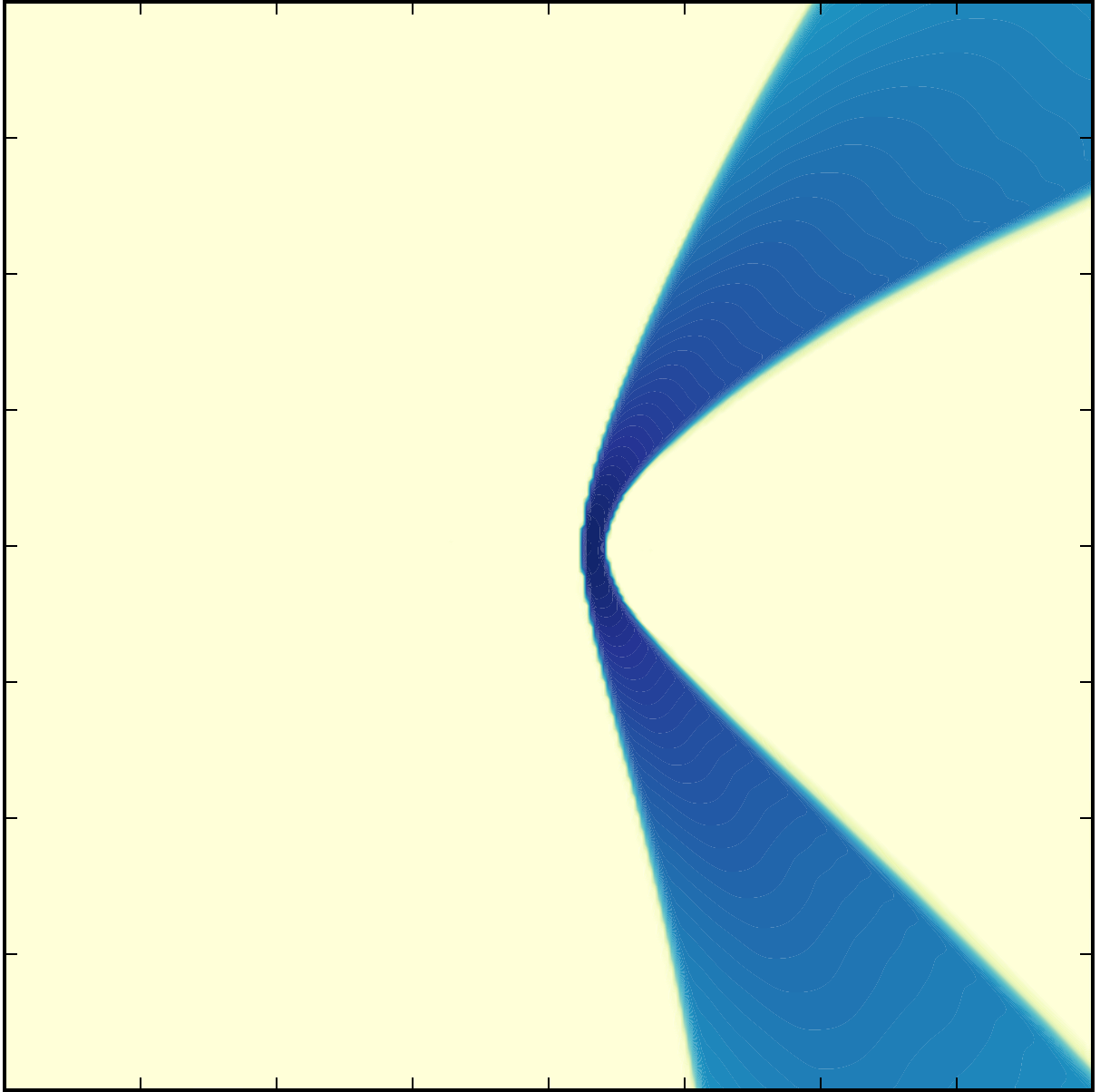}}
				\put(126.7,20){\line(1,0){35.625}}
				\put(126.7,17){\line(0,1){6}}
				\put(162.325,17){\line(0,1){6}}
				\put(126.7,5){\scriptsize{500R$_\odot$}}
			\end{picture}
		\end{subfigure}
		\begin{subfigure}{25\unitlength}
		\begin{picture}(20,290)
				\put(-5,5){\includegraphics[trim=15.9cm 1.4cm 2.4cm 1.3cm, clip=true,height=280\unitlength]{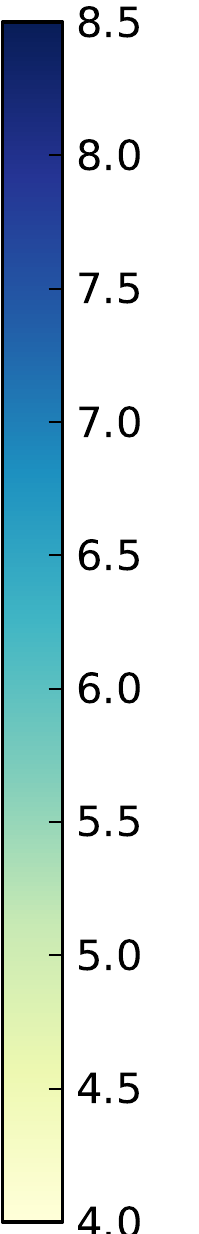}}
			\put(-19,295){\footnotesize{log(K)}}
		\end{picture}
	\end{subfigure}
		\begin{subfigure}{290\unitlength}
			\begin{picture}(290,290)
				\put(0,0){\includegraphics[trim=2.9cm 1.5cm 5.1cm 1.5cm, clip=true,width=290\unitlength]{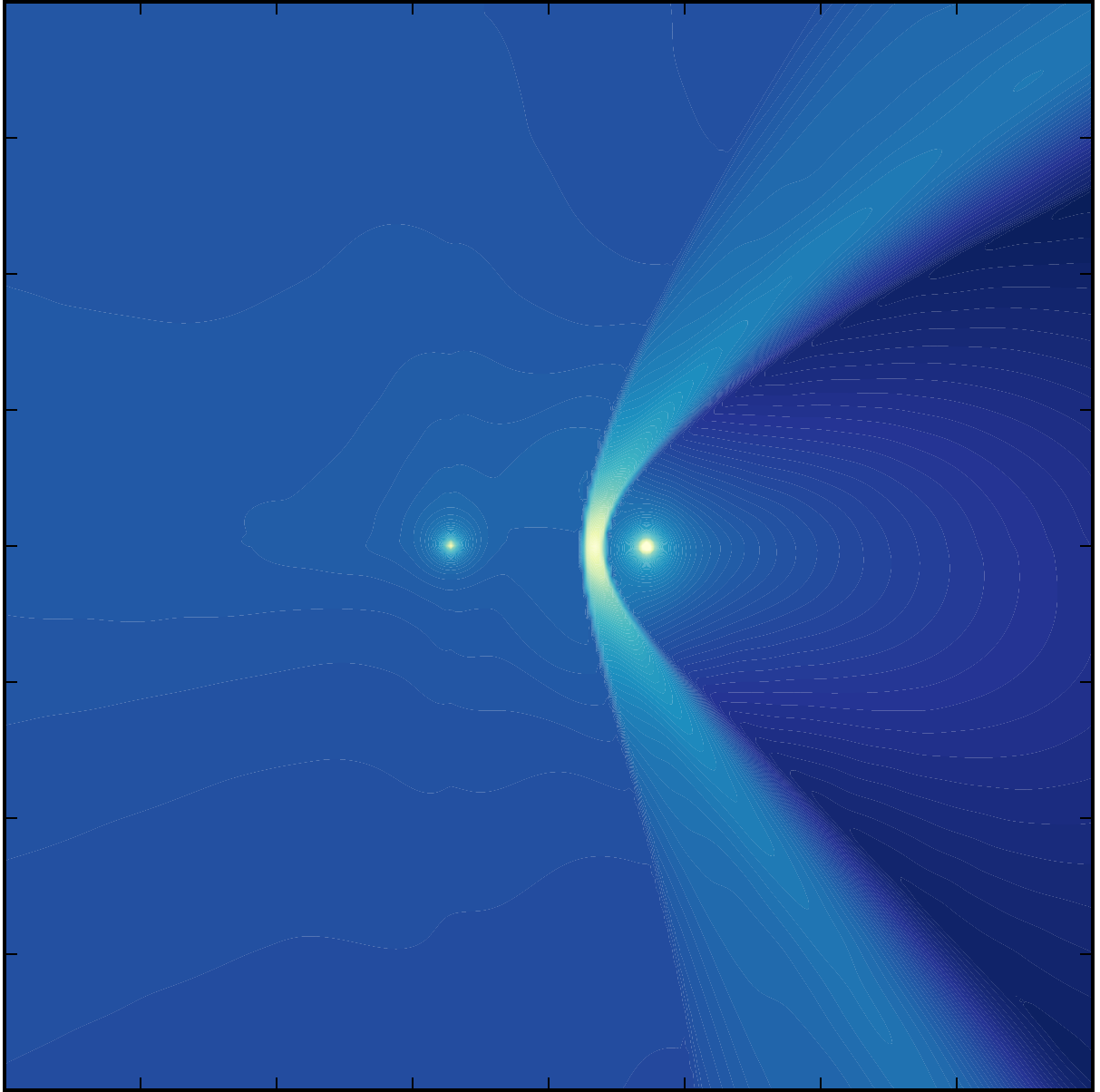}}
				\put(126.7,20){{\color{white}\line(1,0){35.625}}}
				\put(126.7,17){{\color{white}\line(0,1){6}}}
				\put(162.325,17){{\color{white}\line(0,1){6}}}
				\put(126.7,5){\scriptsize{{\color{white}500R$_\odot$}}}
			\end{picture}
		\end{subfigure}
		\begin{subfigure}{20\unitlength}
		\begin{picture}(20,290)
				\put(-5,5){\includegraphics[trim=15.9cm 1.4cm 2.4cm 1.4cm, clip=true,height=280\unitlength]{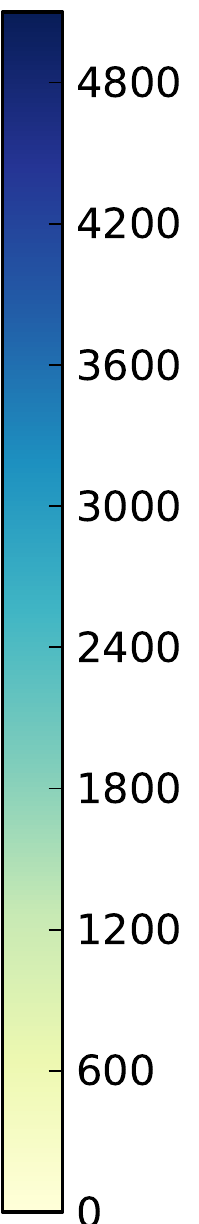}}
			\put(-19,295){\footnotesize{km s$^{-1}$}}
		\end{picture}
	\end{subfigure}
		\begin{subfigure}{300\unitlength}
			\begin{picture}(290,290)
					\put(0,0){\includegraphics[trim=2.9cm 1.5cm 5.1cm 1.5cm, clip=true,width=290\unitlength]{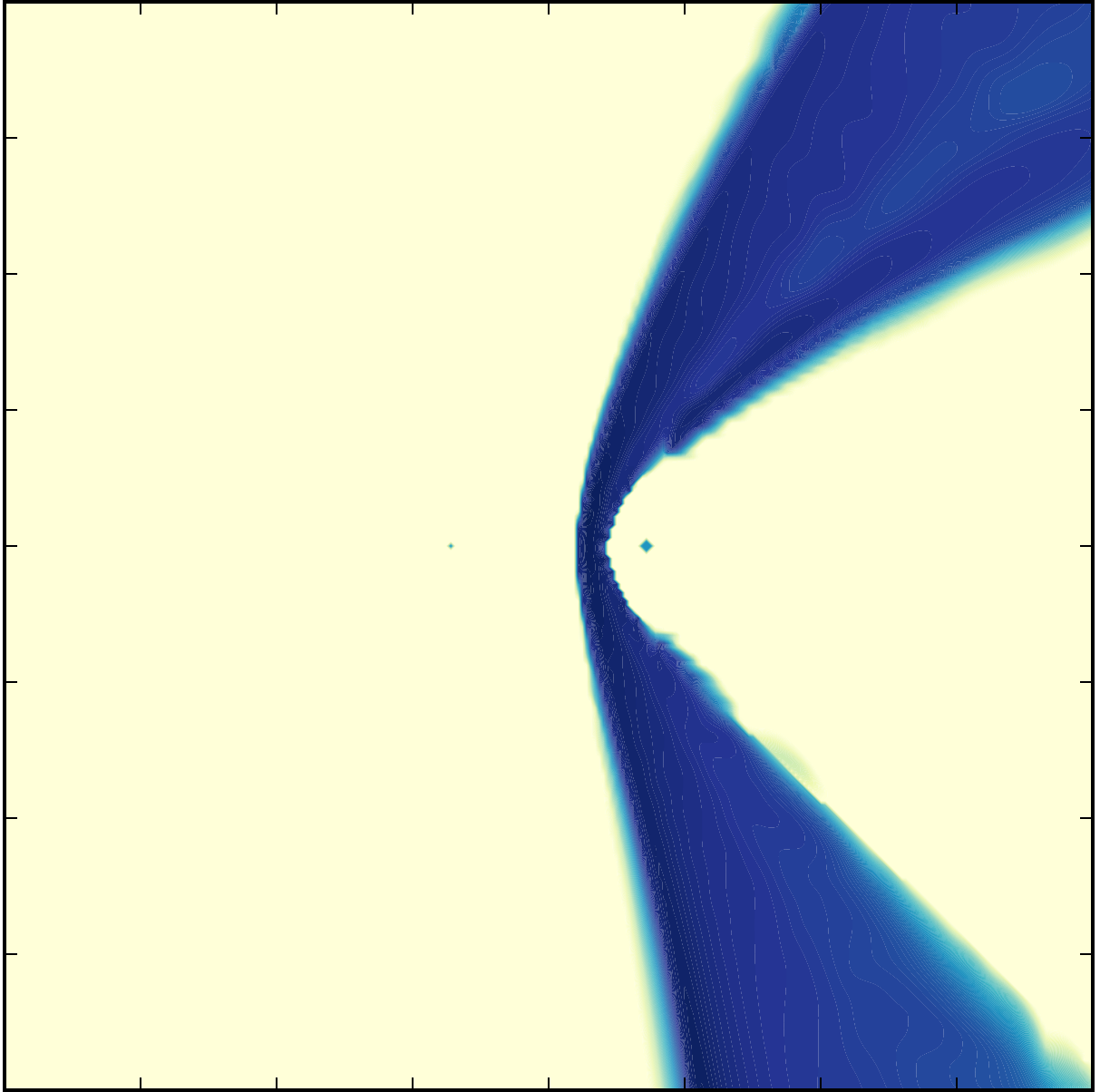}}
				\put(20,250){ 10 MeV}
				\put(126.7,20){\line(1,0){35.625}}
				\put(126.7,17){\line(0,1){6}}
				\put(162.325,17){\line(0,1){6}}
				\put(126.7,5){\scriptsize{500R$_\odot$}}
			\end{picture}
		\end{subfigure}
		\begin{subfigure}{300\unitlength}
			\begin{picture}(290,290)
					\put(0,0){\includegraphics[trim=2.9cm 1.5cm 5.1cm 1.5cm, clip=true,width=290\unitlength]{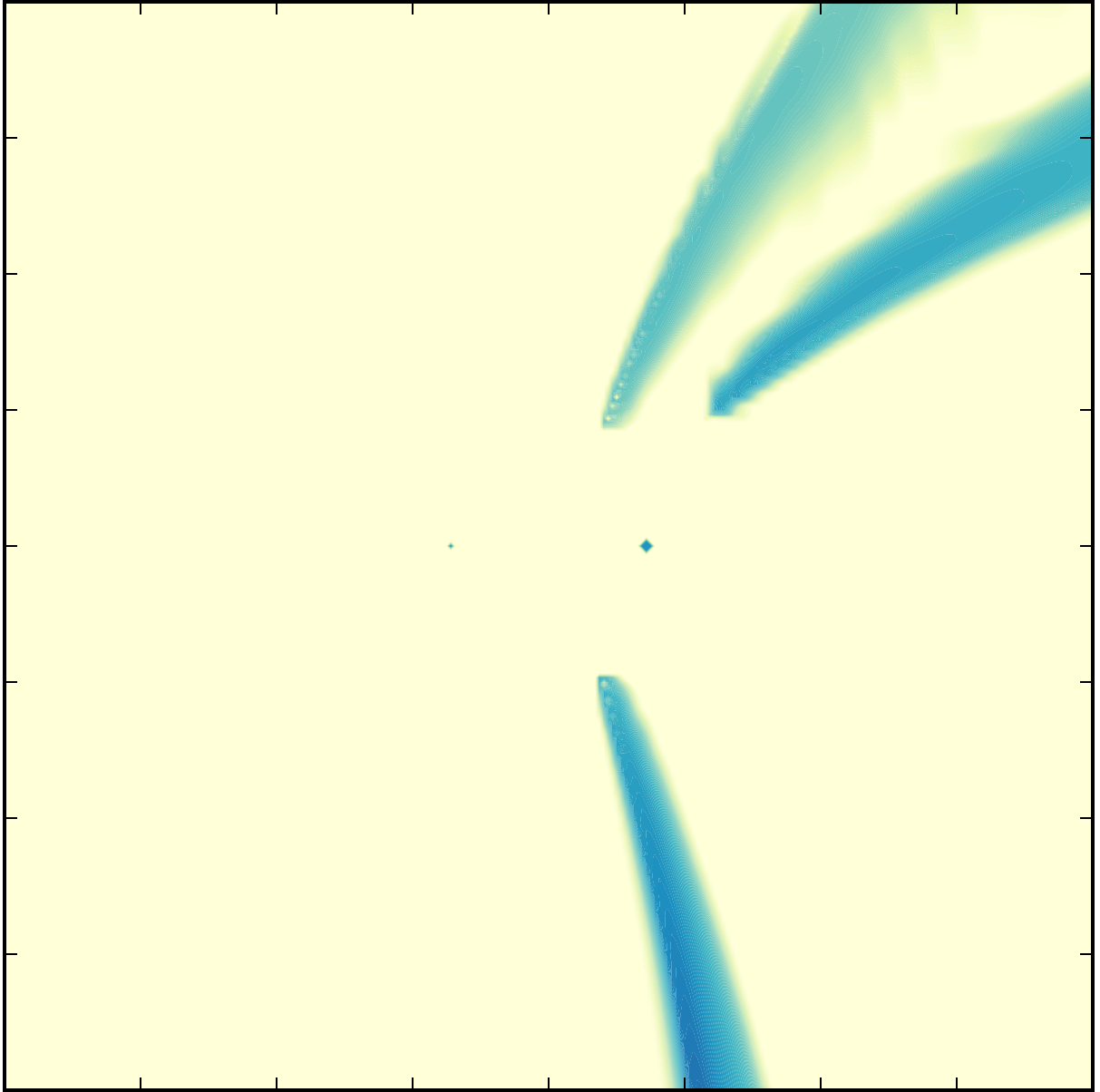}}
				\put(20,250){ 1 GeV}
				\put(126.7,20){\line(1,0){35.625}}
				\put(126.7,17){\line(0,1){6}}
				\put(162.325,17){\line(0,1){6}}
				\put(126.7,5){\scriptsize{500R$_\odot$}}
			\end{picture}
				\end{subfigure}
		\begin{subfigure}{300\unitlength}
			\begin{picture}(290,290)
			\put(0,0){\includegraphics[trim=2.9cm 1.5cm 5.1cm 1.5cm, clip=true,width=290\unitlength]{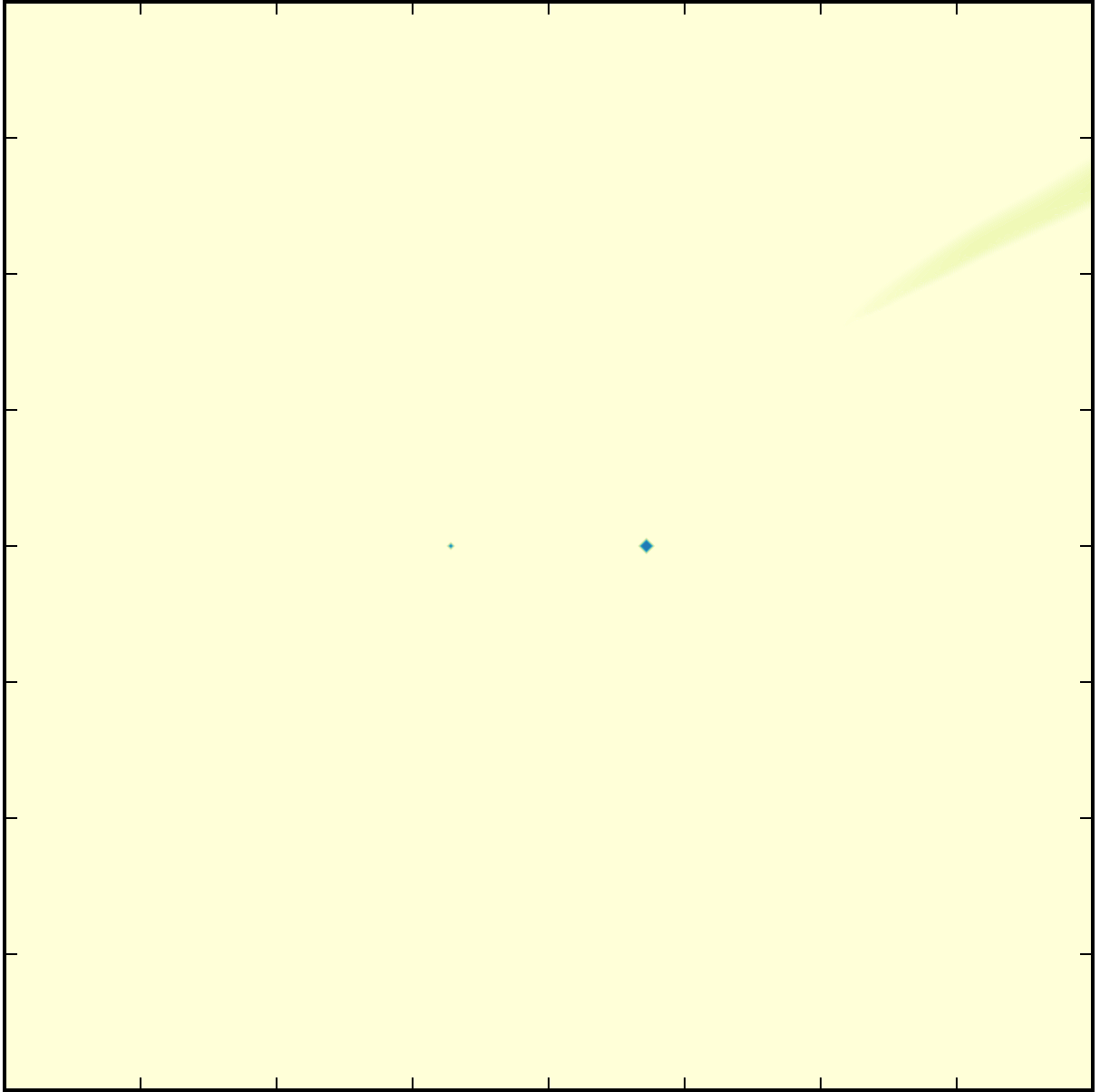}}
				\put(20,250){ 100 GeV}
				\put(126.7,20){\line(1,0){35.625}}
				\put(126.7,17){\line(0,1){6}}
				\put(162.325,17){\line(0,1){6}}
				\put(126.7,5){\scriptsize{500R$_\odot$}}
			\end{picture}
		\end{subfigure}	
		\begin{subfigure}{70\unitlength}
		\begin{picture}(75,290)
				\put(20,5){\includegraphics[trim=15.9cm 1.4cm 2.4cm 1.4cm, clip=true,height=280\unitlength]{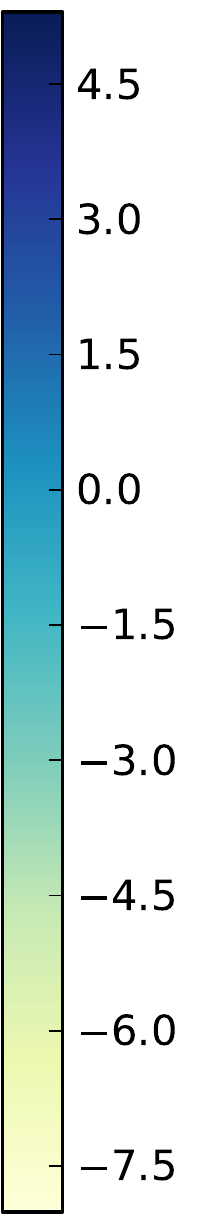}}
			\put(-5,70){\footnotesize{\rotatebox{90}{log(MeV$^{-1}$m$^{-3}$)}}}
		\end{picture}
	\end{subfigure}\\
		\begin{subfigure}{300\unitlength}
			\begin{picture}(290,290)
				\put(0,0){\includegraphics[trim=2.9cm 1.5cm 5.1cm 1.5cm, clip=true,width=290\unitlength]{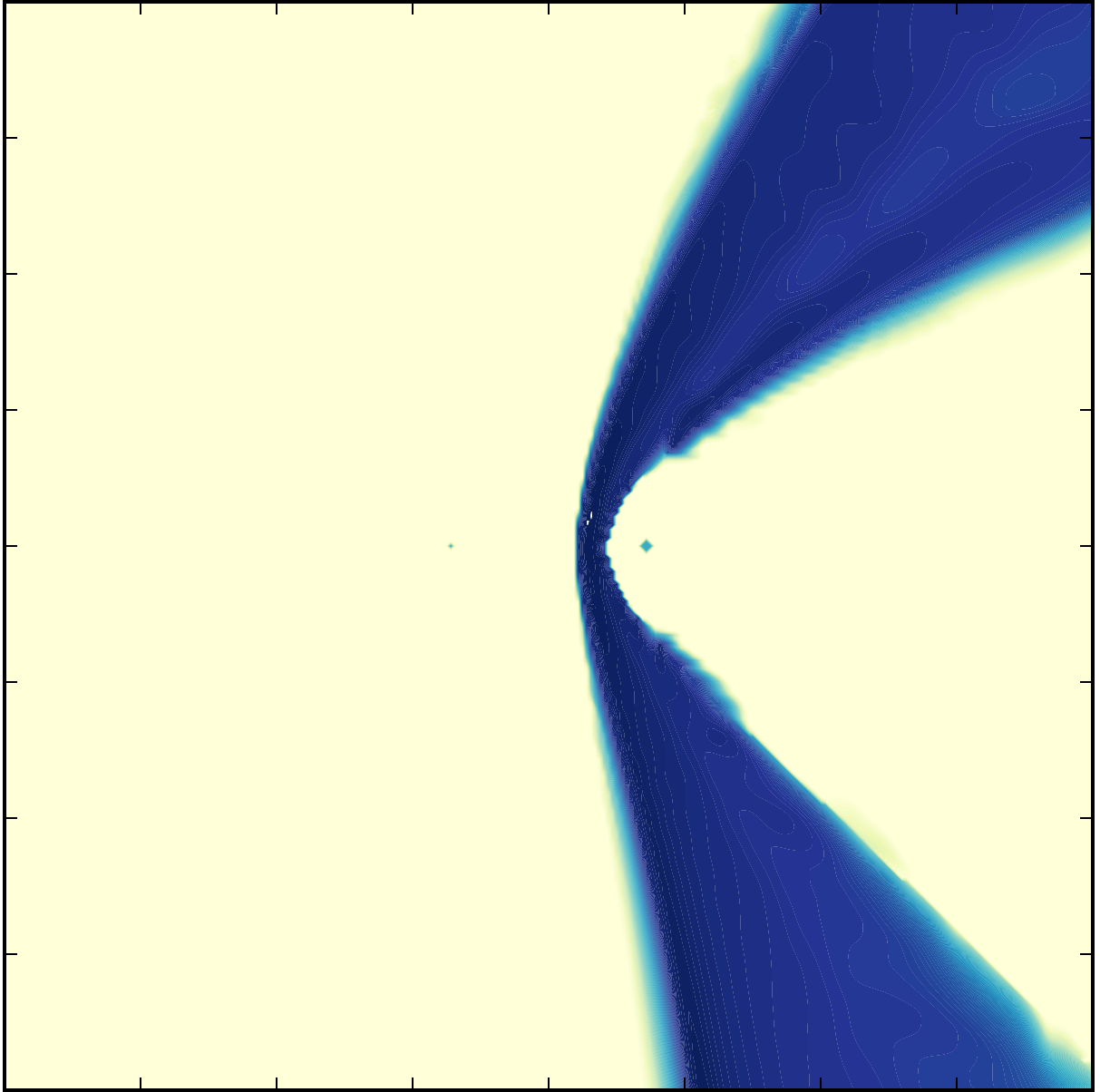}}
				\put(20,250){ 10 MeV}
				\put(126.7,20){\line(1,0){35.625}}
				\put(126.7,17){\line(0,1){6}}
				\put(162.325,17){\line(0,1){6}}
				\put(126.7,5){\scriptsize{500R$_\odot$}}
			\end{picture}
		\end{subfigure}
		\begin{subfigure}{300\unitlength}
			\begin{picture}(290,290)
				\put(0,0){\includegraphics[trim=2.9cm 1.5cm 5.1cm 1.5cm, clip=true,width=290\unitlength]{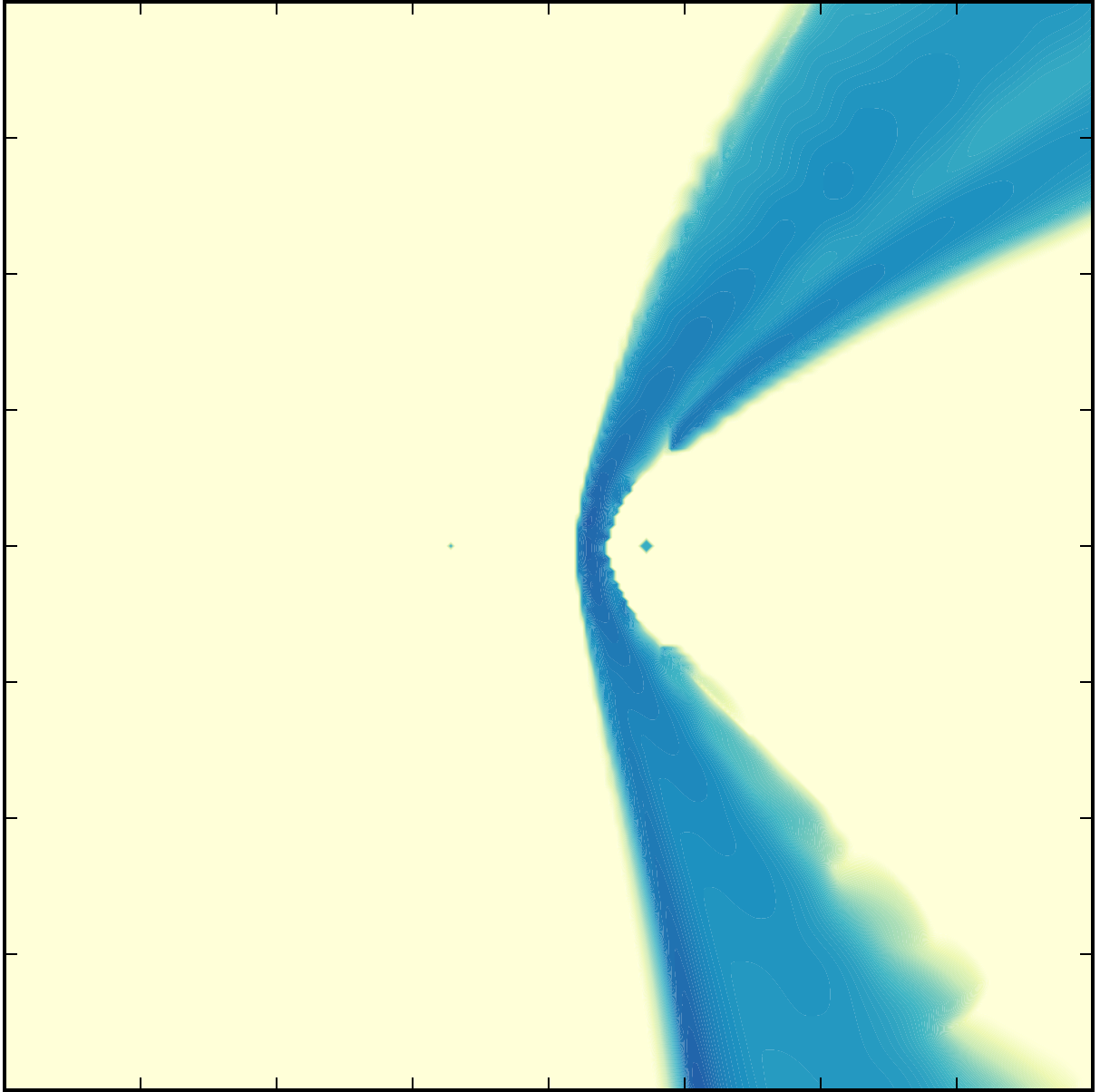}}
				\put(20,250){ 1 GeV}
				\put(126.7,20){\line(1,0){35.625}}
				\put(126.7,17){\line(0,1){6}}
				\put(162.325,17){\line(0,1){6}}
				\put(126.7,5){\scriptsize{500R$_\odot$}}
			\end{picture}
		\end{subfigure}
		\begin{subfigure}{300\unitlength}	
			\begin{picture}(290,290)
				\put(0,0){\includegraphics[trim=2.9cm 1.5cm 5.1cm 1.5cm, clip=true,width=290\unitlength]{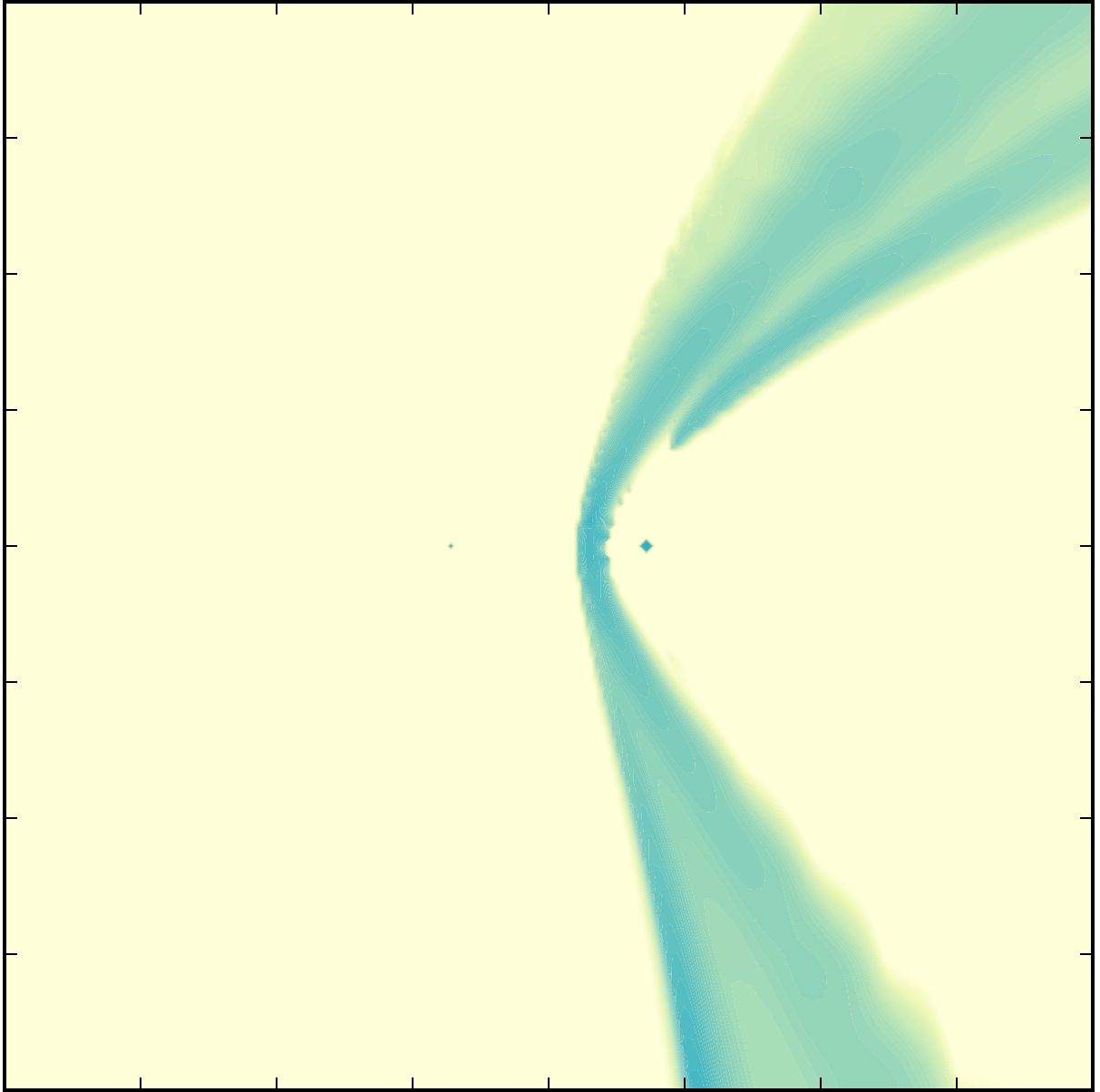}}
				\put(20,250){ 100 GeV}
				\put(126.7,20){\line(1,0){35.625}}
				\put(126.7,17){\line(0,1){6}}
				\put(162.325,17){\line(0,1){6}}
				\put(126.7,5){\scriptsize{500R$_\odot$}}
			\end{picture}
		\end{subfigure}
		\begin{subfigure}{20\unitlength}
		\begin{picture}(75,290)
				\put(20,5){\includegraphics[trim=15.9cm 1.4cm 2.4cm 1.4cm, clip=true,height=280\unitlength]{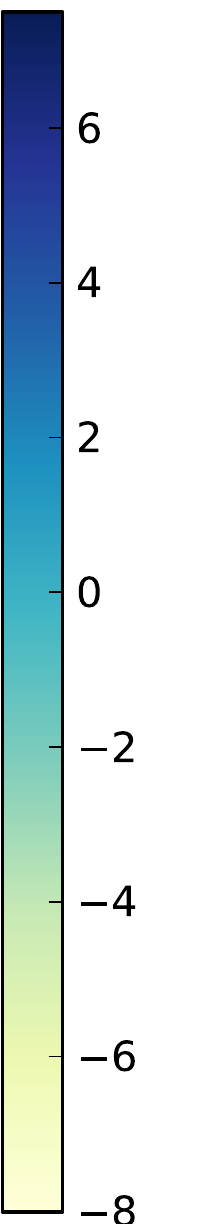}}
			\put(-5,70){\footnotesize{\rotatebox{90}{log(MeV$^{-1}$m$^{-3}$)}}}
		\end{picture}
	\end{subfigure}
\caption{First row: HD quantities particle density, temperature and absolute velocity for case A with orbital motion. Second row: Differential number density of electrons for three different values of kinetic particle energy. Third row: the same for protons. The plots show the $x$--$y$ plane of a $256\times256\times256$ simulation  at z=0.  \label{orbit_hA}}
\end{figure*}

Figure \ref{orbit_hA} shows density, temperature and velocity, as well as particle distributions for different energies on the $x$--$y$ plane (the stars move counter-clockwise). Concerning the hydrodynamic wind variables, the main difference to the stationary case (see Figure \ref{hydros}) lies not in minimal and maximal values but in the now asymmetric shape of the WCR. In paper I, we found that the significant contrasts in the particle distribution between forward and trailing arm are mainly due to the different angle between the stars and the edge of the WCR, resulting in different shock velocities and (as $\dot{E}_\mathrm{DSA}\propto V_\mathrm{Shock}^2$) in different maximum particle energies. For the present case, the ensuing reduction of the DSA rate in the trailing arm of the shock front toward the B star results in a lack of electrons already at 1 GeV (see center plot of Figure \ref{orbit_hA}). In the forward arm, the opposite effect occurs. As wind velocity components are slightly higher close to the edge of the computational domain in the forward arm, an increased DSA rate can accelerate electrons to energies up to 100 GeV. (Note the thin shaded area at $\sim$-6 log(MeV$^{-1}$m$^{-3}$) in the center right plot of Figure \ref{orbit_hA}.) There, electrons reach higher energies than in the case where orbital motion is not taken into account. 
Although we also find a certain degree of asymmetry for protons, the effects of orbital motion are less severe.

\begin{figure*}
	\setlength{\unitlength}{0.001\textwidth}
		\begin{subfigure}{990\unitlength}
			\includegraphics[width=\textwidth]{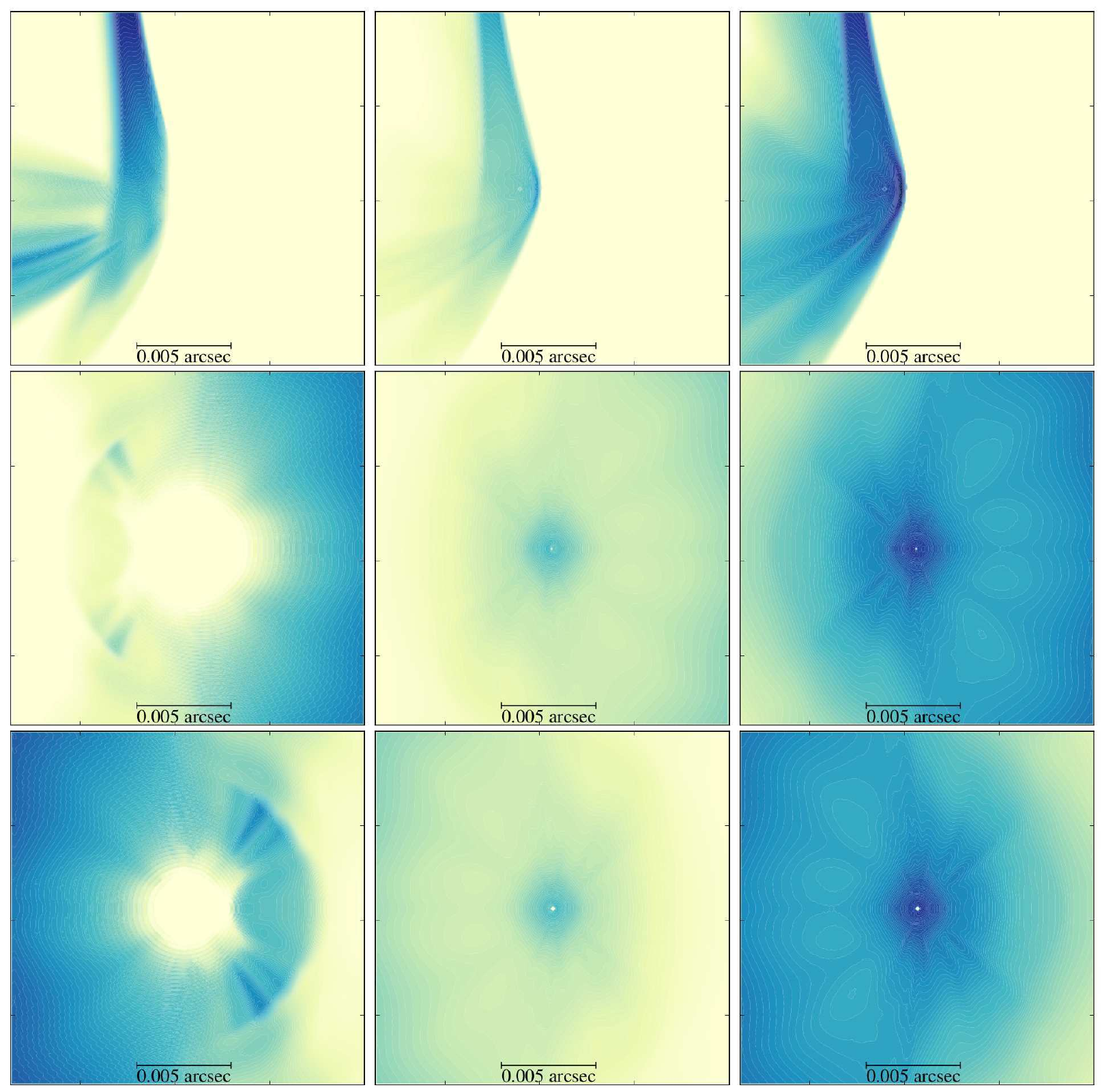}
		\end{subfigure}\\
		\begin{center}
		\begin{subfigure}{500\unitlength}	
			\begin{picture}(500,0)
				\put(0,-20){\includegraphics[trim=2cm 2cm 1cm 11.7cm, clip=true,width=500\unitlength]{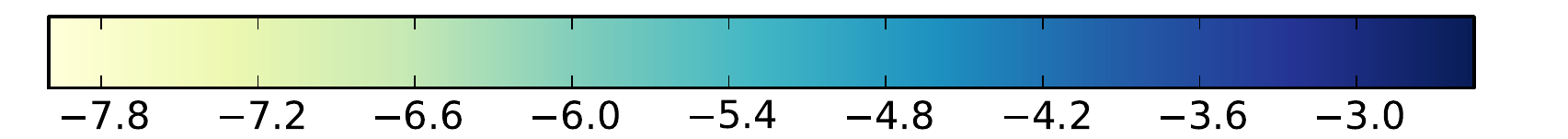}}
				\put(200,25){\footnotesize{ph m$^{-2}$ s$^{-1}$}}
			\end{picture}
		\end{subfigure}
		\end{center}	
		\caption{Projected photon flux above 100 MeV at 1 kpc distance for IC-emission (first column), bremsstrahlung (second column) and neutral pion decay (third column) for case A including orbital motion after 1.5 orbits. The rows represent different orientations. First row: $i=0^\circ$, $\Phi^\prime=0^\circ$ (face-on); second row: $i=$90$^\circ$, $\Phi^\prime$=0$^\circ$ (along--line of--center); third row: $i=$90$^\circ$, $\Phi^\prime=$180$^\circ$ (along line of centers, opposite to second row).
\label{orbit_Pha}}
\end{figure*}

Figure \ref{orbit_Pha} shows the resulting projected photon emission maps for different orientations with respect to the observer. To simplify comparison of the case including orbital motion with the stationary case, we compensate the reversed order of the stars along the $x$ axis by a redefinition of the angle $\Phi$: $\Phi^\prime= \Phi+180^\circ$ and use the same representation of projection maps as discussed previously for $\Phi^\prime$. 

For all three components the face-on orientation (first row) shows considerable asymmetry due to the lack of high-energy particles in the trailing arm of the shock front toward the B star. This produces the apparent lack of emission in the upper left section of the plots. In contrast, a wealth of particles in the trailing arm of the shock front toward the WR star produces significant emission that even dominates the entire emission region for the IC component. 

In the case of alignment of the line of centers and the line of sight (second and third row), the differences of trailing and forward arm manifest themselves most strikingly for the IC component. The circular feature bordering the low-emission center region (most prominent for $i=$90$^\circ$, $\Phi^\prime=$180$^\circ$, third row) is caused by the emission of the forward arm toward the B star touching the edge of the computational domain. If seen from the opposite direction (second row), it is highly suppressed because of an unfavourable scattering angle. 
The lack of emission next to this feature is due to the low number of high-energy electrons in the forward arm towards the WR star. This becomes also apparent in the corresponding plots for bremsstrahlung and neutral pion decay. For the same reasons as in the stationary case, we find that the configuration ($i=$90$^\circ$,$\Phi^\prime=$180$^\circ$) exhibits considerably higher flux for the IC component than the configuration ($i=$90$^\circ$, $\Phi^\prime=$0$^\circ$).

%For bremsstrahlung and neutral pion decay the differences of trailing and forward arm manifest themselves only near the edge of the computational domain where the fluxes are $\sim$2 orders of magnitude below the dominant region in the center. 

%Concerning the IC component, we find that the region where emission is maximal coincides with the trailing arm toward the WR star, where the electrons reach high number densities. There, emission is even higher than for the case when orbital motion is neglected.  The circular shaped region of higher flux 

\begin{figure*}
	\setlength{\unitlength}{0.001\textwidth}
	\begin{subfigure}[c]{500\unitlength}
		\begin{picture}(500,370)
			\put(15,15){
				\includegraphics[width=\textwidth]{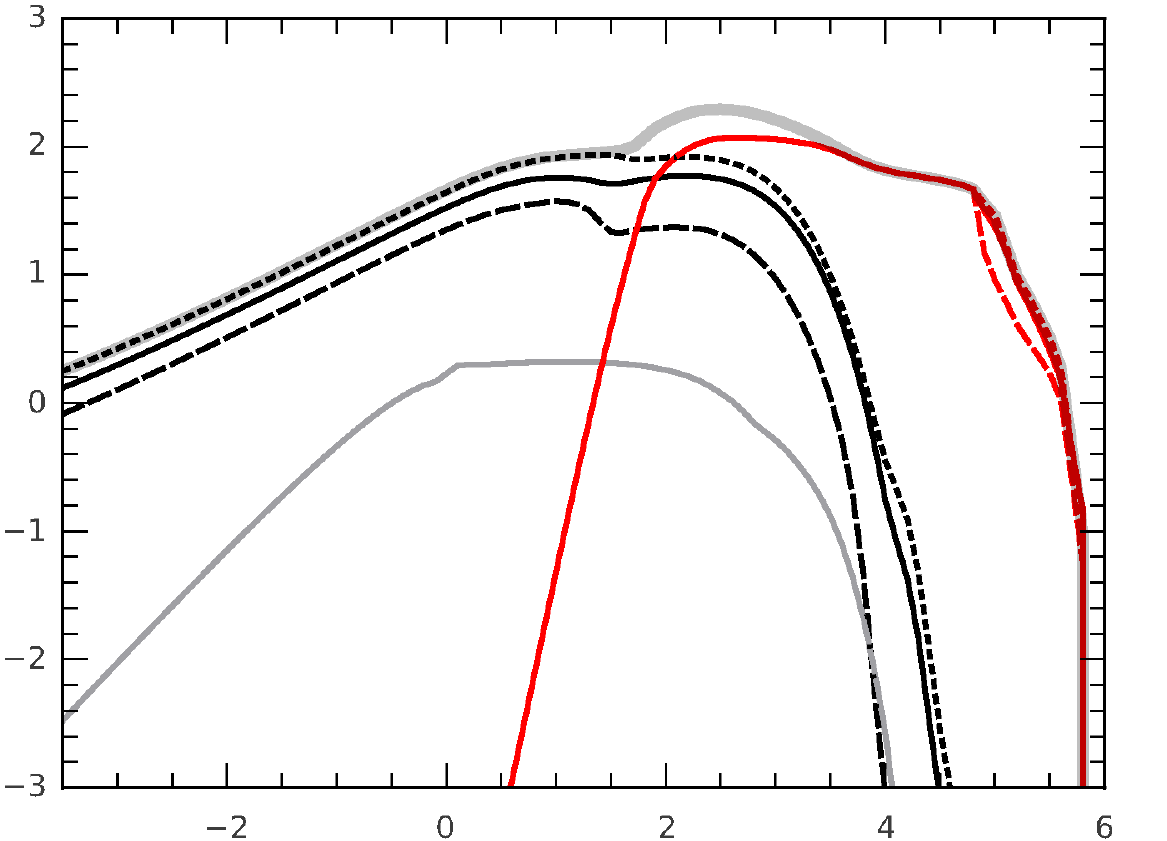}
			}
			\put(200,0){\footnotesize{log( E in MeV)}}
			\put(0,120){\rotatebox{90}{\footnotesize{log( $E^2N$ in MeV m$^{-2}$s$^{-1}$ )}}}
			\put(450,80){a)}
		\end{picture}
	\end{subfigure}
	\begin{subfigure}[c]{500\unitlength}
		\begin{picture}(500,370)
			\put(15,15){
				\includegraphics[width=\textwidth]{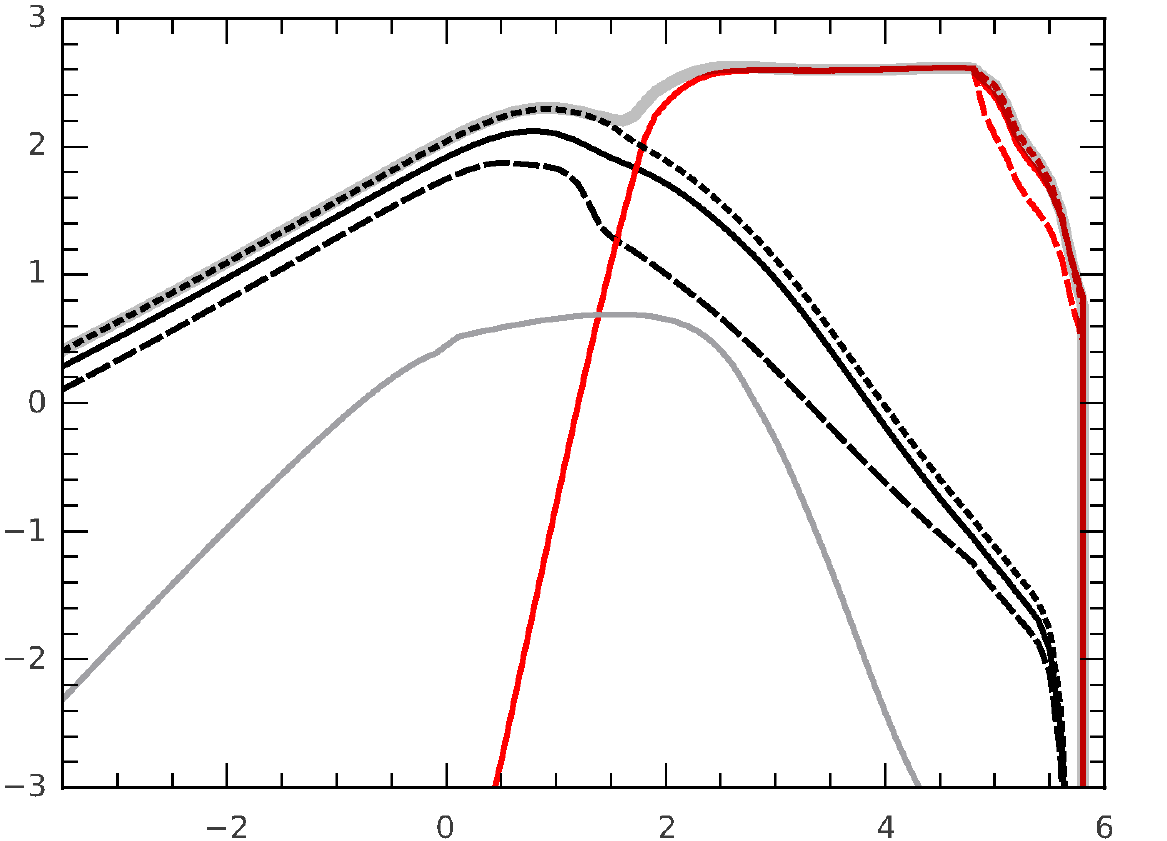}
			}
			\put(200,0){\footnotesize{log( E in MeV)}}
			\put(0,120){\rotatebox{90}{\footnotesize{log( $E^2N$ in MeV cm$^{-3}$ )}}}
			\put(450,80){b)}
		\end{picture}
	\end{subfigure}
		\caption{Same as Figure \ref{gamspec1} for case A with a) and without b) orbital motion. The sum of all three components (for $i=90^\circ$, $\Phi=180^\circ$) is indicated in shaded gray.
\label{orbit_Aspec} }
\end{figure*}
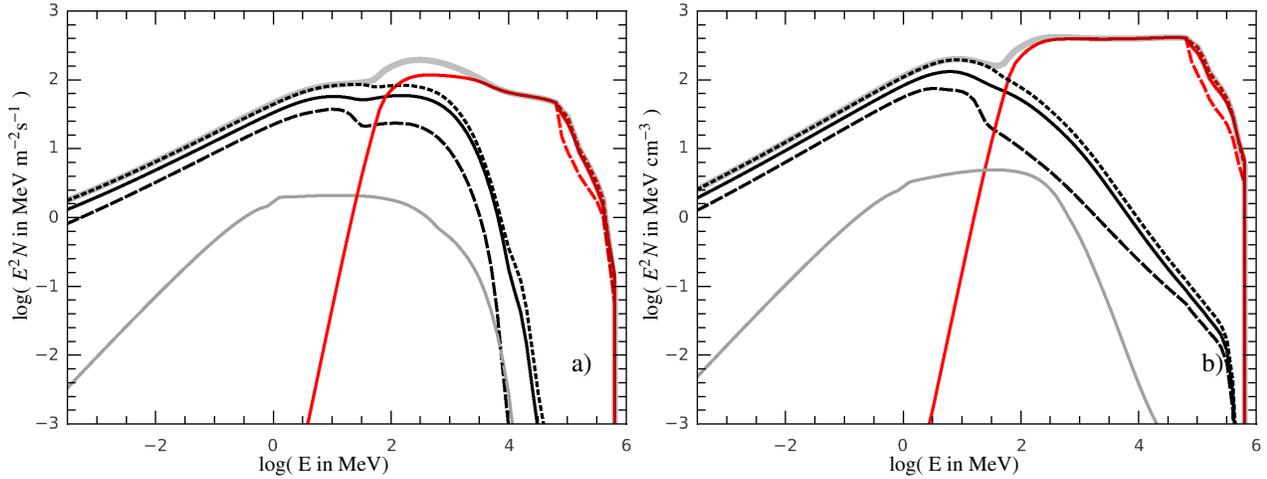

A comparison of the spectra with and without orbital motion is shown in Figure \ref{orbit_Aspec}. We find a decrease in flux for the neutral pion decay and the bremsstrahlung component. Concerning the IC component we find a new spectral feature emerging at $\sim$100 MeV that persists for all three depicted orientations. We interpret this as the influence of the forward arm toward the B star. There, a population of electrons reaching up to 100 GeV is still sufficiently close to the stars in order to exhibit favourable conditions for IC emission. To a lesser degree, the same region also influences the bremsstrahlung component which shows a similar feature at $>$1 GeV. The above shows that orbital motion has considerable impact on the resulting non-thermal high-energy emission.

\section{DISCUSSION}
We have computed different components of nonthermal high-energy photon emission in colliding wind binaries -- for the first time on the basis of a 3D distribution of high-energy particles that was obtained by hydrodynamic simulations and the simultaneous solution of the transport equation. Our approach includes the radiative acceleration of the stellar winds, radiative cooling in the hot shocked gas, diffusive shock acceleration at the shock fronts bordering the WCR and various energy loss mechanisms affecting the high-energy particles. The resulting nonthermal high-energy photon emission has been explored in terms of 2D projection maps, SEDs as well as integrated flux predictions of nonthermal photon emission components based on 3D distributions of high-energy particles in colliding wind binary systems. 

In computing the anisotropic IC-component of the photon emission we take into account the dependence on the scattering angle, as well as the radiation fields of both stars of the binary system. Bremsstrahlung and neutral pion decay components of the photon flux are computed using local wind plasma densities and the present distribution of electrons and protons. The effect of photon photon opacity in the dense radiation fields of the stars has been taken into account for all three nonthermal photon emission mechanisms.
%********** ORBIT case A gammas ***********
Studying three cases of varying stellar separation, we find significant differences in terms of total flux value and also in the identity of the dominant emission component.
We showed that low stellar separations (720 R$_\odot$ for the studied WR-B system, case A) inhibits the acceleration of high-energy electrons at the apex of the WCR due to strong IC and synchrotron losses. This produces an early cutoff in the photon flux ensuing from IC-scattering and bremsstrahlung. Although denser radiation fields and higher plasma densities cause higher photon flux values at lower energies, this early cutoff leads to dominance of the hadronic neutral pion decay component at energies E $\gtrsim$100 GeV for the chosen electron-proton injection ratio of $10^{-2}$. In contrast to that, large stellar separations lead to a dominance of the IC component throughout the studied energy range. As the magnetic and radiation energy density at the WCR are low, the electrons -- and also the ensuing photon fluxes -- reach higher energies. Both, IC and bremsstrahlung components reach up to $\sim$1 TeV if we increase stellar separation by a factor of four (case C). Lower wind plasma densities in the WCR further diminish the significance of the neutral pion decay component. We find that increasing the stellar separation by a factor of four results in an increase of $\sim$ 3 orders of magnitude in the IC and bremsstrahlung components of the photon flux at energies E$>$10~GeV, whereas the neutral pion decay component decreases by about an order of magnitude. For lower energies the importance of high-energy electrons diminishes. Accordingly, the total flux for E$>$1 keV decreases for all three flux components if we increase stellar separation by a factor of four.

The strong dependence on stellar separation suggests that $\gamma$-ray binaries are liable to show a significant variation in their spectra in course of an elliptical orbit with high eccentricity. Spectra can exhibit multiple emission components (as in Figure \ref{gamspec1} a)) for low separations (i.e. during periastron), contrasted by longer periods of one-component spectra (as in Figure \ref{gamspec1} c)) for larger stellar separations. We note that a situation resembling this trend of two apparently distinct emission components has already been observed for the peculiar case of $\eta$ Carinae \citep[see ][]{Takapaper,Walter}. However, this can also be understood in terms of $\gamma$-ray absorption due to the presence a hot X-ray gas \citep[see ][]{Reitberger2012}. The application of our numerical model to other specific candidates for particle-acceleration CWB systems \citep[see catalogue of ][]{Becker2013} will provide predictions for the nonthermal high-energy flux that can then be related to detections and upper limits from observations.

Varying the orientation of the binary system by changing the viewing angles $i$ and $\Phi$, we find a strong dependence of the IC flux component. Photons from bremsstrahlung and neutral pion decay are influenced only indirectly via the changing conditions for photon photon opacity which affect all components at high energies alike. This effect remains small compared to the variation of the IC component due to its dependence on the scattering angle. Since we consider the radiation field of both stars, the IC component does not vanish as the flux from one population of target photons is maximal when the other is minimal. However, since the two radiation fields differ in intensity depending on distance and identity of the star, there remains a significant level of variation. It is largest for the case of small stellar separation where the B star is very close to the WCR. Here, the IC-flux above 100 MeV varies a factor of 8 between the two extreme cases of ($i=90^\circ$, $\Phi=0^\circ$) and ($i=90^\circ$, $\Phi=180^\circ$).
Although this dependence is small for the cases we investigated, it becomes significantly stronger if we have a binary system with greater differences concerning the two stellar radiation fields and the ram pressure of the winds. The IC contribution to the $\gamma$-ray flux will differ by several orders of magnitude in course of an orbital cycle if the thermal photon density of one stellar component is dominant and the orbital inclination is close to the edge-on case of $i~\sim90^\circ$.

% ________________________ spectra  mov case

The effects of photon photon opacity in the radiation field of the star are restricted to incident photons with $E\gtrsim100$ GeV. Given their low prominence in case of low stellar separation, there is no noticeable impact. However, for larger stellar separation, the absorption leaves a clear signature in the spectral components of IC scattering and bremsstrahlung.

Studying the impact of orbital motion for the case of smaller stellar separation, we find a number of effects that are all significantly less severe than the ones obtained for varying the stellar separation by a factor of two. The distortion of the WCR leads to regions of altered wind velocity normal to the shock and, thus, to notable changes in the DSA acceleration rate. Whereas IC emission is greatly reduced in the trailing arm toward the B star (and also in the forward arm toward the WR star), we find regions of higher maximum energies for electrons (than in the case without orbital motion) in the side of the forward arm toward the B star. In terms of SEDs, this leads to an additional bump in the IC component. 
The changes due to orbital motion are less pronounced for the bremsstrahlung and neutral pion decay components.

\section{OUTLOOK}

The presented numerical model can be used to study the high-energy nonthermal flux of specific particle-accelerating CWB systems as a function of time. Possible variations in course of the orbit can be quantitatively assessed and favourable observation periods can be predicted. Compared to observational $\gamma$-ray data (e.g. from the Fermi-LAT instrument), our simulation aims for constraining and refinement of poorly known parameters, such as as the injection fraction of electrons and protons at the shock of the wind collision region. Dependencies on other parameters, such as the magnetic field and the diffusion coefficient, can be studied as well. 

The consequent next step will be the application of our simulation procedure to archetypal nonthermal binary systems. We will confront hardly understood discrepancies between existing flux predictions from analytical models with the present absence of detections of particular enigmatic systems in the \textit{Fermi}-LAT data like e.g. WR104 or $\eta$~Carinae.

\acknowledgments
 This work is supported through the EU FP7 program by Marie Curie IRG grant 248037; K.R. is supported by the Marietta Blau-Stipendium der OeAD-GmbH, financed by the Austrian Ministry of Science BMWF. AR acknowledges financial support through the Austrian Science Fund (FWF) grant P 24926-N27. Additional support was given by the Austrian Ministry of Science BMWF as part of the UniInfrastrukturprogramm of the Research Platform Scientific Computing at the University of Innsbruck and the Austrian Science Fund (FWF) supported Doctorate School DK+ W1227-N16.

%%%%%%%%%%%%%%%%%%%%%%%%%%%%%%%%%%%%%%%
%REFERENCES
%%%%%%%%%%%%%%%%%%%%%%%%%%%%%%%%%%%%%%%

%%%%%%%%%%%%%%%%%%%%%%%%%%%%%%%%%%%%%%%
%FIGURES and TABLES
%%%%%%%%%%%%%%%%%%%%%%%%%%%%%%%%%%%%%%%

\end{document}